\documentclass{article}
\usepackage[dvipsnames]{xcolor}
\usepackage[utf8]{inputenc}
\usepackage{graphicx}
\usepackage[left=2.5cm,right=2.5cm,top=2.5cm,bottom=2.5cm]{geometry}
\usepackage{caption}
\usepackage{subcaption}
\usepackage{lscape}
\usepackage{amsmath}
\usepackage{float}
\allowdisplaybreaks		
\usepackage{amsfonts}
\usepackage{amssymb}
\usepackage{braket}
\usepackage{longtable}
\usepackage{dsfont}
\usepackage{hyperref}
\usepackage{xurl}
\usepackage{enumitem}
\usepackage{stmaryrd}
\usepackage[utf8]{inputenc}
\usepackage{graphicx}
\usepackage[left=2.5cm,right=2.5cm,top=2.5cm,bottom=2.5cm]{geometry}
\usepackage{xcolor}
\usepackage{lscape}
\usepackage{amsthm}
\usepackage{tikz}  
\usepackage{tikzit}
\usetikzlibrary{shapes,decorations,arrows,calc,arrows.meta,fit,positioning}
\usetikzlibrary{arrows.meta}

\theoremstyle{plain}
\newtheorem{Def}{Definition}
\newtheorem{definition}[Def]{Definition}
\newtheorem{theorem}[Def]{Theorem}

\newtheorem{proposition}[Def]{Proposition}
\newtheorem{lemma}[Def]{Lemma}

\newtheorem{remark}[Def]{Remark}
\newtheorem{corollary}[Def]{Corollary}

\newtheorem{example}[Def]{Example}
\newtheorem{examples}[Def]{Examples}

\newcommand{\intv}{\sigma}
\newcommand{\MeasProb}{\cat{Prob}}
\newcommand{\SepMet}{\BorelProb}
\newcommand{\BorelProb}{\cat{BorelProb}}
\newcommand{\MeasStoch}{\cat{Stoch}}
\newcommand{\diagconditioning}{effect conditioning}

\newcommand{\ouralgo}{\texttt{simplify-cf}}
\newcommand{\makecg}{\texttt{make-cg}}
\newcommand{\ouridalgo}{\texttt{id-cf}}
\newcommand{\cfterms}{counterfactual terms}
\newcommand{\stdwts}{\big( s^{(j)}, c^{(j)}, E^{(j)} \big)_{j=1}^k}

\newcommand{\stdwtnoj}{\big( s, c, E \big)}

\newcommand{\ODAGcat}{\cat{ODAG}}
\newcommand{\OCM}{\cat{OCM}}
\newcommand{\ND}{\cat{ND}} %Cat of network diagrams

\newcommand{\OGraph}{\cat{OGraph}}

\newcommand{\ODAG}[1]{{#1}}

\newcommand{\iset}[1]{\mathbf{#1}} %Font for an indexed collection where we do not have a specific indexing set
\newcommand{\mech}[1]{\mathbb{#1}}
\newcommand{\model}[1]{\mech{#1}}

\newcommand{\modelM}{\model{M}}
\newcommand{\modelF}{\model{F}}
\newcommand{\modelN}{\model{N}}
\newcommand{\modelU}{\model{U}}

\newcommand{\normmor}[1]{\normop(#1)}

\DeclareMathOperator{\mechs}{mech} %Mechanisms of an OCM.

\DeclareMathOperator{\intmodel}{int} %Internalise subset of outputs
\DeclareMathOperator{\normop}{norm} %Internalise subset of outputs

\DeclareMathOperator{\openmodel}{open} %Internalise subset of outputs

\DeclareMathOperator{\breakk}{break} %Breaking intervention
\newcommand{\breakint}[2]{\breakk(#1 = #2)}

\DeclareMathOperator{\var}{var} %Variables of a model / set of mechanisms 
\DeclareMathOperator{\invar}{in} %Input Variables of a model / set of mechanisms 
\DeclareMathOperator{\causvar}{caus} %Mechanism Variables of a model / set of mechanisms 
\DeclareMathOperator{\outvar}{out} %Output Variables of a model / set of mechanisms 
\DeclareMathOperator{\intvar}{int} %Hidden Variables of a model / set of mechanisms 
\newcommand{\channel}{\mathrm{channel}}

\DeclareMathOperator{\inputs}{in} 
\DeclareMathOperator{\outputs}{out} 
\DeclareMathOperator{\openDAG}{open} 

\definecolor{darkblue}{rgb}{0.2,0.2,0.7}
\definecolor{darkgreen}{rgb}{0.0, 0.5, 0.0}
\definecolor{ForestGreen}{RGB}{34,139,34}

\definecolor{planning}{rgb}{0.2,0.2,0.7}

\newcommand{\combcolor}{Fuchsia}
\newcommand{\CombLine}{dotted}
\newcommand{\CombThickness}{thick}
\newcommand{\BiarLine}{<->}
\newcommand{\BiarColor}{gray}

\newcommand{\gray}[1]{\textcolor{gray} {#1} }

\newcommand{\rl}[1]{\textcolor{black} {#1}}
\newcommand{\st}[1]{\textcolor{black} {#1}}
\newcommand{\rlb}[1]{\textcolor{black} {#1}}

\newcommand{\rlbcolor}{black}

\newcommand{\etal}{\textit{et al.}}

\newcommand{\Stoch}{\cat{Stoch}}
\newcommand{\FStoch}{\cat{FStoch}}

\newcommand{\MatR}{\cat{Mat}_{\mathbb{R}^+}}

\newcommand{\bigCI}{\mathrel{\text{\scalebox{1.07}{$\perp\mkern-10mu\perp$}}}}

\DeclareMathOperator{\Do}{do}
\DeclareMathOperator{\Cut}{cut}
\DeclareMathOperator{\Pad}{pad}

\newcommand{\Free}{\cat{Free}}

\newcommand{\Pa}{\mathrm{Pa}}
\newcommand{\Nd}{\mathrm{Nd}}
\newcommand{\Ch}{\mathrm{Ch}}
\usepackage{tikz-cd}

\newcommand{\sem}[1]{\ensuremath{\llbracket #1 \rrbracket}}

\DeclareMathOperator{\trim}{trim}
\DeclareMathOperator{\ext}{ext}

\newcommand{\eval}{\mathsf{eval}}

\usepackage{stackengine}

\newcommand{\cat}[1]{\ensuremath{\mathbf{#1}}}
\newcommand{\catC}{\cat{C}}

\newcommand{\id}[1]{\ensuremath{\mathrm{id}_{#1}}}

\newcommand{\hilbH}{\mathcal{H}} %Hilbert space
\newcommand{\hilbK}{\mathcal{K}} %Hilbert space

\usepackage{tikz,xypic}

\tikzstyle{every picture}=[baseline=-0.25em,scale=0.5]

\newenvironment{picc}[1][]
{\begin{aligned}\begin{tikzpicture}[font=\tiny,#1]}
{\end{tikzpicture}\end{aligned}}
\newenvironment{pic}[1][] {\begin{aligned}\begin{tikzpicture}[scale=2.0, font=\tiny,#1]}{\end{tikzpicture}\end{aligned}} %Hard coded picture 

\usetikzlibrary{circuits.ee.IEC}
\pgfdeclarelayer{edgelayer}
\pgfdeclarelayer{nodelayer}
\pgfsetlayers{background,edgelayer,nodelayer,main}

\tikzstyle{whitedott}=[circle, draw=black, fill=white, inner sep=.4ex]
\tikzstyle{greydott}=[circle, draw=black, fill=black!25, inner sep=.4ex] %Added by Sean
\tikzstyle{blackdott}=[circle, draw=black, fill=black, inner sep=.4ex]

\tikzstyle{upgroundsmall}=[circuit ee IEC, thick, ground, rotate=90, scale=1.5, tikzit fill=black, tikzit shape=rectangle]

\newcommand{\discard}[1]{\ensuremath{\tinygroundnew_{#1}}}

\newcommand{\tinygroundnew}{
\smash{
{\hspace{-3pt}
\ensuremath{
\begin{picc}[scale=1.0] 
    \node[upgroundsmall, xscale=0.8, yscale=0.7] (1) at (0,0.16) {};
    \draw (0,0.03) to (0,-0.25);
\end{picc}
}\hspace{-1pt}}}}

\newcommand{\tinymult}[1][whitedott]{
\smash{\raisebox{-2pt}{\hspace{-5pt}\ensuremath{\begin{pic}[scale=0.4,yscale=-1]
    \node (0) at (0,0) {};
    \node[#1, inner sep=1.5pt] (1) at (0,0.55) {};
    \node (2) at (-0.5,1) {};
    \node (3) at (0.5,1) {};
    \draw (0.center) to (1.center);
    \draw (1.center) to [out=left, in=down, out looseness=1.5] (2.center);
    \draw (1.center) to [out=right, in=down, out looseness=1.5] (3.center);
    \node[#1, inner sep=1.5pt] (1) at (0,0.55) {};
\end{pic}
}\hspace{-3pt}}}}

\newcommand{\tinymultflip}[1][whitedott]{
\smash{\raisebox{-2pt}{\hspace{-5pt}\ensuremath{\begin{pic}[scale=0.4,yscale=1]
    \node (0) at (0,0) {};
    \node[#1, inner sep=1.5pt] (1) at (0,0.55) {};
    \node (2) at (-0.5,1) {};
    \node (3) at (0.5,1) {};
    \draw (0.center) to (1.center);
    \draw (1.center) to [out=left, in=down, out looseness=1.5] (2.center);
    \draw (1.center) to [out=right, in=down, out looseness=1.5] (3.center);
    \node[#1, inner sep=1.5pt] (1) at (0,0.55) {};
\end{pic}
}\hspace{-3pt}}}}

\newcommand{\tinycomult}[1][whitedott]{
\smash{\raisebox{-2pt}{\hspace{-5pt}\ensuremath{\begin{pic}[scale=0.4,yscale=1]
    \node (0) at (0,0) {};
    \node[#1, inner sep=1.5pt] (1) at (0,0.55) {};
    \node (2) at (-0.5,1) {};
    \node (3) at (0.5,1) {};
    \draw (0.center) to (1.center);
    \draw (1.center) to [out=left, in=down, out looseness=1.5] (2.center);
    \draw (1.center) to [out=right, in=down, out looseness=1.5] (3.center);
    \node[#1, inner sep=1.5pt] (1) at (0,0.55) {};
\end{pic}
}\hspace{-3pt}}}}

\newcommand{\tinycopy}{\tinycomult[whitedott]}

\newcommand{\tinyunit}[1][whitedott]{
\smash{\raisebox{3pt}{\hspace{-3pt}\ensuremath{\begin{pic}[scale=0.4,yscale=-1]
    \node (0) at (0,0) {};
    \node[#1, inner sep=1.5pt] (1) at (0,0.55) {};
    \draw (0.center) to (1.north);
\end{pic}
}\hspace{-1pt}}}}

\newcommand{\tinycap}{\smash{\raisebox{-3pt}{\hspace{-2pt}\ensuremath{\begin{pic}[scale=0.2, yscale=-1]%,string]
   \pgftransformscale{1.5} \draw[scale = 1] (0,0) to[out=-90,in=-90,looseness=1.5] (1.5,0);
\end{pic}}}}}

\tikzstyle{label}=[font={\footnotesize}, text height=1ex, text depth=0.15ex]
\tikzstyle{map}=[draw, shape=rectangle, inner sep=2pt, minimum height=5mm, fill=white, minimum width=5mm]
\tikzstyle{medium map}=[draw, shape=rectangle, inner sep=2pt, minimum height=5mm, fill=white, minimum width=12mm, tikzit fill=red]
\tikzstyle{large map}=[draw, shape=rectangle, inner sep=2pt, minimum height=5mm, fill=white, minimum width=18mm, tikzit fill=blue]
\tikzstyle{scalar}=[circle, draw, inner sep=2pt, line width=0.7pt]
\tikzstyle{upground}=[circuit ee IEC, thick, ground, rotate=90, scale=1.5, tikzit fill=black, tikzit shape=rectangle]
\tikzstyle{downground}=[circuit ee IEC, thick, ground, rotate=-90, scale=1.5, tikzit fill=black, tikzit shape=rectangle]
\tikzstyle{downgroundnorm}=[circuit ee IEC, thick, ground, rotate=-90, scale=1.5, fill=white, tikzit shape=rectangle, tikzit fill=red]
\tikzstyle{point}=[fill=white, draw, shape=isosceles triangle, shape border rotate=-90, isosceles triangle stretches=true, inner sep=0.2pt, minimum height=0.8mm, yshift=-0.0mm, tikzit shape=rectangle, minimum width=0.5cm]
\tikzstyle{copoint}=[fill=white, draw, shape=isosceles triangle, shape border rotate=90, isosceles triangle stretches=true, inner sep=0.2pt, minimum width=0.5cm, minimum height=0.8mm, yshift=-0.0mm, tikzit shape=rectangle, minimum width=0.5cm]
\tikzstyle{wide point}=[fill=white, draw, shape=isosceles triangle, shape border rotate=-90, isosceles triangle stretches=true, inner sep=0.2pt, minimum height=0.8mm, yshift=-0.0mm, minimum width=12mm, tikzit shape=rectangle, tikzit fill=red]
\tikzstyle{wide copoint}=[fill=white, draw, shape=isosceles triangle, shape border rotate=90, isosceles triangle stretches=true, inner sep=0.2pt, minimum width=0.5cm, minimum height=0.8mm, yshift=-0.0mm, tikzit shape=rectangle, minimum width=12mm, tikzit fill=red]
\tikzstyle{whitedot}=[circle, draw=black, inner sep=.4ex, fill=white]
\tikzstyle{greydot}=[circle, draw=black, inner sep=.4ex, fill=grey]
\tikzstyle{blackdot}=[circle, draw=black, inner sep=.4ex, fill=black]
\tikzstyle{decomp}=[fill=white, draw, shape=isosceles triangle, shape border rotate=-90, isosceles triangle stretches=true, inner sep=0pt, minimum width=0.75cm, minimum height=4mm, yshift=-0.0mm, tikzit shape=rectangle]
\tikzstyle{decompwide}=[fill=white, draw, shape=isosceles triangle, shape border rotate=-90, isosceles triangle stretches=true, inner sep=0pt, minimum width=1.5cm, minimum height=4mm, yshift=-0.0mm, tikzit shape=rectangle, tikzit fill=red]
\tikzstyle{decompflip}=[fill=white, draw, shape=isosceles triangle, shape border rotate=90, isosceles triangle stretches=true, inner sep=0pt, minimum width=0.75cm, minimum height=4mm, yshift=-0.0mm, tikzit shape=rectangle]
\tikzstyle{decompwideflip}=[fill=white, draw, shape=isosceles triangle, shape border rotate=90, isosceles triangle stretches=true, inner sep=0pt, minimum width=1.5cm, minimum height=4mm, yshift=-0.0mm, tikzit shape=rectangle, tikzit fill=red]
\tikzstyle{sharpstate}=[point]
\tikzstyle{sharpeffect}=[copoint]
\tikzstyle{green dot}=[fill=white, draw={rgb,255: red,34; green,139; blue,34}, shape=circle, inner sep=.4ex]
\tikzstyle{greensharpstate}=[point, fill=white, draw={rgb,255: red,34; green,139; blue,34}]
\tikzstyle{grbox}=[fill=white, draw={rgb,255: red,34; green,139; blue,34}, shape=rectangle]
\tikzstyle{split node}=[fill=white, draw={rgb,255: red,255; green,155; blue,255}, shape=circle, dashed]

\tikzstyle{dir}=[->]
\tikzstyle{dashline}=[-, dashed]
\tikzstyle{triangle}=[->, style={{-{Triangle[open]}}}]
\tikzstyle{norm}=[-, dashed, color=blue, draw={rgb,255: red,51; green,0; blue,255}]
\tikzstyle{maptostyle}=[{|->}]
\tikzstyle{dashdir}=[->, dashed]
\tikzstyle{thickline}=[-, thick]
\tikzstyle{FCMBox}=[-, draw={rgb,255: red,191; green,191; blue,191}, thick]
\tikzstyle{green line}=[-, draw={rgb,255: red,34; green,139; blue,34}]

\tikzstyle{tikzfig}=[]

\tikzset{arrow/.style={decoration={
    markings,
    mark=at position #1 with \arrow{>[length=2pt, width=3pt]}},
    postaction=decorate},
    reverse arrow/.style={decoration={
    markings,
    mark=at position #1 with {{\arrow{<[length=2pt, width=3pt]}}}},
    postaction=decorate}
}
\usepackage{subfiles}
\usepackage[symbol]{footmisc}
\title{\textbf{Causal models in string diagrams}}
\author{Robin Lorenz$^{*}$, Sean Tull$^{\dagger}$ \\[0.2cm]
\textit{Quantinuum, 17 Beaumont Street, Oxford, UK}
}
\date{}

\begin{document}
\maketitle

\begin{abstract}
The \emph{framework of causal models}, pioneered by Pearl and his collaborators, as well as Spirtes, Glymour and Scheines, provides a principled approach to causal reasoning which is applied today across many scientific domains.

Here we present the framework of causal models in the language of \emph{string diagrams}, interpreted formally using category theory.  A class of string diagrams, called \emph{network diagrams}, are in 1-to-1 correspondence with \rlb{\emph{directed acyclic graphs}}. A \emph{causal model} is given by interpreting the components of such a diagram in terms of stochastic maps, functions, or more general channels in a 
\rlb{symmetric monoidal category with a `copy-discard' structure (\emph{cd-category}).} 
This represents a model as a single mathematical object that can be reasoned with intuitively and yet rigorously. 

Building on earlier works, most notably by Fong and Jacobs, Kissinger and Zanasi, \rlb{as well as Fritz and Klingler,}  
this work presents diagrammatic definitions of causal models and functional causal models in a cd-category, which generalise causal Bayesian networks and structural causal models, respectively. We formalise the most general kind of \emph{interventions} on a model, including but beyond atomic ones described by do-interventions, and present the natural notion of an \emph{open causal model}, as a  causal model `with inputs'. 

To apply these to causal reasoning, we \rlb{also} give an approach to conditioning based on a \rlb{\emph{normalisation box},} which allows causal inference calculations to be done fully diagrammatically. We use these to define \emph{counterfactuals} in this setup, and to treat the problems of the identifiability of both causal effects and counterfactuals fully diagrammatically. 

The benefits of such a presentation of causal models lie both in foundational questions in causal reasoning, and in particular in their \rlb{clarificatory} role and \emph{pedagogical value}. In fact this manuscript aims to be accessible to different communities, including causal model practitioners as well as researchers in applied category theory. For illustration of the key ideas many examples from the causal model literature are discussed in the diagrammatic language. 

Overall, we argue and demonstrate that \emph{causal reasoning} according to the causal model framework is most naturally and intuitively done as \emph{diagrammatic reasoning}.
\end{abstract}

\footnotetext[1]{\rlb{robin.lorenz@quantinuum.com}}
\footnotetext[2]{\rlb{sean.tull@quantinuum.com}}
\renewcommand*{\thefootnote}{\arabic{footnote}}

\tableofcontents
\vspace*{0.3cm}

%\documentclass[main.tex]{subfiles} \begin{document}

%*****************************************************************
\section{Introduction} \label{Sec_Introduction}
%*****************************************************************

%---------------------
%intro causal models 
%---------------------
Causal reasoning is essential to scientific practice in many areas. 
Despite the omnipresence of problems that are of a causal nature, and thus the need for principled causal reasoning in science, a grounded and fruitful formalisation has come a long way. As well as technical and sociological challenges there were philosophical ones, with controversy around pretty much any epistemological and metaphysical aspect of causation \cite{BeebeeEtAl_2009_OxfordHandbookOfCausation, SEP_MetaphysicsOfCausation, SEP_ProbabilisticCausation}. 
Suspending answers to some of these philosophical issues, the development of the \emph{framework of causal models}, pioneered by Judea Pearl and his collaborators \cite{Pearl_Causality}, as well as Spirtes, Glymour and Scheines \cite{SpirtesEtAL_2000_BookCausationPredictionSearch}, constitutes tremendous technical progress with the clarification and formalisation of causal reasoning in scientific practice. 

In this article, building on an ongoing project from numerous authors \cite{CoeckeEtAl_2012_PicturingBayesianInference,Fong_2013_CausalTheories, ChoEtAl_2019_DisintegrationViaStringDiagrams, JacobsEtAl_2019_CausalInferenceByDiagramSurgery, Fritz_2020_SyntheticApproachToMarkovKernels, JacobsEtAl_2021_CausalInferencesAsDiagramSurgery_DiagramsToCounterfactuals, FritzEtAl_2022_DSeparationInCategoricalProbability, FritzEtAl_2023_DSeparationInCatgeoricalProbabilty}, we present a reformulation of the causal model framework, which aims to bring yet more conceptual clarity, accessibility, and intuition, by recasting causal models and causal reasoning entirely \emph{graphically} in the language of \emph{string diagrams}.

\paragraph{Causal models.}
The causal model framework spells out, given a set of variables $V$, how causal structure and empirical data over (some of) $V$ constrain each other. Empirical data may be either observational or interventional in nature \rl{and causal structure} consists of a set of direct-cause relationships amongst the \rl{variables as represented by a directed graph, most commonly, a} \emph{directed acyclic graph} (DAG). 

\rl{One may distinguish two important and closely related variants of causal models.}
Assuming an underlying determinism, a \emph{structural causal model} (SCM) specifies functional causal mechanisms with dependences according to the causal structure.  At a coarser level, encoding less detailed knowledge, the derived notion of a \emph{causal Bayesian network} (CBN) specifies \emph{stochastic} causal mechanisms, which together define a probability distribution that satisfies the \emph{causal Markov condition} with respect to the DAG. 
These stipulations are then the basis for making causal judgements and answering queries in a principled way. 

Among the achievements are \emph{causal discovery} algorithms in which one infers causal structure to the degree possible given observational data and certain assumptions. Another success is in solutions to \emph{causal inference} problems, where causal knowledge is given and one asks when and how one can predict what would happen if one \emph{intervened} and, e.g. forced some variables to take particular values, as well as unambigously answer \emph{counterfactual} questions. 
The framework paints a precise \emph{causal hierarchy} of three levels of kinds of questions one can answer given what kind of data, going from ones that involve merely associational via interventional to counterfactual reasoning \cite{Pearl_Causality, BareinboimEtAl_2021_PearlsHierarchy}.  

Over the past four decades the framework of causal models has developed into a rich and highly interdisciplinary body of work, extending from the original fields of computer science, statistics and philosophy, to particular sciences and industry, ranging from health to economics and policy making (see, e.g., Refs.~\cite{GlockerEtAl_2021_CausalityInDigitalMedicine, ReynaudEtAl_2022_DArtagnan_CounterfactualVideoGeneration, 
VlontzosEtAl_2022_EstimatingCatgeoricalCFsVisDeepTwinNetworks, 
LeeEtAl_2022_LeveragingDirectedCausalDiscovery} for a tiny selection). 
There is also a recent explosion of works studying causality in the context of machine learning (see Ref.~\cite{SchoelkopfEtAl_2021_TowardCausalRepresentationLearning} and further details below)  and an effort to distil the technical lessons for (classical) causal inference from the study of causality in the foundations of quantum mechanics \cite{WolfeEtAl_2019_InflationTechniqueForCausalInference, BoghiuEtAl_2022_InflationLibrary}.

%---------------------
%transition
%---------------------
The formalism uses both graph and probability theory. A causal model typically involves a DAG $G$ and a probability distribution $P$, but in a sense these stay separate -- one investigates the DAG's properties which may licence certain probabilistic calculations, and conversely one inspects probabilistic equalities and on that basis graphs may be altered. This can lead to fairly involved notation and bookkeeping to keep track of how both objects constrain each other, and so one may wonder whether both aspects can be treated with one mathematical entity. 

%---------------------
%diagrammatic reasoning (and category theory)
%---------------------

\paragraph{The power of diagrammatic reasoning.} 
Over the past twenty years another, seemingly unrelated, success story has been in the applications of \emph{category theory}  and \emph{diagrammatic reasoning} in mathematics, computer science, and quantum computation, \rl{as well as the foundations of quantum theory}. Category theory may be understood as the mathematical language of \emph{composition}, in which one emphasises \emph{processes} or relationships (known as morphisms) between entities rather than entities themselves \cite{mac2013categories}. Depending on the choice of \emph{category} one works in these processes may be of many \rl{kinds, including physical, logical and computational processes} \cite{baez2011physics}.

A major aspect has been the formalisation of scientific topics in terms of 
\emph{monoidal categories}, also known as \emph{process theories} and their associated \emph{graphical calculus}, which allows one to reason about processes rigorously using intuitive \emph{string diagrams} \cite{selinger2011survey}. 

Graphical approaches have been highly successful in applications to the foundations of quantum mechanics \cite{abramsky2004categorical} and quantum computation \cite{coecke2018picturing}, forming the backbone of the \emph{ZX-calculus} \cite{coecke2011interacting} for describing quantum circuits, which is now widely used in industry. In more recent years, the growing field of \emph{applied category theory} \cite{FongEtAl_2019_InvitationACT_SevenSketches} has extended the compositional approach to further scientific areas including probability theory \cite{ChoEtAl_2019_DisintegrationViaStringDiagrams,Fritz_2020_SyntheticApproachToMarkovKernels}, machine learning \cite{shiebler2021category}, game theory \cite{ghani2018compositional} and artificial intelligence \cite{coecke2010mathematical}. 

%---------------------
%prior work
%---------------------
\paragraph{Prior work.} 
A natural question then is whether the framework of causal models can be cast in such a diagrammatic language, and provided with a clarifying category-theoretic formalisation. Over the past decade various works raised this question and undertook important steps towards it. 

Coecke and Spekkens started in Ref.~\cite{CoeckeEtAl_2012_PicturingBayesianInference} by presenting Bayesian inference diagrammatically,  exploiting a special structure a monoidal category may have, called a  
\emph{dagger Frobenius structure}. Closely related to that work, Fong then presented a categorical theory of Bayesian networks intended to formalise causal reasoning ~\cite{Fong_2013_CausalTheories}. In this approach a DAG generates a free category, its \emph{causal theory}, and in particular a string diagram  whose wiring corresponds to the parental sets and which can be given different semantics depending on the category it is interpreted in. 

Since then several authors, notably Jacobs \etal\ \cite{JacobsEtAl_2016_PredicateStateSemanticsBayesianLearning, JacobsEtAl_2017_FormalSemanticsOfInfluenceInBayesianReasoning, ChoEtAl_2019_DisintegrationViaStringDiagrams, JacobsEtAl_2019_LogicalEssentialsOfBayesianReasoning, ChoEtAl_2019_DisintegrationViaStringDiagrams} and Fritz \cite{Fritz_2020_SyntheticApproachToMarkovKernels, FritzEtAl_2022_FreeGSMonoidalCategories}, have developed \emph{categorical probability theory}, which clarified what the structural essence of probabilistic reasoning is in terms of the simple framework of \emph{cd-categories} and \emph{Markov categories} (without requiring the full dagger Frobenius structures). 
Within this setting Fritz and Klingler \cite{FritzEtAl_2022_DSeparationInCategoricalProbability, FritzEtAl_2023_DSeparationInCatgeoricalProbabilty} \rlb{recently furthermore presented categorical \emph{d-separation} -- a diagrammatic notion that generalises the usual notion of d-separation -- and established that it is sound for the corresponding generalisation of conditional independence in Markov categories.}

Building on all these developments, and basically using the string diagrammatic set-up from Fong in Ref.~\cite{Fong_2013_CausalTheories},  
Jacobs, Kissinger and Zanasi in Refs.~\cite{JacobsEtAl_2019_CausalInferenceByDiagramSurgery, JacobsEtAl_2021_CausalInferenceByDiagramSurgery}
go beyond merely Bayesian reasoning and turn to explicitly causal concepts such as interventions, causal effect and counterfactuals in terms of string diagrams. A discussion of their work in more detail will feature throughout this manuscript. 

%---------------------
%This paper
%---------------------
\paragraph{This work.} 
While these efforts established a promising way to treat causal reasoning diagrammatically, they are however somewhat fragmented and tend to be characterised by either a focus on rather formal categorical aspects or only a subset of the framework of causal models. 
What is missing is a comprehensive and accessible presentation of the basic framework of causal models in a string diagrammatic language. 
Here we present such a treatment. % Todo: do we want to say this about Jacobs, and about syntax + semantics?
It can be seen as an extension of the work by Jacobs \etal\ from Ref.~\cite{JacobsEtAl_2021_CausalInferenceByDiagramSurgery}, albeit not insisting on its strict separation of syntax and semantics. 
This work: 
\begin{itemize} 
	\item presents diagrammatic definitions of (generalisations of) \emph{causal Bayesian networks} and \emph{structural causal models};  	
	\item treats \emph{interventions} in full generality, that is the most general kind of intervention, beyond atomic ones, compatible with the definition of causal models, like soft and conditional interventions \cite{CorreaEtAl_2020_CalculusForStochasticInterventions}; 
	\item incorporates \emph{Bayesian conditioning} into the diagrammatic calculus so that calculations for causal inference problems, which often include some conditioning, can be done entirely \rlb{diagrammatically; 
	notably, in a way that accounts for where the given distribution underlying a conditional -- here treated as a \emph{partial channel} -- may lack full support and yet yields a well-behaved notion of conditional independence that satisfies the semi-graphoid axioms and soundness of d-separation;} 
	\item introduces the notion of \emph{open causal models} -- which naturally arise from a certain class of interventions (including do-interventions) and \rlb{are related to models based on `conditional DAGs' from Ref.~\cite{RichardsonEtAl_2022_NestedMarkovForADMGs} and to \emph{open graphs} from Ref.~\cite{OpenGraph1,OpenGraph2}} -- together with \emph{transformations} of (open) causal models, yielding the category of (open) causal models;
	\item discusses how to go from a latent variable model based on an \emph{acyclic directed mixed graph}, which captures latent common causes, to a string diagrammatic representation; 
	\item discusses \emph{causal effect identifiability}, mildly extending the work in Ref.~\cite{JacobsEtAl_2021_CausalInferenceByDiagramSurgery};  
	\item treats \emph{counterfactuals}, where in particular the general definition of counterfactuals in diagrammatic terms is of clarificatory and pedagogical value and also the problem of the identifiability of counterfactuals is treated diagrammatically, including \rl{analogues of the \makecg\ and the \texttt{IDC}$^*$ algorithms \cite{ShpitserEtAl_2008_CompleteIdentificationMethodCausalHierarchy}.} 
\end{itemize}
 
We stress that the goal is not to be overly formal (see Ref.~\cite{FritzEtAl_2022_DSeparationInCategoricalProbability} for more formal foundations), but instead to be accessible for different communities -- some material hence is introductory with a particular reader in mind -- and to showcase the utility of a string diagrammatic representation of causal models.  
Before giving a more detailed exposition of what to expect in this manuscript, we first flesh out some more of the motivations. \\

%---------------------
%Motivation
%---------------------
\paragraph{Motivation} 
A category-theoretic presentation of the causal model framework in string diagrammatic terms is valuable for a number of reasons. 

%-----------
\begin{enumerate}[wide, labelwidth=!,labelindent=0pt,label={(\arabic*)}]
\item \emph{Benefits from an intuitive representation.} The diagrammatic approach has the advantage of allowing one to elegantly combine in one mathematical object all the data that constitutes a causal model -- the causal structure, as well as the stochastic or deterministic maps associated with the causal mechanisms. 
\rlb{Importantly, beyond just providing an intuitive representation of the model, all of what one typically does with a model can be formalised in this language too.} This includes rules of reasoning (rewrite rules) that apply directly to the string diagrams in a way that makes many calculations rather straightforward and avoids otherwise involved bookkeeping. String diagrams appear to be the most natural language for causal reasoning.

Now, given that algorithmic solutions to many causal inference problems exist, why should one care about an elegant representation -- can we not simply let a machine do it all for us?
First, the history of science teaches us that finding the right language facilitates progress and improves our conceptual understanding. Second, it certainly matters for a causal model practitioner who wishes to reason through a concrete example with pen and paper. 
Third, there is a pedagogical value. 	
It is hard to understate the framework's significance -- causal models are here to stay \cite{Pearl_2018_TheBookOfWhy} -- and one may thus hope that in future the basics of causal reasoning would become part of any scientific education, or even be taught at school. To this end one would want the easiest and most intuitive language possible. 
This resonates with the recent effort to use diagrammatics to increase the basic `literacy' in quantum physics and quantum computing as in Ref.~\cite{Coecke_2022_QuantumInPictures}.

\item \emph{Causality and machine learning.} 
A major lesson from the causality literature, stemming from the causal hierarchy, is the fundamental limitation in machine learning (ML) based on purely associational data \cite{Pearl_2018_TheoreticalImpedimentsToML}. After being largely ignored by the ML community, the study of causality in machine learning has gained a prominent role over the past few years and taken promising new directions (see the works by Schölkopf \etal\ \cite{SchoelkopfEtAl_2021_TowardCausalRepresentationLearning, SchoelkopfEtAl_2022_StatisticalToCausalLearning, Schoelkopf_2019_CausalityForMachineLearning} for an overview). The interaction between the fields is, roughly, two-fold. 

First, learning causal mechanisms in a way that combines the success of modern ML since the `statistical turn' with ideas from causal models is argued to help solve major challenges in current ML like (strong) generalisation, robustness and efficiency given limited resources, as well as explainability and fairness \cite{SchoelkopfEtAl_2021_TowardCausalRepresentationLearning}  
(see, e.g., Refs.~\cite{ParascandoloEtAl_2018_LearningIndependentCausalMechanisms, BengioEtAl_2019_MetaTransferObjective, DasguptaEtAl_2019_CausalReasoningFromMetaRL}).  
The hope lies in the potential benefit of models that learn a structured representation where that structure corresponds to (causal) \emph{structure in the world}:  
\begin{center} 
	\begin{minipage}{0.95\textwidth}
	\textit{Future AI models that robustly solve a range of problems in the
	real world will thus likely need to re-use components, which
	requires them to be robust across tasks and environments [...]. 
	An elegant way to do this is to employ a modular structure
	that mirrors a corresponding modularity in the world. In other
	words, if the world is indeed modular, in the sense that components/mechanisms of the world play roles across a range of
	environments, tasks, and settings, then it would be prudent for
	a model to employ corresponding modules.} \cite{SchoelkopfEtAl_2021_TowardCausalRepresentationLearning} 
	\end{minipage}
\end{center}

Second, the causal model framework takes the set of variables that are pertinent to causal analysis as given, saying nothing about how these causal relata are to be determined. 
An idea coined \emph{causal representation learning}\footnote{This has a \rl{precursor in, and is related to,} the study of `causal abstraction' and `causal coarsening' by Chalupka and Eberhardt \etal\  (see, e.g., Refs.~\cite{ChalupkaEtAl_2014_VisualCausalFeatureLearning, ChalupkaEtAl_2016_MultiLevelCauseEffectSystems, ChalupkaEtAl_2016_UnsupervisedDiscoveryOfElNino, Eberhardt_2016_GreenAndGrueCausalVariables, ChalupkaEtAl_2017_CausalFeatureLearning}.} 
in Refs.~\cite{Schoelkopf_2019_CausalityForMachineLearning, SchoelkopfEtAl_2021_TowardCausalRepresentationLearning} 
asks whether a ML model can learn what these causal relata are in the first place, by abstracting away from lower level data (e.g. pixels). 
How exactly this would work and what the necessary conditions are in practice are still unclear, but there is much work exploring this idea 
\cite{BengioEtAl_2017_IndependentlyControllableFeatures, LocatelloEtAl_2019_ChallengingAssumptionsInUnSupervisedDisRep, LeebEtAl_2020_StructureByArchitecture, LocatelloEtAl_2020_WeaklySupervisedDisentanglement, MitrovicEtAl_2020_RepresentationLearningViaInvariantCausalMechanisms, KeEtAL_2020_AmortizedLearningOfNeuralCausalReps, ShenEtAl_2021_DisentangledGenerativeCausalRepLearning, WangEtAl_2021_DesiderataRepLearning_CausalPrespective, TraubleEtAl_2021_DisentangledRepresentationsFromCorrelatedData, LippeEtAl_2022_Citris, BrehmerEtAl_2022_WeaklySupervisedCRL}. 

Now, category theory is the mathematical language in which to study both compositional (modular) structure, as well as structural correspondences, such as between a model and the world. There is also growing interest in the category-theoretic and compositionality focused study of machine learning and AI \cite{ShieblerEtAl_2021_CategoryTheoryInML, Website_CatsForAI}, and specifically the role of causality within it \cite{Cohen_2022_TowardsGroundedTheoryOfCausationForAI, RischelEtAl_2021_CompositionalAbstractionError_CategoryCausalModels, BrehmerEtAl_2022_WeaklySupervisedCRL}. Hence, for both of these ways in which the study of causality and ML may benefit from each other, it can only be beneficial to first cast causal models categorically, with a focus on bringing out causal structure as compositional structure.

%-----------	
\item \emph{Foundational aspects.} A categorical formulation should also be useful within a number of foundational issues in causality. The first is in addressing philosophical questions such as who or what divides the world into causal relata or how to define causal mechanisms and interventions. While suspending such questions and assuming that domain knowledge will typically give uncontroversial answers arguably helped the advent of causal models, trying to find answers \emph{is} important, not only for philosophy of causation, but likely also as a piece of theory to define and judge the scope of causal representation learning (see Refs.~\cite{Cohen_2022_TowardsGroundedTheoryOfCausationForAI, JanzingEtAl_2022_PhenomenologicalCausality}).

Second, to identify the appropriate categorical set-up for causal models means in particular to identify the most general, \emph{abstract structure} relative to which causal reasoning `in its essence' can be defined. This is essential when building further generalisations of the causal model framework. 
The notion of a cd-category, or Markov category, plays such a role -- see below and Refs.~\cite{ChoEtAl_2019_DisintegrationViaStringDiagrams, Fritz_2020_SyntheticApproachToMarkovKernels, FritzEtAl_2022_FreeGSMonoidalCategories}. 

Third, Schmid, Selby and Spekkens develop the framework of \emph{causal-inferential theories} in Ref.~\cite{SchmidEtAl_2020_UnscramblingOmletteOfCausationAndInference}, to `[unscramble] the omelette of causation and inference', that is, capture both formally and identify how they interact to produce causal reasoning as we know it. 
Seeing as their framework is already set in process-theoretic terms, the presentation of causal models here should help
to unscramble causal models by understanding how they relate to causal-inferential theories -- a question already pointed out in Ref.~\cite{SchmidEtAl_2020_UnscramblingOmletteOfCausationAndInference}. 

Fourth, although interesting classically, causal-inferential theories were originally developed to unscramble causation and inference in quantum theory. More broadly, the now large field of `quantum causality' is focused on understanding to what extent causal concepts can be applied to quantum theory, whose foundational controversies make this a pressing task. A promising and well-grounded framework of \emph{quantum causal models} developed in Refs.~\cite{AllenEtAl_2016_QCM, CostaEtAl_2016_QuantumCausalModeling, BarrettEtAl_2019_QCMs, BarrettEtAl_2021_CyclicQCMs, OrmrodEtAl_2022_CausalStructureWithSectorialConstraints} is explicitly concerned with a quantum generalisation of (classical) causal models, but currently lacks the same sort of complete and intuitive diagrammatic presentation. This lack is for technical and, it seems, conceptually deep reasons \cite{LorenzEtAl_2021_CausalAndCompositionalStructure, VanrietveldeEtAl_2021_RoutedQuantumCircuits}. Thus it is imperative to first fully develop the classical case, to allow one to point to precisely which aspects of it do have a quantum generalisation and which do not. This is especially relevant for the case of counterfactuals, where little work has been done for quantum causal models, and the diagrammatic presentation of (classical) counterfactuals here should provide a suitable clarification and reference.

\end{enumerate}

%---------------------
%More details, structure
%---------------------
\subsection{Outline and gentle overview \label{Sec_GentleOverview}}

Let us now give an overview of the diagrammatic treatment of causal models presented in the paper. 

Throughout the work, our basic setting is that of a \emph{symmetric monoidal category} with some extra structure, making what is known as a \emph{cd-category}.  We stress that a reader not familiar with even the basics of category theory need not worry -- all that is essential to the majority of the paper is the intuitive language of string diagrams. A category consists of \emph{objects}, depicted as wires, and \emph{morphisms} or \emph{processes} between them, depicted as boxes. A cd-category also comes with \emph{copy maps} to `share' the values of variables and \emph{discarding} processes to represent marginalisation, \rlb{hence the name `cd'.} % as shown on the right. 
As examples of components of string diagrams, \rlb{which we read bottom up,} the following shows, respectively, a morphism $f$ from $A$ to $B$, another $g$ with \rl{inputs $D, E$ and output $F$,} a copy map and a discard map. 
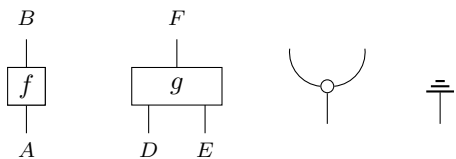
\[
\input{intro-pic.tikz}
\]
An example of a cd-category is $\FStoch$, which has as objects finite sets and as morphisms the stochastic maps between them, where probability distributions are special cases of morphisms without any input. 
We introduce the categorical setup in Section~\ref{sec:catsetup}, which is standard and may be skipped by readers who are familiar with categorical probability theory, apart from perhaps our definition of probabilistic conditioning in Secs.~\ref{sec:normalisation} and \ref{sec:conditioning}. 
The latter builds on \rlb{prior work 
\cite{JacobsEtAl_2016_PredicateStateSemanticsBayesianLearning, JacobsEtAl_2017_FormalSemanticsOfInfluenceInBayesianReasoning,  JacobsEtAl_2019_LogicalEssentialsOfBayesianReasoning, ChoEtAl_2019_DisintegrationViaStringDiagrams, FritzEtAl_2022_DSeparationInCategoricalProbability, FritzEtAl_2023_DSeparationInCatgeoricalProbabilty},} 
but our presentation is tailored to this work and introduces a \emph{normalisation box} that allows one to treat conditioning entirely diagrammatically. 

Section~\ref{sec:causal-models} presents the definition of a \emph{causal model} in a cd-category $\catC$. 
An ordinary causal Bayesian network over discrete variables is an instance and a main example throughout the work.   
Basically, a causal model over a set $\{X_i\}_i$, where each $X_i$ is an object in $\catC$, is a DAG $G$ with vertices $\{X_i\}_i$ and a set of mechanisms $\{ c_i \colon Pa(X_i) \rightarrow X_i \}_i$, \rlb{where $Pa(X_i)$ denotes the set of parents of $X_i$ in $G$}. In fact a DAG can be equivalently represented as a particular kind of string diagram, which we here call a \emph{network diagram}. Roughly, this is a diagram composed of  single-output boxes for the mechanisms and copy maps to feed variables into all their children. A causal model in $\catC$ is then a network diagram with an interpretation of all its constituents in $\catC$. 

\begin{figure}[H]
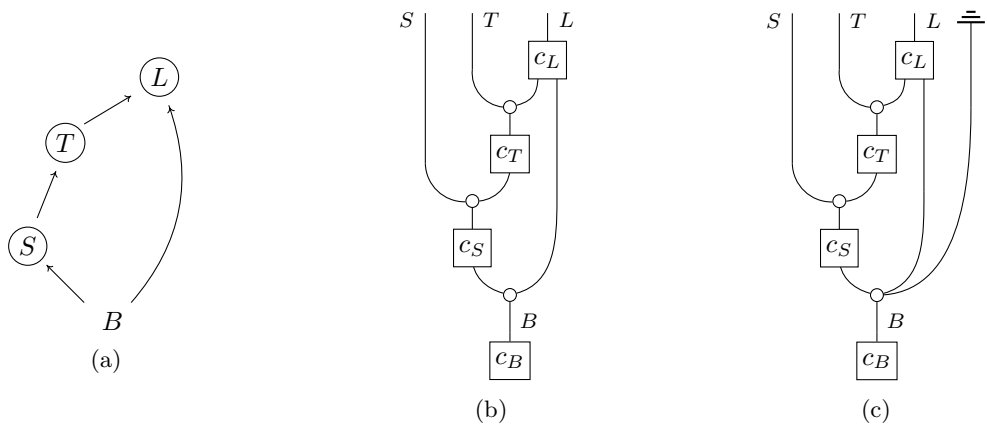

	\centering
	\begin{subfigure}{5cm}
		\centering
		\input{Fig_smoking_example_DAG.tikz}
		\caption{\label{Fig_Intro_example_DAG}}
	\end{subfigure}
	\begin{subfigure}{5cm}
		\centering
		\input{Fig_smoking_example.tikz}
		\caption{\label{Fig_Intro_example_ND}}
	\end{subfigure}
	\begin{subfigure}{5cm}
		\centering
		\input{Fig_Intro_example_marginalised.tikz}
		\caption{\label{Fig_Intro_example_ND_marginalised}}
	\end{subfigure}
	\caption{DAG $G$ in (a) with the encircling of vertices indicating the subset of `observed' variables; 
	in (b) the corresponding network diagram with $\circ$ representing a copy map; 
	in (c) an equivalent diagram with the wire of $B$ `marginalised over' by discarding it, represented by $\protect\discard{}$.   
	Diagrams are read \emph{bottom up}.}
\end{figure}

As a basic, classic example consider a causal model relating the variables $S$ of whether someone smokes, $L$ of whether they develop lung cancer, a genetic condition $B$ which may be a common cause of each of these, and the amount of tar in their lungs $T$. 
The variables $S,T,L$ are observed while $B$ is unobserved (latent). A plausible DAG would be the one in Fig.~\ref{Fig_Intro_example_DAG} and a corresponding causal model is depicted in Fig.~\ref{Fig_Intro_example_ND}.
\noindent 
The choice of output wires of a network diagram \rl{can in particular capture} which variables one considers to be observed. The diagram in Fig.~\ref{Fig_Intro_example_ND_marginalised} has all four variables as output, but then $B$ is discarded and it is a simple rewrite rule in cd-categories that the diagrams in Figs.~\ref{Fig_Intro_example_ND} and \ref{Fig_Intro_example_ND_marginalised} are equivalent. 
Indeed the structure of the diagram in Fig.~\ref{Fig_Intro_example_ND} already expresses that, if interpreted in $\FStoch$, \rl{it represents (the marginal of) a probability distribution over $S, T, L, B$, that} satisfies the \emph{causal Markov condition} relative to $G$, i.e. is of the form
\[ P(S,T,L) \ = \ \sum_B P(L|T,B) \ P(T|S) \ P(S|B) \ P(B) \]
where $P(L|T,B)$, $P(T|S)$ and $ P(S|B)$ are the densities of the stochastic maps $c_L$, $c_T$ and $c_S$ (and these can indeed be derived as Bayesian conditionals from $P$ whenever $P$ has full support). 

A \emph{functional causal model}\footnote{Also known as \emph{structural causal model} and \emph{structural equational model}.} is, in a formal sense, a special case of a causal model and discussed in Sec.~\ref{Sec_FunctionalCausalModel}. Here the notion of a function is generalised to that of a \emph{deterministic} morphism in a cd-category. 

The presence of a causal model, rather than simply a probability distribution, in particular allows one to answer questions about what would happen to the resulting distribution under \emph{interventions} on the variables. Here we allow for a very general kind of intervention in which a model's mechanisms $\{ c_i \}_i$ are replaced with any other set $\{ c_i^* \}_i$ on the same variables and which preserve acyclicity. This agrees with other treatments of general, non-atomic interventions in the literature (see, e.g., Ref.~\cite{CorreaEtAl_2020_CalculusForStochasticInterventions}). In Section~\ref{Sec:Interventions} we discuss interventions and their diagrammatic manifestation (termed `diagrammatic surgery' in Ref.~\cite{JacobsEtAl_2021_CausalInferenceByDiagramSurgery}). 

Most widely considered in practice are the so-called \emph{do-interventions} in which a set of variables are assigned a fixed value. Carrying out a do intervention $\Do(S=s)$ on the above example corresponds to the `breaking' of the $S$ wire, discarding its previous mechanism $c_S$ and feeding forward the sharp state $s$ (i.e. in $\FStoch$ a point distribution centered at $S=s$). 
\rlb{Fig.~\ref{Fig_smoking_example_doS} shows the result $P(S,T,L; \Do(S=s))$.}

\begin{figure}[H]
	\centering
	\input{Fig_smoking_example_doS.tikz} 
	\hspace*{0.1cm} $=$ \hspace*{0.2cm}
	\input{Fig_smoking_example_doS_simplified_2.tikz} 
	\caption{\label{Fig_smoking_example_doS}}
\end{figure}

Now the data that is really required to specify an intervention is often not a full causal model, but only a subset of its mechanisms. This naturally leads one to consider the notion of an \emph{open causal model}, which is essentially a `model with inputs', which we introduce in Section~\ref{sec:openCMs}. An example is the $P(Y|\Do(X))$ when understood as the map that maps $x$ to $P(Y|\Do(X=x))$.    
For the above example this allows one to consider the following:
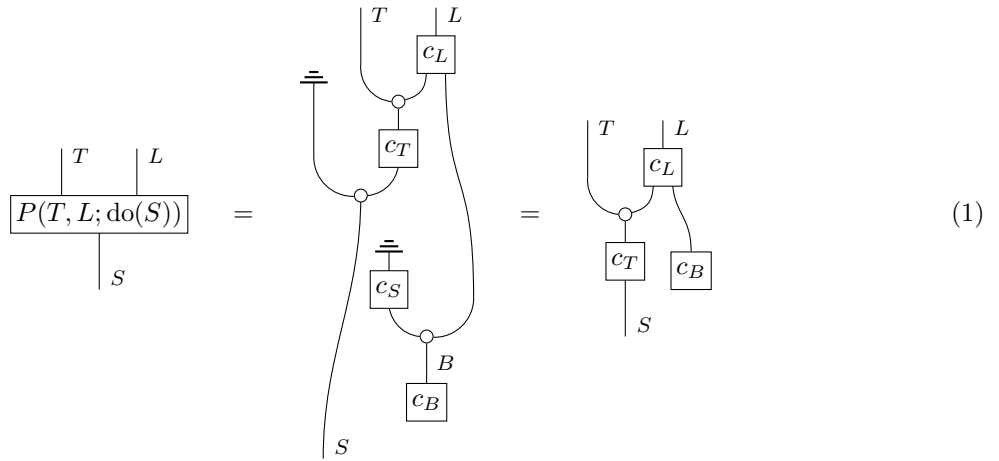
\begin{equation}
	\input{Fig_smoking_example_doS_open_Model_1.tikz} 
	\hspace*{0.5cm} = \hspace*{0.5cm} \input{Fig_smoking_example_doS_open_Model_2.tikz} 
	\hspace*{0.5cm} = \hspace*{0.5cm} \input{Fig_smoking_example_doS_open_Model_3.tikz} 
\end{equation}
The notion of an open model also allows one to formalise the composition of causal models, in parallel and sequentially, as well as transformations between models more general than interventions. 
This is made formal in Section \ref{Sec_ComposingOpenModels}, where we show that open causal models form a symmetric monoidal category themselves. Open models are built \rl{on ideas related} to the \emph{conditional DAGs} from Ref.~\cite{RichardsonEtAl_2022_NestedMarkovForADMGs} and \emph{open graphs} from Ref.~\cite{OpenGraph1,OpenGraph2}. 

The remainder of the article turns to applications of the framework. In such applications, one is not always in possession of `full' knowledge of the causal structure but at times only knowledge that some variables may share a latent common cause. Such scenarios are depicted with an \emph{acyclic directed mixed graph} (ADMG), a DAG which may also have bi-directed edges representing latent causal structure. In Section~\ref{Sec_TreatmentLatentvariables} we discuss ADMGs and the notion of a latent projection from a DAG to an ADMG. We then introduce the \emph{rootification} of an ADMG $G$, a principled way to map it to a DAG such that it has $G$ as its latent projection, through introducing root nodes. The ADMG in Fig.~\ref{Fig_Intro_Example_ADMG} and the DAG in Fig.~\ref{Fig_Intro_example_DAG} stand to each other in such a relationship. 
The network diagrams of rootified ADMGs are the terms in which identifiability problems can then be discussed and, importantly, without loss of generality.\footnote{At least for many problems it is without loss of generality; there are open questions for certain identifiability problems, however these open questions are not specific to our treatment and hold more generally for the treatment of identifiability problems in the causal model literature. See detailed discussion in later sections.} 
\begin{figure}[H]
    \centering
    \input{Fig_smoking_example_ADMG.tikz} 
    \caption{\label{Fig_Intro_Example_ADMG}}
\end{figure}

A major example of an identifiability problem is the problem of \emph{identifiability of causal effects}: when and how we can uniquely determine the distribution resulting from an intervention simply from the observational data and the assumed causal structure, specified by an ADMG. Section~\ref{Sec_CE_Identifiability} briefly recaps the problem and then explores its treatment diagrammatically. We first restate a result by Jacobs \etal\ from Ref.~\cite{JacobsEtAl_2021_CausalInferenceByDiagramSurgery} that casts the c-component condition by Tian and Pearl \cite{TianEtAl_2002_GeneralIdentificationConditionForCausalEffects}, at least for a special case of when that condition is applicable, in diagrammatic terms.  Here our setup allows this result to be extended in the obvious way to interventions more general than atomic ones. 
Second, we discuss several pedagogical examples from the literature, showing how easy these concrete examples become when represented in string diagrammatic language.  We will also pay attention to how the diagrammatic normalisation box makes explicit the interplay between when one would have to condition on events with vanishing probability and concerns of identifiability. 

A final central aspect to the causal model framework are \emph{counterfactuals}, questions such as \textit{would Mary have had a headache, had she taken an aspirin?} In Section~\ref{Sec_CF_TheNotion} we give a general formal definition of counterfactuals in diagrammatic terms, which allows for arbitrarily many parallel worlds in the counterfactual statement. While it is in keeping with the literature (see, e.g., Refs.~\cite{Pearl_Causality, ShpitserEtAl_2008_CompleteIdentificationMethodCausalHierarchy, Pearl_2011_AlgorithmizationOfCounterfactuals}), we argue that the diagrammatic presentation and our precise conditions on data that defines a counterfactual have a clarificatory value. In particular it makes clear the difference between counterfactual questions and `what if one were to intervene' kind of questions. 

We find that it is easy and natural to treat counterfactuals diagrammatically, as demonstrated with an example in Sec.~\ref{Sec_CF_id_example}. This leads to the statement of an algorithm, \ouralgo, in Sec.~\ref{Sec_CF_makeCG_algorithm_1}, which simplifies the diagram of a counterfactual by applying rewrite rules \rl{in a way that makes its output correspond to that of the \makecg\ algorithm from Ref.~\cite{ShpitserEtAl_2008_CompleteIdentificationMethodCausalHierarchy}, but is arguably much more straightforward. 
Sec.~\ref{Sec_CF_id_criteria} then presents an algorithm, \ouridalgo, which on the basis of \ouralgo's output assesses the identifiability of a counterfactual and outputs the corresponding identifying expression in diagrams. This can be seen as the translation of the main ideas of the algorithmic solution from  Ref.~\cite{ShpitserEtAl_2008_CompleteIdentificationMethodCausalHierarchy} into diagrammatic terms. We argue the algorithm to be sound and discuss completeness.} 

Finally, Sec.~\ref{Sec_CF_Further_Generalisation} touches on a generalisation of the notion of counterfactuals that becomes natural in this work's setup, namely a notion where worlds may be defined by changes in causal structure more general than through do-interventions and where one may condition upon fuzzy facts. 

Section~\ref{Sec_Conclusions} closes with a list of promising directions for future work.

% \bibliographystyle{utphys} \bibliography{CauseComp.bib} % Uncomment while working on file standalone, if needed 
%\end{document}
%\documentclass[main.tex]{subfiles} \begin{document}

%*****************************************************************
\section{The categorical setup} \label{sec:catsetup}
%*****************************************************************
We will begin by introducing the category-theoretic approach to probability theory developed by numerous \rl{authors, see, e.g., Refs.~\cite{CoeckeEtAl_2012_PicturingBayesianInference,ChoEtAl_2019_DisintegrationViaStringDiagrams,fritz2020synthetic}.} Formally, this involves working in a `symmetric monoidal category' or more specifically a `cd-category'. In practice this amounts to working with intuitive but formal diagrams known as \emph{string diagrams} \cite{selinger2011survey}, which describe probabilistic and (here causal) processes. For the purposes of this article it suffices to consider the category $\MatR$ of $\mathbb{R}^+$-valued finite matrices, described in Example \ref{ex:MatR+} below, but the categorical approach is much more general. 

%***********************************
\subsection{String diagrams and symmetric monoidal categories \label{Sec_SMCsAndStringDiagrams}}
%***********************************

\rl{Let us start then with the basic notion of symmetric monoidal categories, as well as the string diagrammatic language used for their representation.}
Recall that a \emph{category} $\catC$ consists of a collection of objects $X, Y, \dots$ and morphisms $f \colon X \to Y$ between them, which we can compose in sequence. Such a morphism is also referred to as a \emph{process} $f$ from $X$ to $Y$. An object $X$ is depicted as a wire labelled by $X$, and a morphism $f \colon X\to Y$ as a box with lower input $X$ and upper output $Y$, read from bottom to top.
\[
\begin{tikzpicture}
	\begin{pgfonlayer}{nodelayer}
		\node [style=map] (0) at (0, 0) {$f$};
		\node [style=none] (1) at (0, 1) {};
		\node [style=none] (2) at (0, -1) {};
		\node [style=label] (3) at (0, 1.5) {$Y$};
		\node [style=label] (4) at (0, -1.5) {$X$};
	\end{pgfonlayer}
	\begin{pgfonlayer}{edgelayer}
		\draw (2.center) to (1.center);
	\end{pgfonlayer}
\end{tikzpicture}
 
\]
Given another morphism $g \colon Y \to Z$ we can form their sequential composite $g \circ f \colon X \to Z$, depicted as follows.
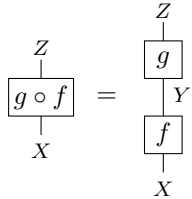
\[
\input{composite-1.tikz}
\]
Note that we may only compose morphisms sequentially when their types match in this way. Each object $X$ in a category also comes with an \emph{identity} morphism $\id{X} \colon X \to X$ depicted as a blank wire:
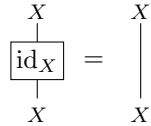
\[
\input{identity.tikz}
\]
which acts as a unit for composition: $\id{Y} \circ f = f = f \circ \id{X}$ for any $f \colon X \to Y$. Note that this rule is trivial in the graphical language. 

Formally, a \emph{symmetric monoidal category} $(\catC, \otimes, I)$, is a category $\catC$ coming with a functor $\otimes \colon \catC \times \catC \to \catC$, distinguished object $I$ and natural transformations which express that $\otimes$ is suitably associative and symmetric, with $I$ as a unit \cite{coecke2006introducing}. These features are most naturally expressed in string diagrams, as follows. 

Firstly, for any pair of objects $X, Y$ we can form their parallel composite or `tensor' $X \otimes Y$, depicted by placing their wires side-by-side. 
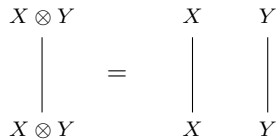
\[
\input{tensor-ob.tikz}
\]
Similarly, given morphisms $f \colon X \to W$ and $g \colon Y \to Z$ we can form their parallel composite $f \otimes g \colon X \otimes Y \to W \otimes Z$, depicted as below. 
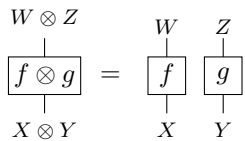
\[
\input{tensor.tikz}
\]
The tensor is symmetric meaning that we can also `swap' pairs of wires past each other, such that swapping twice leaves the wires alone, and boxes can move along the swaps as below. 
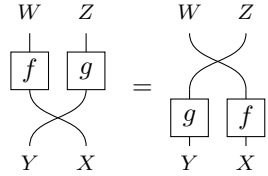
\[
\input{symmetry.tikz}
\]
A monoidal category also comes with a distinguished \emph{unit object} $I$, with (the identity on) $I$ depicted simply as empty space. 
\begin{equation} \label{eq:empty-space}
\input{idI.tikz}  \ \ \quad =   \ \ \quad 1 
\end{equation} 
Intuitively, tensoring any object by $I$ simply leaves it invariant. Formally this is expressed via coherence isomorphisms $X \otimes I \simeq X \simeq I \otimes X$. 

The unit object allows us to give meaning to morphisms without inputs and/or outputs. A morphism $\omega \colon I \to X$ is called a \emph{state} of $X$ and is depicted with `no input'. Similarly an \emph{effect} is a morphism of the form $e \colon X \to I$, and is depicted with no output.   
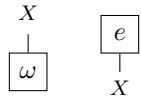
\[
\begin{tikzpicture}
	\begin{pgfonlayer}{nodelayer}
		\node [style=map] (0) at (0, 0) {$\omega$};
		\node [style=none] (1) at (0, 1) {};
		\node [style=label] (2) at (0, 1.5) {$X$};
	\end{pgfonlayer}
	\begin{pgfonlayer}{edgelayer}
		\draw (1.center) to (0);
	\end{pgfonlayer}
\end{tikzpicture}
 \qquad \begin{tikzpicture}[tikzfig]
	\begin{pgfonlayer}{nodelayer}
		\node [style=map] (0) at (0, 1) {$e$};
		\node [style=none] (1) at (0, 0) {};
		\node [style=label] (2) at (0, -0.5) {$X$};
	\end{pgfonlayer}
	\begin{pgfonlayer}{edgelayer}
		\draw (1.center) to (0);
	\end{pgfonlayer}
\end{tikzpicture}
 
\]
A morphism $r \colon I \to I$ is called a \emph{scalar} and is depicted with no inputs or outputs. Scalars can move `freely' around diagrams, and also can be multiplied together via $r \cdot s = r \otimes s = r \circ s = s \cdot r$, i.e.: 
\[ 
\begin{tikzpicture}
	\begin{pgfonlayer}{nodelayer}
		\node [style=scalar] (0) at (-0.25, 0) {$r$};
		\node [style=scalar] (1) at (1, 0) {$s$};
		\node [style=none] (2) at (2.25, 0) {$=$};
		\node [style=scalar] (3) at (4, 0) {$r \cdot s$};
	\end{pgfonlayer}
\end{tikzpicture}
 
\]
We denote the `empty space' scalar in \eqref{eq:empty-space} by $1 = \id{I}$, satisfying $1 \circ r = r$ for all scalars $r$.  

The composition operations $\circ, \otimes$ in a (monoidal) category satisfy numerous axioms which we omit here but which are self-evident in the graphical language. For example, in any category associativity of composition $(h \circ g) \circ f = h \circ g \circ f = h \circ (g \circ f)$ is automatic in diagrams, as below.
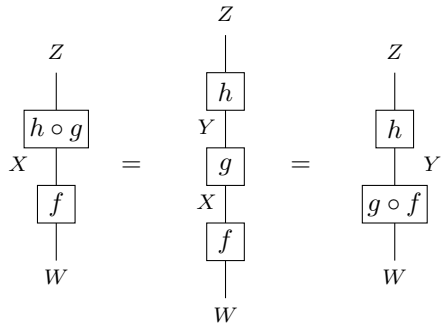
\[
\input{associativity.tikz}
\]
Functoriality of $\otimes$ means that $(f \otimes g) \circ (f' \otimes g') = (f \circ f') \otimes (g \circ g')$ which is also automatic from the diagrams (left below). A consequence is the `interchange law' $(f \otimes \id{}) = f \otimes g = (\id{} \otimes g)$ which lets us freely slide boxes along wires (right below). 
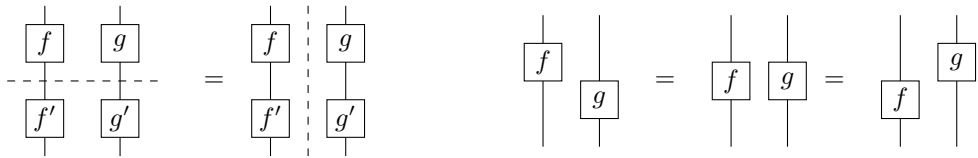
\[
\input{functoriality2.tikz} \qquad \qquad \qquad 
\input{interchange2.tikz} 
\]
At times we often omit labelling certain wires (objects) in diagrams as in the above. Let us now introduce our primary example category in this article. 

\begin{example} \label{ex:MatR+}
In the category $\MatR$ of positive matrices, the objects are finite sets $X,Y,\dots$ and the morphisms $M \colon X \to Y$ are functions $M \colon X \times Y \to \mathbb{R}^+$ where $\mathbb{R}^+ := \{r \in \mathbb{R} \mid r \geq 0 \}$. We think of such a function as an `$X \times Y$ matrix' with entries $M(y \mid x) := M(x,y) \in \mathbb{R}^+$ for $x \in X$, $y \in Y$.  
\[
\begin{tikzpicture}
	\begin{pgfonlayer}{nodelayer}
		\node [style=none] (0) at (0, 1) {};
		\node [style=none] (1) at (0, -1) {};
		\node [style=map] (2) at (0, 0) {$M$};
		\node [style=label] (3) at (0, 1.5) {$Y$};
		\node [style=label] (4) at (0, -1.5) {$X$};
	\end{pgfonlayer}
	\begin{pgfonlayer}{edgelayer}
		\draw (1.center) to (0.center);
	\end{pgfonlayer}
\end{tikzpicture}
 
\quad   :: \   (x,y) \  \mapsto \ M(y \mid x)
\]
We compose morphisms $N \circ M$ via the matrix product, given by summation over internal wires.  
\[
\input{mat-comp.tikz}
\quad   :: \   (x,z) \  \mapsto \ \sum_{y \in Y} N(z \mid y) M(y \mid x)
\]
The tensor $\otimes$ is given on objects by $X \otimes Y = X \times Y$, and on morphisms by the Kronecker product
\[
\input{mat-tens.tikz} 
\quad :: \  ((x,y),(w,z)) \  \mapsto \ M(w \mid x)N(z \mid y)
\]
The symmetry is simply the obvious isomorphism $X \times Y \simeq Y \times X$. The unit object is \rlb{the} singleton set $I=\{\star\}$. A state $\omega$ of $X$ is then equivalent to a positive function on $X$:
\[
\begin{tikzpicture}
	\begin{pgfonlayer}{nodelayer}
		\node [style=map] (0) at (0, -0.25) {$\omega$};
		\node [style=none] (1) at (0, 0.75) {};
		\node [style=label] (2) at (0, 1) {$X$};
	\end{pgfonlayer}
	\begin{pgfonlayer}{edgelayer}
		\draw (1.center) to (0);
	\end{pgfonlayer}
\end{tikzpicture}
  \quad   :: \    x \  \mapsto \ \omega(x)
\]
where $\omega(x) := \omega(x \mid \star)$. Special cases, discussed later, are probability distributions over $X$.
In just the same way, an effect $e$ on $X$ is also equivalent to a positive function on $X$ via $e(x) := e(\star \mid x)$.
\[
\begin{tikzpicture}[tikzfig]
	\begin{pgfonlayer}{nodelayer}
		\node [style=map] (0) at (0, 1) {$e$};
		\node [style=none] (1) at (0, 0) {};
		\node [style=label] (2) at (0, -0.5) {$X$};
	\end{pgfonlayer}
	\begin{pgfonlayer}{edgelayer}
		\draw (1.center) to (0);
	\end{pgfonlayer}
\end{tikzpicture}
  \quad   :: \    x \  \mapsto \ e(x)
\]
 A scalar $r \colon I \to I$ is precisely a positive real $r \in \mathbb{R}^+$, and composing scalars amounts to multiplication $r \otimes s = r \circ s = r \cdot s$ in $\mathbb{R}^+$. 
\end{example}

\subsection{Cd-categories}

Categories such as $\MatR$ come with further structure, which allow one to describe many aspects of probability theory entirely diagrammatically \cite{ChoEtAl_2019_DisintegrationViaStringDiagrams,fritz2020synthetic}. 

\begin{definition} \cite{ChoEtAl_2019_DisintegrationViaStringDiagrams} \label{def:cd_category}
A \emph{cd-category} (\emph{copy-discard category}) is a symmetric monoidal category in which each object comes with a specified pair of morphisms
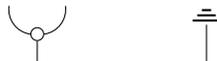
\[
\input{copy-delete.tikz}
\]
called \emph{copying} and \emph{discarding}, respectively, which satisfy the following:
\[
\input{markov-axioms.tikz}
\]
Formally, these say that copying and discarding form a \emph{commutative comonoid}. The choice of these morphisms is moreover `natural' in that the following hold for all objects $X, Y$.
\begin{equation} \label{eq:nat-rules}
\input{copy-nat.tikz} \qquad \qquad \input{disc-nat.tikz} \qquad \qquad \input{disc-I.tikz}
\end{equation}
\end{definition}

Thanks to the axioms for copying, we can unambiguously define a copying morphism with $n$ output legs, for any $n \geq 1$, as follows.
\[
\input{copy-n2.tikz}
\]
Formally we also define the copying morphism with $n=0$ output legs to be the discarding effect $\discard{}$. 

In a cd-category the processes of a truly `probabilistic' or \rlb{`stochastic'} nature are those satisfying the following. 

\begin{definition} \label{Def_Channel}
A morphism $f$ is a \emph{channel} when it preserves discarding:
\[
\input{causal2.tikz}
\]
In particular, we call a state $\omega$ \emph{normalised} when the following holds.
\[
\begin{tikzpicture}
	\begin{pgfonlayer}{nodelayer}
		\node [style=map] (0) at (-0.5, -0.75) {$\omega$};
		\node [style=upground] (1) at (-0.5, 0.75) {};
		\node [style=none] (2) at (0.75, 0) {$=$};
		\node [style=none] (3) at (2, 0) {$1$};
	\end{pgfonlayer}
	\begin{pgfonlayer}{edgelayer}
		\draw (1) to (0);
	\end{pgfonlayer}
\end{tikzpicture}

\]
A cd-category in which every morphism is a channel, or equivalently $\discard{}$ is the unique effect on any object, is called a \emph{Markov category} \cite{fritz2020synthetic}. Given any cd-category $\catC$, its subcategory $\catC_\channel$ of channels always forms a Markov category. 
\end{definition}

A useful rule, which follows from \eqref{eq:nat-rules}, is that any channel with multiple inputs $f$ satisfies:
\[
\input{mech-disc.tikz}
\]
The presence of discarding allows one to `ignore' certain outputs of morphisms. Given any morphism $f$ from $X$ to $Y, Z$, its \emph{marginal} $X \to Y$ is the following morphism. 
\[
\input{marginal.tikz} 
\]

\begin{example} 
$\MatR$ is a cd-category. The copy map on $X$ is given $\tinymultflip[whitedot](y,z \mid x) = \delta_{x,y,z}$ with value $1$ iff $x=y=z$ and $0$ otherwise. Discarding $\discard{}$ on $X$ is given by the function with $x \mapsto 1$ for all $x \in X$. 
A state $\omega$ is normalised precisely when it forms a normalised probability distribution over $X$.
\[
\begin{tikzpicture}
	\begin{pgfonlayer}{nodelayer}
		\node [style=map] (0) at (0.5, -0.5) {$\omega$};
		\node [style=none] (1) at (0.5, 0.75) {};
		\node [style=label] (2) at (0.5, 1) {$X$};
	\end{pgfonlayer}
	\begin{pgfonlayer}{edgelayer}
		\draw (1.center) to (0);
	\end{pgfonlayer}
\end{tikzpicture}
 \ \ \text{normalised} 
\quad 
 \iff 
 \quad 
\sum_{x \in X} \omega(x) = 1
\]
More generally, $M \colon X \to Y$ is a channel precisely when it forms a \emph{probability channel}, or equivalently the matrix is \emph{Stochastic}, meaning that it sends each $x \in X$ to a normalised distribution.  
\[
\begin{tikzpicture}
	\begin{pgfonlayer}{nodelayer}
		\node [style=none] (0) at (0, 1) {};
		\node [style=none] (1) at (0, -1) {};
		\node [style=map] (2) at (0, 0) {$M$};
		\node [style=label] (3) at (0, 1.5) {$Y$};
		\node [style=label] (4) at (0, -1.5) {$X$};
	\end{pgfonlayer}
	\begin{pgfonlayer}{edgelayer}
		\draw (1.center) to (0.center);
	\end{pgfonlayer}
\end{tikzpicture}
 \ \ \text{a channel} 
 \iff 
\sum_{y \in Y} M(y \mid x) = 1 \ \ \forall x \in X.
\]
Such a probability channel from $X$ to $Y$ is often also referred to as a \emph{conditional probability distribution} and called $P$ or $P(Y \mid X)$, with values denoted $P(y \mid x) := P(Y=y \mid X =x)$ for $x \in X$, $y \in Y$.  

The subcategory of channels in $\MatR$ is the Markov category $\FStoch$ of finite Stochastic matrices, consisting of probability channels between finite sets. Often we will restrict attention to this subcategory.

Diagrams can now help us express a few basic features of probability theory. For example, given probability distributions $\omega, \sigma$ over $X, Y$, their tensor corresponds to the resulting independent distribution over $X \times Y$.
\[
\input{prod-state.tikz}
\]
An arbitrary normalised state on $X \otimes Y$ corresponds to a joint distribution over $X, Y$ (left-hand below). Similarly a channel as on right-hand below represents a probability channel \rl{$P(Y_1,\dots,Y_m \mid X_1,\dots,X_n)$. }
\[
\input{joint-state.tikz} \qquad \qquad \qquad  \input{MXY.tikz} 
\]
Given any state over $X$ and $Y$ in $\MatR$, the marginal on $X$ corresponds to taking the marginal in the usual probabilistic sense, by summation over $Y$. In general, marginalisation of any morphism is given by summation over the discarded object.
\[
\input{marginal2.tikz} :: x \mapsto \sum_{y \in Y} \omega(x,y)
\qquad \qquad \qquad 
\input{margM.tikz} :: (x,y) \mapsto \sum_{z \in Z}  M(y,z \mid x)
\]
Finally observe that for any effect $e \colon X \to \mathbb{R}^+$ and normalised state $\omega$ the scalar $e \circ \omega$ corresponds to the expectation value of $e$ according to the probability distribution $\omega$. 
\[
\begin{tikzpicture}
	\begin{pgfonlayer}{nodelayer}
		\node [style=map] (0) at (0, -1) {$\omega$};
		\node [style=map] (1) at (0, 1) {$e$};
		\node [style=label] (2) at (-0.5, 0) {$X$};
	\end{pgfonlayer}
	\begin{pgfonlayer}{edgelayer}
		\draw (1) to (0);
	\end{pgfonlayer}
\end{tikzpicture}
  \ \ = \ \ \sum_{x \in X} e(x) \omega(x)
\]
\end{example}

When defining causal models we are largely concerned with channels and so will work in the Markov category $\FStoch$ of finite probabilistic processes, the subcategory of channels in $\MatR$. However for causal reasoning (such as in the calculation in Sections \ref{Sec_CE_Identifiability} and \ref{Sec_Counterfactuals}) it can be helpful to work in a broader category, in particular containing non-trivial effects, such as $\MatR$.  

\subsection{Deterministic processes and caps}

Amongst general processes of a `probabilistic' nature it can be helpful to identify those processes which behave `deterministically', like (partial) functions. In fact these are precisely those that respect copying, as follows.

\begin{definition} \label{def:deterministic} 
A morphism $f$ is called \emph{deterministic} when the following holds.
\[
\input{deterministic.tikz}
\]
\end{definition}

An important special case in practice are deterministic states, which here we also call \emph{sharp}. We depict a sharp state $x$ of $X$ as:
\[
\begin{tikzpicture}[tikzfig]
	\begin{pgfonlayer}{nodelayer}
		\node [style=sharpstate] (0) at (0, 0) {$x$};
		\node [style=none] (1) at (0, 1.25) {};
		\node [style=label] (2) at (0, 1.75) {$X$};
	\end{pgfonlayer}
	\begin{pgfonlayer}{edgelayer}
		\draw (1.center) to (0);
	\end{pgfonlayer}
\end{tikzpicture}

\]
By definition these are the states which are directly copied by the copy morphisms:
\begin{equation} \label{eq:copy-points}
\input{copy-points.tikz}
\end{equation}

A feature we will find useful in causal reasoning later is a correspondence between deterministic states and effects. For this we require an extra property of our cd-category. 

\begin{definition} 
We say that a cd-category $\catC$ has \emph{caps} when each object $X$ comes with a distinguished effect on $X \otimes X$ depicted as $\tinycap$ and satisfying the following: 
\begin{equation}\label{eq:mult-map}
\input{cap-sym.tikz} \qquad \qquad \ \ \ 
\input{cap-2.tikz}  \ \ \ \qquad \qquad  \input{mult-map2.tikz} 
\end{equation}
The choice of caps is moreover `natural' in that the following hold for all objects $X, Y$. 
\[
\input{capXYrule.tikz} 
% \qquad \qquad \qquad \tikzfig{capIrule}
\]
\end{definition} 
It follows that each cap is deterministic.\footnote{One may verify that the right-hand morphism in \eqref{eq:mult-map} is commutative and associative, forming a commutative \emph{semigroup} multiplication $\tinymult[whitedot]$ in $\catC$.} Now in any cd-category with caps, for any sharp state $x$ there is a corresponding deterministic effect denoted $x^\dagger$ and depicted as `flipping $x$ upside-down':
\[
\input{cap-3.tikz}
\]
We call an effect \emph{sharp} when it is of the form $x^\dagger$. One may verify that for any sharp state $x$, the effect $x^\dagger$ is the unique effect satisfying the following. 
\begin{equation} \label{eq:sharp-state-eff}
\begin{tikzpicture}
	\begin{pgfonlayer}{nodelayer}
		\node [style=sharpeffect] (0) at (-3, 0.75) {$x$};
		\node [style=sharpstate] (1) at (-3, -0.75) {$x$};
		\node [style=none] (2) at (-1.5, 0) {$=$};
		\node [style=none] (4) at (-0.25, 0) {$1$};
	\end{pgfonlayer}
	\begin{pgfonlayer}{edgelayer}
		\draw (1) to (0);
	\end{pgfonlayer}
\end{tikzpicture}
 \qquad \qquad 
\input{sharp-2.tikz}
\end{equation}
Caps are particularly useful in diagrammatic reasoning when they are \emph{cancellative}, meaning that: 
\[ 
\input{capcancel.tikz} \label{eq:cancel-caps}
\]
for all morphisms $f, g$.

\begin{example}
$\MatR$ has cancellative caps. Each point $x \in X$ corresponds to a normalised sharp state on $X$ which we again denote by $x$, given by the point probability distribution $\delta_x$ \rlb{on $X$.} The sharp effect $x^\dagger$ is then given by the same function $\delta_x$ on $X$ also. 
\[
\begin{tikzpicture}[tikzfig]
	\begin{pgfonlayer}{nodelayer}
		\node [style=sharpstate] (0) at (0.5, -0.75) {$x$};
		\node [style=none] (1) at (0.5, 0.5) {};
		\node [style=label] (2) at (0.5, 1) {$X$};
	\end{pgfonlayer}
	\begin{pgfonlayer}{edgelayer}
		\draw (1.center) to (0);
	\end{pgfonlayer}
\end{tikzpicture}
  \ \ , \ \ \begin{tikzpicture}[tikzfig]
	\begin{pgfonlayer}{nodelayer}
		\node [style=sharpeffect] (0) at (0.5, 0.75) {$x$};
		\node [style=none] (1) at (0.5, -0.5) {};
		\node [style=label] (2) at (0.5, -1) {$X$};
	\end{pgfonlayer}
	\begin{pgfonlayer}{edgelayer}
		\draw (1.center) to (0);
	\end{pgfonlayer}
\end{tikzpicture}
 
\quad   :: \ y \mapsto 
\begin{cases} 
1 & x=y \\ 
0 & \text{otherwise} 
\end{cases} 
\]
The cap is given by \ $\tinycap(x,y) = \delta_{x,y}$. Hence, for all $x, y \in X$ the following holds. 
\[
\begin{tikzpicture}[tikzfig]
	\begin{pgfonlayer}{nodelayer}
		\node [style=none] (0) at (-2, -0.25) {};
		\node [style=none] (1) at (-0.25, -0.25) {};
		\node [style=sharpstate] (2) at (-2, -0.5) {$x$};
		\node [style=sharpstate] (3) at (-0.25, -0.5) {$y$};
	\end{pgfonlayer}
	\begin{pgfonlayer}{edgelayer}
		\draw [bend left=90, looseness=1.75] (0.center) to (1.center);
	\end{pgfonlayer}
\end{tikzpicture}
\quad = \quad 
\begin{tikzpicture}[tikzfig]
	\begin{pgfonlayer}{nodelayer}
		\node [style=none] (0) at (4.75, -0.75) {};
		\node [style=sharpeffect] (1) at (4.75, 1) {$y$};
		\node [style=sharpstate] (2) at (4.75, -0.75) {$x$};
	\end{pgfonlayer}
	\begin{pgfonlayer}{edgelayer}
		\draw (1) to (0.center);
	\end{pgfonlayer}
\end{tikzpicture}
  \ \ = \ \ \begin{cases} 1 & x = y \\ 0 & \text{otherwise} \end{cases}
\]
Every sharp state on $X$ is of the above form for some $x \in X$, or else given by the \emph{zero state} $0$ defined by $0(x) = 0$ for all $x \in X$. Then $0^\dagger$ is given by the constant zero function on $X$ also. In particular the only sharp scalars are $0$ and $1$. We note the useful observation that for any morphism $M \colon X \to Y$ its scalar values $M(y \mid x) \in \mathbb{R}^+$ can be given diagrammatically by composing with $x$ and $y^\dagger$ as below. 
\[
\ M(y \mid x)  \ \ \ =  \ \ \ \begin{tikzpicture}[tikzfig]
	\begin{pgfonlayer}{nodelayer}
		\node [style=sharpeffect] (0) at (0.5, 1.5) {$y$};
		\node [style=none] (1) at (0.5, 1.5) {};
		\node [style=sharpstate] (2) at (0.5, -1.5) {$x$};
		\node [style=map] (3) at (0.5, 0) {$M$};
	\end{pgfonlayer}
	\begin{pgfonlayer}{edgelayer}
		\draw (2) to (1.center);
	\end{pgfonlayer}
\end{tikzpicture}
  
\]

Since $\tinycopy(y,z \mid x) = \delta_{x,y,z}$ it is indeed the case that $\tinycopy$ copies each state given by $x \in X$ as in \eqref{eq:copy-points}. In contrast, a general state $\omega$ is not copyable. 
\[
\begin{tikzpicture}
	\begin{pgfonlayer}{nodelayer}
		\node [style=whitedot] (0) at (0, 0) {};
		\node [style=none] (1) at (-1, 1) {};
		\node [style=none] (2) at (1, 1) {};
		\node [style=none] (3) at (0, -1.25) {};
		\node [style=label] (4) at (-1, 1.5) {$X$};
		\node [style=label] (5) at (1, 1.5) {$X$};
		\node [style=map] (6) at (0, -1.5) {$\omega$};
		\node [style=none] (7) at (2.5, 0) {$\neq$};
		\node [style=map] (8) at (4.25, -0.75) {$\omega$};
		\node [style=map] (9) at (5.75, -0.75) {$\omega$};
		\node [style=none] (10) at (4.25, 0.5) {};
		\node [style=none] (11) at (5.75, 0.5) {};
		\node [style=label] (12) at (4.25, 1) {$X$};
		\node [style=label] (13) at (5.75, 1) {$X$};
	\end{pgfonlayer}
	\begin{pgfonlayer}{edgelayer}
		\draw [bend right=45] (1.center) to (0);
		\draw [bend right=45] (0) to (2.center);
		\draw (3.center) to (0);
		\draw (11.center) to (9);
		\draw (10.center) to (8);
	\end{pgfonlayer}
\end{tikzpicture}
\] 
The left-hand side above is given by $(x,y) \mapsto \omega(x) \delta_{x,y}$, while the right is $(x,y) \mapsto \omega(x) \omega(y)$, which differ unless $\omega$ is zero or $\omega = x$ for some $x \in X$. \rl{The left-hand side defines a distribution, where the copies of $X$ are perfectly correlated, while on the right-hand side it is a product distribution with both $X$ independent from each other.}

Deterministic morphisms in $\MatR$ correspond to (partial) functions. Any partial function $f \colon X \to Y$ defines a morphism $X \to Y$ which we again denote by $f$, given on sharp states $x \in X$ by $f \circ x = f(x)$ whenever $f(x)$ is defined, and $f \circ x = 0$ otherwise. Equivalently, 
% $f(y \mid x) = \delta_{x, f(x)}$ whenever $f(x)$ is defined, and zero otherwise. Equivalently, 
\[
\input{pfunc.tikz} \ = \ 
\begin{cases} 
1 & f(x) \text{ defined and } y = f(x)  \\ 
0 & \text{otherwise}
\end{cases}
\]
\end{example}

The framework of cd-categories is much more general than simply the finite discrete setting of $\MatR$. In Appendix \ref{sec:measure-theory} we detail a cd-category for the treatment of full measure-theoretic probability, though this is not required to read the remainder of the paper. 

The remainder of this section introduces further properties of a cd-category useful for causal reasoning, \rl{which while not essential for defining causal models as such, will be very useful in Sections \ref{Sec_CE_Identifiability} and \ref{Sec_Counterfactuals}.}  

\subsection{Normalisation} \label{sec:normalisation}

In causal reasoning one often needs to normalise states and processes. Intuitively, for a general state $\omega$, its normalisation should be another state $\normmor{\omega}$, depicted with a dashed box around $\omega$, such that the following holds:
\begin{equation} \label{eq:state-norm}
\input{state-norm-cond.tikz}
\end{equation}
\rl{In $\MatR$ this corresponds to writing a state as a product of a non-negative constant and a `correctly normalised' probability distribution.}
More generally, for any process $f$ its normalisation $\normmor{f}$ is a process depicted:
\begin{equation} \label{eq:norm-box}
\input{norm-box3.tikz}
\end{equation}
which intuitively is given on each sharp state input $x$ by normalising $f \circ x$.
\begin{equation} \label{eq:norm-on-states}
\input{norm-det.tikz}
\end{equation}
Another way to express the latter  without reference to states is the following.
\begin{equation} \label{eq:norm-sup-cond}
\input{normsupcond1.tikz}
\end{equation}
Plugging in a sharp state $x$ \rl{into} \eqref{eq:norm-sup-cond} shows that \eqref{eq:state-norm} holds for $\omega = f \circ x$, with $\normmor{\omega} = \normmor{f} \circ x$. When $f = \omega$ is already a state this reduces to \eqref{eq:state-norm}.

\rl{One can often think of a normalised process} as a channel, so that for a state $\omega$ its normalisation $\normmor{\omega}$ is normalised in our earlier sense. But this is not always possible. For example in $\MatR$ if $\omega$ is the zero state then $\normmor{\omega} = 0$. More generally if $f(x) = 0$ for some input $x$, then $\normmor{f}(x) = 0$ also. 

Instead, a normalisation $g = \normmor{f}$ will in general only be a \emph{partial channel}, meaning that the following holds. 
\[
\input{pchannelg.tikz}
\]
We can now define normalisation formally, while stating some of its essential properties.

\begin{definition} \label{def:norm}
Let $\catC$ be a cd-category. We say $\catC$ has \emph{normalisation} when for every morphism $f$ there is a distinguished partial channel $\normmor{f}$, depicted with a dashed box as in \eqref{eq:norm-box}, such that \eqref{eq:norm-sup-cond} holds and:
\begin{enumerate}
\item 
Whenever $f$ is a partial channel we have $\normmor{f} = f$. 
\item 
For all morphisms $f, g$:
\begin{equation} \label{eq:mon-cond}
\input{norm-monoidal.tikz}
\end{equation}
\item 
For all morphisms $f$: 
\begin{equation} \label{eq:disc-cond}
\input{norm-discard.tikz}
\end{equation}
\item  
For all morphisms $f$:
\begin{equation} \label{eq:copy-cond}
\input{copynorm.tikz}
\end{equation}
\end{enumerate}
\end{definition} 

In Appendix \ref{sec:normalisation-appendix} we in fact prove that a normalisation structure on a cd-category $\catC$ is unique whenever it exists, being totally fixed by the above axioms. We also \rl{establish in App.~\ref{sec:normalisation-appendix}} further conditions, which ensure that a cd-category has normalisation, and the following further properties of normalisation.

\begin{lemma} \label{lem:norm-lemma}
Let $\catC$ be a cd-category with normalisation. 
\begin{enumerate}[resume]
\item \label{enum:chancond}
Whenever $f$ is a channel, for all morphisms $g$ we have: 
\[
\input{normchan3.tikz}
\]
\item \label{enum:Extracopy}
For all morphisms $f$ we have:
\[
\input{copynorm2.tikz}
\]
\item \label{enum:detstatecond}
If $\catC$ has caps then \eqref{eq:norm-on-states} holds and more generally for every morphism $f$ and sharp state $x$ we have: 
\begin{equation} \label{eq:norm-on-states-wide}
\input{norm-det-wide.tikz}
\end{equation}
\end{enumerate}
\end{lemma}

\begin{example}
$\MatR$ has normalisation. On each object $X$ the zero state $0$, given by $0(x) = 0$ for all $x \in X$, satisfies $\normmor{0} = 0$. For a non-zero state $\omega$ we have \[
\normmor{\omega}(x) = \frac{\omega(x)}{\sum_{x' \in X} \omega(x')}
\]
For a general morphism $M \colon X \to Y$ the normalisation $\normmor{M}$ sends $x \in X$ to $\normmor{M(x)}$. That is:
\[
\normmor{M}(y \mid x) = 
\begin{cases} 
\frac{M(y \mid x)}{\sum_{y' \in Y} M(y' \mid x)} & \text{ if }

\sum_{y \in Y} M(y \mid x) \neq 0  \\ 
0 & \text{otherwise}
\end{cases}
\]
\end{example}

\subsection{\rlb{Conditioning and conditional independence}} \label{sec:conditioning}

For our purposes the main use of normalisation will be to define probabilistic conditioning. Here we give a diagrammatic account of conditioning which extends previous treatments by Cho and Jacobs \cite{ChoEtAl_2019_DisintegrationViaStringDiagrams} and Fritz \cite{Fritz_2020_SyntheticApproachToMarkovKernels} from channels to general processes, when in the presence of normalisation and well-behaved caps. 

\begin{definition} \label{def:conditional-morphism}
We say a cd-category $\catC$ has \emph{\diagconditioning} when it has normalisation and cancellative caps. Then for any morphism $f \colon X \to Y \otimes Z$ we define the \emph{conditional} $f|_Z \colon X \otimes Z\to Y$ to be the partial channel 
\begin{equation} \label{eq:disintegration}
\input{disint-norm.tikz}
\end{equation}
For any deterministic state $z$ of $Z$ we then define $f|_z \colon X \to Y$ by composing with $z$, so that 
\begin{equation} \label{eq:manyzs}
\input{cond-by-z4short.tikz}
\end{equation}
\end{definition}

The following useful property generalises the \emph{chain rule} or \emph{product rule} \rlb{$P(x,y) = P(y | x) P(x)$} in probability theory, and for the case where $\omega$ is normalised has been referred to by Cho and Jacobs as the \emph{disintegration} of a joint state \cite{ChoEtAl_2019_DisintegrationViaStringDiagrams}.

\begin{lemma}[Disintegration] \label{lem:chain-rule}
If $\catC$ has \diagconditioning, 
any state $\omega$ of $X \otimes Y$ satisfies 
\begin{equation} \label{eq:chainrule}
\input{disint-states.tikz}
\end{equation}
\end{lemma}

 We prove an extension of this result in Appendix \ref{sec:normalisation-appendix}, which establishes in what sense our definition of $f|_Z$ satisfies the categorical definition of a `conditional' from \rlb{Refs.~\cite{ChoEtAl_2019_DisintegrationViaStringDiagrams, Fritz_2020_SyntheticApproachToMarkovKernels}.} Note that this only holds when cancellative caps are present. The latter approaches provide a more general account of conditioning (for channels) which we do not use here.

\begin{example} \label{ex:conditioning-in-Mat}
$\MatR$ has \diagconditioning\ as we have seen. For a morphism $M \colon X \to Y \otimes Z$ the partial channel $M|_Z$ is given by 
\[
M|_Z(y \mid x, z) = \frac{M(y, z \mid x)}{\sum_{z' \in Z}M(y,z' \mid x)}
\]
whenever the sum in the denominator is non-zero, and $M|_Z(y \mid x, z) = 0$ for all $y$ otherwise. In standard probability theory notation, a channel $M \colon X \to Y \otimes Z$ is described by a density $P(Y, Z \mid X)$, and $M|_Z$ corresponds to the conditional density $P(Y \mid X, Z)$. 
\[
\input{prob-notation-1.tikz}
\]
For any $z \in Z$, the partial channel $M|_z$ corresponds to the conditional density $P(Y \mid X, Z=z)$. 
\[
\input{prob-notation-4.tikz}
\]
To see the relation of \eqref{eq:chainrule} to the product rule in probability theory, observe that for any joint distribution $P(X,Y)$ we have the following.
\[
P(x,y) \ \ = \ \ \input{chain-rule-connection.tikz}
 \ \ 
 \ \ =  \ \ P(x) P(y \mid x) 
 \]
\end{example}

\color{\rlbcolor}

\begin{remark} 
The significance of \emph{partial} channels in the definition of normalisation is that it allows our conditionals to be defined even for distributions without full support, which are important for various aspects of causal reasoning.  As above, in $\MatR$ a conditional $P(X | Z=z)$ outputs the \emph{zero state} whenever $P(Z=z) = 0$, interpreted as an `undefined' distribution over $X$, and otherwise yields the expected conditional distribution. 
\end{remark}

We will now establish that conditionals behave in keeping with our intuitions, with all proofs found in Appendix \ref{sec:cond-appendix}. 

\begin{proposition} \label{Prop_Commutativity_of_conditional}
	Let $\catC$ be a cd-category with \diagconditioning\ and $\omega$ a state of $X \otimes Y \otimes Z$. 
	Then 
	\begin{equation}
		\input{Fig_Commutativity_of_conditional.tikz}
	\end{equation}
\end{proposition}

This, together with~\eqref{eq:disc-cond}, means that relative to a given state $\omega$ one can drop the labels in boxes of conditionals -- their identity is unambiguously determined by the input and output types. Whenever the context makes the reference state $\omega$ clear, it is therefore convenient to employ the following notation.  
Given a state $\omega$ of $X_1 \otimes \dots \otimes X_n$ for any disjoint subsets $S,T \subseteq \{X_1, ..., X_n\}$ and  $R:= \{X_1, ..., X_n\} \setminus (S \cup T)$ we write: 
\begin{equation}
	\input{Fig_notation_for_conditionals.tikz}
\end{equation}
In particular we say that $\omega$ has \emph{full support} over $S \subseteq \{X_1,\dots,X_n\}$ when we have: 
\[
\input{full-supps.tikz}
\]
% This box \rlb{also generally, that is when $\omega$ may \emph{not} have full support on $S$,} induces a morphism on $S$ that leaves all conditionals invariant under composition, as follows.  
\st{Even when $\omega$ lacks full support on $S$, this box induces a morphism on $S$ that leaves all conditionals invariant under composition, as follows.}  

\begin{lemma} \label{Lem_support_projector}
	Let $\catC$ be a cd-category with \diagconditioning\ and $\omega$ a state of $X \otimes Y$. Then the following hold.  
	\begin{center}
		(1) \ $ \input{proj-idem.tikz}$
		\hspace*{0.2cm}  \qquad (2) \ \ $ \input{Fig_Projector_proposition_one.tikz} $ \ \ 
		\hspace*{0.1cm} \qquad (3) \ $ \input{Fig_Projector_proposition_three.tikz} $ .
	\end{center}
\end{lemma}

Note that $X$ may of course have the structure of a product over a subset $S$ of objects as above. 

We can now define conditional independence, which lets the same idea as in Refs.~\cite{ChoEtAl_2019_DisintegrationViaStringDiagrams, Fritz_2020_SyntheticApproachToMarkovKernels} manifest in terms of conditionals in the above sense. 

\begin{definition} \label{def:cond_ind}
\textnormal{(Conditional independence):} 
Let $\catC$ be a cd-category with \diagconditioning. 
Given a state $\omega$ of $X \otimes Y \otimes Z$, we 
say $X$ is independent from $Y$ relative to $Z$, and write $(X \bigCI Y | Z)_{\omega}$, iff
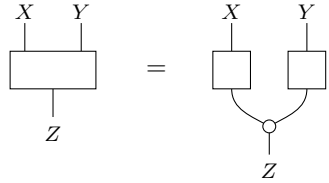
\[ \input{Fig_Def_CI_v2.tikz} \]
\end{definition}

It is instructive to observe the following equivalences. 

\begin{lemma} \label{Lem_CI_equivalences}
Let $\catC$ be a cd-category with \diagconditioning. 
Given a state $\omega$ of $X \otimes Y \otimes Z$, conditional independence $(X \bigCI Y | Z)_{\omega}$ is equivalent to the following: 
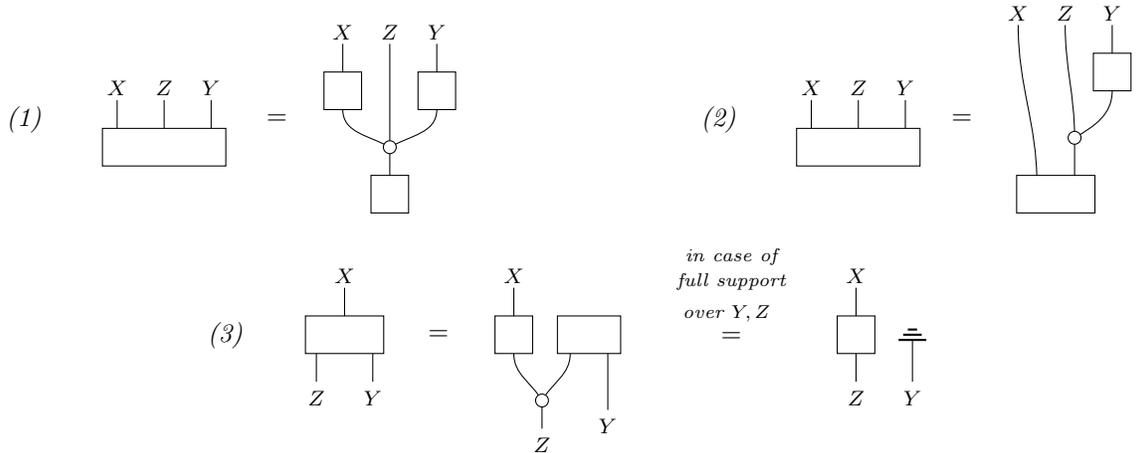
\begin{center}
(1) \hspace*{0.5cm} $\input{Fig_CI_prop_2.tikz}$ \hspace*{3cm}
(2) \hspace*{0.5cm} $\input{Fig_CI_prop_3.tikz}$ \\[0.4cm]
(3) \hspace*{0.5cm} $\input{Fig_CI_prop_4.tikz}$ \hspace*{0.3cm} 
$\stackrel{\begin{minipage}{1.7cm} \centering {\scriptsize in case of \\[-0.1cm] full support over $Y, Z$} \vspace*{0.1cm} \end{minipage}}{=}$ 
\hspace*{0.3cm} $\input{Fig_CI_prop_4_b.tikz}$ 
\end{center}
\end{lemma}

Similarly to the treatment of the full support case in \cite{ChoEtAl_2019_DisintegrationViaStringDiagrams}, we are able to establish string diagrammatic forms of the well-known \emph{semi-graphoid axioms} for reasoning about conditional independence, recalled and proven in Appendix~\ref{App_Proof_Thm_SG_axioms}. 

\begin{theorem} \label{Thm_SG_axioms}
	Let $\catC$ be a cd-category with \diagconditioning\ and $\omega$ some state. 
	The conditional independence relation $( \_\_ \bigCI \_\_ | \_\_ )_{\omega}$ from Def.~\ref{def:cond_ind}, understood as a 3-place relation for arbitrary triples of disjoint subsets (of the objects of $\omega$'s codomain) satisfies the \emph{semi-graphoid axioms}. 
\end{theorem}

\begin{example}
In $\MatR$ the condition $(X \bigCI Y | Z)_{P}$ for some $P(X,Y,Z)$ according to Def.~\ref{def:cond_ind} indeed reads 
\[ P(X,Y | Z) \ = \ P(X | Z) \ P(Y | Z) \ , \]
understood as $P(X,Y | Z=z) = P(X | Z=z) P(Y | Z=z)$ whenever $P(Z=z) \neq 0$ and it is set to the zero state in $\MatR$ for all other $z$. 
Conditions $(1)$, and $(2)$ of Lem.~\ref{Lem_CI_equivalences}, respectively, are:
\[ P(X,Y,Z) \ = \ P(Y | Z) \ P(X|Z) \ P(Z) \quad \text{and} \quad P(X,Y,Z) \ = \ P(Y | Z) \ P(X,Z) \ , \] 
both understood as above concerning the dependence on full support. 
Condition $(3)$, in the full support case, reads
\[ P(Y| X, Z) \ = \ P(Y | Z) \ . \]
\end{example}

\color{black}

% \bibliographystyle{utphys} \bibliography{CauseComp.bib} % Uncomment while working on file standalone, if needed 
%\end{document}
%\documentclass[main.tex]{subfiles} \begin{document}

%*****************************************************************
\section{Causal models} \label{sec:causal-models}
%*****************************************************************

%*****************************************************************
\subsection{Definition} \label{sec:def-causal-models}
%*****************************************************************

Let us now turn to the main object of study of this work and give the general definition of a causal model in a cd-category. Recall that a finite \emph{directed graph} $G=(V,E)$ consists of a finite set $V=\{X_1,\dots,X_n\}$ of \emph{vertices} and a subset $E \subseteq V \times V$ of directed \emph{edges}. An edge $(X_i,X_j) \in E$ is usually denoted $X_i \to X_j$. A finite \emph{directed acyclic graph (DAG)} furthermore has no directed cycles, i.e. no sequence of edges of the form $X_{i_1} \to X_{i_2} \to \dots \to X_{i_n} \to X_{i_1}$ of length $n \geq 1$.
\rlb{For a vertex $X \in V$ write $\Pa(X)\subseteq V$ and $\Ch(X) \subseteq V$ for the sets of parents and children of $X$, respectively.}

A causal model is essentially given by specifying a DAG, whose edges describe direct-cause relations between variables, along with channels that describe the mechanisms by which each variable is caused by its parents in the DAG. 

\begin{definition} \label{def:causal-model}
Let $\catC$ be a cd-category. A \emph{causal model} $\modelM$ in $\catC$ is given by: 
\begin{enumerate} 
\item 
a finite DAG $G$ with vertices $V=\{X_1,\dots,X_n\}$; 
\item 
a specified subset $O \subseteq V$  of \emph{output} vertices;
\item 
for each vertex $X_i \in V$ an associated object of $\catC$ also denoted $X_i$, and a channel in $\catC$ of the form:
\begin{equation} \label{eq:channels} 
% c_i \colon \bigotimes \Pa(X_i) \to X_i
\input{cismall.tikz}
\end{equation} 
where $\Pa(X_i)$ denotes the parents of $X_i$ in $G$. We call the channel $c_i$ the \emph{mechanism} for $X_i$.
\end{enumerate}
The pair $\ODAG{G}_\modelM = (G,O)$ is called the \emph{causal structure} of $\modelM$. The indexed objects $\var(\modelM) = (X_i)^n_{i=1}$ of $\catC$ are called the \emph{variables} of the model. The variables corresponding to output vertices are called the \emph{output} variables and denoted $\outvar(\modelM) \subseteq \var(\modelM)$.  
\end{definition} 

Note that this definition allows for only a subset of variables to be considered as `outputs' of the model. This specification does not necessarily come with the interpretation that these are the only `observed' variables; it is a piece of data that is useful for different purposes.

Rather than specify the DAG directly, we can also simply specify a causal model via its collection of mechanisms. More precisely, we can equivalently define a causal model  $\modelM$ in $\catC$ as consisting of an indexed collection of objects $\var(\modelM) = (X_i)^n_{i=1}$ with a subset $\outvar(\modelM)$, and a collection of channels $(c_i)^n_{i=1}$ as in \eqref{eq:channels}, such that the directed graph $G$ with vertices $X_i$ and an edge $X_i \to X_j$ whenever $X_i$ is an input to $c_j$ is acyclic.

When we take our category to describe probabilistic mappings between finite sets, we immediately obtain the more standard notion of a causal model. 

\begin{example} \label{ex:CBN}
A \emph{Causal Bayesian Network (CBN)} is a causal model in $\catC = \FStoch$. Thus it consists of:
\begin{itemize}
\item 
a finite DAG $G=(V,E)$ with a subset $O \subseteq V$ of outputs; 
\item 
for each vertex $X_i$ an associated variable with a finite set of values also denoted $X_i$, and a mechanism $c_i$ given by a probability channel with density 
\begin{equation} \label{eq:mechanism-density}
P(X_i \mid \Pa(X_i))
\end{equation}
\end{itemize} 
A CBN defines a joint distribution over all the variables with density 
\begin{equation} \label{eq:markov}
P(X_1,\dots,X_n) := \prod^n_{i=1} P(X_i \mid \Pa(X_i))
\end{equation}
As the notation suggests, the mechanisms can often be thought of as conditional densities for $P(X_1,\dots,X_n)$, though we emphasise that they are given channels -- see Rem.~\ref{Rem_MechanismsFirst}.
In general any joint density $P$ is said to satisfy the \emph{causal Markov condition} for $G$ when there exist probability channels such that \eqref{eq:markov} holds. 
Often one is particularly interested in the \emph{output distribution} of the CBN, given by the marginal distribution over the output variables $O$.
\end{example}

\begin{remark} \label{Rem_MechanismsFirst}
It is perhaps more common to define a CBN
as given by a DAG $G$ and corresponding variables, along with a joint distribution $P(X_1,\dots,X_n)$ that satisfies the Markov condition \eqref{eq:markov}, where one defines the mechanisms as corresponding conditional densities \eqref{eq:mechanism-density}. 
% Todo: Make sure this point about support is clear. 
However, these conditional channels are not uniquely determined if the joint distribution lacks full support, and their precise choice can make a practical difference when considering, for instance, interventions (Sec.~\ref{Sec:Interventions}). 
In contrast, we consider the specific channels $c_i$ to be \emph{the} key aspect of a causal model, since they come with an interpretation as causal mechanisms. We therefore take a CBN to be defined by its mechanisms rather than the joint distribution. 
\end{remark} 

\begin{example} \label{Ex_CBN} 
	As a basic, more concrete example of a CBN, which will also serve as a running example throughout, consider the following example, taken from Ref.~\cite{CorreaEtAl_2020_CalculusForStochasticInterventions}. 
	Similarly to the example in Sec.~\ref{Sec_GentleOverview}, it is about smoking and lung cancer, though this time with different further variables that will be relevant to studying policy making later. 
	Let $S$ be a person's choice to smoke, $L$ whether or not they develop lung cancer, $A$ their age and $B$ a set of relevant background conditions like socio-economic status, education etc. 
A plausible causal model consists of the DAG below, where the vertices corresponding to output variables are circled, along with the specification of each of the probability channels listed to the right, which give the mechanisms.
	\begin{eqnarray}
		\input{Fig_Example_CBN_DAG.tikz} 
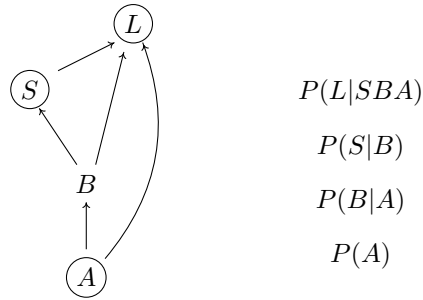
		\begin{minipage}{5cm}
			\centering
			$P(L | SBA)$ \\[0.3cm]
			$P(S | B)$ \\[0.3cm]
			$P(B | A)$ \\[0.3cm]
			$P(A)$	
		\end{minipage}
		\nonumber
	\end{eqnarray}
	As emphasised in Rem.~\ref{Rem_MechanismsFirst}, the CBN is not taken to be defined by conditional distributions obtained from a joint distribution, but by causal mechanisms. This data though of course defines a marginal distribution over the output variables:
	\begin{equation}
		P(SLA) = \sum_{B} \ P(L | SBA) \ P(S | B) \ P(B | A) \ P(A) \nonumber
	\end{equation}
\end{example}

%*************************************************
\subsection{Causal models as diagrams} \label{subsec:NetDiagramsCMs}
%*************************************************

Let us now see how to describe CBNs and more general kinds of causal models in diagrams. In fact there is an exact correspondence between DAGs and a certain class of string diagrams which lies at the heart of the diagrammatic approach to causal models. This construction was first given by Fong \cite{Fong_2013_CausalTheories} and made still more explicit by Jacobs et al.~\cite{JacobsEtAl_2019_CausalInferenceByDiagramSurgery}. 

\begin{definition} \label{def:DGO}
Let $G$ be a DAG over vertices $V=\{X_1,\dots,X_n\}$, and $O \subseteq V$ a subset of output vertices. We define a string diagram $D_{G,O}$ with no inputs and a single output for each $X_i \in O$, as follows. 
\begin{enumerate} \item 
For each vertex $X_i$ we draw a box denoted $c_i$ from the parents $Y_1,\dots,Y_m$ of $X_i$ to $X_i$ as below. We then copy the output of the box $k + 1$ times if $X_i$ is an output variable, or $k$ times otherwise, where $k$ is the number of children of $X_i$. 
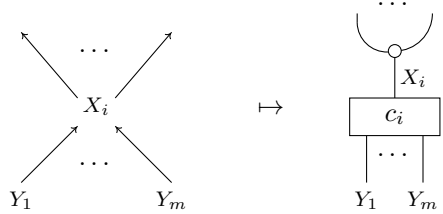
\[
\input{DAG-extract.tikz}
\qquad 
\mapsto
\qquad 
\input{DAG-extractsd.tikz}
\]
In particular if a variable $X_i$ has no parents then the box $c_i$ has no inputs, and if $X_i$ has no children and is not an output then $c_i$ is proceeded by the discarding effect $\discard{}$. 
\item 
We then wire these pieces together by plugging an $X_i$ output from (the copy maps following) $c_i$ into the $X_i$ input of $c_j$, whenever there is an edge $X_i \to X_j$ in $G$. 
\end{enumerate} 
\end{definition} 

Since $G$ is acyclic, this produces a well-defined string diagram of the specified type. In particular by construction each vertex $X_i$ appears as an output only if $X_i \in O$, and in this case precisely once. In practice we allow any names to be given to the boxes $c_i$ in the diagram.

\begin{example} \label{ex:DAG}
For the DAG $G$ on the left below with outputs $O=\{X_2, X_3\}$ circled, the diagram $D_{G,O}$ is shown to the right. 
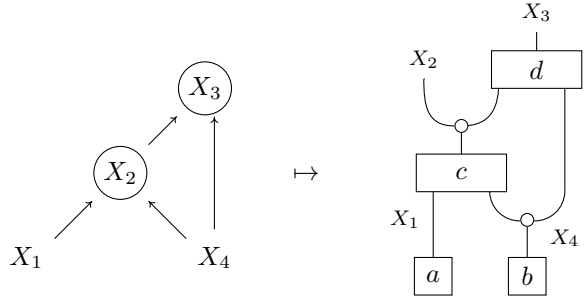
\[
\input{Graph-obs2.tikz}
\qquad 
\mapsto
\qquad 
\input{state-GO.tikz}
\]
\end{example} 

By construction, each string diagram $D_{G,O}$ contains only boxes with a single output, and copy maps. More precisely, we can identify the kinds of string diagrams that arise from DAGs in this way as follows. 

\begin{definition} \label{Def_NetworkDiagram}
A \emph{network diagram} is a string diagram $D$ built from single-output boxes, copy maps and discarding effects:
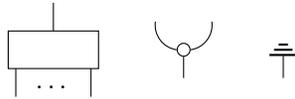
\[
\input{nd-box.tikz}
\qquad 
\input{nd-copy.tikz}
\qquad 
\begin{tikzpicture}[tikzfig]
	\begin{pgfonlayer}{nodelayer}
		\node [style=upground] (0) at (1.5, 0) {};
		\node [style=none] (1) at (1.5, -0.75) {};
	\end{pgfonlayer}
	\begin{pgfonlayer}{edgelayer}
		\draw (1.center) to (0);
	\end{pgfonlayer}
\end{tikzpicture}

\] 
along with labellings on the wires, such that any wires not connected by a sequence of copy maps are given distinct labels, and each label appears as an output at most once and as an input to any given box at most once. A \emph{state network diagram} is a network diagram with no inputs. 
\end{definition}

\st{Causal models as defined so far will only require state network diagrams, but general network diagrams with inputs will be used  in Sec.~\ref{sec:openCMs} and throughout Secs.~\ref{Sec_CE_Identifiability} and \ref{Sec_Counterfactuals}}.

We consider two network diagrams to be equivalent when they are equal according to the general axioms of a cd-category, up to relabellings of the wires and boxes.  

\st{
\begin{remark} 
Network diagrams form a subset of the `gs-monoidal string diagrams' considered by Fritz and Liang \cite{FritzEtAl_2022_FreeGSMonoidalCategories}, which allow boxes with multiple outputs. The latter are not required here since each variable in a causal model is the unique output of its mechanism.
\end{remark} 
}

\begin{proposition}  
\label{prop:DagsNDs}
Up to equivalence and relabellings of diagrams, the following are equivalent: 
\begin{itemize} 
\item a DAG $G=(V,E)$ with outputs $O \subseteq V$; 
\item a state network diagram 
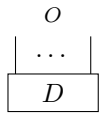
\[
\input{stateNDO.tikz}
\]
with wires $V$ and outputs $O$;
\end{itemize} 
via $(G,O) \mapsto D_{G,O}$ and $D \mapsto (G_D, O_D)$. 
\end{proposition} 

\begin{proof} 
Given $(G,O)$ we define $D_{G,O}$ as in Def.~\ref{def:DGO}. Conversely, given a state network diagram $D$, we define the DAG $G_D$ with a vertex $X$ for each output of a non-copy box in $D$, and set $X \in O_D$ iff it is an output of the diagram. We include a directed edge $X \to Y$ whenever $X$ is an input to the unique box with output $Y$. Up to relabellings, it is straightforward to see that these mappings are inverse to each other. 
\end{proof} 

The correspondence so far relates DAGs and state network diagrams, which are purely syntactic in nature, outlining only a causal structure. To provide an actual causal model, we must interpret them in a category, identifying the wires and boxes with specific objects and channels in the category. 

\begin{definition} 
An \emph{interpretation} $\sem{-}$ of a network diagram $D$ in a cd-category $\catC$ consists of specifying in $\catC$ an object $\sem{X_i}$ for each wire $X_i$ and channel $\sem{f} \colon \sem{X_1} \otimes \dots \otimes \sem{X_k} \to \sem{X}$ for each box $f$ in $D$ with inputs $X_1,\dots, X_k$ and output $X$.
\end{definition}

Thus an interpretation $\sem{-}$ of $D$ involves specifying objects for all wires and channels for all boxes in $D$. \st{For brevity, we at times also call a network diagram $D$ along with an interpretation $\sem{-}$ in $\catC$ a \emph{network diagram $D$ in $\catC$}}.

\begin{theorem}  \label{Thm:Equivalences_CausalModel_Defs}
Let $\catC$ be a cd-category. Specifying a causal model $\modelM$ in $\catC$ is equivalent to specifying a state network diagram $D$ with an interpretation $\sem{-}$ in $\catC$.
\end{theorem} 

\begin{proof}
Thanks to Proposition \ref{prop:DagsNDs} a causal structure $(G,O)$ is equivalent to a state network diagram $D = D_{G,O}$. Interpreting the network diagram $D$ is then precisely the same as specifying a mechanism \eqref{eq:channels} for each variable, i.e. specifying a causal model with this causal structure. 
\end{proof}

Thus, in short, a causal model $\modelM$ in $\catC$ is the same as a state network diagram in $\catC$. This allows one to specify a full causal model by simply drawing its network diagram, which encodes the causal structure $(G,O)$ and names the mechanisms, and then interpreting the mechanisms in $\catC$. \rlb{In practice and especially later in Sections~\ref{Sec_CE_Identifiability} and \ref{Sec_Counterfactuals},} given a model $\modelM$ we typically omit writing $\sem{-}$ and so write simply $X$, $f$ for both the labels and boxes in the network diagram and their corresponding interpretation in $\catC$, rather than denoting the latter by $\sem{X}, \sem{f}$.     

\st{
\paragraph{The output state defined by a model.} Given any network diagram $D$ with an interpretation $\sem{-}$ in $\catC$, we can compose the interpreted channels in the diagram according to $D$ to yield a single overall channel in $\catC$ from the inputs to the outputs of the diagram, which we denote by $\sem{D}$. In particular, for any causal model $\modelM$ with network diagram $D$ and outputs $O$ we obtain a normalised state in $\catC$ over the outputs $O$: 
\[
\input{output-state.tikz}
\]
which we call the \emph{output state} of the model.  
}

\st{Conversely, in practice one often starts from a normalised state $\omega$ over a set of objects $O$, and asks whether $\omega$ is compatible with a causal structure given by some DAG $G$, or equivalently a network diagram. We say that $\omega$ \emph{factorises according to} a network diagram $D$ if there exists an interpretation $\sem{-}$ of $D$ with output objects $O$ the same as those of $\omega$ (i.e. a choice of objects for the remaining non-output wires in the diagram, and channels for all the boxes in the diagram) such that $\sem{D} = \omega$. Similarly we say that $\omega$ \emph{factorises according to} a DAG $G$ over $V$ with specified outputs $O \subseteq V$ if it factorises according to $D_{G,O}$.}

\begin{example}
An interpretation of a network diagram in $\FStoch$ consists of specifying finite sets of values for each label $X_i$, and a probability channel $c_i \colon Y_1 \times \dots \times Y_n \to X_i$ for each box with these inputs and output in the diagram. 
Hence, we can specify a CBN via a network diagram along with interpretations for each of its boxes as finite probability channels, or for short, a network diagram in $\FStoch$. 

\st{ 
Given such a CBN $\modelM$, the output state $\sem{\modelM}$ over $O$ in $\catC$ describes the resulting output distribution $P(O)$ over the output variables $O$.}

\st{In general a distribution $\omega = P(O)$ factorises according to a network diagram $D=D_{G,O}$ \rlb{whenever $\omega$ satisfies the causal Markov condition for $G$ (see Ex.~\ref{ex:CBN}).}} 
\end{example}

\begin{example} \label{Ex_CBN_ND}
	\rl{Recall the CBN from Ex.~\ref{Ex_CBN}. 
	Its representation as a network diagram $D$ is given by the below diagram together with interpreting 
	$c_L$, $c_S$, $c_B$ and $c_A$ as the channels $P(L | SBA)$, $P(S | B)$, $P(B | A)$ and $P(A)$, respectively.}
	\begin{equation} \label{eq:CBN-ndag-ex}
		\input{Fig_Example_CBN_DAG_ND.tikz} 
	\end{equation}
\st{Viewing the above diagram as a single state over $S, L , A$ yields precisely the output state of the model, i.e. the output distribution $P(S,L,A)$.}
\end{example} 

\rl{It is worth pointing out that, given} a state network diagram $D=D_{G,O}$ corresponding to DAG $G$ with vertices $V$ and outputs $O$, we can always extend it to the diagram $D_G := D_{G,V}$ on the same DAG, but in which now every wire appears as an output. In fact $D_{G,O}$ can be obtained as a marginal of the latter: 
\[
\input{DGomarginal.tikz}
\]
and conversely we can obtain $D_G$ by copying each non-output wire an extra time: 
\[
\input{copies.tikz}
\] 

\st{ 
\begin{example} 
For a CBN, the normalised state $\sem{D_G}$ over $V$ in $\catC$ describes the probability distribution $P(V)$ over the full set of variables $V$ of the model, rather than merely the outputs $O$. For the CBN with diagram $D$ in Example \eqref{eq:CBN-ndag-ex} with variables $V=\{A, B, S, L\}$ this is given by: 
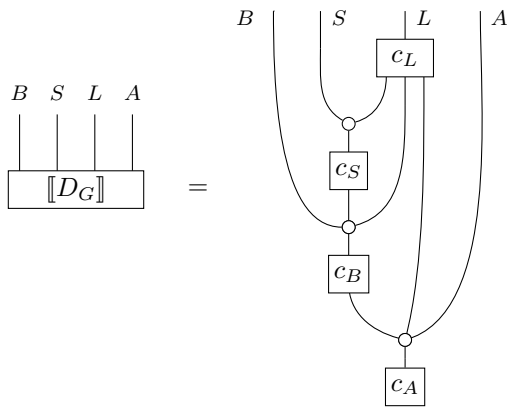
\[
\input{CBN-nd-full.tikz}
\]
\end{example} 
}

\begin{remark} \label{rem:JKZ_SynG_Perspective}
As observed in \cite{JacobsEtAl_2019_CausalInferenceByDiagramSurgery}, an interpretation of a network diagram $D$ can be equivalently defined as a cd-functor 
\[
\Free(D) \to \catC
\]
where $\Free(D)$ is the \emph{free cd-category} generated by the labels and boxes in the diagram. Thus a causal model in $\catC$ over a DAG $G$ is determined by (its outputs $O$ and) a cd-functor $\Free(D_G) \to \catC$. This is the original perspective on causal models in \cite{Fong_2013_CausalTheories}, where Fong calls the category $\Free(D_{G})$ the \emph{causal theory} of $G$. \st{This perspective is compatible with our approach, but here we place less emphasis on the view of a model as a functor.}
 \end{remark}

%*************************************************
\subsection{Functional causal models \label{Sec_FunctionalCausalModel}}
%*************************************************

In the Pearlian causal model framework there is a further central notion of a causal model, in addition to that of a CBN, known as a \emph{Functional Causal Model} (FCM), or perhaps more commonly a \emph{Structural Causal Model} (SCM). 
This kind of model is considered to encode a more `refined' form of causal knowledge and constitutes the basis for more detailed causal explanations and answering different kinds of causal queries, most notably counterfactual ones -- it is the model that is associated with the third and highest level of the \emph{causal hierarchy} \cite{Pearl_Causality, BareinboimEtAl_2021_PearlsHierarchy}.

In an SCM each mechanism is factored in terms of a deterministic functional component, representing the underlying causal process, and an additional noise variable with no parents, which encodes our uncertainty about the underlying state of the world. Conceptually, the notion of SCMs may be seen to underlie that of CBNs in that the former formalise causal reasoning on the basis of causal relations defined as functional dependencies, and then give rise to the latter after marginalising over the `local noise variables'.  
Formally, though, an SCM can be seen as a special case of a CBN over a larger set of variables, and the following presentation shall make use of this fact. 

In Definition \ref{def:deterministic} we saw that functions generalise to deterministic channels in a cd-category. 
This naturally yields the following definition, where in our category-theoretic context we prefer to use the term `functional causal model', saving the term `structural causal model' for their usual instantiation in $\FStoch$.

\begin{definition} \label{def:FCM}
Let $\catC$ be a cd-category. A \emph{Functional Causal Model (FCM)} in $\catC$ is a causal model $\model{M}$ whose variables are partitioned into $V = \{X_i\}^n_{i=1}$, called the \emph{endogenous} variables, and $U = \{U_i\}^n_{i=1}$,  called the \emph{exogenous} variables, with output variables $O \subseteq V$ and such that:
\begin{enumerate}
\item 
each $U_i$ has no parents; 
\item 
each $X_i$ has a mechanism of the form  
\begin{equation} \label{eq:SCM-2}
\input{SCM-2.tikz}
\end{equation}
where $\Pa'(X_i) \subseteq V$ and $f_i$ is a deterministic channel.
\end{enumerate}
\end{definition} 

That is, an FCM is a causal model in which every endogenous variable $X_i$ -- the modelled, or explained ones -- has a deterministic mechanism $f_i$ with one non-output parent $U_i$, and every such $U_i$ variable --  thought of as a `local disturbance' -- has no parents and only one child. Since each $U_i$ has no parents, its mechanism will, by the definition of a causal model, simply be a state  
\[
\begin{tikzpicture}[tikzfig]
    \begin{pgfonlayer}{nodelayer}
        \node [style=none] (0) at (0, 0.25) {};
        \node [style=map] (1) at (0, -0.75) {$\lambda_i$};
        \node [style=label] (2) at (0, 0.75) {$U_i$};
    \end{pgfonlayer}
    \begin{pgfonlayer}{edgelayer}
        \draw (0.center) to (1);
    \end{pgfonlayer}
\end{tikzpicture}
\]

The usual kinds of FCMs, which are considered in the literature, are the following. 

\begin{example}
A \emph{Structural Causal Model (SCM)} is an FCM in $\FStoch$. 
Hence, it is given by finite sets $X_1,\dots,X_n$, $U_1,\dots,U_n$, describing values of the endogenous and exogenous variables, respectively, along with a distribution $\lambda_i$ over each $U_i$ and for each $X_i$ a function 
\[
f_i \colon \Pa'(X_i) \times U_i \to X_i
\] 
 for a subset $\Pa'(X_i) \subseteq \{X_1,\dots, X_n\}$. 
\end{example}

Typically an FCM with variables $V \cup U$ is considered to represent the deterministic mechanisms $f_i$ that underlie a (non-functional) causal model on $V=\{X_1, \dots, X_n\}$, where all randomness comes from our ignorance about the $U_i$, captured in the `noise' distributions $\lambda_i$. 
Given any FCM $\model{M}$ we can define a causal model $\model{M}|_{V}$ over just $V$ by assigning each $X_i$ the mechanism
\begin{equation} \label{eq:deSCM}
\input{de-SCM.tikz}
\end{equation} 
Conversely, every causal model in $\FStoch$, i.e. every CBN, is in fact of this form for some SCM $\model{M}$, since we can always extend any probabilistic mechanism to a functional part along with an additional `noise' distribution, thanks to the following well-known fact. 
\begin{proposition}
For every channel $c$ in $\Stoch$ there exists an object $U$ with a state $\lambda$ and a deterministic channel $f$ such that 
\begin{equation}
	\input{purif.tikz} \label{Eq_FunctionalDilation}	
\end{equation}
\end{proposition} 
Crucially, the latter direction of decomposing a given mechanism $c$ in terms of suitable $f, U, \lambda$ is highly non-unique. 
Hence, there are a great many SCMs with variables $V \cup U$ that give rise to the \emph{same} CBN with variables $V$. \rlb{This fact essentially} is behind the separation between the second and third level of the causal hierarchy. 
The significance of FCMs to formalise counterfactuals will be discussed in Sec.~\ref{Sec_Counterfactuals}.

\begin{example} \label{Ex_CBN_FCM_ND}
An SCM with variables $V=\{B,S,L,A\}$ and $U=\{U_B, U_S,U_L,U_A\}$ and the same causal structure amongst $V$ as in Ex.~\ref{Ex_CBN_ND} is given by the below network diagram. 
The (interpretations of the) pairs $f_X, \lambda_X$ for $X \in V$ may be chosen such that Eq.~\eqref{eq:deSCM} holds with the left-hand side being the corresponding channel $c_X$ from our example CBN in Ex.~\ref{Ex_CBN_ND}, with these channels indicated by the grey boxes.
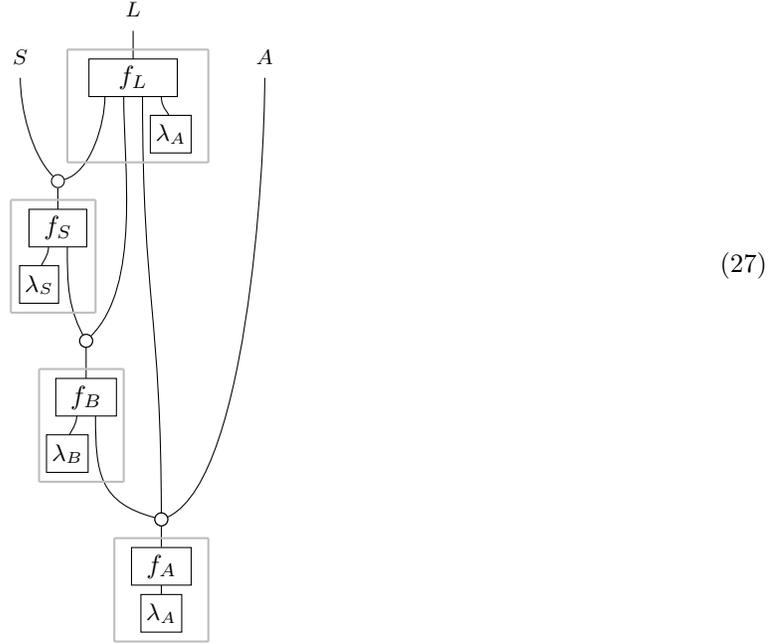
\begin{equation}
	\input{Fig_Example_FCM_DAG_ND_boxed_up.tikz}
\end{equation}
\end{example} 

\rlb{Another significance of viewing causal relations as ultimately defined in terms of dependency relations relative to some, even if unknown, true function (deterministic morphism), is that} it justifies the otherwise ad-hoc stipulation in the definition of a causal model that each mechanism has only a single output. While the details of this argument can be read up elsewhere 
\rlb{-- see, e.g., Refs.~\cite{Pearl_Causality, SEP_ProbabilisticCausation, SEP_ReichenbachsPrincple, AllenEtAl_2016_QCM, Lorenz_2022_MeritsOfTheSpirit_Synthese, otsuka2022process} --} we note that this fact becomes visually manifest in the diagrammatic representation of FCMs.

Given an FCM like Ex.~\ref{Ex_CBN_FCM_ND}, by regarding the pairs $f_X, \lambda_{X}$ for $X \in V$ as mechanisms $c_X$ according to Eq.~\eqref{eq:deSCM} we obtain a causal model in the sense of Def.~\ref{def:causal-model}. However, the difference between asserting the existence of mechanisms as single-output channels $c_X \colon \Pa(X) \rightarrow X$ versus single-output \emph{deterministic} channels  
$f_X \colon \Pa(X) \rightarrow X$ lies in their factorisation properties. Any function with multiple outputs, say $f \colon A \to B \times C$ factorises in terms of component functions  $f_B, f_C$ with outputs $B, C$. More generally any deterministic morphism in a cd-category factorises in this way:
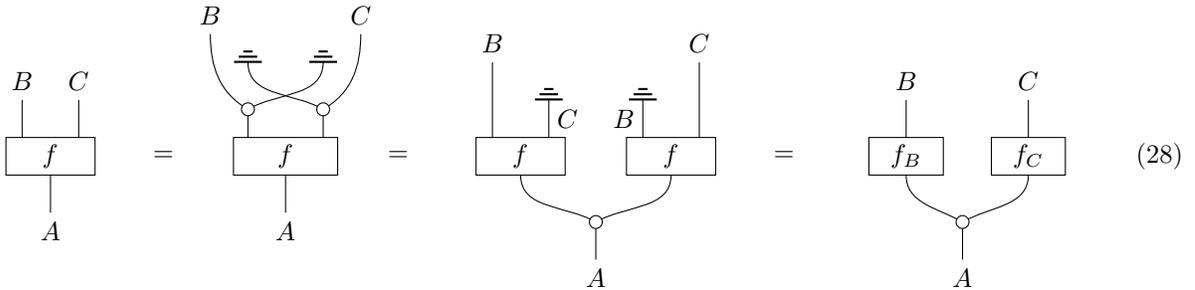
\begin{equation} \label{Eq_FunctionsFactorise}
	\input{Fig_functions_factorise_b.tikz} 
\end{equation}  

The same property holds iteratively for more than two factors in the same way.\footnote{If the input has a product structure corresponding to distinct variables, one may then combine this factorisation property with rewriting the deterministic channels $f_B$ and $f_C$ according to their actual dependency structure so that $f_B$ and $f_C$ have only those wires as input on which $B$ and $C$ actually depend through $f_B$ and $f_C$, respectively.} 
Crucially, generic channels $c \colon A \to B \otimes C$ in a cd-category do \emph{not} factorise in this way, just like generic probabilistic channels in $\FStoch$ do not.

Thus viewing causal models as simply defined via probability channels one may be tempted to allow for channels with multiple outputs, but when basing \rl{causal relations} ultimately on functional dependencies one arrives at the view that mechanisms should have single outputs only.

\color{\rlbcolor}

\subsection{Notions of faithfulness\label{subsec:discussion_faithfulness}}

An important aspect of the causal model framework is the precise link it provides between causal assertions and conditional independence relations. 
A probability distribution has to respect all those conditional independences that are implied by a causal structure for it to be compatible with that causal structure.  
The link thus also is at the basis of algorithmic solutions to \emph{causal discovery} \cite{Pearl_Causality, SpirtesEtAL_2000_BookCausationPredictionSearch, PetersEtAl_2017_ElementsOfCausalInference} -- the epistemological problem of inferring (constraints on) the causal structure given a probability distribution.  

This link is formalised in the \emph{d-separation theorem} \cite{Verma&Pearl_1990_CausalNetworks, GeigerEtAl_1990_IdentifyingIndependence} in terms of the purely graphical notion of \emph{d-separation}. 
Given a DAG $G$, for disjoint subsets \rlb{of vertices $Y$, $Z$ and $W$,  write $(Y \bigCI Z | W)_G$ if $Y$ and $Z$ are d-separated by $W$  (Def.~\ref{Def_DSeparation} in Appendix ~\ref{App_Proof_Thm_DSeparation_Theorem}). The following establishes that d-separation is in general sound for conditional independence in our sense, proven in Appendix~\ref{App_Proof_Thm_DSeparation_Theorem}.\footnote{Recently, Fritz and Klingler presented a generalised notion of d-separation in Markov categories in purely string diagrammatic terms \cite{FritzEtAl_2023_DSeparationInCatgeoricalProbabilty}. However, since this work uses a distinct notion of conditional independence (involving partial channels), we must independently verify soundness of d-separation.}}

\begin{theorem} \label{Thm_DSeparation_Theorem}
	Let $\catC$ be a cd-category with \diagconditioning{} and $\omega$ a normalised state over $V$ which factorises according to a DAG $G$ with variables $V$. Then for any disjoint subsets $Y,Z,W \subset V$ we have
	\[ (Y \bigCI Z | W)_{G} \quad \Rightarrow \quad (Y \bigCI Z | W)_{\omega} \ . \]
\end{theorem}

Conversely, \rlb{we call} a causal model $\modelM$ in $\catC$ \emph{faithful} iff the only conditional independences that its induced state $\omega=\sem{\modelM}$ satisfies are those that are implied by the causal structure, that is: $(Y \bigCI Z | W)_{\omega} \implies (Y \bigCI Z | W)_{G_{\modelM}}$. The significance of notions of faithfulness generally is as Occham's razor type of desiderata in causal discovery  \cite{Pearl_Causality}. 

Now since an actual causal model comes with the specification of causal mechanisms for each variable, we also obtain a second, logically distinct notion of faithfulness, which is most easily stated by first defining the following property of a channel. 
\begin{definition} \textnormal{(No-signalling):} \label{Def_NoSignalling}
	Given a channel $c : S \rightarrow X$ from inputs $S$ to $X$, say there is \emph{no signalling} from $T \subseteq S$ to $X$ iff there exists a channel $d$ such that 
	\begin{equation} 
		\input{nosignal.tikz} \label{eq:nosignal}
	\end{equation}
\end{definition}

The terminology refers to the fact that the above condition ensures that any input $\rho$ to the systems in $T$ does not lead to a change in $X$ as determined by $c$, that is:

\begin{equation} 
	\input{nosignal_ImplicationForStates.tikz} \label{eq:nosignal_ImplicationForStates}
\end{equation}

Given a causal model $\modelM$ with variables $V$ and DAG $G_{\modelM}$, let us call the  mechanism $c_X \colon \Pa(X) \rightarrow X$ of $X \in V$ \emph{faithful} iff there does not exist any non-empty $T \subseteq \Pa(X)$ such that Eq.~\eqref{eq:nosignal} holds. We then call a causal model $\modelM$ \emph{mechanism faithful} if all of its mechanisms are faithful. Thus for each variable $X$ there is the potential to signal to $X$ from each of its parents. The following result helps us to connect \rlb{the discussed notions of faithfulness.} 

\begin{lemma} \label{lem:con-indep-mechs}
Let $\modelM$ be a causal model whose induced state $\omega=\sem{\modelM}$ has full support over $\Pa(X)$, for some variable $X$. Then 
\begin{equation}
	\input{Fig_mechanism_and_conditional.tikz} \label{Eq_mechanism_and_conditional}
\end{equation}
\end{lemma}
\begin{proof} 
Appendix~\ref{app:mechcon}. 
\end{proof} 

It follows from this result that if $\modelM$ is a causal model whose induced state has full support over all variables, 
\rlb{its mechanisms are uniquely determined (by conditioning) and faithfulness of $\modelM$ implies mechanism faithfulness.} 
In general, however, Eq.~\eqref{Eq_mechanism_and_conditional} fails and it is not hard to convince oneself that neither notion of faithfulness implies the other.\footnote{We note that Ref.~\cite{BarrettEtAl_2019_QCMs} introduces a further notion of conditional independence, therein referred to as strong relative independence.  In light of the split-node models from App.~\ref{Sec:Split_node_models} strong relative independence naturally generalises from $\MatR$ in Ref.~\cite{BarrettEtAl_2019_QCMs} to causal models in cd-categories. This notion, logically stronger than $(Y \bigCI Z | W)_{\omega}$, also induces a corresponding notion of faithfulness in terms of d-separation relations. We leave exploring the relations between all three notions of faithfulness in more detail to future work.} 

One way to understand the conceptual significance of non-faithful causal models is in the context of how causal models relate to functional causal models. Suppose that we have a causal model with a mechanism $c_X$ ultimately deriving from a FCM via a functional dependency $f_X$ and extra `noise' state $\lambda_X$ as \rlb{in Eq.~\eqref{eq:deSCM}. Then it is perfectly conceivable} that there actually is causal influence through $f_X$ from $Y$ to $X$ for some $Y \in \Pa(X)$, albeit for some particular (fine-tuned) state $\lambda_{X}$ on $U_X$ this influence is `washed out', yielding an effective mechanisms $c_X$, which is \emph{not} faithful. For this reason, we must allow for causal models which are not mechanism faithful.

\color{black}

% \bibliographystyle{utphys} \bibliography{CauseComp.bib} % Uncomment while working on file standalone, if needed 
%\end{document}
%\documentclass[main.tex]{subfiles} \begin{document}

\section{Interventions} \label{Sec:Interventions}

\subsection{Definition \label{Subsec:Definition_Intervention}}

One of the major advantages of a causal model is the ability to describe what would happen if one were to \emph{intervene} on the variables, by `reaching in' to modify the causal mechanisms and thereby also the causal structure in general. 
The majority of the causal model literature focuses on so called \emph{do-interventions}, also called \emph{atomic} interventions -- so much so that `intervention' is often synonymous with `do intervention'. 
But also more general kinds of interventions like \emph{soft}, or \emph{conditional} interventions have received attention, especially more recently. 
Here, we begin with a very general definition of intervention in our setting, which matches the general notion in the context of CBNs and SCMs in the literature (see, e.g., Ref.~\cite{CorreaEtAl_2020_CalculusForStochasticInterventions}). 
We then recover the more common, particular kinds of interventions as special cases.

\begin{definition} \label{Def_general_Intervention}
Let $\catC$ be a cd-category. An \emph{intervention} $\intv$ on a causal model $\modelM$ in $\catC$ with variables $\var(\modelM)$ is a modification of the mechanisms to yield a new causal model $\modelM'$ in $\catC$ over the same variables $\var(\modelM)$. 

\end{definition}
We often refer to the intervention $\intv$ via the function sending each mechanism $c_i$ in $\modelM$ to the corresponding mechanism $c'_i$ in $\modelM'$. 
\begin{equation} \label{eq:intervention} 
\input{int-map.tikz}
% \intv ::
% \tikzfig{cismall}
% \mapsto \ \ 
% \tikzfig{cidsmall} 
\end{equation} 

While the vertices, variables and output variables remain fixed under an intervention, the edges in the DAG may change and so $G_\modelM \neq G_{\modelM'}$ in general. In Section \ref{sec:transforms} we consider more general kinds of transformations of causal models which may modify \emph{any} aspect of them.

The above definition of intervention is so general that it includes the transition between one causal model to any other causal model so long as the set of variables is unchanged. 
However, in practice interventions tend to concern only one or a few variables. 
We will therefore often refer to a specific intervention by writing $\intv_X$, where $X$ denotes the subset of the variables whose mechanisms do change with respect to the original model, i.e. it is understood implicitly that for all $X_i \in V\setminus X$, the mechanisms are unchanged, $c'_i = c_i$. For the case of $\FStoch$, given a CBN $\modelM$ with corresponding distribution $P(V)$ over its variables $V$, we will denote the resulting distribution of the causal model $\modelM'$ following an intervention $\intv_X$ by $P(V ; \intv_X)$.

\subsection{The zoo of kinds of interventions \label{Subsec:Zoo_of_Interventions}}

We now list some noteworthy classes of interventions and give examples to illustrate their use in diagrammatic causal reasoning -- when we specify a causal model by a network diagram, interventions intuitively amount to doing `string diagram surgery' \cite{JacobsEtAl_2019_CausalInferenceByDiagramSurgery}.

\paragraph{Breaking interventions.}
A \emph{breaking intervention at $X_i$} is an intervention of the form 
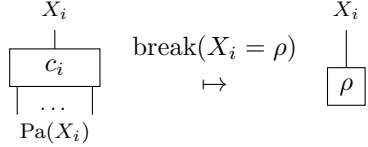
\[
\input{breakint2.tikz}
\]
In the modified DAG $X_i$ has no parents; it is turned into a root node, seeing as the previous mechanism $c_i$ is overwritten and replaced with a state $\rho$. 
In $\FStoch$ this corresponds to assigning $X_i$ a probability distribution and captures a special case of `soft interventions'. 
For instance, one may think of a `fuzzy do-intervention', where for practical limitations one \rlb{can only fix $X_i$} to some value with a certain probability. 

Indeed, a do-intervention arises for the special case of $\rho$ a point distribution, or more generally in any cd-category, 
a \emph{do intervention} $\Do(X_i = x)$ is a breaking intervention with $\rho = x$ a sharp state $x$ of $X_i$.
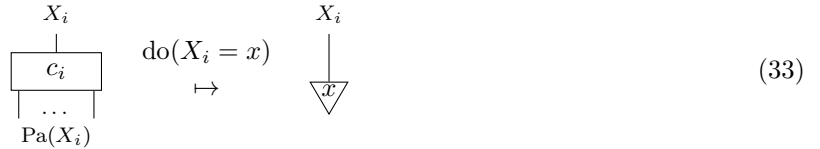
\begin{equation} \label{eq:do}
\input{doint.tikz}
\end{equation} 
Given a CBN with variables $V$ we denote the corresponding post-intervention distribution of the new model by $P(V ; do(X_i = x))$. 
Do-interventions are prominent in questions of causal effect identifiability (see Sec.~\ref{Sec_CE_Identifiability}). 
As another special case of a breaking intervention, a \emph{cut intervention} $\Cut(X_i)$ is one where $\rho = \tinyunit[blackdot]$ is a distinguished normalised \emph{uniform state} on $X_i$ \cite{JacobsEtAl_2019_CausalInferenceByDiagramSurgery}. 
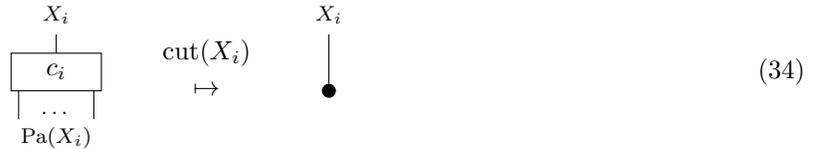
\begin{equation} \label{eq:cut}
% \tikzfig{cutint}
\input{cutintgraydot.tikz}
\end{equation} 
Intuitively, a cut intervention replaces the original mechanism for a variable with a uniformly random distribution over its values. Cut interventions feature prominently in \cite{JacobsEtAl_2019_CausalInferenceByDiagramSurgery}. 
See the subsequent example for how they may be seen to relate to do-interventions.

\begin{examples} \label{Ex_Do_Intervention}
	Returning to the CBN from Ex.~\ref{Ex_CBN_ND}\hspace*{0.05cm}, consider the do-intervention $\Do(S=s)$ that forces $S$ to be in the deterministic state $s$. This might correspond to, for instance, an idealised piece of legislation that forces everyone to smoke, or not smoke, depending on the value of $s$. \st{The effect on the output distribution $P(S,L,A)$ of the model is as follows. }
	\begin{equation}
		\input{Fig_Example_CBN_DAG_ND.tikz}  
		\hspace*{0.6cm} \stackrel{\begin{minipage}{1.5cm} \small $\Do(S = s)$ \vspace*{0.2cm} \end{minipage} }{\mapsto} \hspace*{0.8cm}	
		\input{Fig_Ex_Do_Intervention.tikz} 
		\hspace*{0.3cm} = \hspace*{0.3cm}
		\input{Fig_Ex_Do_Intervention_factored.tikz} 
		\label{Eq_Ex_Do_Intervention} 
	\end{equation}
	where we \rlb{have used the fact that} deterministic states `copy through', see Eq.~\eqref{eq:copy-points}.	
	Let us use this example also to demonstrate the relationship between a $\Do(S = s)$ and a $\Cut(S)$ intervention together with conditioning on $S=s$ in the diagrammatic fashion of Sec.~\ref{sec:catsetup}\hspace*{0.05cm}:
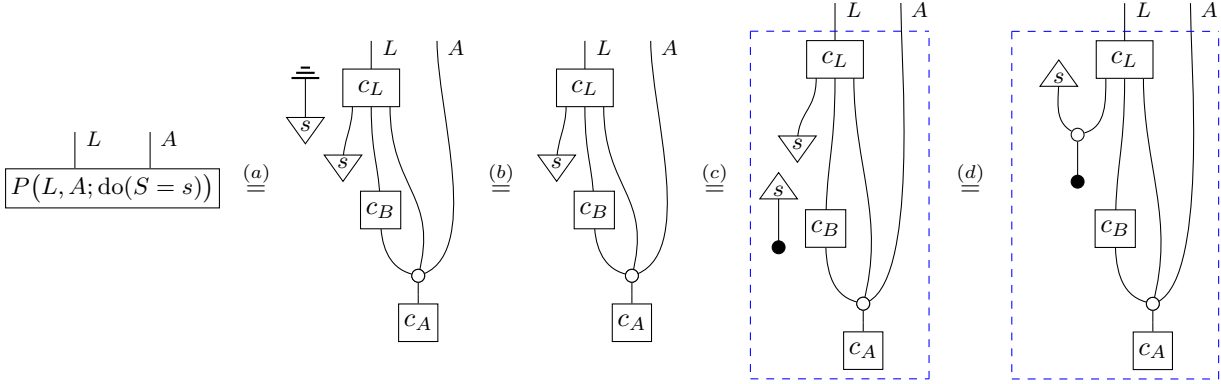
\hspace*{0.05cm}
	\begin{equation}
		\input{Fig_Ex_Do_Intervention_Equalities_1.tikz} \hspace*{0.2cm} 
		\stackrel{(a)}{=}
		\begin{minipage}{3cm}
			\centering
			\input{Fig_Ex_Do_Intervention_Equalities_2.tikz}
		\end{minipage}
		\hspace*{-0.3cm} \stackrel{(b)}{=} \hspace*{0.2cm}
		 \begin{minipage}{2.3cm}
			\centering
			\input{Fig_Ex_Do_Intervention_Equalities_3.tikz}
		\end{minipage}
		\hspace*{-0.2cm} \stackrel{(c)}{=}
		\begin{minipage}{3cm}
			\centering
			\input{Fig_Ex_Do_Intervention_Equalities_4.tikz}
		\end{minipage}
		\hspace*{-0.15cm} \stackrel{(d)}{=}
		\begin{minipage}{3.5cm}
			\centering
			\input{Fig_Ex_Do_Intervention_Equalities_5.tikz}
		\end{minipage} 
%		\label{Eq_Ex_Cut_Intervention} 
\nonumber
	\end{equation}
	The above equalities are representative of the sort of diagrammatic rewriting that will make frequent appearance throughout this work:  
	$(a)$ holds by definition of marginalisation, $(b)$ uses that the state $s$ is normalised, $(c)$ uses Eq.~\eqref{eq:mon-cond} and that the state{ } $\tinyunit[blackdot]$ has full support and $(d)$ uses Eq.~\eqref{eq:sharp-state-eff}. 
	Note that the right-hand side of $(d)$ is nothing but conditioning on $S=s$ in the model that underwent $\Cut(S)$,\footnote{This is the sense in which Jacobs \etal\ used cut interventions to model do-interventions, though their set-up in Ref.~\protect\cite{JacobsEtAl_2021_CausalInferencesAsDiagramSurgery_DiagramsToCounterfactuals} did not allow one to treat the conditioning diagrammatically.}	  
	which one can see as modelling a randomised controlled trial, where the treatment that $S$ receives -- ignoring how ethically problematic that would be in this case -- is randomised, followed by conditioning on $S=s$ in the distribution obtained from such experiment.
	
    Importantly, for a general breaking intervention the (not necessarily deterministic) state $\rho$ does not `copy through' the copy map to yield a product distribution as above -- there will be non-trivial correlation between $S$ and $L,A$:
	\begin{equation}
 \input{break-int-ex.tikz}
 % \tikzfig{Fig_Ex_Do_Intervention_Equalities_1} \quad  
		\qquad = \qquad 
		% \begin{minipage}{3cm}
		% 	\centering
  %  		\end{minipage}
		\input{Fig_Ex_Breaking_Intervention.tikz} 
		\label{Eq_Ex_Breaking_Intervention}
	\end{equation}
\end{examples} 

\st{
Steps $(a)$ and $(b)$ of the above example illustrate how, for any subset $S \subseteq V$ of the variables $V$ of a CBN $\modelM$, the result of a do-intervention diagrammatically manifests the so-called \emph{truncated factorisation formula} for $P(V\setminus S ; \Do(S=s))$. This instructs one to drop all factors corresponding to $S$ from $P(V)$ in Eq.~\eqref{eq:markov} and to fix $S_X=s_X$, wherever $X \in S $ appears as a parent of some variable not in $S$.}

\st{Indeed in general this distribution arises from doing `surgery' on the corresponding network diagram $D_{\modelM}$, by replacing the mechanism $c_X$ with deterministic state $s_X$ for each $X \in S$, then `copying through' $s_X$ and discarding the output wires of $D_{\modelM}$ that correspond to $S$. One is left with the overall state representing $P(V\setminus S ; \Do(S=s))$ which is indeed in keeping with the formula.}

The diagrammatic approach also makes it easy to go beyond breaking interventions, such as the following. 

\paragraph{Local interventions.}
A \emph{local intervention at $X_i$} is an intervention of the form 
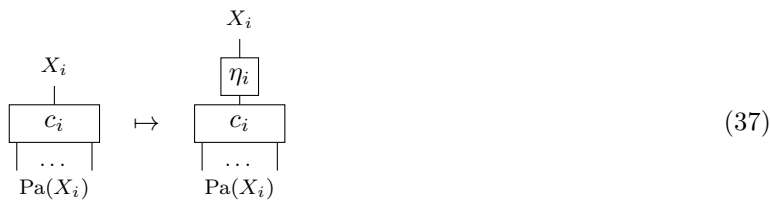
\begin{equation}
\input{cismall.tikz} 
\quad 
\mapsto
\quad 
\input{etai.tikz} 
\end{equation}
for some channel $\eta_i \colon X_i \to X_i$. More generally, a \emph{wide local intervention at $X_i$} is of the form 
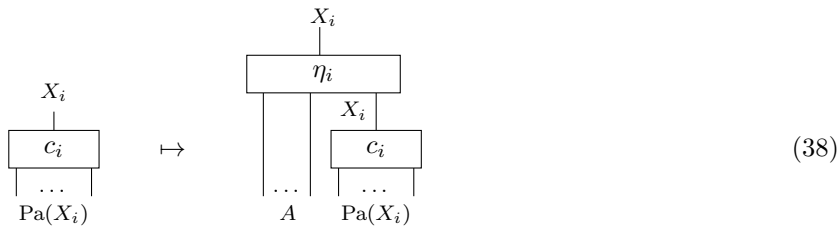
\begin{equation} \label{eq:DeepIntervention}
\input{cismall.tikz} 
\qquad 
\mapsto
\qquad 
\input{non-deep-wide.tikz}
\end{equation} 
for some channel $\eta_i \colon A \otimes X_i \to X_i$ with $A \subset V \setminus Pa(X_i)$, where `wide' indicates the bringing in of new direct causes to $X_i$.  
For either kind of local intervention one can imagine an agent stationed at $X_i$, who implements the intervention locally. The channel $\eta_i$ is an effective description of the `post-processing' done to the system she receives as determined by the mechanism $c_i$.
\rlb{Local interventions are also closely related to the class of interventions considered in the so called `split-node' representation of causal models \cite{BarrettEtAl_2019_QCMs}, see App.~\ref{Sec:Split_node_models}.}

\begin{example} \label{Ex_Wide_Local_Intervention}
	Once more, let us turn to our running example CBN, for which  %from Ex.~\ref{Ex_CBN_ND}. 
	Ex.~\ref{Ex_Do_Intervention} considered a do-intervention $\Do(S=s)$, as well as a generic breaking intervention, which is a non-atomic intervention and already accounts for one aspect of soft interventions. 
	A yet more realistic policy to improve public health may be such that, effectively, all people under 21 years have a, say, 90\% probability to not smoke, while all those above 21 years of age are unaffected by the policy \cite{CorreaEtAl_2020_CalculusForStochasticInterventions}.  
	This can be modelled by a wide local intervention that brings in $A$ as a new direct cause of $S$ and is of the following form:
	
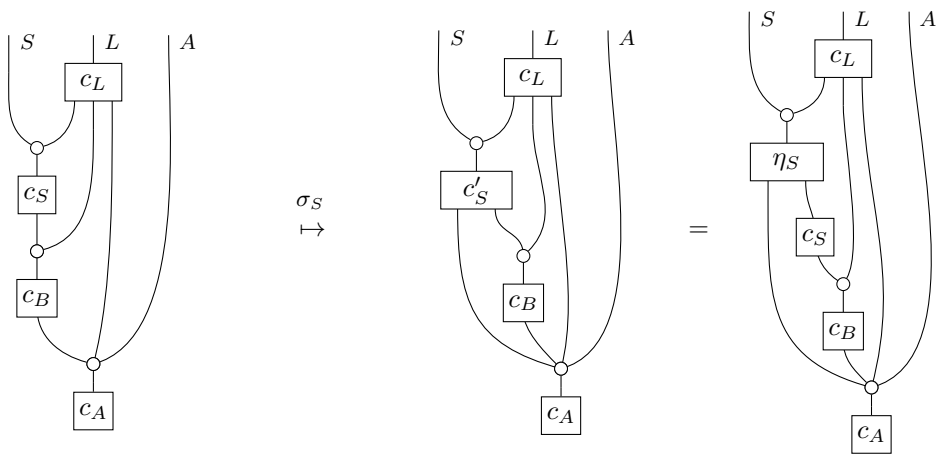
\begin{equation}  
		\begin{minipage}{3cm}
			\centering
			\input{Fig_Example_CBN_DAG_ND.tikz} 
		\end{minipage}
		\hspace*{0.6cm} \stackrel{\begin{minipage}{1cm} \small $ \ \hspace{0.05cm} \ \sigma_S$ \vspace*{0.2cm} \end{minipage} }{\mapsto} \hspace*{0.8cm}	
		\begin{minipage}{3.5cm}
			\centering
			\input{Fig_Ex_Wide_Local_Intervention.tikz} 
		\end{minipage}
		=
		\begin{minipage}{4cm}
			\centering
			\input{Fig_Ex_Wide_Local_Intervention_b.tikz} 
		\end{minipage}
		\nonumber 
	\end{equation}
	where the new mechanism for $S$ can be expressed in terms of the old one as:
	\begin{equation}
		c_S'(B,A) := \begin{cases}	0.9 \ \delta_{S,0} + 0.1 \ \delta_{S,1}  &  \forall A \leq 21 \\  c_S(B) &  \forall A > 21 \end{cases} 
		\label{Eq_WideLocalIntervention_NewChannel}
	\end{equation}
	Eq.~\eqref{Eq_WideLocalIntervention_NewChannel} also allows one to read off the appropriate channel $\eta_S$, which when pre-composed with $c_S$ yields $c'_S$.
\end{example}

At a more formal level, we can consider as interventions the process of simply removing certain, `effectively irrelevant' causal parents from a variable, or adding superfluous ones, 
which we refer to, respectively, as `trimming' and `padding'.

\st{ 
\paragraph{Trimming interventions.}
Consider a mechanism $c_i$ for $X_i$, and suppose that each object $X_j \in \Pa(X_i)$ has at least one normalised state. In Appendix \ref{app:trim} we show that there is then a unique subset $T \subseteq S = \Pa(X_i)$ and morphism $d$ such that $c_i$ factors as $\discard{T} \otimes d$ as in Eq.~\eqref{eq:nosignal} and $d$ is faithful when viewed as a mechanism with inputs $T$. 
We call $d=\trim(c_i)$ the \emph{trimming} of $c_i$ and define the intervention $\trim(X_i)$ by $c_i \mapsto \trim(c_i)$.
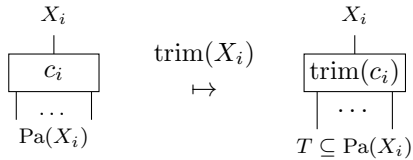
\[
\input{trim.tikz}
\]
At the level of the DAG, this involves removing all edges $Y \to X_i$ for which $Y$ is in fact non-signalling to $X_i$ through the mechanism $c_i$ \rlb{(see Def.~\ref{Def_NoSignalling})}. A model is invariant under all trimming iff \rlb{the model satisfies mechanism faithfulness (see Sec.~\ref{subsec:discussion_faithfulness}).}
}

\paragraph{Padding interventions.}
Conversely, for any subset of variables $S$ and variable $X_i$ we define a \emph{padding intervention} $\Pad(X_i, S)$ as
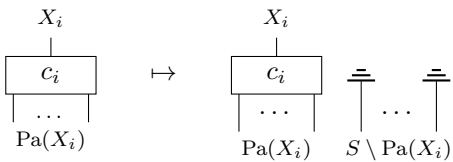
\[
\input{padding.tikz}
\]
This  formally sets the variables in $S$ to be parents of $X_i$, adding any edges $S \to X_i$ that are not already present in the DAG.  By construction if $S \not \subseteq \Pa(X_i)$ the new mechanism for $X_i$ will not be faithful. Padding is inverse to trimming in that whenever $c_i$ is faithful and $S$ and $\Pa(X_i)$ are disjoint we have that $\trim(\Pad(X_i,S))$ leaves $c_i$ invariant.
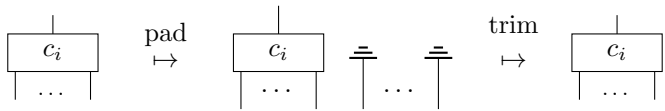
\[
\input{padding2.tikz}
\]

\st{We emphasise that trimming and padding as such are neither of an operational nature, nor is their role to dilute the conceptual significance of non-faithful models. The latter are important to be able to assert that there is a direct-cause relation $X \rightarrow Y$, even when there is no way to make a difference to $Y$ by altering $X$, given that the effective description of $Y$'s mechanism happens to be a particular $c_Y$. \rlb{(See discussion in Sec.~\ref{subsec:discussion_faithfulness}.)}
Trimming and padding are nonetheless convenient pieces of formalism to state some obvious, but useful facts, including the following.}

\begin{remark} \textnormal{(\st{Breaking} interventions as discarding and trimming):} \label{Rem_Do_As_CompositionPlusTrimming}
A breaking intervention  $c_i \to \rho_i$ may be seen as a composite of a standard local intervention,  composing $c_i$ with $\rho_i \circ \discard{}$, followed by a trimming intervention, as follows:
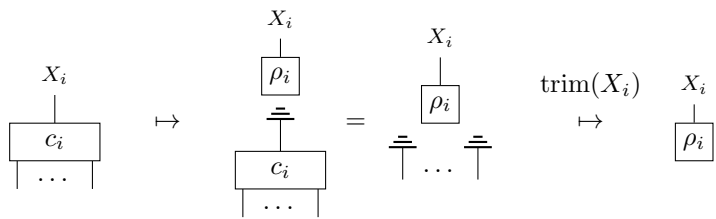
\[
\input{break-as-trim.tikz}
\]
It is in this sense that the example in Fig.~\ref{Fig_smoking_example_doS} in the introduction represented a do-intervention. Also when discussing examples in later sections we will often depict do-interventions in this manner of discarding the original mechanism, followed by simplifying rewrites that include, in particular, the trimming.
\end{remark}

In the framework of causal models, the fact that an intervention can alter one mechanism without affecting any other mechanism is the case simply by definition of the formalism. The assumption that one can \emph{actually do} that in practice is often referred to as the \emph{autonomy of mechanisms}.\footnote{This may also be phrased as the assumption that one can choose the causal relata in an appropriate manner such that their mechanisms are indeed autonomous and the formalism of causal models is rendered applicable -- understanding when and in what precise sense this assumption applies is the subject of ongoing works, see, e.g., Ref.~\cite{janzing2022phenomenological, Cohen_2022_TowardsGroundedTheoryOfCausationForAI} }
The specific classes of interventions we have met so far can be described as `product interventions' in that the resulting mechanism for each variable does not depend on the mechanisms for the other variables.  
\st{For illustration that interventions do not have to be of a `product form' -- without changing the status of mechanisms as autonomous -- consider the following final class of interventions.}

\paragraph{Rewiring interventions.}
Given mechanisms $c_1,\dots,c_n$ for $X_1,\dots,X_n$, a permutation $\phi$ of $\{1,\dots,n\}$ and morphisms $f_i \colon X_{\phi(i)} \to X_i$ we can define a \emph{rewiring intervention} by 
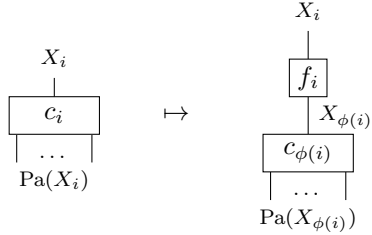
\[
\input{swap-mechanism.tikz}
\]
for $i=1,\dots,n$. At the level of the DAG this amounts to permuting the labels of the vertices according to $\phi$, while leaving the edges `as they were'. 
More precisely, the parental sets in the new model are given by $\Pa'(X_i) = \Pa(X_{\phi(i)})$. At the level of mechanisms the variable $X_i$ is now given by the mechanism for $X_{\phi(i)}$, up to the transformation $f_i$. This intervention is most natural at the level of the network diagram, where it amounts to `swapping wires' and then composing with the $f_i$. 

\begin{example} 
Consider when we have two variables on the same object $X_1=X_2$ and consider a rewiring which swaps these, leaving others invariant, with $f_1 = f_2 = \id{X_1} = \id{X_2}$. This amounts to inserting a swap into the network diagram as below: 
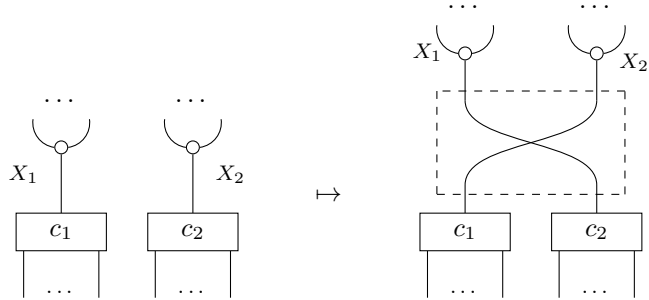
\[
\input{swap-before.tikz} 
\qquad 
\mapsto
\qquad 
\input{swap-mechanisms.tikz}
\]
so that now $X_1$ has the mechanism $c_2$ and $X_2$ the mechanism $c_1$.
\end{example} 

Many more examples of how the listed types of interventions manifest as `diagrammatic surgery' in concrete causal models can be found in Sec.~\ref{Sec_CE_Identifiability}.

\color{black} 

% \bibliographystyle{utphys} \bibliography{CauseComp.bib} % Uncomment while working on file standalone, if needed 
%\end{document}
%\documentclass[main.tex]{subfiles} \begin{document}

\section{Open causal models} \label{sec:openCMs}

\subsection{Definition} \label{sec:definition_openCMs}

When defining interventions, we saw that they often  really only act on a subset of mechanisms, rather than an entire causal model. Formally, such a subset of mechanisms can be viewed as a `causal model with inputs', as well as outputs, in that it may include some variables whose mechanisms are not specified, and which may be interpreted as inputs to the model. 

In this section we expand the treatment from Section \ref{sec:causal-models} to include such `open' causal models. 
Though less widely studied, these are in fact very natural from the categorical perspective, particularly in the correspondence with network diagrams from Sec.~\ref{subsec:NetDiagramsCMs}, when dealing with (breaking) interventions, and in our treatment of counterfactuals in Section \ref{Sec_Counterfactuals}. 
We begin by giving a suitable notion of a DAG with `inputs and outputs'.

\begin{definition}
An \emph{open DAG} $(G,I,O)$ is given by a finite DAG $G$ with vertices $V$, along with a subset $O \subseteq V$ of \emph{outputs} and a subset $I \subseteq V$ of \emph{inputs}, such that every input $i \in I$ has no parents in $G$. 
\end{definition}

Note that we do not require $I$ and $O$ to be disjoint, so that a vertex can be both an input and output. We often refer to such an open DAG simply by $G$ for short. 

We can depict an open DAG $(G,I,O)$ just as for a DAG $G$,  with the output variables $O$ highlighted with circles, and with a special input arrow with no source entering each input vertex $i \in I$. 
\rl{See Rem.~\ref{Rem_ODAGsvsOPenGraphs} for how open DAGs are related to the `open graphs' from Ref.~\cite{OpenGraph1,OpenGraph2}.}

\begin{example} \label{ex:ODAG}
The following depicts an open DAG over $V = \{X_1,\dots,X_5\}$ with inputs $I = \{X_2,X_3\}$ and outputs $O=\{X_3, X_5\}$. 
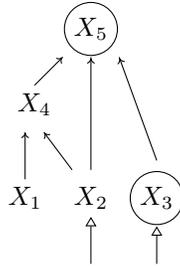
\[
\input{open-DAG-rotate3.tikz}
\]
\end{example}

Open DAGs specify the causal structure of our corresponding notion of `causal model with inputs'.

\begin{definition}
Let $\catC$ be a cd-category. An \emph{open causal model} $\modelM$ in $\catC$ is given by: 
\begin{enumerate}
\item an open DAG $\ODAG{G}=(G,I,O)$ with vertices $V$; 
\item for each vertex $X_i \in V$ an associated object of $\catC$ also denoted $X_i$; 
\item 
for each vertex $X_i \in V \setminus I$ a channel in $\catC$ of the form:
\begin{equation} \label{eq:channels-open} 
 \input{cismall.tikz} 
\end{equation} 
called the \emph{mechanism} for $X_i$. 
\end{enumerate}
The open DAG $\ODAG{G}=(G,I,O)$ is called the \emph{causal structure} of $\modelM$ and also denoted $\ODAG{G}_\modelM$. The objects of $\catC$ indexed by $V$ are called the \emph{variables} $\var(\modelM)$ of $\modelM$. The objects indexed by $I$ and $O$ are called the \emph{input} and \emph{output} variables and denoted $\invar(\modelM), \outvar(\modelM) \subseteq \var(\modelM)$, 
respectively. We call an open causal model $\modelM$ \emph{closed} when it has no inputs, $\invar(\modelM) = \emptyset$, i.e. forms a causal model in our earlier sense.
\end{definition}

In other words, an open causal model $\modelM$ with variables $\var(\modelM)$ is defined just like a (closed) causal model with the same variables $\var(\modelM)$ except that no mechanisms are specified for the input variables $\invar(\modelM)$. 
We call the remaining (non-input) variables the \emph{caused} variables, and the variables which are not output the \emph{internal} variables, denoting these as follows.
\begin{align*}
\causvar(\modelM) &:= \var(\modelM) \setminus \invar(\modelM) \\ 
\intvar(\modelM) &:= \var(\modelM) \setminus \outvar(\modelM)
\end{align*}
We denote the collection of mechanisms \eqref{eq:channels-open} of a model $\modelM$ by $\mechs(\modelM) := (c_i)_{X_i \in \causvar(\modelM)}$.

As before, we can define an open causal model $\modelM$ without specifying the causal structure $G$ explicitly, simply by giving an indexed collection of objects $\var(\modelM)$ with subsets \rl{$\invar(\modelM)$ and $\outvar(\modelM)$, as well as a channel (see Eq.~\eqref{eq:channels-open}) for each variable $X_i$ that is not an input}. To be a valid model the induced directed graph $G$ must be acyclic, where $G$ has a vertex for each variable and \rl{edges $X_i \to X_j$ whenever $X_j \not\in \invar(\modelM)$ and $X_i$ is an input to the mechanism $c_j$.} 
We then define the \rl{corresponding induced open DAG $\ODAG{G}_{\modelM} = (G,\invar(\modelM),\outvar(\modelM))$.}

Our correspondence between causal models and network diagrams now extends straightforwardly to include those with inputs, as follows.   

\begin{proposition}  
\label{prop:ODagsNDs}
Up to equivalence and relabellings of diagrams, the following are equivalent: 
\begin{itemize} 
\item an open DAG $\ODAG{G}=(G,I,O)$ over vertices $V$; 
\item a network diagram 
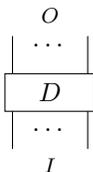
\[
\input{ONDag.tikz}
\]
with wire labels $V$, inputs $I$ and outputs $O$. 
\end{itemize} 
\end{proposition} 

\begin{proof}
Given an open DAG $\ODAG{G} = (G,I,O)$ we define the network diagram $D_\ODAG{G}$ just as in Definition \ref{def:DGO}, except that each input $i \in I$ is not assigned a mechanism, but is simply input to the diagram and copied to each of the mechanisms of its children, and additionally as an output if $i \in O$. 

Conversely, given a network diagram $D$ with inputs $I$ and outputs $O$ we define the open DAG $\ODAG{G}_D = (G,I,O)$ just as for $G_D$ and $O$ in Proposition \ref{prop:DagsNDs}, except that we also include an input vertex $i \in I$ for each input wire to $D$.  As before these assignments are easily seen to be inverse to one another up to relabellings and equivalence of diagrams. 
\end{proof} 

\begin{example}
The open DAG in Example \ref{ex:ODAG} corresponds to the following network diagram. 
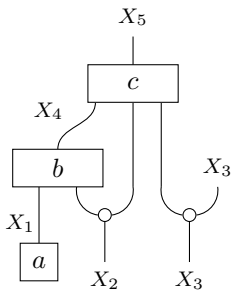
\[
\input{open-DAG-sd3.tikz}
\]
An open causal model of this form amounts to specifying objects \rlb{$X_1,\dots,X_5$} and channels $a,b,c$ as above. 
\end{example}

We immediately obtain a diagrammatic characterisation of open causal models as before, now allowing for network diagrams with inputs. 

\begin{theorem}  \label{Thm:Equivalences_OpenCausalModel_Defs}
Let $\catC$ be a cd-category. Specifying an open causal model $\modelM$ in $\catC$ is equivalent to specifying a network diagram $D$ with an interpretation $\sem{-}$ in $\catC$.
\end{theorem} 

Thus for short we can say that an open causal model $\modelM$ in $\catC$ is a network diagram in $\catC$. For any open model we write $\sem{\modelM} := \sem{D_\modelM}$ for the resulting channel in $\catC$ from its inputs to its outputs. 

\begin{remark}
Given a (closed) causal model $\modelM$ in $\catC$ given by a state network diagram $D$ with interpretation $\sem{-}$, restricting to a sub-network diagram $D'$ (now with inputs) yields an open causal model again via $\sem{-}$. In \cite{Fong_2013_CausalTheories}, Fong calls such an induced morphism $\sem{D'}$ between the causal variables a \emph{causal conditional}. 
\end{remark}

A main intuition and use case associated with open models comes from dealing with the interventions from Sec.~\ref{Sec:Interventions}.  One may think of the input wire $X$ of an open model as corresponding to a `do-intervention $\Do(X=x)$ up to specifying the state $x$', or in fact more generally a `breaking intervention $\breakint{X}{\rho}$ up to specifying $\rho$'.

More precisely, given a (closed) causal model $\modelM$ in $\catC$ with variables $V$ and outputs $O \subseteq V$, we can define an open model $\modelM'$ with variables $V$, input $X$ and outputs $O$ by removing the mechanism for $X$. The common expression $P(Y ; \Do(X))$ for $Y \subseteq O$ is precisely the (marginal on $Y$ of) the overall channel $\sem{\modelM'}$ induced by this open model $\modelM'$, which when fed an input state $x$ yields the distribution $P(Y; \Do(X=x))$ resulting from a do-intervention $\Do(X=x)$ on $\modelM$. 
\begin{equation}
	\input{Fig_DoIntAsOpencausalModel_Y.tikz} 
	\hspace*{1.0cm} \text{where} \hspace*{1.0cm}
	\input{Fig_DoIntAsOpencausalModel_2_Y.tikz} 
	\label{Eq_DoIntAsOpencausalModel}
\end{equation}

We will often use the notation $P(Y ; \Do(X))$ in this sense when discussing do-interventions. For the sake of concreteness consider the following example.

\begin{example} \label{Ex_Do_As_OpenModel}
Returning to Ex.~\ref{Ex_Do_Intervention}, 
we may now write:
% for the channel \rlb{defined by the corresponding} open causal model: 
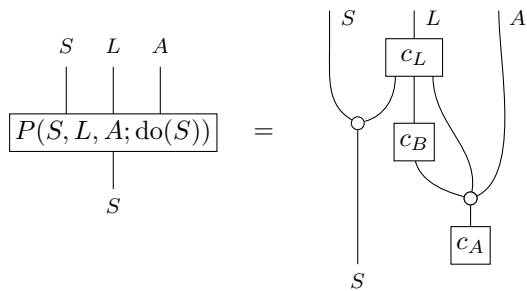
\[
\input{Fig_Example_OpencausalModel.tikz}
\]
That is, $P(S,L,A;\Do(S)) = \sem{\modelM}$ for the open causal model $\modelM$ with network diagram given by the right-hand side. Feeding in the deterministic state $s$ or generic state $\rho$, recovers Eq.~\eqref{Eq_Ex_Do_Intervention} and Eq.~\eqref{Eq_Ex_Breaking_Intervention}, respectively.
\end{example}

\subsection{Composing open models \label{Sec_ComposingOpenModels}}

Allowing inputs as well as outputs \rl{now allows for the composition of} open models, network diagrams and open DAGs, and in fact each of these form categories in their own right. 

Composition is perhaps most readily understood for network diagrams. Given network diagrams $D, D'$ we can compose them in parallel $D \otimes D'$ by simply drawing them side-by-side, and in sequence $D' \circ D$ by plugging them together whenever the outputs $O$ of $D$ match the inputs $I$ of $D'$.
\begin{equation} \label{eq:nd-compose}
\input{compose-NDs-2.tikz}
\end{equation}
\rlb{where $S \amalg T$ denotes the disjoint union of sets $S$ and $T$.} 
Formally, we have the following, \rl{where by} a \emph{relabelling} of indexed collections $(X_i)_{i \in I} \simeq (Y_j)_{j \in J}$ we mean a bijection $f \colon I \to J$ such that $X_i = Y_{f(i)}$ for all $i \in I$.

\begin{proposition} \label{prop:cat-NDs}
Network diagrams form a symmetric monoidal category $\ND$, where the objects are finite sets and morphisms $D \colon I \to O$ are network diagrams equipped with relabellings of their inputs $I_D \simeq I$ and outputs $O_D \simeq O$.
\end{proposition} 
\begin{proof} 
We define composition as in \eqref{eq:nd-compose}, with the identity diagram on $I$ consisting entirely of identity wires, and $X \otimes Y = X \amalg Y$ on objects, with $I = \emptyset$. Explicitly, write $L_D$ for the wire labels of a diagram $D$. Then  $L_{D \otimes D'} = L_D \amalg L_{D'}$ is given by disjoint union, while $L_{D' \circ D} = L_D \amalg L_{D'} / \sim$ is its quotient after identifying $I'$ with $O$. 
\end{proof}

Thanks to the correspondence from Proposition \ref{prop:ODagsNDs} between network diagrams and open DAGs, we immediately obtain an isomorphic category of open DAGs. 

\begin{corollary}
Open DAGs form a symmetric monoidal category $\ODAGcat$, isomorphic to $\ND$, where the objects are finite sets and morphisms $\ODAG{G} \colon I \to O$ are open DAGs with relabellings $I_G \simeq I$, $O_G \simeq O$. 
\end{corollary} 
\begin{proof} 
Again on objects $X \otimes Y = X \amalg Y$, with $I=\emptyset$, and $G \otimes G'$ is given by disjoint union of inputs, outputs and DAGs. Explicitly, for $G \colon I \to O$ and $G' \colon I' = O \to O'$, $G' \circ G$ is again the quotient of the disjoint union after identifying the corresponding elements of $O=I'$. The identity on $X$ is the open DAG with $V=I=O=X$, and no edges. 
\end{proof} 

\begin{example} 
Using our earlier notation, an example composite of open DAGs is the following. 
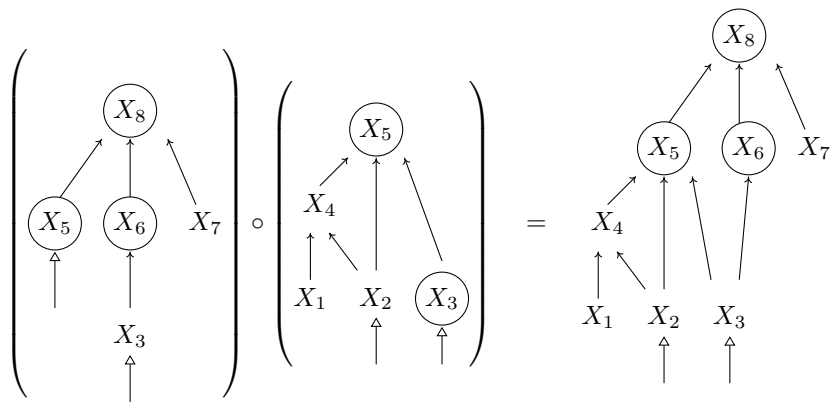
\[
\left( \input{ODAG2.tikz} \right) \circ 
\left( \input{ODAG1.tikz} \right) 
\quad 
=
\quad 
\input{ODAG3.tikz}
\]
\end{example} 

\begin{remark} \label{Rem_ODAGsvsOPenGraphs}
Implicitly in \cite{OpenGraph1}, and explicitly in \cite{OpenGraph2}, an \emph{open graph} $G \colon I \to O$ is defined to be a directed graph $G=(V,E)$ with functions `input' $i \colon I \to V$ and `output' $o \colon O \to V$.  An open DAG is then \rlb{a finite open graph} such that $i$ and $o$ are injective, identified with subsets $I, O \subseteq V$, and every $i \in I$ has no parents in $G$. Open graphs form an SMC $\OGraph$, with composition defined again by identifying corresponding outputs and inputs (formally, via pushout). This makes $\ODAGcat$ a monoidal subcategory of $\OGraph$. 
\end{remark} 

Since an open causal model is just a network diagram along with an interpretation, we can compose open models just like network diagrams, and they form their own category also. 

\begin{corollary}
Open causal models in $\catC$ form a symmetric monoidal category $\OCM(\catC)$, where objects are finite indexed collections $\iset{X} = (X_i)^n_{i=1}$ of objects of $\catC$, and morphisms 
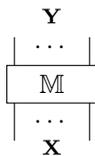
\[
\begin{tikzpicture}[tikzfig]
	\begin{pgfonlayer}{nodelayer}
		\node [style=medium map] (0) at (0, 0) {$\modelM$};
		\node [style=none] (1) at (-1, 0.25) {};
		\node [style=none] (2) at (1, 0.25) {};
		\node [style=none] (3) at (1, 1.25) {};
		\node [style=none] (4) at (-1, 1.25) {};
		\node [style=label] (5) at (0, 1.75) {$\iset{Y}$};
		\node [style=label] (6) at (0, -1.75) {$\iset{X}$};
		\node [style=none] (7) at (-1, -1.5) {};
		\node [style=none] (8) at (1, -1.5) {};
		\node [style=none] (9) at (1, -0.25) {};
		\node [style=none] (10) at (-1, -0.25) {};
		\node [style=none] (11) at (0, 1) {$\dots$};
		\node [style=none] (12) at (0, -1) {$\dots$};
	\end{pgfonlayer}
	\begin{pgfonlayer}{edgelayer}
		\draw (4.center) to (1.center);
		\draw (3.center) to (2.center);
		\draw (10.center) to (7.center);
		\draw (9.center) to (8.center);
	\end{pgfonlayer}
\end{tikzpicture}
\]
are open causal models $\modelM$ in $\catC$, equipped with relabellings  $\invar(\modelM) = \iset{X}$ and $\outvar(\modelM) = \iset{Y}$. Here $\iset{X} \otimes \iset{Y} = \iset{X} \amalg \iset{Y}$ and $\iset{I} = \emptyset$.
\end{corollary}

Explicitly, given models $\modelM \colon \iset{X} \to \iset{Y}$ and $\modelN \colon \iset{Y} \to \iset{Z}$, their sequential composite: 
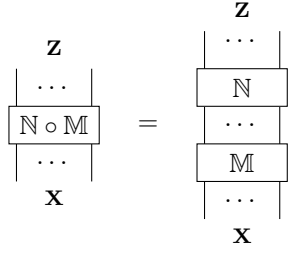
\[
\input{sequential-models.tikz}
\]
is given by composing their corresponding network diagrams, so that $D_{\modelN \circ \modelM} = D_\modelN \circ D_\modelM$, and taking the (disjoint) union $\mechs(\modelM) \amalg \mechs(\modelN)$ of mechanisms from each model. 
The variables $\var(\modelN \circ \modelM)$ form the quotient of the disjoint union of the variables of $\modelM$ and $\modelN$ by the equivalence relation that identifies where they match in $\iset{Y}$. Similarly, given models $\modelM \colon \iset{X} \to \iset{Y}$ and $\modelM' \colon \iset{X'} \to \iset{Y'}$ their parallel composite 
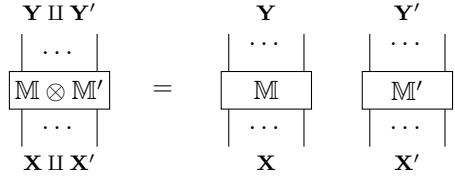
\[
\input{parallel-models.tikz}
\]
is given by drawing diagrams side by side, with $D_{\modelM \otimes \modelM'} = D_\modelM \otimes D_{\modelM'}$, and the disjoint union of variables and mechanisms from each model. 

\begin{remark} 
We can summarise the categories in this section in the following commutative diagram of monoidal functors, where $\simeq$ denotes an isomorphism of categories:
 \[
 \input{comm-diag2.tikz}
 \]
 Here $\sem{-}$ is the functor given by $\iset{X} \mapsto \bigotimes^n_{i=1} X_i$ and $\modelM \mapsto \sem{\modelM} := \sem{D_\modelM}$,  sending an open causal model to a single channel in $\catC$ from its inputs to its outputs.
\end{remark} 

Composing open models provides a way to build new causal models from old ones; here are a few basic examples. 
First of all note that any individual channel $c \colon X_1 \otimes \dots \otimes X_n \to Y$ in $\catC$ forms an open causal model $ c \colon \iset{X} = (X_i)^n_{i=1} \to (Y)$ in $\catC$ which has $c$ as its only mechanism. In particular any normalised state $\omega$ of $Z$ in $\catC$ defines a closed causal model $\omega \colon \emptyset \to (Z)$. Thus we can view the following as morphisms either in $\catC$ or $\OCM(\catC)$.
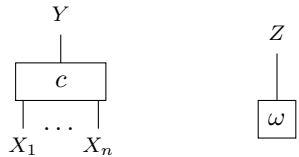
\[
\begin{tikzpicture}[tikzfig]
	\begin{pgfonlayer}{nodelayer}
		\node [style=none] (0) at (0, 1) {};
		\node [style=none] (1) at (-1, -0.5) {};
		\node [style=none] (2) at (1, -0.5) {};
		\node [style=none] (3) at (1, -1.5) {};
		\node [style=none] (4) at (-1, -1.5) {};
		\node [style=none] (5) at (0, -1.5) {$\dots$};
		\node [style=medium map] (6) at (0, -0.25) {$c$};
		\node [style=none] (7) at (0, 0) {};
		\node [style=label] (8) at (0, 1.5) {$Y$};
		\node [style=label] (9) at (-1, -2) {$X_1$};
		\node [style=label] (10) at (1, -2) {$X_n$};
		\node [style=map] (11) at (5.75, -1.25) {$\omega$};
		\node [style=none] (12) at (5.75, 0.5) {};
		\node [style=label] (13) at (5.75, 1) {$Z$};
	\end{pgfonlayer}
	\begin{pgfonlayer}{edgelayer}
		\draw (4.center) to (1.center);
		\draw (3.center) to (2.center);
		\draw (7.center) to (0.center);
		\draw (12.center) to (11);
	\end{pgfonlayer}
\end{tikzpicture}
\]

\begin{example}
\st{As a simple case of parallel composition, consider adding to a causal model $\modelM$ an additional output $Z$  with no parents. The case in which we assign $Z$ no mechanism, making it an input, is captured on the left below, while assigning it a mechanism given by a normalised state $\omega$ is on the right.} 
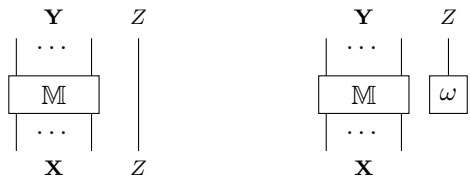
\[
\input{newvar1.tikz} 
\qquad 
\qquad 
\qquad 
\input{newvar2.tikz}
\]
\end{example}
 
\begin{example} \label{Ex_FCM_as_composition}
Let us call an open causal model \emph{deterministic} when it has only deterministic mechanisms. A functional causal model (Def.~\ref{Sec_FunctionalCausalModel}) can now be alternatively defined as a marginal of a (closed) causal model of the form 
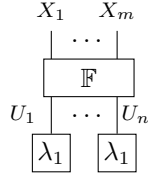
\[
\input{FCM.tikz}
\]
for some normalised states $\lambda_i$ of $U_i$, for $i=1,\dots, n$, where $\model{F}$ is a deterministic model with variables $X_1,\dots,X_n,U_1,\dots,U_n$ such that each $X_i$ has only one of the inputs $U_i$ as a parent and each $U_i$ has one child. Here $m \leq n$ with $X_1,\dots,X_m$ the output variables.  
\end{example}

Beyond sequential and parallel composition there is a final notion that, while not a composition of models in $\OCM(\catC)$, still combines several open models into a new one. We call this the \emph{sharing of variables}. 
This is particularly useful in the construction of a `parallel worlds' model when treating counterfactuals in Sec.~\ref{Sec_CF_GeneralDef}. 
\begin{definition} \label{Def_SharingOfVariables}
Given open causal models with identical inputs $\modelM_i \colon \iset{X} \to \iset{Y}^{(i)}$ for $i=1,\dots,n$ we can build a new open model denoted $\modelM := \bigotimes^{\iset{X}}_i \modelM$, \rlb{which shares the inputs, most naturally defined} via its network diagram, which composes the network diagrams for each model with the input variables $\iset{X}$ copied and passed to each model: 
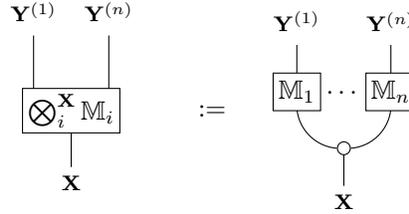
\[
\input{OCM-copyin.tikz}
\]
\end{definition}
Note that the diagram on the right-hand side above is not itself a diagram in $\OCM(\catC)$, seeing as the latter lacks copying.

\subsection{Transformations of causal models} \label{sec:transforms}

We have seen how one can build new causal models through composing them in the sense of the SMC that they form and through sharing variables. 
In this section we describe various further examples of \emph{transformations} of open causal models and open DAGs which allow us to construct new ones. 
Each form of transformation makes sense both as an operation on open models and on open DAGs, and we specify both. 
These transformations include many `operations' one would commonly do to and with causal models in practice and which thereby become part of the formalism in an explicit way.

\subsubsection{Internalising and externalising}

As a first kind, one can transform a model simply by altering which variables are considered as outputs.  
Section~\ref{subsec:NetDiagramsCMs} already mentioned how discarding and copying-out are inverses to each other at the level of network diagrams. Their manifestations at the level of models in $\catC$ correspond to transformations in $\OCM(\catC)$. 

Firstly we note that while $\OCM(\catC)$ lacks copying, it has discarding. For any set of variables $\iset{X}$ the discarding effect:
\[
\begin{tikzpicture}[tikzfig]
	\begin{pgfonlayer}{nodelayer}
		\node [style=none] (0) at (0, 0) {};
		\node [style=upground] (1) at (0, 1) {};
		\node [style=label] (2) at (0, -0.5) {$\iset{X}$};
	\end{pgfonlayer}
	\begin{pgfonlayer}{edgelayer}
		\draw (0.center) to (1);
	\end{pgfonlayer}
\end{tikzpicture}

\]
is the open causal model with inputs $\iset{X}$, no outputs and no mechanisms. 

Now, given a causal model $\modelM$ and subset $\iset{A}$ of its outputs, we can form a new causal model $\intmodel_{\iset{A}}(\modelM)$, which we call  \emph{internalising} $\iset{A}$, as follows.\footnote{We stress again that this diagram is not a network diagram in $\catC$, but a diagram in $\OCM(\catC)$, where the discards are formally distinct from those in $\catC$, but have the same effect as if one discarded the wires corresponding to $\iset{A}$ given the network diagram in $\catC$.}  
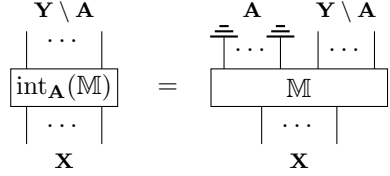
\[
\input{model-discA.tikz}
\]
The model $\intmodel_{\iset{A}}(\modelM)$ has the same variables, inputs and mechanisms as $\modelM$, but with the variables $\iset{A}$ no longer considered as outputs. 
This corresponds to replacing the open DAG $\ODAG{G} = (G,I,O)$ of $\modelM$ by $\ODAG{G}' = (G,I, O \setminus A)$.

Note that $\OCM(\catC)$ has more structure than just that of an SMC. 
In particular, we can `reach inside' morphisms to perform extra operations. Given a causal model $\modelM$ and subset $\iset{A} \subseteq \var(\modelM)$ we define the \emph{externalisation} of $\iset{A}$:
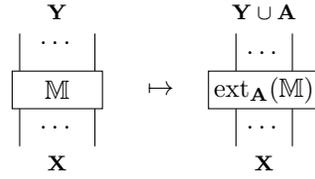
\[
\input{externalising.tikz}
\]
as the model with the same inputs and mechanisms as $\modelM$ and with outputs given by $\outvar(\modelM) \cup \iset{A}$. At the level of open DAGs we replace $\ODAG{G} = (G,I,O)$ by $(G,I,O \cup A)$. 

Whenever $\iset{A}$ is disjoint from $\iset{Y} = \outvar(\modelM)$ it makes sense to depict externalising as follows. 
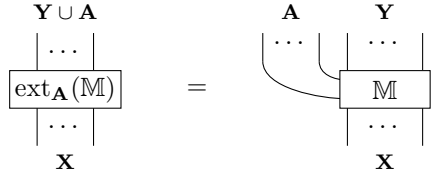
\[
\input{externalising-2.tikz}
\]
Note that externalising and interalising are inverse to each other in the sense that whenever $\iset{A}$ is disjoint from $\iset{Y}$ we have: 
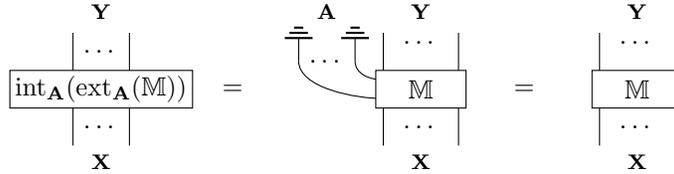
\[
\input{int-ext.tikz}
\]
and whenever $\iset{A} \subseteq \iset{Y}$ we have: 
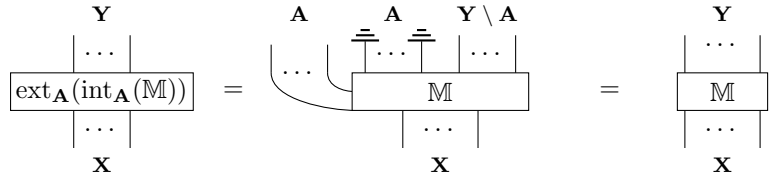
\[
\input{ext-int.tikz}
\] 
By definition both operations are idempotent: $\ext_\iset{A}(\ext_\iset{A}(\modelM)) =  \ext_\iset{A}(\modelM)$ and similarly for $\intmodel_\iset{A}$.

\begin{remark}
$\OCM(\catC)$ has the following `dilation' property. Let us call $\modelM$ \emph{maximal} if whenever 
\begin{equation} \label{eq:dilation}
\input{maximal.tikz}  \quad 
\end{equation}
then in fact $\iset{Z} = \emptyset$ and $\modelN = \model M$. Equivalently, $\modelM$ is maximal iff $\outvar(\modelM) = \var(\modelM)$. Then every $\modelM$ is of the form \eqref{eq:dilation} for a unique maximal $\modelN$, namely $\modelN = \ext_{\var(\modelM)}(\modelM)$, given by making every variable of $\modelM$ an output. 
\end{remark}

\begin{remark}
Note that \eqref{eq:dilation} holds only when $\modelM$ and $\modelN$ have precisely the same variables $\var(\modelM) = \var(\modelN)$, inputs and mechanisms, and only differ in which variables are considered outputs, with $\outvar(\modelM) \subseteq \outvar(\modelN)$. So besides the inputs and outputs of a morphism (model) there are generally many more variables living `inside'. In particular $\OCM(\catC)$ has many scalars: 
\[
\begin{tikzpicture}
	\begin{pgfonlayer}{nodelayer}
		\node [style=map] (0) at (0, 0) {$\modelM$};
	\end{pgfonlayer}
\end{tikzpicture}

\]
since these are in bijection with closed causal models with no outputs. 
\end{remark}

\subsubsection{Opening and interventions}

The transformations considered so far have either combined models or simply adjusted which variables are considered the outputs. 
The following transformation instead alters the mechanisms, more specifically by removing some.  

Given a model $\modelM$ from $\iset{X}$ to $\iset{Y}$ and a subset  $\iset{S} \subseteq \var(\modelM)$ we define the
transformation of \emph{opening}
\[
\input{open.tikz}
\]
where the model $\openmodel_{\iset{S}}(\modelM)$ is obtained by making each $S \in \iset{S}$ an input, thus removing its mechanism from the model if it has one. Thus $\openmodel_\iset{S}(\modelM)$ has mechanism set 
$\{c_X \in \mechs(\modelM) \ | \  X \in \causvar(\modelM) \setminus \iset{S} \}$ with inputs $\invar(\modelM) \cup \iset{S}$, and the same variables and output sets. 

This corresponds to replacing the open DAG $\ODAG{G} = (G,I,O)$ by the open DAG $(G_{\overline{S}}, I \cup S, O)$, where $G_{\overline{S}}$ is the DAG given by removing from $G$ all those edges \rlb{that point to vertices in $S$.} In the resulting open DAG each vertex in $S$ therefore has no parents and is labelled as an input.

\begin{remark}
Opening is closely related to the \emph{fixing} operation defined for so-called CADMGs in \cite{RichardsonEtAl_2022_NestedMarkovForADMGs}, in which a variable $S$ has its mechanism removed, and is then usually set to some pre-defined value $s$. 
For the models considered here fixing is equivalent to performing a do intervention $\Do(S=s)(\modelM)$. 
\end{remark}

The relation between open causal models and do-interventions, and breaking interventions more generally, was already mentioned at the end of Sec.~\ref{sec:definition_openCMs}. We can now describe this relation in a way that has the intervention itself be a transformation of models, using opening. 

Given an open model $\modelM$ we can define a breaking intervention $\breakint{S}{\rho}$ for $S \subseteq \causvar(\modelM)$ just as for closed models. Then it is given by the following open model:
\[
\input{open-break.tikz}
\]
Thus breaking the mechanism for $S$ and replacing it with $\rho$ amounts to opening the model $\modelM$ at $S$ to make an input, and then composing it with $\rho$. 
The special case of $\Do(S=s)$ and $\Cut(S)$ are: 
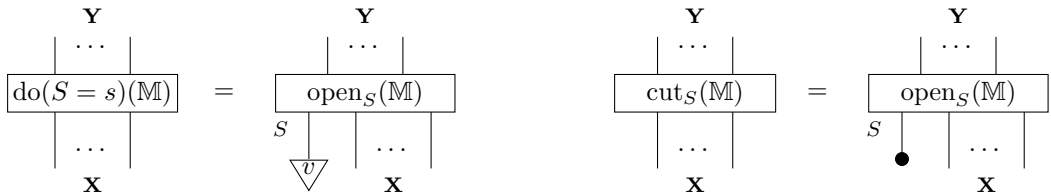
\[
\input{open-do.tikz}
\qquad \qquad \qquad 
\input{open-cut.tikz}
\]
The overall channel of a do-intervention $P(Y; \Do(X))$ up to specifying the state is given as follows (so that in the discussion in Sec.~\ref{sec:definition_openCMs}  we had $\modelM' = \openmodel_X(\modelM)$ in Eq.~\eqref{Eq_DoIntAsOpencausalModel}).
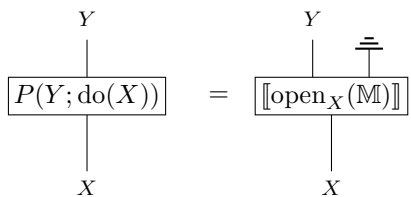
\[
\input{do-from-open_v2.tikz}
\]

In the account of breaking interventions above, we alluded to a general notion of interventions on open models. 
Similarly to closed causal models, we define an \emph{intervention on an open causal model} $\modelM $ to be any reassignment to a new open causal model $\intv(\modelM)$ with the same variables $\var(\intv(\modelM)) = \var(\modelM)$, inputs and outputs:
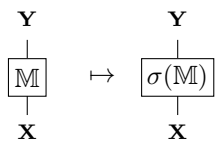
\[
\input{openintervention.tikz}
\]
Thus $\intv(\modelM)$ differs only in its choice of mechanisms and as a result in its underlying open DAG. All of our earlier classes of interventions from Section \ref{Sec:Interventions} can equally be considered as interventions on open causal models in the same way. 
Again, interventions amount to doing `surgery' on the network diagram describing an open causal model.

Beyond those we have listed, there will be many more natural variants of transformations of open causal models. Some can be stated simply at the level of the category $\OCM(\catC)$, such as composition of models and internalising, while others such as externalising and general interventions require `reaching inside' the morphism to the underlying causal model within. The ability to do so \rlb{reflects the fact that} open causal models form much more than simply a category. Determining precisely what further formal structure they do satisfy we leave for future work. 

% \bibliographystyle{utphys} \bibliography{CauseComp.bib} % Uncomment while working on file standalone, if needed 
%\end{document}

%\documentclass[main.tex]{subfiles} \begin{document}

%*****************************************************************
\section{Treatment of causal models with latent variables \label{Sec_TreatmentLatentvariables}}
%*****************************************************************

In practice a common situation is the following. 
Observational data over some set of \emph{observed} variables $O$ is available so that one can estimate the corresponding probability distribution $P(O)$, and one also has some, but \emph{incomplete}, knowledge of the direct cause relations amongst the variables in $O$. This knowledge is incomplete in that either one knows about, or simply cannot exclude the possibility of, the presence of a latent (unobserved) confounder variable $L$, which is a common cause of some variables in $O$. In short, one has $P(O)$, but $O$ cannot be assumed to be a \emph{causally sufficient} set of variables.  

It is therefore useful to be able to represent causal knowledge in a way that also encodes where there may or may not be latent common causes and understand how such partial knowledge constrains the observed distribution $P(O)$. 
There is extensive literature on various kinds of graphical models accounting for latent variables (see, e.g., Refs.~\cite{Pearl_Causality, Evans_2018_MarginsOfDiscreteBN, RichardsonEtAl_2022_NestedMarkovForADMGs}). 

Our definition of a (closed) causal model in Def.~\ref{def:causal-model} involves causal mechanisms for all variables $V$ and a subset \rlb{$O\subseteq V$} of \emph{output} variables. Though the term `output' is a formal and deliberately neutral notion, the subset $O$ will often indeed correspond to the `observed' or accessible variables in the usual sense in causal modelling. In this section we show how the otherwise studied latent variable models and our diagrammatic definition of causal models relate to one another.

%*************************************************
\subsection{Latent projections and ADMGs \label{Sec_ADMGs}}
%*************************************************

A useful graph theoretic notion is the following. 

\begin{definition} \textnormal{(Acyclic directed mixed graph (ADMG)\footnote{In the causal model literature ADMGs are sometimes also referred to as `causal diagrams'.}):} 
An acyclic directed mixed graph (ADMG) $G$ is a tuple $(V,E,B)$ with set of vertices $V$, directed edges $E \subseteq V \times V$ such that $(V,E)$ is a DAG and $B$ a set of unordered pairs of vertices (i.e. $B \subseteq \mathcal{P}(V)$ such that $|S| = 2$ for all $S \in B$). 
Graphically, $G$ is represented as the DAG $(V,E)$ with, additionally, bi-directed edges $a \leftrightarrow b$ for all $\{a,b\} \in B$. 
\end{definition}

\begin{figure}[H]
	\centering
	\input{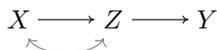}
	\caption{Example of an ADMG. \label{Fig_Example_ADMG}}
\end{figure}

\begin{figure}[H]
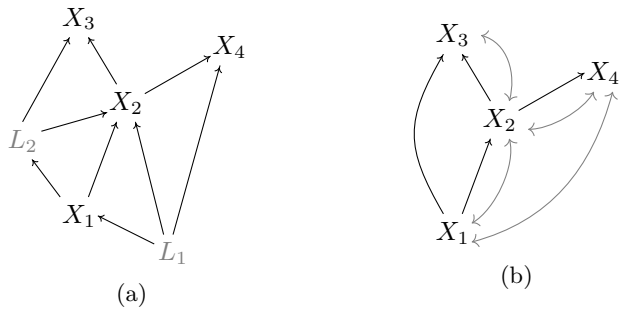

	\centering
	\begin{subfigure}{5cm}
		\centering
		\input{Fig_ExampleDAG_2.tikz}
		\caption{\label{Fig_ExampleDAG_2}}
	\end{subfigure}
	\begin{subfigure}{5cm}
		\centering
		\input{Fig_ExampleDAG_2_LP.tikz}
		\caption{\label{Fig_ExampleDAG_2_LP}}
	\end{subfigure}
	\caption{Example of a DAG in (a) and its latent projection to a subset of the vertices in (b).}
\end{figure}

The idea is that bi-directed edges indicate where there may be latent common causes not included in $V$. 
Suppose the DAG $G$ in Fig.~\ref{Fig_ExampleDAG_2} represents the correct causal structure of a causally sufficient set of variables $V = O \cup L$ with $O=\{X_1,X_2,X_3,X_4\}$ and $L=\{L_1,L_2\}$. 
The DAG $G_O$, the mere restriction of $G$ to $O$, is missing the causal relevance of $L$ to $O$. 
A graph with vertices $O$, but that does `remember enough' from $G$, at least for certain purposes, is given by the ADMG that is the latent projection of $G$ to $O$ (see, e.g., Ref.~\cite{RichardsonEtAl_2022_NestedMarkovForADMGs}).  
\begin{definition} \textnormal{(Latent projection)\footnote{For the sake of simplicity, here defined only for a DAG, but the extension to the latent projection of an ADMG to a subset of its vertices is straightforward \cite{RichardsonEtAl_2022_NestedMarkovForADMGs}.}:} 
Given a DAG $G=(V,E)$ its \emph{latent projection} onto $O \subseteq V$ is the ADMG $\pi_O(G)=(O,E',B)$, where 
\begin{eqnarray}
	E'&:=& \{ (a,b) \in O \times O | \ \exists \ \text{directed path} \ a \rightarrow ... \rightarrow b \ \text{with all intermediate vertices in} \ V \setminus O \} \nonumber \\
	B &:=& \{ \{a,b\} \subseteq O | \ \exists \ \text{path of the form}\footnotemark \ a \leftarrow ... \rightarrow b \ \text{with all intermediate vertices in} \ V \setminus O \}	\ . \nonumber
\end{eqnarray}
\end{definition}
\footnotetext{A path, that is a \emph{not necessarily} directed path, is a sequence of edges no matter their orientation and the condition here is that the two edges at the ends point into $a$ and $b$, respectively.}

Note $E'$ contains at least all directed edges in $E$ between vertices in $O$. 
Fig.~\ref{Fig_ExampleDAG_2_LP} shows the latent projection of the DAG in Fig.~\ref{Fig_ExampleDAG_2} to $O=\{X_1,X_2,X_3,X_4\}$.

There are other kinds of graphs that allow projecting to a subset of a DAG's vertices in a way that captures more details of the DAG. 
In particular the notion of \emph{marginalised DAGs} (mDAGs) generalises ADMGs by essentially allowing hyperedges that connect more than two vertices of a graph. 
It was introduced by Evans \cite{Evans_2016_GraphsForMarginsOfBNs, Evans_2018_MarginsOfDiscreteBN},  
in order to study the full set of constraints on a distribution $P$ -- including inequality constraints -- that are implied by $P$ arising from some DAG based model. 
The present work will however not be concerned with mDAGs. 

The significance of ADMGs for the causal model framework stems from two facts. 
First, an ADMG $A$ with vertices $O$ encodes all the equality constraints on a distribution $P(O)$, in particular all conditional independence relations, that have to be satisfied if $P(O)$ is the marginal of a distribution that is Markov for a DAG $G$ such that $A = \pi_O(G)$.  
The associated compatibility condition of a distribution $P(O)$ relative to an ADMG $A$ with vertices $O$ is called the \emph{nested Markov condition} \cite{RichardsonEtAl_2022_NestedMarkovForADMGs}. 
For its definition we refer the reader to the literature (see Ref.~\cite{RichardsonEtAl_2022_NestedMarkovForADMGs}), since it will not be needed for the present purposes.  
Second, the significance of ADMGs stems from the fact that they turn out to encode precisely the sufficient level of detail to assess questions of causal effect identifiability \cite{RichardsonEtAl_2022_NestedMarkovForADMGs}, which we explore in Section~\ref{Sec_CE_Identifiability}.

%*************************************************
\subsection{ADMGs and network diagrams \label{Sec_ADMGsAndNetworkDiagrams}}
%*************************************************

Let $\modelM$ be a CBN, i.e. a closed causal model in $\FStoch$ according to Def.~\ref{def:causal-model}, with causal structure $\ODAG{G}_\modelM = (G,O)$.  
The model defines a distribution $P(O)$, represented by the corresponding network diagram $D_{\modelM}$ with output $O$ (once interpreted as a state in $\FStoch$). 
By construction this $P(O)$ satisfies the nested Markov condition relative to the ADMG $\pi_O(G)$, the latent projection of $G$ to $O$. 
Suppose one has observational data only over $O$, but has specific knowledge about the latent common causes as encoded in the DAG $G$. 
Then the network diagram $D_{\modelM}$ is indeed the natural object to work with. Though not every channel in the diagram will be uniquely determined by the observational data over $O$, the mechanisms exist by assumption and can be reasoned with elegantly via $D_{\modelM}$. 
\rl{Recall Example~\ref{Ex_CBN_ND}, where the (interpreted) network diagram defines a state (in $\FStoch$) with outputs $S,L$ and $A$, but the mechanism $c_B$ is included explicitly -- while $B$ may not be observed, the network diagram expresses a concrete and plausible assumption about the direct-cause relations that obtain amongst all four variables.} 

Conversely, suppose an ADMG $G$ with vertices $O$ is given and a distribution $P(O)$, which satisfies the nested Markov condition relative to $G$. This is the kind of given data relative to which causal query problems are often stated in the causal model literature. 
How, though, might one then associate a network diagram with this data seeing as state network diagrams are in 1-to-1-correspondence with DAG-based causal structures (see Prop.~\ref{prop:DagsNDs})? 
There are in fact two natural choices for how to associate a DAG with a given ADMG in a principled way\footnote{For the more general mDAGs, mentioned in Sec.~\ref{Sec_ADMGs}, there is one canonical DAG for each mDAG \cite{Evans_2016_GraphsForMarginsOfBNs}.}. Given an ADMG $G=(V,E,B)$, we consider causal structures $(D,V)$ where $D$ is either of the following DAGs:

\begin{enumerate}[label={(\arabic*)},leftmargin=1.2cm]

	\item $\rho(G) := (V \cup \{R_b\}_{b\in B}, E \cup E')$ is the DAG obtained by adding an extra root node $R_b$ for each bi-directed edge $b$, and edges $E'= \{ R_b \rightarrow v \ | \ \text{whenever} \ b \in B  \ \text{and} \  v \in b \}$ from these nodes to their two corresponding endpoints;

\item $\tilde{\rho}(G) := (V \cup \{R_m\}_{m\in M}, E \cup E')$ where 
	$M= \{ m \subseteq V \ | \ m \ \text{maximal such that for all distinct} \ a,b\in m, \ \{a,b\} \in B \ \}$ and 
	$E'= \{ R_m \rightarrow v \ | \ \text{whenever} \ m \in M  \ \text{and} \  v \in m \}$, i.e. obtained by introducing a root node $R_m$ per maximal subset $m \subseteq V$ such that all pairs of vertices in $m$ share a bi-directed edge in $G$,  with $m$ as its children.
\end{enumerate}

\begin{figure}[H]
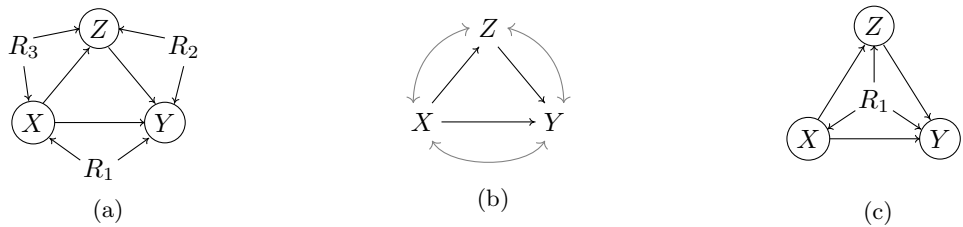

	\centering
	\begin{subfigure}{5cm}
		\centering
		\input{Fig_Example_ADMG_InducedDAG_1_circled.tikz}
		\caption{\label{Fig_Example_ADMG_InducedDAG_1}}
	\end{subfigure}
	\begin{subfigure}{5cm}
		\centering
		\input{Fig_Example_ADMG_For_InducedDAG.tikz}
		\caption{\label{Fig_Example_ADMG_For_InducedDAG}}
	\end{subfigure}
	\begin{subfigure}{5cm}
		\centering
		\input{Fig_Example_ADMG_InducedDAG_2_circled.tikz}
		\caption{\label{Fig_Example_ADMG_InducedDAG_2}}
	\end{subfigure}
	\caption{Example of an ADMG $G$ in (b) and its rootifications $\rho(G)$ in (a) and $\tilde{\rho}(G)$ in (c).}
\end{figure}

\rl{Figs.~\ref{Fig_Example_ADMG_InducedDAG_1} and \ref{Fig_Example_ADMG_InducedDAG_2} show the induced DAGs $\rho(G)$ and $\tilde{\rho}(G)$, respectively,} for the ADMG $G$ in Fig.~\ref{Fig_Example_ADMG_For_InducedDAG}. 

\noindent We refer to $\rho \colon G \mapsto \rho(G)$ and $\tilde{\rho} \colon G \mapsto \tilde{\rho}(G)$ as \emph{rootification maps} -- they turn an ADMG into a DAG by associating root nodes to (sets of) bi-directed edges. 

For any given ADMG $G$, the induced causal structures $\rho(G)$ and $\tilde{\rho}(G)$ induce corresponding network diagrams $D_{\rho(G)}$ and $D_{\tilde{\rho}(G)}$, which when interpreted in $\FStoch$ define corresponding causal models.  
There are many other DAGs that all have $G$ as their latent projection and whose induced network diagram one could consider, too. Importantly however, as long as one is concerned with studying the equality constraints on $P(O)$ that are associated with the given ADMG $G$, or with questions of the identifiability of causal effects involving variables in $O$ (see Sec.~\ref{Sec_CE_Identifiability}), it does not matter which such DAG and corresponding network diagram one chooses, provided the DAG's latent projection onto $O$ is indeed $G$ \cite{RichardsonEtAl_2022_NestedMarkovForADMGs}. 

It is for this reason that many causal inference problems can hence be studied diagrammatically by moving from some ADMG $G$ to a network diagram induced by the rootification map $\rho$ (or $\tilde{\rho}$). The latter is a recipe that uniquely specifies a DAG from any given ADMG. To interpret $D_{\rho(G)}$ in $\FStoch$ then means to regard the additional root nodes introduced by the rootification as some unspecified and unconstrained variables of the same kind as $O$, i.e. discrete finite variables. 
Similarly, the channels appearing in  $D_{\rho(G)}$ are simply regarded as any channels  such that the marginalisation over the additional root nodes gives back the distribution $P(O)$, while their exact identity often is irrelevant. 
For instance, we may associate a causal model of the form as in Eq.~\eqref{Fig_Example_ND_From_ADMG} with the ADMG $G$ in Fig.~\ref{Fig_Example_ADMG_For_InducedDAG} via the rootification $\rho(G)$ from Fig.~\ref{Fig_Example_ADMG_InducedDAG_1}.   

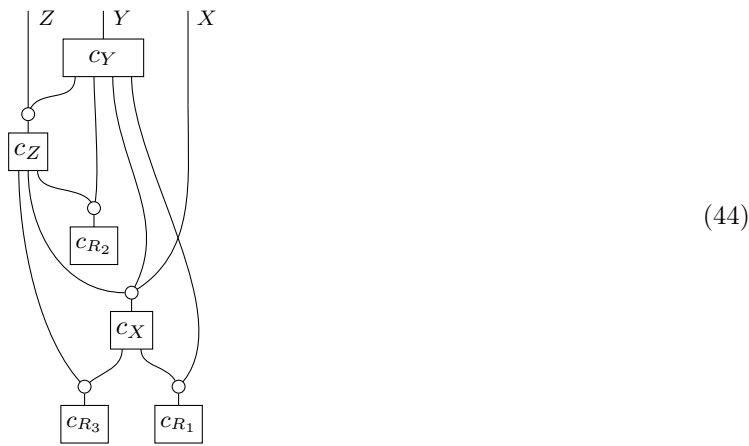
\begin{equation}
	\input{Fig_Example_ND_From_ADMG.tikz}  \label{Fig_Example_ND_From_ADMG}
\end{equation}
 
In summary, given a CBN $\modelM$ with \rlb{DAG $G$,} one can consider the induced \rlb{ADMG $\pi_O(G)$} for some subset $O$ of vertices, but it is the \rl{DAG, or the network diagram $D_{\modelM}$ that} represents \emph{the actual} causal structure. In contrast, given an \rlb{ADMG $G'$} with vertices $O$ and some compatible $P(O)$ given, we may consider the DAG and corresponding causal model induced \rlb{via $\rho(G')$} in order to reason diagrammatically about the given data, but that DAG is \emph{not} asserted to be \emph{the} causal structure. Rather, it is \emph{a plausible} causal structure with the right features to licence reasoning \rlb{with $D_{\rho(G')}$} for certain purposes.  
The context of the given data always makes clear which status is to be assigned to a causal model.

% \bibliographystyle{utphys} \bibliography{CauseComp.bib} % Uncomment while working on file standalone, if needed 
%\end{document}

%\documentclass[main.tex]{subfiles} \begin{document}

%*****************************************************************
\section{Identification of causal effects \label{Sec_CE_Identifiability}}
%*****************************************************************

%*************************************************
\subsection{The problem \label{Sec_CEI_TheProblem}}
%*************************************************

There are many situations in which one would to like to know the effect of an intervention. However, often \rl{one} is not able to obtain this information experimentally, by actually doing the intervention -- it might be technically impossible (like fixing planets' positions or people's socio-economic status) or ethically unacceptable (like forcing people to smoke). 
Pearl referred to \rl{$P(Y;\Do(X))$\footnote{Recall that this is the family of post-intervention distributions for different values $x$ in $\Do(X=x)$, or equivalently, the channel defined by the corresponding open causal model as explained in Sec.~\ref{sec:openCMs}.} as the \emph{causal effect} of $X$ on $Y$. 
So an important question is this:} 
%The general problem of \emph{identifiability of causal effects} asks, 
given what sort of data, and in which circumstances, can one predict what would happen if we were to intervene on some variables, \rl{such as determining a causal effect $P(Y;\Do(X))$}?   

Just given observational data, i.e. some distribution $P(O)$, one cannot say anything about any post-intervention distribution. 
At the other extreme, given a causal model $\modelM$ over variables $V$, which includes a causal mechanism for every variable in $V$,
and given any intervention whatsoever (Def.~\ref{Def_general_Intervention}), atomic or not, then the resultant model describing the post-intervention scenario is also fully known. 
In particular, the marginal distribution over any subset of variables is known.  For an atomic intervention $\Do(X)$ this is known as the \rl{`truncated factorisation formula' \cite{Pearl_Causality} (see the discussion in Sec.~\ref{Sec:Interventions})}. 

A quantity $\phi$ is said to be \emph{identifiable} relative to some data $E$ iff all causal models that agree on $E$ necessarily agree also on $\phi$. 
Understanding when and how causal effects are identifiable if the given data comprises $P(O)$ and \emph{partial} causal knowledge is the problem of \emph{identifiability of causal effects}. More precisely, it is defined as:  
given $P(O)$, an ADMG $G$ with vertices $O$ and disjoint subsets $X,Y,Z \subset O$, when is $P(Y|Z;\Do(X))$ identifiable?\footnote{The set $Z$ may of course be empty, but note that in general identifiability depends on whether the query involves some variables that are conditioned on -- the identifiability of $P(Y|Z;\Do(X))$ is not equivalent to that of $P(Y,Z;\Do(X))$ or $P(Y;\Do(X))$. 
Also, while usually left implicit in the literature, it is of course assumed that $P(O)$ is compatible with the given ADMG $G$, that is it satisfies the nested Markov condition.}

\begin{figure}[h]
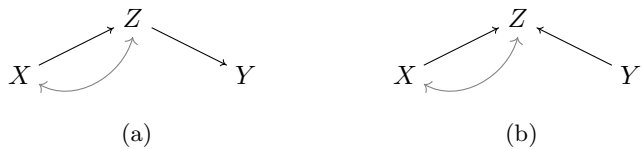

	\centering
	\begin{subfigure}{5cm}
		\centering
		\input{Fig_Example_ADMG_ForCEID_1.tikz}
		\caption{\label{Fig_Example_ADMG_ForCEID_1}}
	\end{subfigure}
	\begin{subfigure}{5cm}
		\centering
		\input{Fig_Example_ADMG_ForCEID_2.tikz}
		\caption{\label{Fig_Example_ADMG_ForCEID_2}}
	\end{subfigure}
	\caption{Examples of causal effect identifiability from~\cite{ShpitserEtAl_2008_CompleteIdentificationMethodCausalHierarchy}. If $P(X,Y,Z)$ is given, then with respect to the ADMG in (a) $P(Y | Z; \Do(X))$ is identifiable, whereas $P(Y; \Do(X))$ is not; while the converse holds with respect to the ADMG in (b). 
	 Note that $P(Y,Z; \Do(X))$ is not identifiable in either case, due to the failure of the c-component criterion in Sec.~\protect\ref{Sec_JKZResult}. See Sec.~\ref{Sec_CEI_ExamplesGeneralCase} for a diagrammatic treatment of these two cases. \label{Fig_Examples_ADMG_ForCEID}}
\end{figure}

The problem of causal effect identifiability has been studied intensely over the past 30 years and since been solved. See, e.g., Shpitser and Pearl \cite{ShpitserEtAl_2008_CompleteIdentificationMethodCausalHierarchy} for an introduction and overview, including a presentation of the algorithmic solution that is sound and complete for this problem. See Fig.~\ref{Fig_Examples_ADMG_ForCEID} for a couple of pedagogical examples.  

More generally, one may ask about the identifiability of $P(Y| Z; \sigma_X)$ for some general, non-atomic intervention $\sigma_X$ if given $P(O)$ and an ADMG $G$ with vertices $O$ (possibly also given some other set of post-intervention distributions). 
This more general version of the problem was defined and studied by Correa and Bareinboim in Ref.~\cite{CorreaEtAl_2020_CalculusForStochasticInterventions}, where they present an algorithmic solution and prove it to be sound. 

Now, if the problem is solved -- algorithmic solutions exist and necessary and sufficient graphical conditions on the tuple ($G$, $X$, $Y$, $Z$) for identifiability are known -- why should one discuss it here at any length?
First, these necessary and sufficient graphical conditions are not particularly straightforward to assess (involving the absence of so called hedges 
\cite{ShpitserEtAl_2008_CompleteIdentificationMethodCausalHierarchy}, or equivalently, the `intrinsicness' of certain districts \cite{RichardsonEtAl_2022_NestedMarkovForADMGs}). 
Second, even once one knows that an instance of the problem is identifiable, the algorithmic solution involves the application of a sequence of do-calculus rules \cite{ShpitserEtAl_2008_CompleteIdentificationMethodCausalHierarchy} (or, more generally, $\sigma$-calculus rules for non-atomic interventions \cite{RichardsonEtAl_2022_NestedMarkovForADMGs}), where knowing which rules will get one to the solution is not always easy.  
So, while automatisation due to the algorithms exists, if a more intuitive and easier to reason with representation of the problem in terms of diagrams existed, then this would be of value for any practitioner who wishes to reason about a specific case with pen and paper, as well as for pedagogical purposes. 
In the following two subsections we argue, echoing the work by Jacobs \etal\ in Ref.~\cite{JacobsEtAl_2019_CausalInferenceByDiagramSurgery}, that this is indeed the case.

%*************************************************
\subsection{The c-component condition diagrammatically -- the JKZ result \label{Sec_JKZResult}}
%*************************************************

A special case of the problem occurs for $|X|=1$ and $Y=V \setminus X$, i.e. the intervention concerns only a single variable and one is asking about the identifiability of $P(V \setminus X ; \Do(X))$, the post-intervention distribution over all other variables.  
Tian and Pearl showed that in this case there is a simple graphical condition, which is necessary and sufficient for identifiability \cite{TianEtAl_2002_GeneralIdentificationConditionForCausalEffects}.  
Given an ADMG $G=(V,E,B)$, for a vertex $X$ its \emph{c-component}  is $C(X) := \{ v \in V \ | \ \exists \ \text{confounded path between} \ X \ \text{and} \ v \}$, 
where a \emph{confounded path} is one that consists solely of bi-directed edges. 
Tian and Pearl's condition, the \emph{c-component condition}, then reads $Ch(X) \cap C(X) = \emptyset$, i.e. $X$ mustn't have a counfounded path to any of its children.
 
Jacobs, Kissinger and Zanasi showed in Ref.~\cite{JacobsEtAl_2019_CausalInferenceByDiagramSurgery} that the c-component condition has a neat equivalent formulation as a structural constraint on a network diagram, and moreover that the solution, i.e. expressing $P(V \setminus X ; \Do(X))$ in terms of the given distribution $P(O)$, can be obtained diagrammatically, too. 
We restate this result here for later reference and in slightly more general form seeing as the present set-up \rlb{allows the treatment of arbitrary} interventions diagrammatically, going beyond the treatment of only the \emph{cut} intervention in~\cite{JacobsEtAl_2019_CausalInferenceByDiagramSurgery}.  

\begin{theorem} \label{Thm_JKZResult} \textnormal{(JKZ theorem \cite{JacobsEtAl_2019_CausalInferenceByDiagramSurgery}\footnote{The proof of this result only relies on the uniqueness of `comb-disintegration', also shown in Ref.~\cite{JacobsEtAl_2019_CausalInferenceByDiagramSurgery} and, hence, the proof of the present statement concerning more general interventions goes through in just the same way as for the version proven in  Ref.~\cite{JacobsEtAl_2019_CausalInferenceByDiagramSurgery}.}):} 
Let $D$ be a network diagram in $\FStoch$ over variables $V$ and with output $O \subseteq V$. 
Let $X \in O$ and suppose there exists a partitioning of the remaining variables into disjoint subsets $O \setminus \{X\} = A \cup B \cup C$ such that $D$ can be seen to be of the below form (in the sense that $f_1,f_2,g$ are network diagrams):\hspace*{0.05cm}\footnote{Equivalently, $f_1,f_2,g$ are morphisms in $\Free(D)$, see Rem.~\ref{rem:JKZ_SynG_Perspective}.}
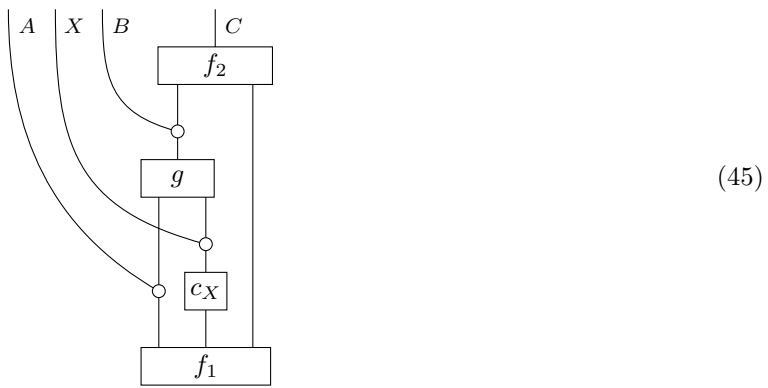
	\begin{equation}
		\input{Fig_JKZResults_condition.tikz}  \label{Eq_JKZResult_cond}
	\end{equation}	
	Then for $\eta_X$ an intervention of the below form, 
	$P(O ; \eta_X)$ is identifiable from any $P(O)$ that can arise from an interpretation of $D$ in \FStoch\ and is given by:
	
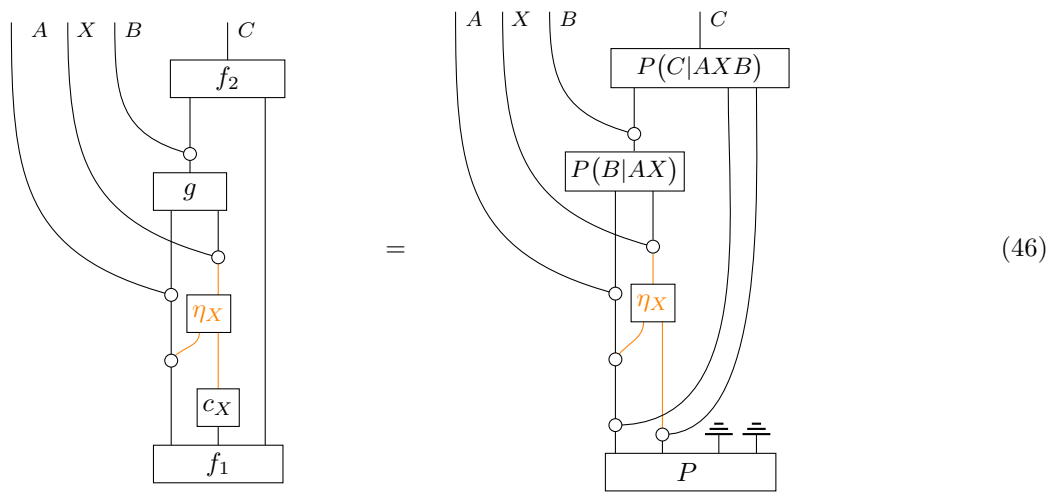
\begin{eqnarray}
		\begin{minipage}{4.0cm} \centering 
		\input{Fig_JKZResults_LHS_withoutComb.tikz} \end{minipage}
		 \hspace*{1.0cm} = \hspace*{0.5cm} 
		\begin{minipage}{4.55cm} \centering 
		\input{Fig_JKZResults_RHS_b_withoutComb.tikz} \end{minipage}   \label{Eq_JKZResult}
	\end{eqnarray}
\end{theorem}

Crucially, the right-hand side of Eq.~\eqref{Eq_JKZResult} contains only marginals and conditionals that are, ignoring issues of full support, computable from $P(O)$. Note that the special case of a $\Do(X=x)$-intervention is obtained for $\eta_X = x \circ \big( \discard{A} \otimes \discard{X} \big)$.\footnote{\rl{And trimming of course, see Rem.~\ref{Rem_Do_As_CompositionPlusTrimming}.}} 

As Jacobs \etal\ argue in Ref.~\cite{JacobsEtAl_2019_CausalInferenceByDiagramSurgery} (put in our terms), given an ADMG $G$ with vertices $O$, if for some $X \in O$ it holds that $Ch(X) \cap C(X) = \emptyset$ then $O \setminus X $ can be partitioned into disjoint subsets $A,B,C$ such that the network diagram $D_{\rho(G)}$ induced by $G$ is of the form~\eqref{Eq_JKZResult_cond}. 
Conversely, starting from a network diagram $D$ that is of this form, it is not hard to see that $\pi_O(G_{D})$, the latent projection of the DAG induced by $D$ to $O$, satisfies the corresponding c-component condition for $X$. 
The intuition behind this equivalence is that any confounding path connecting to $X$ -- that is, a wire not copied out into an output -- may connect to descendants of $X$, but not the children of $X$, which in Eq.~\eqref{Eq_JKZResult_cond} have to be contained in $B$. 
 
\begin{remark} \label{Rem_JKZResult}
Note that, contrary to the claim in Ref.~\cite{JacobsEtAl_2019_CausalInferenceByDiagramSurgery}, Theorem~\ref{Thm_JKZResult} is not more general than Tian and Pearl's original result \rlb{in that the former would cover cases beyond the assumption of `semi-Markovianity' underlying the latter result.  Semi-Markovianity is} an assumption on the nature of the unobserved variables, namely that they be only root nodes.\footnote{At times semi-Markovianity is understood as the condition that the unobserved root nodes furthermore have at most two children. Thanks to Thm.~\protect\ref{Thm_TheFolkloreResult} neither the weaker, nor stronger notion of semi-Markovianity constitute an assumption for causal effect identifiability results, so long as the given ADMG is understood as the latent projection of the correct causal structure of a causally sufficient set of variables.} 
This is a subtle issue to do with the role of latent projections, and the fact that the status of semi-Markovianity as an assumption in causal effect identifiability seems to have not always been clear in the literature. In Ref.~\cite{RichardsonEtAl_2022_NestedMarkovForADMGs} Richardson \etal\ prove a result to the effect of such clarification -- see Thm.~\ref{Thm_TheFolkloreResult} below and the discussion in the subsequent section for the details. 
\end{remark}

%*************************************************
\subsection{The general case diagrammatically -- examples \label{Sec_CEI_ExamplesGeneralCase}}
%*************************************************

Beyond the case when the c-component condition is applicable, let us return \rl{to the general case of the identifiability of causal effects $P(Y |Z ; \Do(X))$} for arbitrary disjoint subsets $X,Y,Z \subset O$, given $P(O)$ and an ADMG $G$ with vertices $O$. 
Also in this general case a necessary and sufficient graphical condition for identifiability is known 
and one may thus wonder about a corresponding generalisation of Thm.~\ref{Thm_JKZResult}. 
However, that graphical condition 
-- the absence of so called \emph{hedges} 
\cite{ShpitserEtAl_2008_CompleteIdentificationMethodCausalHierarchy}\footnote{Equivalently, the graphical condition from \cite{RichardsonEtAl_2022_NestedMarkovForADMGs} in terms of the intrinsicness of certain `districts'.} -- is a fairly involved graph-theoretic notion, which we will not discuss here and instead refer the reader to the literature.  
We note though that, upon close inspection of its definition, it seems clear that there cannot exist one single form into which a network diagram could be cast whenever the condition holds, as was the case for the c-component condition in Thm.~\ref{Thm_JKZResult}.\footnote{For what it is worth, one may wonder about a potentially insightful analogy with the graphical condition of \rlb{\emph{d-separation} (see Sec.~\ref{subsec:discussion_faithfulness} and App.~\ref{App_Proof_Thm_DSeparation_Theorem}).} 
The latter also does not have a single, simple equivalent diagrammatic condition as a particular structural constraint on the network diagram induced by a DAG. Recently, however, Fritz \etal\ showed that there is a neat condition indeed equivalent to d-separation, but it is one in terms of surgery operations on the network diagram, rather than one particular form of the diagram as such \protect\cite{FritzEtAl_2022_DSeparationInCategoricalProbability}. 
Similarly, it is then not inconceivable that by allowing diagrammatic operations like `surgery' on a network diagram, it might be possible to identify a diagrammatic condition that is equivalent to the identifiability of causal effects in the general case.  
We leave exploring this possibility for future work. \label{Footnote_Categorical_DSeparation_ForCEid}} 

Importantly, even with no general result yet that generalises Thm.~\ref{Thm_JKZResult}, a diagrammatic approach to studying the identifiability of causal effects has benefits. 
Network diagrams are a particularly natural and suitable language for reasoning in this context; they combine causal knowledge with rules from probability theory in a way that makes many steps rather straightforward that otherwise can seem involved and hard to spot.   
In order to substantiate this impression, the following will discuss four examples, which all have appeared in conventional, that is non-diagrammatic treatment in the literature before: 
Examples~\ref{Ex_CEID_0_a} and \ref{Ex_CEID_0_b} are the simple examples from Fig.~\ref{Fig_Examples_ADMG_ForCEID}, which also appear as pedagogical examples in Ref.~\cite{ShpitserEtAl_2008_CompleteIdentificationMethodCausalHierarchy}; 
Example~\ref{Ex_CEID_1} is a more involved example from Ref.~\cite{ShpitserEtAl_2008_CompleteIdentificationMethodCausalHierarchy}; 
Example~\ref{Ex_CEID_2} is from Ref.~\cite{CorreaEtAl_2020_CalculusForStochasticInterventions} and involves non-atomic interventions. 
None of these examples are covered by Thm.~\ref{Thm_JKZResult}.

\rl{Discussing these examples diagrammatically will follow the expected approach in light of Sec.~\ref{Sec_TreatmentLatentvariables}. 
Given an ADMG $G$ with vertices $O$ and three disjoined subsets $X,Y,Z \subseteq O$, consider the rootification $\rho(G)$ and the induced network diagram $D_{\rho(G)}$. 
Then let $\modelM$ be a corresponding CBN with $D_{\rho(G)}$ and an interpretation in $\catC=\MatR$ and consider the channel $P(Y |Z ; \Do(X))$ based on the thus defined causal model with output state $P(O)$. 
The goal is to simplify and rewrite $P(Y |Z ; \Do(X))$ so as to express it in terms of $P(O)$, that is, $P(O)$ itself, its marginals and conditionals. 
Now, three things may happen, exposing an important interplay between syntax and semantics:} 
\begin{enumerate}[label=(\arabic*)]
	\item \rl{One succeeds to rewrite $P(Y |Z ; \Do(X))$ in terms of $P(O)$ in a way that only uses the diagrammatic reasoning that holds in any cd-category with \diagconditioning\ and hence did not depend on the concrete category $\catC$, nor on the contingent properties of $P(O)$. 
	One concludes identifiability.}
	\item \rl{One succeeds to rewrite $P(Y |Z ; \Do(X))$ in terms of $P(O)$, but only subject to some assumptions on the properties of a specific $P(O)$ (like whether certain scalars vanish or not, see examples below). Again, though now subject to the contingent properties of $P(O)$, one concludes identifiability.}
	\item \rl{One fails to rewrite $P(Y |Z ; \Do(X))$ in terms of $P(O)$. Then arguing for non-identifiability on the basis of the simplified diagram relies on stronger assumptions than $\catC$ being a cd-category with \diagconditioning\ -- essentially one needs enough maps in the category to be guaranteed to be able to construct suitable counterexamples. 
Below there are two examples claimed to be non-identifiable -- we understand that conclusion in $\catC=\MatR$ and, beyond giving the intuitive reason, we simply refer to the literature for suitable counterexamples.}
\end{enumerate}

 \rl{Now, the fact that we \emph{can} conclude identifiability in case of a successful rewriting, i.e. case $(1)$ and $(2)$, is owed to the following} result by Richardson \etal\ in Ref.~\cite{RichardsonEtAl_2022_NestedMarkovForADMGs} (Cor.~50 therein) -- one that they refer to as a `folklore result' seeing as it had never been formally shown beforehand despite its importance and implicit use in many earlier works: 
\begin{theorem} \textnormal{(\cite{RichardsonEtAl_2022_NestedMarkovForADMGs}):} \label{Thm_TheFolkloreResult} 
Suppose $G$ and $G'$ are two DAGs such that their respective sets of vertices $V_G$ and $V_{G'}$ overlap on some set $O$, i.e. $O \subseteq V_G \cap V_{G'}$ and such that their latent projections to $O$ coincide, $\pi_O(G) = \pi_O(G')$.  
Then for disjoint subsets $X,Y \subset O$ the causal effect $P(Y ; \Do(X))$ is identifiable from $P(O)$ assuming $G$ as the causal structure iff it is identifiable from $P(O)$ assuming $G'$ as the causal structure and, moreover, if indeed identifiable both DAGs imply the same $P(Y ; \Do(X))$. 
\end{theorem} 

This result's significance for the present work of course is that the process of going from an ADMG to a network diagram through rootification (see Sec.~\ref{Sec_ADMGsAndNetworkDiagrams}) to then reason on the basis of that network diagram does not constitute any loss of generality concerning the identifiability of the corresponding causal effect.\footnote{Note that the identifiability of causal effects of the form $P(Y |Z=z ; \Do(X))$ for a \emph{triple} of disjoint subsets $X,Y,Z \subset O$ reduces to an `unconditional' case for a \emph{pair} of new sets of vertices defined in terms of the sets $X,Y,Z$ and, importantly, relative to the same ADMG $G$ \cite{ShpitserEtAl_2008_CompleteIdentificationMethodCausalHierarchy}. 
Hence, while the result in Thm.~\ref{Thm_TheFolkloreResult} only mentions `unconditional' causal effects, it nonetheless allows one to conclude that there is no loss of generality in studying identifiability of $P(Y |Z=z ; \Do(X))$ by rootifying a given ADMG.}

\rl{As the probabilistic notation suggests, we have $\catC=\MatR$ in mind, but one may allow for $P(O)$ and $P(Y |Z ; \Do(X))$ to stand just as well for the state on outputs and the channel of an open causal model, respectively, when defined by a corresponding causal model in some cd-category $\catC$ with \diagconditioning. 
As long as $\catC$ is such that Thm.~\ref{Thm_TheFolkloreResult} extends appropriately so that considering the rootification $\rho(G)$ does not bring in any causal assumption, in addition to the given ADMG $G$, then our approach allows one to conclude identifiability in the cases $(1)$ and $(2)$ for any such $\catC$.}

\begin{example} \label{Ex_CEID_0_a} 
Consider the ADMG $G$ in Fig.~\ref{Fig_Example_ADMG_ForCEID_1_again} (reproduced here from Fig.~\ref{Fig_Example_ADMG_ForCEID_1} for convenience; also see Ref.~\cite{ShpitserEtAl_2008_CompleteIdentificationMethodCausalHierarchy} for a discussion of this example), together with its rootification $\rho(G)$ in Fig.~\ref{Fig_Example_ADMG_ForCEID_1_rootified} and the corresponding network diagram $D_{\rho(G)}$ in Fig.~\ref{Fig_Example_ADMG_ForCEID_1_asND}. 
\begin{figure}[H]
	\centering
	\begin{subfigure}{5cm}
		\centering
		\input{Fig_Example_ADMG_ForCEID_1.tikz}
		\caption{\label{Fig_Example_ADMG_ForCEID_1_again}}
	\end{subfigure}
	\begin{subfigure}{5cm}
		\centering
		\input{Fig_Example_ADMG_ForCEID_1_rootified.tikz}
		\caption{\label{Fig_Example_ADMG_ForCEID_1_rootified}}
	\end{subfigure}
	\begin{subfigure}{4cm}
		\centering
		\input{Fig_Example_ADMG_ForCEID_1_asND.tikz}
		\caption{\label{Fig_Example_ADMG_ForCEID_1_asND}}
	\end{subfigure}
	\vspace*{-0.5cm}
	\caption{\label{Fig_Ex_CEID_0_a_fig1}}
\end{figure}
By reasoning on the basis of $D_{\rho(G)}$ \rl{let us study the claims} from the caption of Fig.~\ref{Fig_Examples_ADMG_ForCEID} that $P(Y \mid Z; \Do(X))$ is identifiable while $P(Y ; \Do(X))$ is not. 
Firstly, note that:
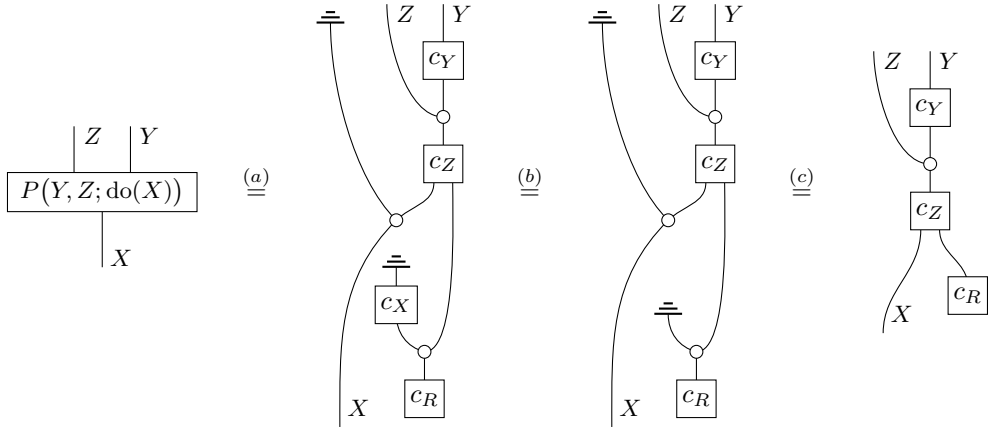
\begin{eqnarray}
	\input{Fig_Example_CEI_0_0.tikz}
	\hspace*{0.5cm} \stackrel{(a)}{=} \hspace*{0.5cm}  \input{Fig_Example_CEI_0_1.tikz} 
	\hspace*{0.5cm} \stackrel{(b)}{=} \hspace*{0.5cm}  \input{Fig_Example_CEI_0_2.tikz}  
	\hspace*{0.5cm} \stackrel{(c)}{=} \hspace*{0.5cm}  \begin{minipage}{2cm} \centering \input{Fig_Example_CEI_0_3.tikz} \end{minipage}	\nonumber
\end{eqnarray}
Step (a) simply applies the definition of $\Do(X)$ as defining a corresponding open model (without feeding in a particular state $x$), see Sec.~\ref{sec:openCMs};
step (b) just \rlb{uses the fact that} $c_X$ is a channel; step (c) uses how copying and discarding interact (see Def.~\ref{def:cd_category}).  Next, by simply marginalising over $Z$ one finds:
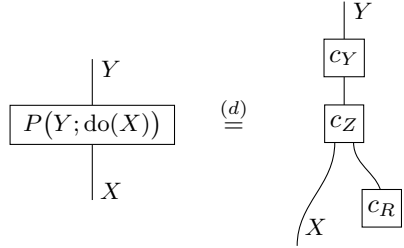
\begin{eqnarray}
	\input{Fig_Example_CEI_2_0.tikz}
	\hspace*{0.5cm} \stackrel{(d)}{=} \hspace*{0.5cm} \begin{minipage}{1.5cm} \centering \input{Fig_Example_CEI_2_1.tikz}  \end{minipage} \nonumber
\end{eqnarray}
and it is obvious that this is not identifiable from $P(X,Y,Z)$ due to the presence of $R$. We omit repeating the standard proof, this being a paradigmatic example of an unobserved confounder $R$ between $X$ and $Z$, but give the intuition of how to see it in the diagrams. In general 
different choices of $c_Z$ and $c_R$ will clearly make a difference to $P\big(Y;\Do(X) \big)$. However, as visible in the network diagram in Fig.~\ref{Fig_Example_ADMG_ForCEID_1_asND}, one can find fine-tuned choices of these two channels such that it makes no difference to $P(X,Y,Z)$ because that difference is compensated by a suitable choice  of the channel $c_X$. 

Note also that the Bayesian conditional given by the extracted (partial) channel $P(Y|X)$ (see Sec.~\ref{sec:conditioning}), is identifiable, but is generally distinct from the sought channel, \rlb{where the below simplification uses the standard steps employed throughout Sec.~\ref{sec:conditioning} and App.~\ref{sec:cond-appendix}:}
\rlb
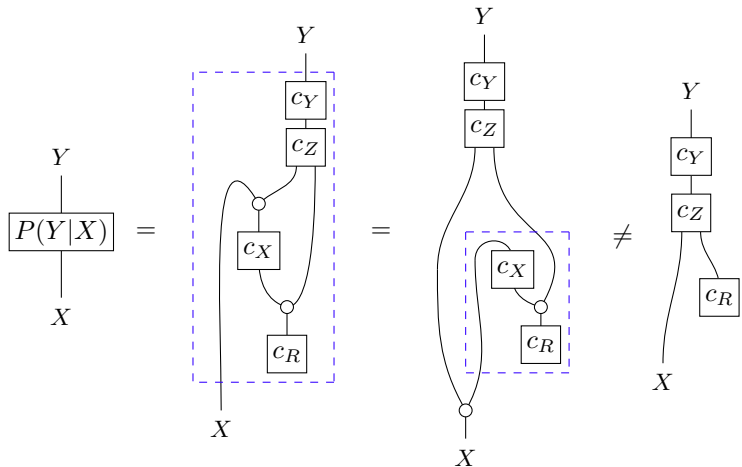
{
\begin{equation}
	\begin{minipage}{8.0cm}
		\centering
		\input{Fig_extractedChannel_simplified_b.tikz} 
	\end{minipage}	
	\neq
	\begin{minipage}{1.6cm}
		\centering
		\input{Fig_soughtChannel.tikz} 
	\end{minipage}	
	\nonumber 
\end{equation}
}

Next, consider the conditional causal effect
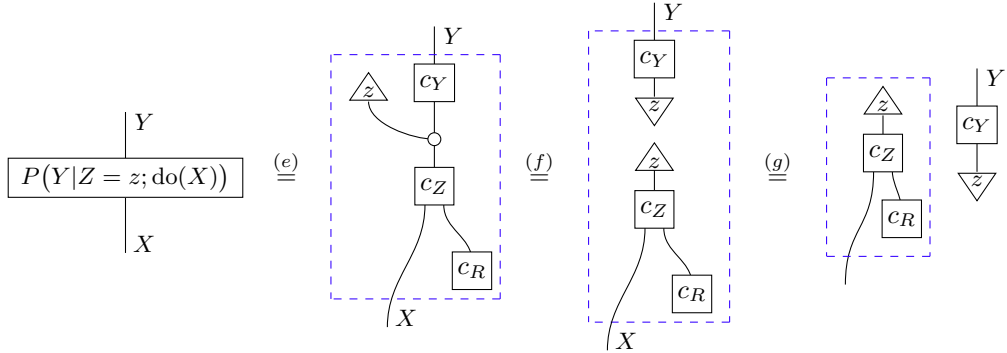
\begin{eqnarray}
	\input{Fig_Example_CEI_1_0.tikz}
	\hspace*{0.3cm} \stackrel{(e)}{=} \hspace*{0.3cm} \begin{minipage}{2cm} \centering \input{Fig_Example_CEI_1_1_a.tikz}  \end{minipage}
	\hspace*{0.5cm} \stackrel{(f)}{=} \hspace*{0.3cm} \begin{minipage}{2cm} \centering\input{Fig_Example_CEI_1_1_b.tikz}  \end{minipage}
	\hspace*{0.3cm} \stackrel{(g)}{=} \hspace*{0.3cm} \begin{minipage}{2.5cm} \centering \input{Fig_Example_CEI_1_2.tikz}  \end{minipage}
%	\hspace*{0.3cm} \stackrel{(g)}{=} \hspace*{0.3cm} \begin{minipage}{1.5cm} \centering \tikzfig{Fig_Example_CEI_1_3}  \end{minipage} 
	\nonumber
\end{eqnarray}
Step (e) uses the diagrammatic definition for conditionals (Sec.~\ref{sec:conditioning}); 
step (f) uses Eq.~\eqref{eq:sharp-state-eff}; 
step (g) uses Eq.~\eqref{eq:mon-cond} and that $c_Y$ and $z$ are already normalised. 
Now, note that in case the effect inside the normalisation box has full support,
% i.e. \rlb{for every sharp state $x$ of $X$ it gives a non-vanishing scalar,} 
 then the normalisation box turns this effect simply into discarding, so that the channel of interest, $P\big(Y|Z=z;\Do(X) \big)$, is equal to
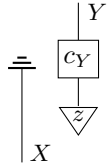
\begin{center}
	\input{Fig_Example_CEI_1_3.tikz}
\end{center}
In that case the conditional causal effect $P\big(Y|Z=z;\Do(X=x) \big)$ is independent of $x$ and 
with the channel $c_Y$ identical to the Bayesian conditional $P(Y|Z)$ (wherever it is defined),  
the causal effect $P\big(Y|Z=z;\Do(X) \big)$ is identifiable, that is whenever $P(Z=z)\neq 0$, and given by $P\big(Y|Z=z\big)$. 
This is also the answer one finds in Ref.~\cite{ShpitserEtAl_2008_CompleteIdentificationMethodCausalHierarchy}. 
Looking at the ADMG $G$ in Fig.~\ref{Fig_Example_ADMG_ForCEID_1_again} this is of course expected since $Y$ depends on $X$ only through $Z$, which here is conditioned on $Z=z$.  

However, when interested in a conditional distribution, there is a crucial difference between two situations. 
One is asserting that a particular state is identifiable, but may, depending on the specific given data, be the zero state, which in the literature would usually be referred to as that the conditional is not defined (see Sec.~\ref{sec:conditioning}).  
Another situation is that it is not identifiable. 
In the above example $P\big(Y|Z=z;\Do(X=x) \big)$ appears to be identifiable, but this is, strictly speaking, not true. 
Whether or not the scalar 
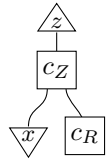
\begin{center}
	\input{Fig_Example_CEI_1_TheScalar.tikz}
\end{center}
is non-vanishing for all $x$ -- and only then does one find independence of $P\big(Y|Z=z;\Do(X=x) \big)$ from $x$ -- 
depends on a quantity which actually is \emph{not} identifiable, namely for the very same reason that $P\big(Y;\Do(X) \big)$ is not identifiable. 
This is a subtlety, which is easily glossed over, as appears to be the case in Ref.~\cite{ShpitserEtAl_2008_CompleteIdentificationMethodCausalHierarchy} -- the diagrammatic treatment helps to prevent this.
\end{example}

\begin{example} \label{Ex_CEID_0_b} 
Consider the ADMG $G$ in Fig.~\ref{Fig_Example_ADMG_ForCEID_2_again} (again reproduced here from Fig.~\ref{Fig_Example_ADMG_ForCEID_2} for convenience and see Ref.~\cite{ShpitserEtAl_2008_CompleteIdentificationMethodCausalHierarchy} for a discussion of this example), together with its rootification $\rho(G)$ in Fig.~\ref{Fig_Example_ADMG_ForCEID_2_rootified} and the corresponding network diagram $D_{\rho(G)}$ in Fig.~\ref{Fig_Example_ADMG_ForCEID_2_asND}.
\begin{figure}[H]
	\centering
	\begin{subfigure}{5cm}
		\centering
		\input{Fig_Example_ADMG_ForCEID_2.tikz}
		\caption{\label{Fig_Example_ADMG_ForCEID_2_again}}
	\end{subfigure}
	\begin{subfigure}{5cm}
		\centering
		\input{Fig_Example_ADMG_ForCEID_2_rootified.tikz}
		\caption{\label{Fig_Example_ADMG_ForCEID_2_rootified}}
	\end{subfigure}
	\begin{subfigure}{4cm}
		\centering
		\input{Fig_Example_ADMG_ForCEID_2_asND.tikz}
		\caption{\label{Fig_Example_ADMG_ForCEID_2_asND}}
	\end{subfigure}
	\caption{\label{Fig_Ex_CEID_0_b_fig1}}
\end{figure}
By reasoning with $D_{\rho(G)}$ it is again straightforward to \rl{study} the claims from the caption of Fig.~\ref{Fig_Examples_ADMG_ForCEID} that $P(Y \mid Z; \Do(X))$ is  not identifiable while $P(Y ; \Do(X))$ is. The below steps are analogous to Example~\ref{Ex_CEID_0_a}. 
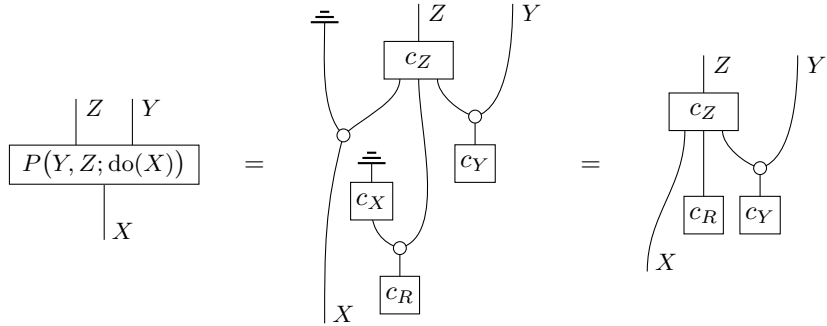
\begin{eqnarray}
	\input{Fig_Example_CEI_0_0.tikz}
	\hspace*{0.5cm} = \hspace*{0.5cm}  \begin{minipage}{3cm} \centering \input{Fig_Example_2_CEI_0_1.tikz} \end{minipage}
	\hspace*{0.5cm} = \hspace*{0.5cm}  \begin{minipage}{2cm} \centering \input{Fig_Example_2_CEI_0_2.tikz} \end{minipage}	\nonumber
\end{eqnarray}
Hence, conditioning on $Z=z$ yields an expression with quantities non-identifiable from $P(X,Y,Z)$:\footnote{We will again not prove this fact; see analogous intuitive argumentation in Ex.~\ref{Ex_CEID_0_a}.} 
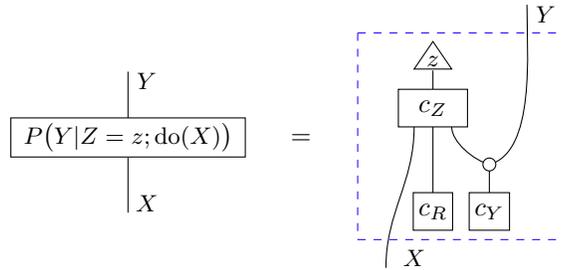
\begin{eqnarray}
	\input{Fig_Example_CEI_1_0.tikz}
	\hspace*{0.5cm} = \hspace*{0.5cm} \begin{minipage}{2cm} \centering \input{Fig_Example_2_CEI_1_1.tikz}  \end{minipage}\nonumber
\end{eqnarray}
In contrast, marginalising over $Z$ separates $Y$ from any dependence on $X$ and $P\big(Y;\Do(X)\big)$ turns out identifiable -- it is equal to the marginal $P(Y)$: 
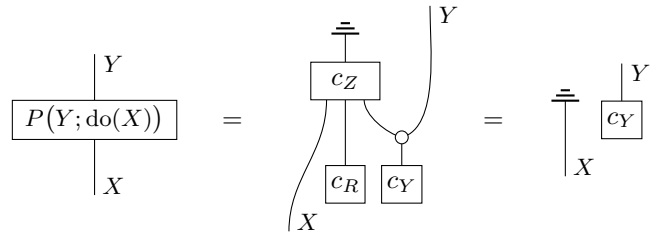
\begin{eqnarray}
	\input{Fig_Example_CEI_2_0.tikz}
	\hspace*{0.5cm} = \hspace*{0.5cm} \begin{minipage}{2.0cm} \centering \input{Fig_Example_2_CEI_2_1.tikz}  \end{minipage} 
	\hspace*{0.5cm} = \hspace*{0.5cm} \begin{minipage}{1.5cm} \centering \input{Fig_Example_2_CEI_2_2.tikz}  \end{minipage} \nonumber
\end{eqnarray}
Note that, in contrast to above Example~\ref{Ex_CEID_0_a}, this equality in the last step holds unconditionally and there is no subtlety to identifiability in this case. 
\end{example}

\begin{example} \label{Ex_CEID_1} 
\rlb{Consider the ADMG in Fig.~\ref{Fig_Ex_Shpit_3a_ADMG} and suppose $P(Y_1, Y_2,W_1,W_2,X)$ is given.} 
Is $P(Y_1, Y_2; \Do(X))$ then identifiable? 
Shpitser and Pearl discuss this example in Ref.~\cite{ShpitserEtAl_2008_CompleteIdentificationMethodCausalHierarchy} and answer the question positively using the \emph{IDENTIFY} algorithm. 
\begin{figure}[H]
	\centering
	\begin{subfigure}{4cm}
		\centering
		\resizebox{0.55\textwidth}{!}{\input{Fig_Ex_Shpit_3a_ADMG.tikz}} 
		\caption{\label{Fig_Ex_Shpit_3a_ADMG}}
	\end{subfigure}
	\begin{subfigure}{4cm}
		\centering
		\resizebox{0.75\textwidth}{!}{\input{Fig_Ex_Shpit_3a_rhoOfG.tikz}} 
		\caption{\label{Fig_Ex_Shpit_3a_rhoOfG}}
	\end{subfigure}
	\begin{subfigure}{7cm}
		\centering
		\resizebox{0.7\textwidth}{!}{\input{Fig_Ex_Shpit_3a_ND.tikz}} 
		\caption{\label{Fig_Ex_Shpit_3a_ND}}
	\end{subfigure}
	\caption{ }
\end{figure}
Consider the network diagram $D_{\rho(G)}$ in Fig.~\ref{Fig_Ex_Shpit_3a_ND}, induced by the rootification $\rho(G)$ in Fig.~\ref{Fig_Ex_Shpit_3a_rhoOfG}. 
Straightforward computation yields: 
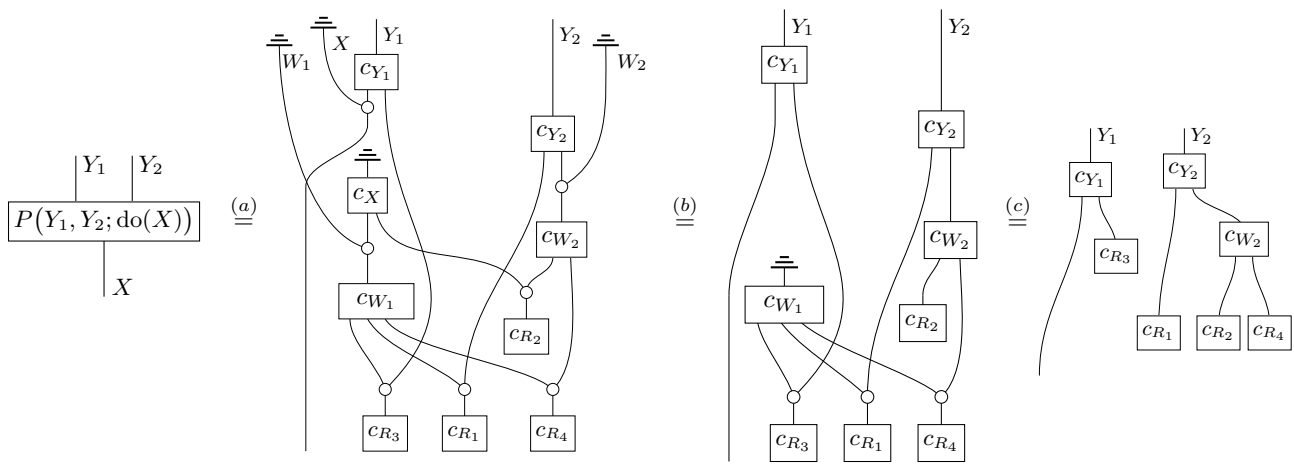
\begin{eqnarray}
	\input{Fig_Ex_Shpit_3a_step0.tikz} \hspace*{0.3cm} \stackrel{(a)}{=} \resizebox{0.32\textwidth}{!}{\input{Fig_Ex_Shpit_3a_step1.tikz}}  
	\stackrel{(b)}{=} \hspace*{0.3cm} \resizebox{0.21\textwidth}{!}{\input{Fig_Ex_Shpit_3a_step4.tikz}} 
	\hspace*{0.1cm} \stackrel{(c)}{=} \resizebox{0.21\textwidth}{!}{\input{Fig_Ex_Shpit_3a_step5.tikz}} \nonumber
\end{eqnarray}
where step $(a)$ just applies the corresponding \emph{opening} operation (and marginalisation) to $D_{\rho(G)}$ and $(b)$ and $(c)$ are the obvious simplifications (see analogous steps in Example~\ref{Ex_CEID_0_a}). 
Hence, the question of the identifiability of $P(Y_1, Y_2; \Do(X))$ boils down to the identifiability of the two factors on the right-hand side of $(c)$. 
The second factor is just the marginal $P(Y_2)$ as can easily be seen from discarding all outputs in $D_{\rho(G)}$ apart from $Y_2$.   
But what of the first factor?  In order to `extract' it from $P$, consider the marginal $P(Y_1,W_1,X)$: 
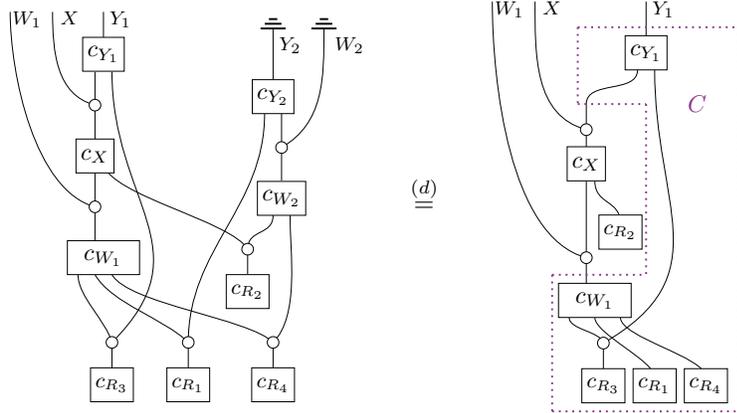
\begin{eqnarray}
	\begin{minipage}{5cm} \centering \resizebox{5cm}{!}{\input{Fig_Ex_Shpit_3a_step6.tikz}} \end{minipage} 
	\hspace*{0.3cm} \stackrel{(d)}{=} \hspace*{0.3cm} 
	\begin{minipage}{4cm} \centering \resizebox{3.5cm}{!}{\input{Fig_Ex_Shpit_3a_step7.tikz}} \end{minipage} \nonumber
\end{eqnarray}
where $(d)$ only did the same sort of simplifications as always involving the discards and sliding the boxes around. 
Now, observe that the violet dotted box is identifiable from $P(Y_1,W_1,X)$ -- provided that distribution has full support -- via comb disintegration \cite{JacobsEtAl_2019_CausalInferenceByDiagramSurgery}, which in turn yields
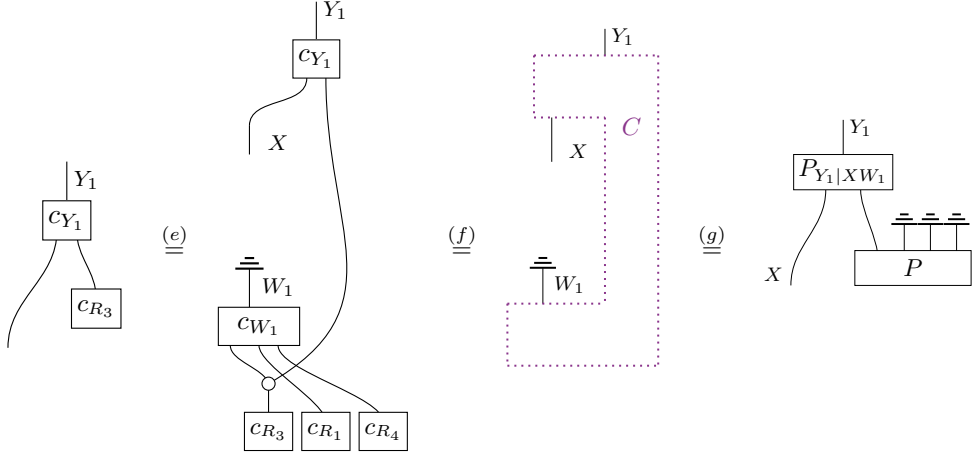
\begin{eqnarray}
	\resizebox{0.1\textwidth}{!}{\input{Fig_Ex_Shpit_3a_step8.tikz}}
	\hspace*{0.3cm} \stackrel{(e)}{=} \hspace*{0.3cm} \resizebox{0.16\textwidth}{!}{\input{Fig_Ex_Shpit_3a_step9.tikz}} 
	\hspace*{0.3cm} \stackrel{(f)}{=} \hspace*{0.3cm} \resizebox{0.13\textwidth}{!}{\input{Fig_Ex_Shpit_3a_step10.tikz}} 
	\hspace*{0.3cm} \stackrel{(g)}{=} \hspace*{0.3cm} \resizebox{0.18\textwidth}{!}{\input{Fig_Ex_Shpit_3a_step11.tikz}} \nonumber
\end{eqnarray}
where $(e)$ took the left-hand side of $(d)$ and opened $X$, as well as marginalised all outputs apart from $Y_1$; step $(f)$ just used $(d)$ and step $(g)$  used said result from Ref.~\cite{JacobsEtAl_2019_CausalInferenceByDiagramSurgery}. 
Thus we can conclude identifiability with the identifying expression being: 
\begin{eqnarray}
	\begin{minipage}{5cm} \centering \resizebox{3.5cm}{!}{\input{Fig_Ex_Shpit_3a_step12.tikz}} \end{minipage}
	\hspace*{0.3cm} = \hspace*{0.5cm} 
	\begin{minipage}{5cm} \centering  \resizebox{6.5cm}{!}{\input{Fig_Ex_Shpit_3a_step13.tikz}} \end{minipage} \nonumber
\end{eqnarray}
which agrees with the answer from Ref.~\cite{ShpitserEtAl_2008_CompleteIdentificationMethodCausalHierarchy}. 
\end{example}

For our final example we suppose that caps in $\catC$ are cancellative \eqref{eq:cancel-caps}, as is the case in $\MatR$. 

\begin{example} \label{Ex_CEID_2} 
As an example involving non-atomic interventions let us consider \rl{the following one, taken} from Ref.~\cite{CorreaEtAl_2020_CalculusForStochasticInterventions}. 
Let $G$ be the ADMG in Fig.~\ref{Fig_Ex_Corr_ADMG} and suppose $P(R,W,X,Y,Z)$, as well as $P(R,W,X,Y,Z; \sigma_Z)$ are given, where the intervention $\sigma_Z$ replaces $c_Z$ with some $c_Z^* : X \rightarrow Z$. 
Now, letting $\sigma_X$ be the intervention that replaces $c_X$ with some mechanism \rlb{$c_X^* : R \rightarrow X$,} is $P(Y | R,Z; \sigma_{X})$  identifiable from the given data?
Note that in this case the given data includes not just observational data, but also interventional data for $\sigma_Z$. 
\begin{figure}[H]
	\centering
	\begin{subfigure}{4cm}
		\centering
		\resizebox{0.5\textwidth}{!}{\input{Fig_Ex_Corr_ADMG.tikz}} 
		\caption{\label{Fig_Ex_Corr_ADMG}}
	\end{subfigure}
	\begin{subfigure}{4cm}
		\centering
		\resizebox{0.68\textwidth}{!}{\input{Fig_Ex_Corr_ADMG_rootified.tikz}} 
		\caption{\label{Fig_Ex_Corr_ADMG_rootified}}
	\end{subfigure}
	\begin{subfigure}{7cm}
		\centering
		\resizebox{0.7\textwidth}{!}{\input{Fig_Ex_Corr_ND.tikz}} 
		\caption{\label{Fig_Ex_Corr_ND}}
	\end{subfigure}
	\caption{ }
\end{figure}
\noindent On the basis of the network diagram $D_{\rho(G)}$ in Fig.~\ref{Fig_Ex_Corr_ND}, induced by the rootification $\rho(G)$ in Fig.~\ref{Fig_Ex_Corr_ADMG_rootified}\hspace*{0.05cm},\footnote{Note that Thm.~\ref{Thm_TheFolkloreResult} does not, at least not in an immediate sense, apply to interventions more general than do-interventions. 
While to our knowledge no analogous `folklore result' to the effect that latent projection ADMGs come without loss of generality for causal effect identifiability also with general interventions has been proven yet, there is no loss of generality for this specific example. 
This is because any DAG with $G$ as its latent projection is a special case of $\rho(G) \ (=\tilde{\rho}(G))$ and it is indeed an identifiable example on the basis of $D_{\rho(G)}$.} the quantity of interest can be expressed as: 
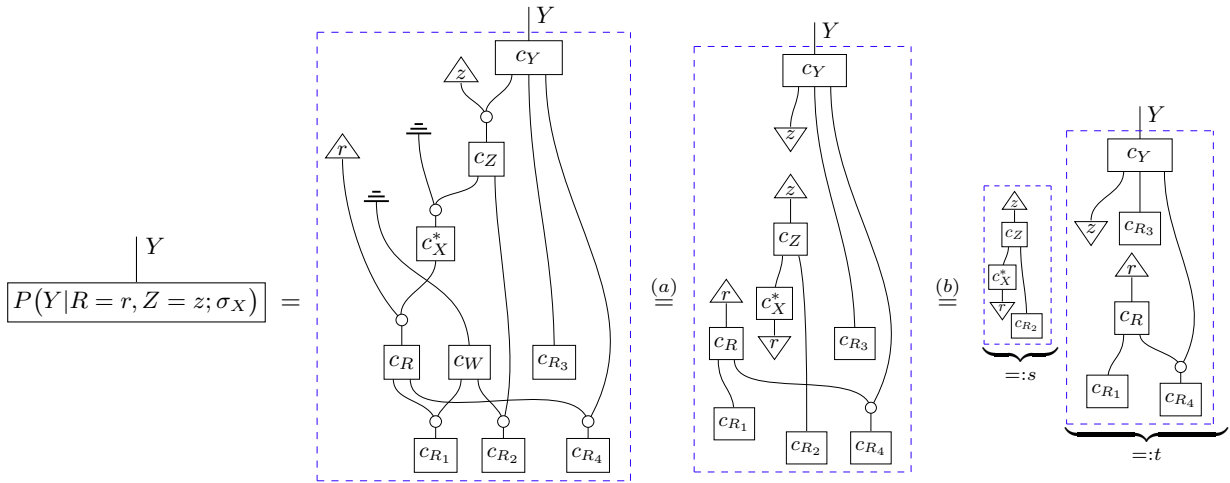
\begin{eqnarray}
	\input{Fig_Ex_Corr_step0.tikz} \hspace*{0.1cm} =  \hspace*{0.1cm}
	\resizebox{0.28\textwidth}{!}{\input{Fig_Ex_Corr_step1.tikz}} 
	\hspace*{-0.3cm} \stackrel{(a)}{=} \hspace*{0.1cm} \resizebox{0.2\textwidth}{!}{\input{Fig_Ex_Corr_step2.tikz}} 
	\hspace*{-0.2cm} \stackrel{(b)}{=}  \hspace*{0.2cm} 
	\begin{minipage}{0.06\textwidth} 
	\[ \underbrace{ \resizebox{1.0\textwidth}{!}{\input{Fig_Ex_Corr_step3_scalar.tikz}}}_{=: s} \]
	 \vspace*{0.5cm} \end{minipage} \hspace*{0.1cm} 
	 \begin{minipage}{0.13\textwidth} 
	\[ \underbrace{ \resizebox{1.0\textwidth}{!}{\input{Fig_Ex_Corr_step3.tikz}}}_{=: t} \]
	 \vspace*{0.5cm} \end{minipage} \nonumber
\end{eqnarray}
where $(a)$ and $(b)$ just do the usual simplifications. 

First of all note that, similarly to Ex.~\ref{Ex_CEID_0_a}, there is a subtlety to do with the identifiability of scalars. 
Whenever the scalar inside the normalisation box of $s$ is non-vanishing, the normalised $s$ is the real number one and can be dropped. 
However, in case the scalar is zero, the normalised $s$ is still zero and it can evidently not be dropped. 
Judging which of the cases obtain requires knowledge of the identity of 
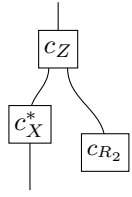
\begin{center}
	\input{Fig_Ex_Corr_step3_MapForScalar.tikz}
\end{center}
While $c_X^*$ is given by assumption, the composite of $c_Z$ and $c_{R_2}$ is in fact \emph{not} identifiable from the given data (see analogous arguments in Ex.~\ref{Ex_CEID_0_a}).  
Hence, strictly speaking, $P(Y | R,Z; \sigma_{X})$ is not identifiable either -- again, a subtlety that is easily spotted in the diagrammatic treatment (and seemingly glossed over in Ref.~\cite{CorreaEtAl_2020_CalculusForStochasticInterventions}). 
Under the assumption that $c_X^*$, $c_Z$ and $c_{R_2}$ -- whatever their precise identities may be -- are such that $s$ is equal to one for all $z$ and $r$, then one may continue assessing the identifiability of the remaining state $t$. 

Now, in order to see whether that state $t$ is identifiable from the given data, consider the given distribution $P(R,W,X,Y,Z; \sigma_Z)$. 
Seeing as the $c_Z^*$ of $\sigma_Z$ does not appear in $t$, the goal is to get rid of its appearance. So, let us actually consider the conditional $P(Y |R=r, X=x, Z=z; \sigma_Z)$ to `separate off' the channel $c_Z^*$ as follows:
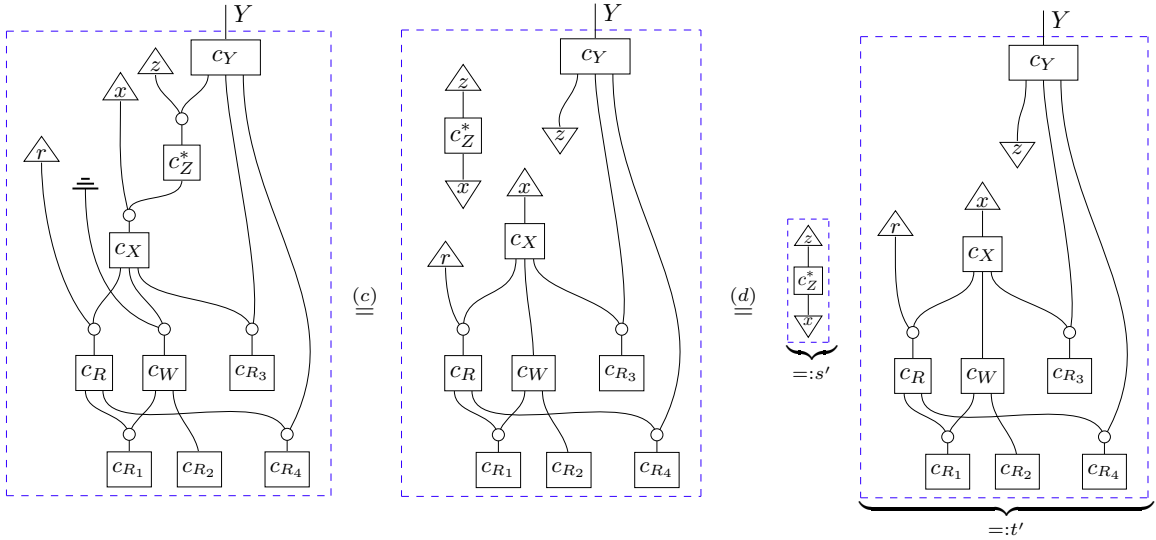
\begin{eqnarray}
	\hspace*{0.6cm} 
	\resizebox{0.29\textwidth}{!}{\input{Fig_Ex_Corr_step4a.tikz}} 
	\hspace*{-0.3cm} \stackrel{(c)}{=}  \hspace*{0.2cm} \resizebox{0.27\textwidth}{!}{\input{Fig_Ex_Corr_step4b.tikz}}
	\hspace*{-0.2cm} \stackrel{(d)}{=}  \hspace*{0.3cm} 
	\begin{minipage}{0.04\textwidth} 
	\[ \underbrace{ \resizebox{1.0\textwidth}{!}{\input{Fig_Ex_Corr_step4c_scalar.tikz}}}_{=: s'} \]
	 \vspace*{0.3cm} \end{minipage} \hspace*{0.1cm} 
	 \begin{minipage}{0.26\textwidth} 
	\[ \underbrace{ \resizebox{1.0\textwidth}{!}{\input{Fig_Ex_Corr_step4c.tikz}} \hspace*{-0.4cm} }_{=: t'} \]
	 \vspace*{1.0cm} \end{minipage} \nonumber 
\end{eqnarray}
\noindent where $(c)$ and $(d)$ used Eqs.~\eqref{eq:sharp-state-eff} and \eqref{eq:mon-cond}. Once again, one can drop the scalar whenever $z \circ c_Z^* \circ x \neq 0$. 
Since $c_Z^*$ is given by assumption one can always evaluate $z \circ c_Z^* \circ x$. 

In order to see how the state $t$ relates to that of $t'$, note that one can of course `copy out discards'. 
By the latter we mean to replace a wire by a copy map with one of its outputs discarded again as done (multiple times) in step (e) below. 
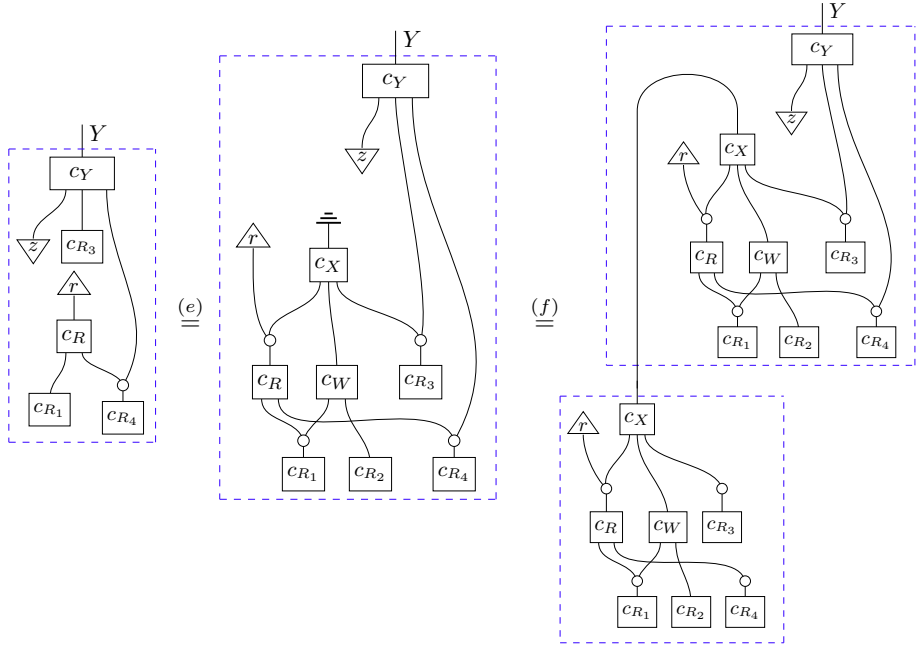
\begin{eqnarray}
	\begin{minipage}{0.13\textwidth} \resizebox{1.0\textwidth}{!}{\input{Fig_Ex_Corr_step3.tikz}} \vspace*{1cm} \end{minipage}
	\hspace*{0.0cm} \stackrel{(e)}{=}  \hspace*{0.1cm} \resizebox{0.25\textwidth}{!}{\input{Fig_Ex_Corr_step5b.tikz}}
	\hspace*{-0.1cm} \stackrel{(f)}{=}  \hspace*{-0.1cm} 
	\begin{minipage}{0.3\textwidth} \resizebox{1.0\textwidth}{!}{\input{Fig_Ex_Corr_step5c.tikz}}  \end{minipage} 
	\nonumber
\end{eqnarray}
Step $(f)$ is simply an application of the diagrammatic chain rule (Lemma \ref{lem:chain-rule}).

In conclusion, without the assumption concerning $s$, that is the corresponding assumption on $c_X^*$, $c_Z$ and $c_{R_2}$, 
the quantity $P(Y | R=r, Z=z; \sigma_{X})$ is not identifiable. 
Under the assumption that $s=1$  \emph{and} $s'=1$, the quantity $P(Y | R, Z; \sigma_{X})$ is identifiable and given by 
\begin{equation} \label{Eq_Ex_CEID_2_result}
	\input{Fig_Ex_Corr_step0.tikz} \hspace*{0.3cm} =  \hspace*{0.2cm} 
	\begin{minipage}{0.21\textwidth} \resizebox{1.0\textwidth}{!}{\input{Fig_Ex_Corr_step5d_v2.tikz}} \vspace*{1cm} \end{minipage}  
\end{equation}
We note that -- apart from the qualification stemming from the scalars -- this is precisely the expression also found by Correa \etal\ \cite{CorreaEtAl_2020_CalculusForStochasticInterventions}. 
However, also given the assumption that $s=1$, but in case $s'=0$, which as pointed out above is computable, then Eq.~\ref{Eq_Ex_CEID_2_result} does obviously \emph{not} hold and the present discussion is inconclusive.
\end{example}

\rl{Although the diagrams may take more space on paper than a conventional treatment, the required steps of reasoning in these examples are easy to spot when looking at the diagrams and the representation actually helps reveal subtleties.} 

% \bibliographystyle{utphys} \bibliography{CauseComp.bib} % Uncomment while working on file standalone, if needed 
%\end{document}

%\documentclass[main.tex]{subfiles} \begin{document}

%*****************************************************************
\section{Counterfactuals \label{Sec_Counterfactuals}}
%*****************************************************************

\begin{center}
\textit{What if Martin Luther King had died when he was stabbed in 1958? \cite{Byrne_2007_RationalImagination} \\
Today Mary woke up with a headache, but had she taken an aspirin last night, would she have woken up with a headache today? \cite{ShpitserEtAl_2008_CompleteIdentificationMethodCausalHierarchy}} 
\end{center}

Questions of a \emph{counterfactual} nature, such as the above examples, are abundant, be they in our everyday life, legal contexts or in science.  In a question such as whether Mary would have had a headache had she taken an aspirin, one means, implicitly, with \emph{all else the same} and \emph{given the fact} that she did actually \emph{not} take an aspirin and did develop a headache. However, we cannot in fact go back and give an aspirin to the very same Mary who did not take one, and in this lies the basic difficulty in providing an empirical basis for counterfactuals. Despite their prevalence in human reasoning, counterfactuals have led to much controversy and philosophical debates concerning all of their aspects -- semantic, epistemic and metaphysical \cite{SEP_Counterfactuals}. 

Here we will put these questions to the side and simply focus on the formalisation of counterfactuals offered within the causal model framework \cite{BalkeEtAl_1994_CounterfactualProbabilities, HalpernEtAl_2005CausesAndExplanationsI, HalpernEtAl_2005CausesAndExplanationsII, Pearl_Causality, Pearl_2011_AlgorithmizationOfCounterfactuals, ShpitserEtAl_2008_CompleteIdentificationMethodCausalHierarchy, BareinboimEtAl_2021_PearlsHierarchy}. 
This consists of a definition of counterfactual concepts and variables and then on that basis a `logic', or set of rules, for reasoning probabilistically about counterfactual statements.  The formalisation rests on two features of the framework of causal models. 

\begin{enumerate} 
	\item 
	Interventions allow one to precisely contemplate a hypothetically different world, which was altered in some way, in particular when `fixing' some variable to a value distinct from what actually occurred;
	\item 
	A functional causal model, with its deterministic relationships between endogenous variables, allows one to give a precise meaning to \emph{`with all else the same'}.
\end{enumerate} 

In the following we will give a basic diagrammatic presentation of this formalisation, which turns out to be a very natural language to define and clarify counterfactual concepts.

%**********************************************
\subsection{The notion of a counterfactual \label{Sec_CF_TheNotion}}
%**********************************************

%**********************************************
\subsubsection{Introductory example \label{Sec_CF_IntroductoryExample}}
%**********************************************

Let us begin by considering again our example question: ``today Mary woke up with a headache, but had she taken an aspirin last night, would she have woken up with a headache today?" (also discussed in Ref.~\cite{ShpitserEtAl_2008_CompleteIdentificationMethodCausalHierarchy}). 
Let $X\in \{1,0\}$ denote whether or not Mary takes an aspirin, $Y \in \{1,0\}$ whether or not she develops a headache. 
\rl{The question implicitly includes that in the actual world Mary did not take an aspirin ($X=0$) and had a headache this morning ($Y=1$).}

\rl{In order to answer the question, one has to make assumptions. 
Let $\modelM$ be a structural causal model as in Fig.~\ref{Fig_CF_Intro_Example_step1} with functional mechanisms $f_X$ and $f_Y$ and $X \rightarrow Y$ the causal structure among the endogenous variables, which we assume correctly models the causal dependencies.} 
\rl{First, consider a situation, in which we furthermore} know that the precise values of the background variables that gave rise to $X=0$ and $Y=1$ were $U_X = a$ and $U_Y = b$, respectively, i.e. the equation in Fig.~\ref{Fig_CF_Intro_Example_step2} holds. 	
\begin{figure}[H]
	\centering
	\begin{subfigure}{4cm}
		\centering
		\input{Fig_CF_Intro_Example_1.tikz}
		\caption{\label{Fig_CF_Intro_Example_step1}}
	\end{subfigure}
	\hspace*{1cm}
	\begin{subfigure}{6cm}
		\centering
		\input{Fig_CF_Intro_Example_2.tikz}
		\hspace*{0.1cm}	$=$  \hspace*{0.1cm}
		\input{Fig_CF_Intro_Example_3.tikz}
		\caption{\label{Fig_CF_Intro_Example_step2}}
	\end{subfigure}
	\caption{ }
\end{figure}
Given such perfect knowledge of both the causal mechanisms $f_X$, $f_Y$ and the values $a$ and $b$ of the exogenous variables in the actual world, there is a clear meaning and answer for the question of whether Mary would have had a headache had she taken aspirin. 
Eq.~\eqref{Fig_CF_Intro_Example_step3} 
encapsulates this: the left-hand side shows the same model as before with states $a$ and $b$ as in the actual world, but additionally the ``had she taken aspirin" is formalised through a do-intervention \rlb{that fixes $X=1$;  
then copying through the `1'-state establishes, as shown on the right-hand side, that the answer to the question is given by $f_Y(1,b)$.} 
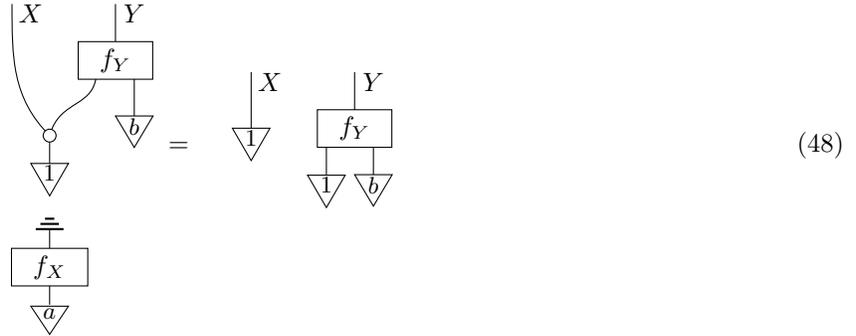
\begin{equation}
		\rlb{\input{Fig_CF_Intro_Example_4.tikz}}
		\hspace*{0.1cm}	=  \hspace*{0.1cm} 
		\begin{minipage}{3cm} \centering \rlb{\input{Fig_CF_Intro_Example_5.tikz}} \end{minipage}	
		\label{Fig_CF_Intro_Example_step3} 
\end{equation}
However, even if we are in possession of the correct structural causal model -- knowing $f_X$, $f_Y$, $P(U_X)$ and $P(U_Y)$ -- we typically will not be able to \emph{observe the particular values} $a$ and $b$ of the exogenous variables. 
\rl{The fact that $X=0$ and $Y=1$ obtained in the actual world, however, constrains what $U_X$ and $U_Y$ could have been. 
Hence, in order to answer the counterfactual question -- assuming $\modelM$, but not knowing $a$ and $b$ -- the following stipulation is natural.} 

First, we are reasoning about two worlds. One is the actual world, the other the counterfactual one in which $X=1$ is fixed by fiat.  
Whatever the values of $U_X$ and $U_Y$ are, they should be the same in both worlds. The diagram in Fig.~\ref{Fig_CF_Intro_Example_step4} represents the two worlds stitched together through $U_X$ and $U_Y$ in this way, where the (copies of the) variables $X$ and $Y$ in the counterfactual world are denoted $X^*$ and $Y^*$ for distinction. This is \rl{the} formalisation of the \emph{`all else the same'}. 

Second, the way in which the facts of the actual world constrain what $U_X$ and $U_Y$ may have been is obtained simply by conditioning on $X=0$ and $Y=1$ in that actual world -- see Fig.~\ref{Fig_CF_Intro_Example_step5}. This conditioning makes an inferential influence go from $X$ and $Y$ in the actual world to their counterparts $X^*$ and $Y^*$ in the counterfactual world, mediated via $U_X$ and $U_Y$. The \rl{distribution $P(Y^*)$ on the} right-hand side of Fig.~\ref{Fig_CF_Intro_Example_step5}, obtained through straightforward simplification,\footnote{Note that \rlb{$\delta_0 \circ f_X \circ P(U_X) \neq 0$ by assumption (after all, $X=0$ was observed)} and that scalar can hence be dropped due to the normalisation box.} thus \rl{is the answer to the counterfactual question assuming $\modelM$.} 
\begin{figure}[H]
	\centering
	\begin{subfigure}{4.7cm}
		\centering
		\rlb{\input{Fig_CF_Intro_Example_6.tikz}}
		\caption{\label{Fig_CF_Intro_Example_step4}}
	\end{subfigure}
	\hspace*{1.5cm}	
	\begin{subfigure}{10.0cm}
		\centering
		\input{Fig_CF_Intro_Example_7_noX.tikz} $=$  \begin{minipage}{3.2cm} \centering \input{Fig_CF_Intro_Example_8_b_noX.tikz} \end{minipage}	
		\caption{\label{Fig_CF_Intro_Example_step5}}
	\end{subfigure}
	\caption{ }
\end{figure}

The next section will present the general definition of counterfactuals based on the same ideas as this example. 
\rl{This} diagrammatic treatment closely follows the \emph{twin network approach} from the conventional literature (see, e.g, Ref.~\cite{ShpitserEtAl_2008_CompleteIdentificationMethodCausalHierarchy}). 
The twin network graph for the previous example, shown in Fig.~\ref{Fig_CF_Intro_Example_twin_network}, represents how the causal structures of the actual and counterfactual worlds are linked. But in this approach one then performs the Bayesian updating and calculations on the given data separately, rather than all in one diagram as in Fig.~\ref{Fig_CF_Intro_Example_step5}.  
Also note that the latter diagram is more informative than the twin network graph -- it makes explicit the identity of the causal mechanisms, while the \rl{twin network graph} only shows where there are causal dependences.  
\begin{figure}[H]
	\centering
	\input{Fig_CF_Intro_Example_twin_network.tikz}
	\caption{ \label{Fig_CF_Intro_Example_twin_network}}
\end{figure}
Note that already in Ref.~\cite{JacobsEtAl_2021_CausalInferencesAsDiagramSurgery_DiagramsToCounterfactuals} Jacobs \etal\ have used a similarly simple example to illustrate a string diagrammatic approach to counterfactuals. 
Apart from the greater generality which we offer below in Secs.~\ref{Sec_CF_GeneralDef} and \ref{Sec_CF_Further_Generalisation}, 
another difference is that our formal set-up allows for the Bayesian conditioning to be done diagrammatically, too, so that the entire calculation of the counterfactual is contained in Fig.~\ref{Fig_CF_Intro_Example_step5}. 
That is, all three steps of the so called \emph{3-step recipe} -- abduction, intervention and prediction -- with which the treatment of counterfactuals is also often introduced in the literature (see, e.g., Ref.~\cite{Pearl_2011_AlgorithmizationOfCounterfactuals}) are captured in one diagram.

%**********************************************
\subsubsection{General definition \label{Sec_CF_GeneralDef}}
%**********************************************

This section presents a general definition of counterfactuals, formalising the ideas behind the above example. 
Conceptually speaking, this definition matches how the term tends to be understood in the causal model literature, including that counterfactuals may make reference to more than just two worlds, while formally speaking, it is more general in that it is not restricted to a concrete category such as $\MatR$. Further generalisations, of a more conceptual kind, are postponed to Sec.~\ref{Sec_CF_Further_Generalisation}. 

Starting from a formal definition helps to disentangle two things; on the one hand, the \emph{definition} of counterfactuals, in particular making explicit relative to what kind of data counterfactuals are well-defined, and on the other hand, the study of epistemological questions such as given what kind of data can one \emph{in practice} infer anything unambiguously about some counterfactual. 
As the example from Sec.~\ref{Sec_CF_IntroductoryExample} suggests, the definition is relative to a given functional causal model and it will be helpful to employ the open causal model perspective from Sec.~\ref{sec:openCMs}. %======

Henceforth, let $\catC$ be a cd-category with \diagconditioning\ (see Section \ref{sec:catsetup}). Let $\model{M}$ be a functional causal model in $\catC$ with endogenous variables $V$, output variables $O\subseteq V$, deterministic mechanisms $(f_X)_{X \in V}$ and exogenous variables $(U_X)_{X \in V}$, with corresponding states \rl{$(\lambda_X)_{X \in V}$} (see Def.~\ref{def:FCM}). 
Following Example~\ref{Ex_FCM_as_composition}, let then $\model{F}$ be the respective deterministic part with input $U$ and output $O$, seen as an open causal model, \rl{and $\model{L}$ be the overall product state $\bigotimes_{X} \lambda_X$ on $U$ such that $\model{M} = \model{F} \circ \model{L}$.}
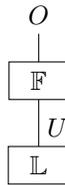
\begin{figure}[H]
	\centering
	\input{Fig_Model_For_ParallelWorldsDiagram_single_wire.tikz}
\end{figure}

With this notation at hand, one can straightforwardly state the general version of a twin network kind of stitching together of more than two worlds. 

\begin{definition} \textnormal{(Parallel worlds model\footnote{This terminology reflects the similarity with the `parallel worlds graph' treatment of counterfactuals (see, e.g., Refs.~\protect\cite{ShpitserEtAl_2008_CompleteIdentificationMethodCausalHierarchy, AvinETAl_2005_IdentifiabilityPathSpecificEffects}).}):} \label{Def_ParallelWorldsModel}
Let \rl{$\model{M} = \model{F} \circ \model{L}$} be a functional causal model in $\catC$ with output variables $O$. 
For each $j=1,...,k$ let $\sigma_j$ be an intervention on $\model{F}$ 
such that $\sigma_j\big( \model{F} \big)$ is still deterministic.   
The induced causal model \rl{$\model{M}^k_{\sigma_j} := \big( \bigotimes_j^{U} \sigma_j\big( \model{F} \big)\hspace*{0.05cm}\big) \circ \model{L}$}, i.e. the model with network diagram below, is called a \emph{parallel worlds model}.\footnote{\rl{Recall the notation of $\bigotimes_j^{U}$ for the `sharing' of the variables $U$ from Def.~\ref{Def_SharingOfVariables}.}} 
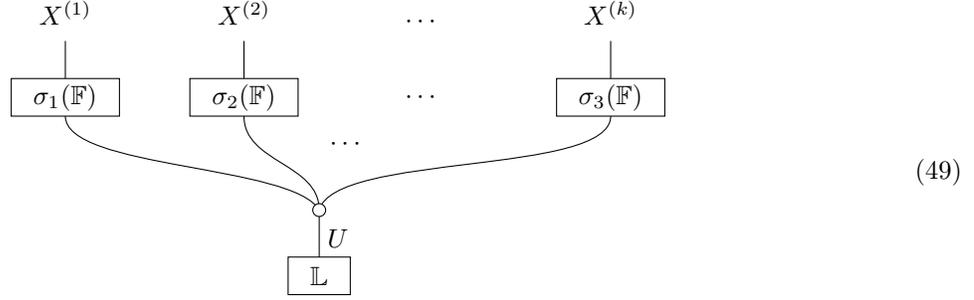
\begin{equation} \label{eq:parallel-worlds}
	\centering
	\input{Fig_Def_ParallelWorldsModel_single_wires_b.tikz}
\end{equation}
For distinction the non-input variables of $\sigma_j\big( \model{F} \big)$ are relabelled with a superscript $(j)$.  
\end{definition}

The notion of a counterfactual involves such parallel worlds, now specifically using do-interventions, along with conditioning.

\begin{definition} \textnormal{(Counterfactual):} \label{Def_Counterfactual}
Let \rl{$\model{M} = \model{F} \circ \model{L}$} be a functional causal model in $\catC$ with endogenous variables $V$ and output variables $O\subseteq V$.  A \emph{counterfactual} is a state of the form: 
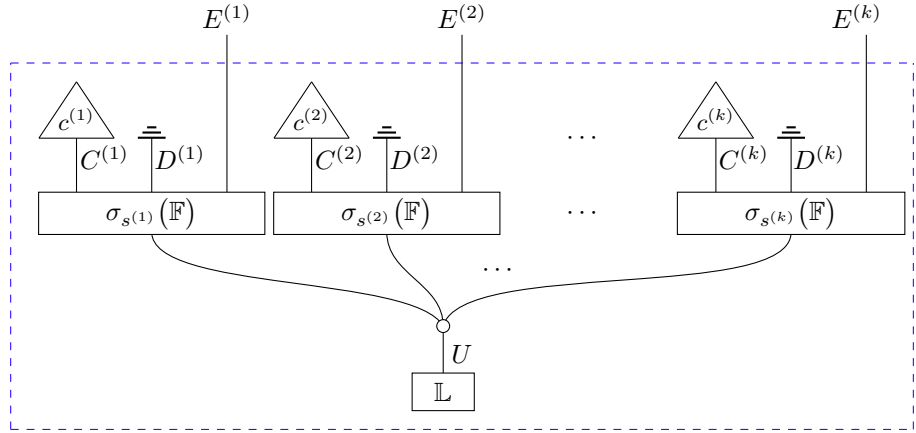
\[
	\input{Fig_Def_Counterfactual_new.tikz}
\]
arising from a parallel worlds model where each $\sigma_{s^{(j)}}$ carries out a do-intervention $\Do(S^{(j)}=s^{(j)})$
for some sharp state $s^{(j)}$ of $S^{(j)} \subseteq O$. Here each $O^{(j)}$ is partitioned as $C^{(j)} \cup D^{(j)} \cup E^{(j)}$ and each $c^{(j)}$ is a sharp effect on $C^{(j)}$. Furthermore, we require for \rl{some $j,j'$ with $j \neq j'$} that
\begin{equation} \label{eq:CFcond}
C^{(j)} \neq \emptyset \neq E^{(j')}
\end{equation}
\end{definition}

In other words, a counterfactual is a model of $k$ parallel worlds defined by do-interventions, composed with sharp effects on each $C^{(j)}$, each $D^{(j)}$ discarded, and then all normalised. 
Typically, one of the sets $S^{(j)}$ will be empty, i.e. one of the parallel worlds is described by the unaltered model $\model{F}$ and understood as the `actual world'. 
In order to be able to link with conventional notation, but \emph{without} implying any particular category like $\catC = \MatR$, we at times denote the state defined by a counterfactual as in Def.~\ref{Def_Counterfactual} also as  
\begin{equation}
	P_{\model{M}^k_{s}} \big(E^{(1)},...,E^{(k)} \ \allowbreak | \allowbreak \ C^{(1)} = c^{(1)}, ..., C^{(k)} = c^{(k)} \big)	\label{Eq_CF_Prob_Notation}
\end{equation}
with $s$ collectively standing for the family of $s^{(j)}$ and, for yet more brevity, we also write just 
$P_{\model{M}^k_{s}} \big( E \ \allowbreak | \allowbreak \ C=c \big)$. 

The following terminology is helpful in allowing us to succinctly refer to the key data in a counterfactual.

\begin{definition} \label{Def_CfTerms}
Given a set $O$ of objects in $\catC$, a \emph{world term} $\stdwtnoj$ is given by a sharp state $s$ on $S \subseteq O$, sharp effect $c$ on $C \subseteq O$ and subset $E$ disjoint from $C$.  A finite family $W^k = \stdwts$ of world terms on the same set of objects $O$ are called \emph{\cfterms} if they satisfy \eqref{eq:CFcond} \rlb{for some $j,j'$ with $j \neq j'$.} 
\end{definition}

While the data in a family of world terms makes no reference to an FCM, the notation is of course with Def.~\ref{Def_ParallelWorldsModel} in mind and the superscripts $(j)$ will typically label the worlds of a parallel worlds model.  Thus a counterfactual is determined by an FCM $\modelM$ along with \cfterms.

\begin{remark} \label{Rem_CF_Notion} \textnormal{(Relation to conventional notation):} 
	In the causal model literature the discussion of counterfactuals tends to use expressions like $P(\gamma | \delta)$,  
where $\gamma$ and $\delta$ are conjunctions of \emph{counterfactual expressions}, which are -- written in our notation -- value assignments of the form 
\begin{eqnarray}
	\gamma &=& \big( E^{(1)} = e^{(1)} \ \wedge ... \wedge \ E^{(k)} = e^{(k)} \big) \nonumber \\ 
	\delta &=& \big( C^{(1)} = c^{(1)} \ \wedge ... \wedge \ C^{(k)} = c^{(k)} \big) \nonumber
\end{eqnarray}
that is, $\delta$, $\gamma$ is basically the same data as $C=c$, $E=e$ but understood as a logical expression. 
Note that:
\begin{itemize}
	\item This conventional treatment insists on a counterfactual as a probability $P(\gamma | \delta) \in [0,1]$. 
	A consequence is that $P(\gamma | \delta)$  may not be defined seeing as $P(\delta)$ might vanish \cite{ShpitserEtAl_2008_CompleteIdentificationMethodCausalHierarchy}. 

	In contrast, a counterfactual defined as a state over $E^{(1)},...,E^{(k)}$ as in Def.~\ref{Def_Counterfactual} is always well-defined in this sense. For example in  $\MatR$ the normalisation box gives either the conditional probability distribution or else the zero \emph{state}  if $P(\delta)$ vanishes. The latter is not a probability distribution, as it assigns zero to all values in the domain, so cannot be expressed at the level of individual probabilities like $P(\gamma | \delta)$. 

Of course, even if $P_{\model{M}^k_{s}} \big( E \ \allowbreak | \allowbreak \ C=c \big)$ is a probability distribution, it may not have full support, i.e. $P_{\model{M}^k_{s}} \big( E=e  \ \allowbreak | \allowbreak \ C=c \big)$ may vanish for some $e$. This can be computed by plugging the corresponding sharp effects into the open wires in Def.~\ref{Def_Counterfactual}, which are crucially outside the normalisation box. 

\item In the literature the different worlds and their variables are usually indexed by a subscript indicating the variables on which the world's  do-intervention acts, like $Y_x$ instead of $Y^*$ for the introductory example. This notation would lead to clutter in Def.~\ref{Def_Counterfactual}, hence the corresponding superscripts for the $j$th world.

	\item The subscript in  $P_{\model{M}^k_{s}}$ makes explicit the functional model relative to which the counterfactual is defined. 

\end{itemize}
\end{remark}

On a more conceptual side, there are a few points that deserve attention and discussion, including the separation of the two Definitions~\ref{Def_ParallelWorldsModel} and \ref{Def_Counterfactual}, and the \rlb{particular condition} \eqref{eq:CFcond} on \emph{\cfterms}. 

A counterfactual relates worlds that differ in two respects; the sense of factuality counter to which an alternative situation is being considered is two-fold: 

\begin{enumerate}
	\item The worlds differ by causal structure. 
The `had Mary taken an aspirin, while all else the same', requires fiddling with the causal structure, else all deterministic causes of some variable $X$ being the same would simply make for the same value of $X$. 
Such `reaching in' to consider some variable taking a different value is formally a do-intervention, hence their appearance in Def.~\ref{Def_Counterfactual}. 
However, the way to think about it is not an operational one -- it is a formalisation of what corresponds to the \textit{small miracle} in Lewis' possible worlds semantics of counterfactuals \cite{Lewis_1973_Counterfactuals, Lewis_1973_Causation}.\footnote{Note that there are crucial differences though compared to Lewis' treatment of counterfactuals and his wider programme \protect\cite{SEP_Counterfactuals, Pearl_Causality}.} 

\item Each world involves its own event-like facts -- as opposed to facts about causal structure -- captured by (conditioning on) the particular values $c^{(j)}$ for variables $C^{(j)}$. The `had Mary taken an aspirin' turns the question about her headache in an alternative world only into a counterfactual question relative to the fact that in actuality she did not take aspirin and did develop a headache. 
\end{enumerate}

Next let us discuss the condition \ref{eq:CFcond}. This requires that at least in some world, i.e. for some $j$, the subset $C^{(j)}$ is non-empty so that some variables in some world are being conditioned on. 
In contrast, consider a state as in Def.~\ref{Def_Counterfactual}, but without the conditioning on the effects $c^{(j)}$, that is, the state that represents the distribution $P_{\model{M}^k_{s}}( E, C)$ on all the $C^{(j)}, E^{(j)}$: 
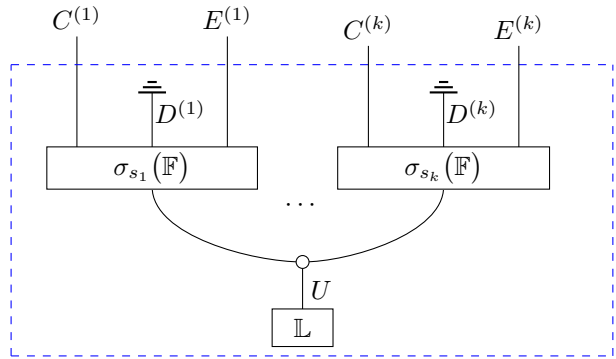
\[
\input{CF-no-cond.tikz}
\]
The output variables of the worlds will be correlated, since they are connected up through the background variables $U$. 
However, by construction, whether or not some particular variables $S_j$ have been set to particular values in some $j$th world -- the very thing that defines the respective world -- does \emph{not} influence the probabilities in some other world. 
It is only the sharp effects $c^{(j)}$ that make an inferential influence `travel' from one world to the other through $U$, which arguably is the whole point of a counterfactual. 

This is particularly evident diagrammatically. Considering the marginal on, say, the $k$th world, one is left with just that -- the respective interventional model making predictions for $P\big( E^{(k)} | C^{(k)}=c^{(k)} ; \Do(S^{(k)}=s^{(k)}) \big)$, which has nothing to do with a counterfactual: 
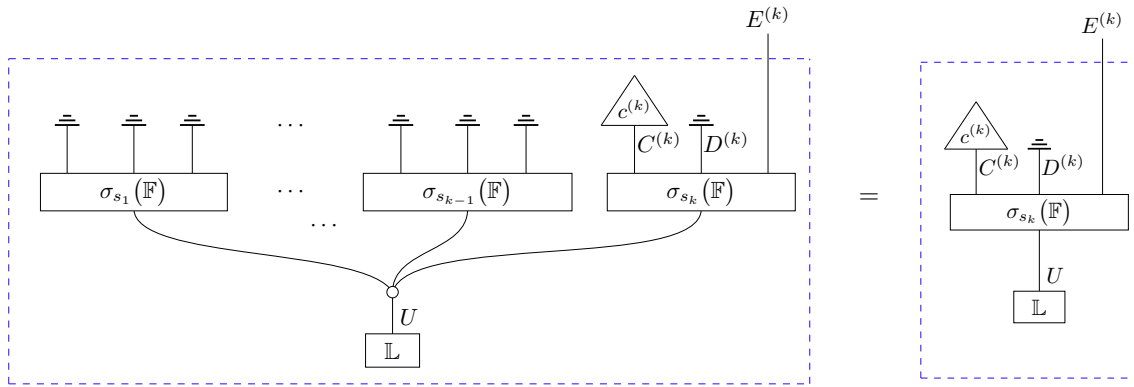
\begin{figure}[H]
	\centering
	\begin{minipage}{0.65\textwidth} \resizebox{1.0\textwidth}{!}{\input{Fig_NotACounterfactual.tikz}} \end{minipage}
	\hspace{0.3cm}	$=$ \hspace{0.3cm}
	\begin{minipage}{0.185\textwidth} \resizebox{1.0\textwidth}{!}{\input{Fig_NotACounterfactual_2.tikz}} \end{minipage}
\end{figure}
\noindent Needless to say, parallel world states and probabilities $P(\gamma)$, are important ingredients when computing a counterfactual. 
What our Def.~\ref{Def_Counterfactual} makes manifest though is that a counterfactual can \emph{not} be faithfully paraphrased as contemplating \emph{what if one were to intervene} -- it is more than that.\footnote{\rlb{In the literature one sometimes finds a counterfactual paraphrased as a `what if' question. This is misleading since there is nothing necessarily counterfactual about a question `what if I were to reach in and set $X=x$, i.e. perform a do-intervention on $X$?'. One may in principle answer such a question with an experiment, or in some cases without any intervention at all, given the right kind of data (see Sec.~\ref{Sec_CE_Identifiability}). In contrast, it is impossible in principle  to empirically verify a counterfactual claim. The diagrammatic presentation makes this distinction precise via the notions of worlds and of factuality that counterfactuality relies upon.}} 
At the same time Def.~\ref{Def_Counterfactual} is broad in that it does not demand that conditioning happen in all worlds, or that it happen only in the actual world $\model{F}$.  
A concrete example, which is more involved than the aspirin and headache one, will be discussed in the subsequent section.

%**********************************************
\subsection{Identifiability of counterfactuals \label{Sec_CF_Identifiability}}
%**********************************************

%**********************************************
\subsubsection{Problem statement and an example \label{Sec_CF_id_example}}
%**********************************************

The notion of a counterfactual is defined relative to a given functional causal model. However, at times the identity of a particular counterfactual can be inferred unambiguously from data that does not single out a functional causal model, which in practice, of course, one basically never has. 

This yields the epistemological problem of the \emph{identifiability of counterfactuals}, which in conventional terms (c.f. Rem.~\ref{Rem_CF_Notion}) is as follows:  
given an ADMG $G$ with vertices $O$ and the set of all do-interventional distributions,  
$P_*(O):= \{P\big(Y ; \Do(X) \big) \ | \ X,Y \subseteq O, \ \text{disjoint} \}$, is $P(\gamma | \delta)$ identifiable? 
Note that the \emph{given data} includes all \emph{do-interventional} distributions.\footnote{Note that $P(O)$ is included in this set $P_*(O)$ for $X= \emptyset$.} 
If a particular counterfactual is identifiable according to this definition, the relevant interventional distributions can either be obtained from corresponding experimental data, or else the problem is reduced to the problem of the identifiability of these interventional distributions -- this is a manifestation of the causal hierarchy \rl{\cite{Pearl_Causality, BareinboimEtAl_2021_PearlsHierarchy}}. 

Here we will discuss the problem in the following terms.  %Todo: Again see if want to name such a category
Let $\catC$ be a cd-category with \diagconditioning, and $O$ a set of objects in it -- to be thought of as the `observed' variables. Given an ADMG $G$ with vertices $O$, some set $P_*(O)$ and \cfterms\ $W^k = \stdwts$ concerning the set $O$, the problem asks: is the corresponding counterfactual:
\begin{equation} 
P_{\model{M}_{s}^k} \big(E \ | \ C=c \big) \label{eq_cf_id}
\end{equation}
uniquely determined, and thus expressable in terms of $P_*(O)$?

It may seem odd to write the expression \rlb{in Eq.~\eqref{eq_cf_id}} given that the data does not enforce any particular FCM $\model{M}$. However, 
the formalisation of counterfactuals relies on an underlying determinism, and so implicitly that there is a correct such model. In case a counterfactual \emph{is} identifiable, it does not matter which $\model{M}$ compatible with $G$ and $P_*$ is considered, since any such models imply \rlb{the same counterfactual.} This is no different than the role of $P(\gamma | \delta)$ in the literature, which in case of non-identifiability is not a well-defined quantity. 

The diagrammatic treatment of the problem will consider a diagram $C$ that makes reference to an FCM $\model{M}$ that is in keeping with the given data and such that on the basis of $C$ one \emph{can} establish whether the \rlb{counterfactual is identifiable.} Hence in either case the specific $\model{M}$ doesn't matter -- it was merely an intermediate step to employ the diagrammatic reasoning. 

\paragraph{Formal setup of the problem.} Given an ADMG $G$ with vertices $O$, consider the rootification $\tilde{\rho}(G)$ with vertices $V = O \cup R$, where $R$ are the additional root nodes introduced by $\tilde{\rho}$,  and 
let \rl{$\model{M} = \model{F} \circ \model{L}$} be a corresponding FCM with background variables $\{U_X\}_{X\in V}$ compatible with $P_*$ and such that its causal structure is compatible \rl{with $G$, i.e. $\pi_O(G_{\model{M}}) = G$.} We will then construct and inspect the counterfactual diagram $C$ as induced by this $\model{M}$ and the \cfterms\ $W^k$.

As a notational convention, for any $X \in V$ we will write $c_X$ for the effective causal mechanism defined by the pair $(f_X, \lambda_X)$ according to Eq.~\eqref{Eq_CBN_SCM_Relation}. 
The set $\{c_X\}_{X \in V}$ thus defines a causal model with DAG $\tilde{\rho}(G)$. 
\begin{equation}
		\centering
		\input{Fig_Alg_3_RHS.tikz} \hspace{0.2cm} := \hspace{0.4cm} \input{Fig_Alg_3_LHS.tikz}
	\label{Eq_CBN_SCM_Relation}
\end{equation}

\begin{remark} \textnormal{(Role of rootification):} \label{Rem_Rootification_issue_CF} 
	Note that we explicitly specified the rootification method $\tilde{\rho}$ defined in Sec.~\ref{Sec_ADMGsAndNetworkDiagrams}. 
This is because for any given ADMG $G$ the alternative rootification $\rho(G)$ can be regarded as a  special case of $\tilde{\rho}(G)$, where some root nodes with more than two children actually are sets of root nodes each of which has only two children. 
For causal effect identifiability in Sec.~\ref{Sec_CEI_TheProblem} we appealed to a result from Ref.~\cite{RichardsonEtAl_2022_NestedMarkovForADMGs}, namely Thm.~\ref{Thm_TheFolkloreResult}, to argue that it does \emph{not} matter how one rootifies an ADMG. 
For counterfactuals, this result is however not directly applicable. 
Although a priori it seems not obvious why the way one treats bi-directed edges in an ADMG should not matter to the identifiability of counterfactuals, works like \cite{ShpitserEtAl_2008_CompleteIdentificationMethodCausalHierarchy} unfortunately are silent on this question. 
We too will leave this question open, but use $\tilde{\rho}$ to be on the safe side.\footnote{This means that in case of identifiability it is \rlb{as strong a claim as possible}, whereas in case of non-identifiability it does a priori not imply non-identifiability if one assumed all bi-directed edges corresponded to two-children root nodes as the rootification method $\rho$ would have it. We will come back to this question in Rem.~\ref{Rem_ADMGsAndNonID}. \label{Footnote_RootificationForCF}}  
Finally, we note that for any example in \cite{ShpitserEtAl_2008_CompleteIdentificationMethodCausalHierarchy}, many of which will be discussed below, $\tilde{\rho}$ and $\rho$ lead to the same DAG anyway.  
\end{remark}

Just as for  the identifiability of causal effects, the problem of the identifiability of counterfactuals also has been solved -- an algorithm exists that is sound and complete for it \cite{ShpitserEtAl_2008_CompleteIdentificationMethodCausalHierarchy}. 
The subsequent sections will discuss a diagrammatic treatment of the solution.

Before that, however, it is instructive to consider an example from the literature and treat it in our diagrammatic set-up. This serves several purposes.  
First, it illustrates our otherwise rather abstract definition of counterfactuals in Def.~\ref{Def_Counterfactual}. 
Second, it builds the intuitions needed for the general algorithmic solution that follows. 
Third, the very fact that one \emph{can} discuss this example and present the solution in a straightforward way without having to prove any facts about counterfactuals, only through diagrammatic reasoning -- employing existing rewrite rules in the category -- speaks for itself. 

\begin{example} \label{Example_CF_identifiability}
Shpitser and Pearl study the following example in Ref.~\cite{ShpitserEtAl_2008_CompleteIdentificationMethodCausalHierarchy}.  
Consider the ADMG $G$ in Fig.~\ref{Fig_CFid_Example_ADMG} and suppose $P_*(O)$ is given with $O=\{X,Y,Z,W,D\}$. 
Is the counterfactual $P_{\model{M}^3_{s}} \big(Y^{(1)} \ | \ X^{(2)} = \tilde{x}, D^{(2)} = d, Z^{(3)} = z \big)$, 
\rl{with the worlds-defining $s$ standing for  
$\Do(X^{(1)} =x)$ and $\Do(D^{(3)}=d)$}\footnote{That is, the \cfterms\ $W^3$ with 
$S^{(1)} = \{X^{(1)}\}$, $s^{(1)} =x$, $E^{(1)} = \{Y^{(1)}\}$, $C^{(1)} = \emptyset$, 
$S^{(2)} = \emptyset$, $E^{(2)} =  \emptyset$, $C^{(2)} = \{X^{(2)}, D^{(2)} \}$, $c^{(2)} = \tilde{x} \otimes d$,
$S^{(3)} = \{D^{(3)}\}$, $s^{(3)}=d$, $E^{(3)} =  \emptyset$, $C^{(3)} = \{Z^{(3)}\}$,  $c^{(3)} = z$.} 
identifiable? 
They also describe a medical scenario, which gives rise to this query with its reference to three worlds: 
``For instance, we might be interested in how likely the patient would be to have a symptom $Y$ given a certain dose $x$ of drug $X$, assuming we
know that the patient has taken dose $\tilde{x}$ of drug $X$, dose $d$ of drug $D$, and we know how an intermediate symptom $Z$ responds to treatment $d$." 
\begin{figure}[H]
	\centering
	\begin{subfigure}{2.5cm}
		\centering
		\input{Fig_CFid_Example_ADMG.tikz} 
		\caption{ \label{Fig_CFid_Example_ADMG}}
	\end{subfigure}
	\hspace*{0.7cm}
	\begin{subfigure}{2.5cm}
		\centering
		\input{Fig_CFid_Example_ADMG_rootified.tikz} 
		\caption{ \label{Fig_CFid_Example_ADMG_rootified}}
	\end{subfigure}
	\hspace*{0.3cm}
	\begin{subfigure}{5cm}
		\centering
		\input{Fig_CFid_Example_CBN_ND.tikz} 
		\caption{ \label{Fig_CFid_Example_CBN_ND}}
	\end{subfigure}
	\begin{subfigure}{5cm}
		\centering
		\input{Fig_CFid_Example_SCM_ND.tikz} 
		\caption{ \label{Fig_CFid_Example_SCM_ND}}
	\end{subfigure}
	\caption{ }
\end{figure}
For the diagrammatic treatment consider the rootification $\tilde{\rho}(G)$ in Fig.~\ref{Fig_CFid_Example_ADMG_rootified} and the corresponding network diagram $D_{\tilde{\rho}(G)}$ in Fig.~\ref{Fig_CFid_Example_CBN_ND}. 
Fig.~\ref{Fig_CFid_Example_SCM_ND} in turn shows the corresponding FCM, relative to which we will consider the counterfactual of interest. This counterfactual is 
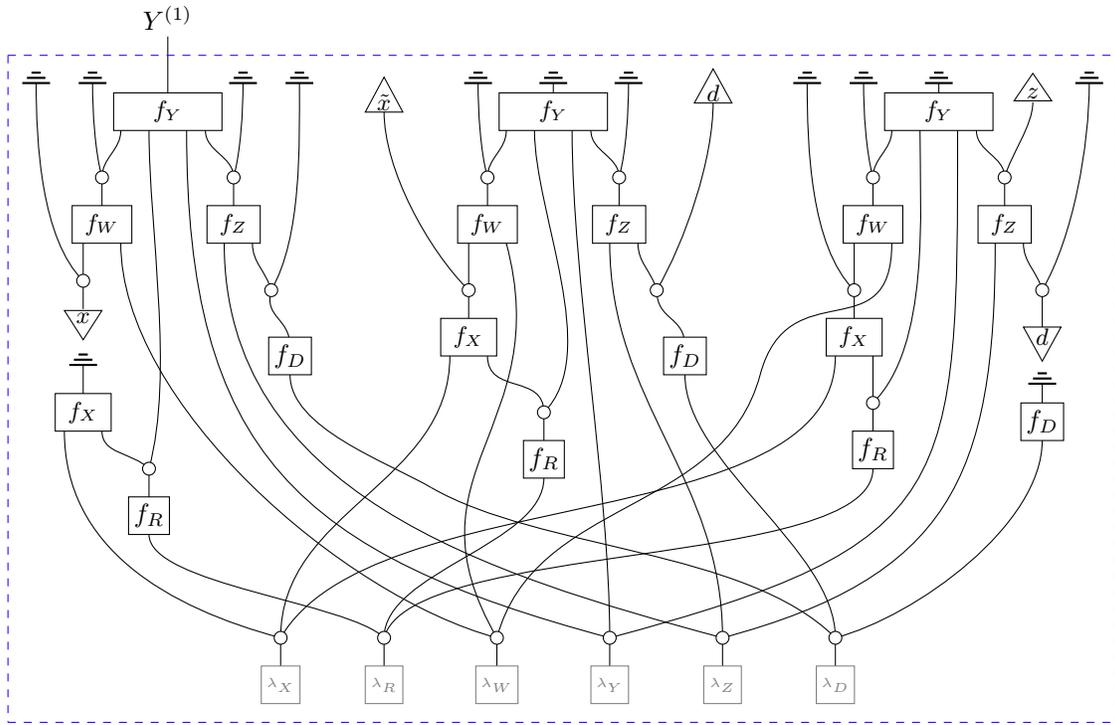
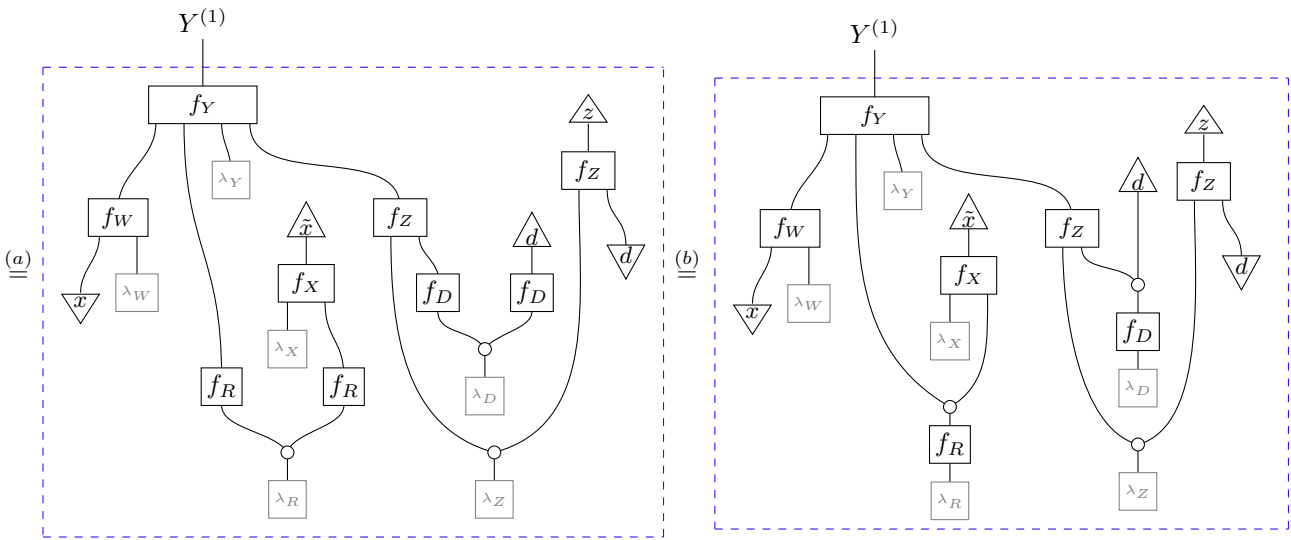
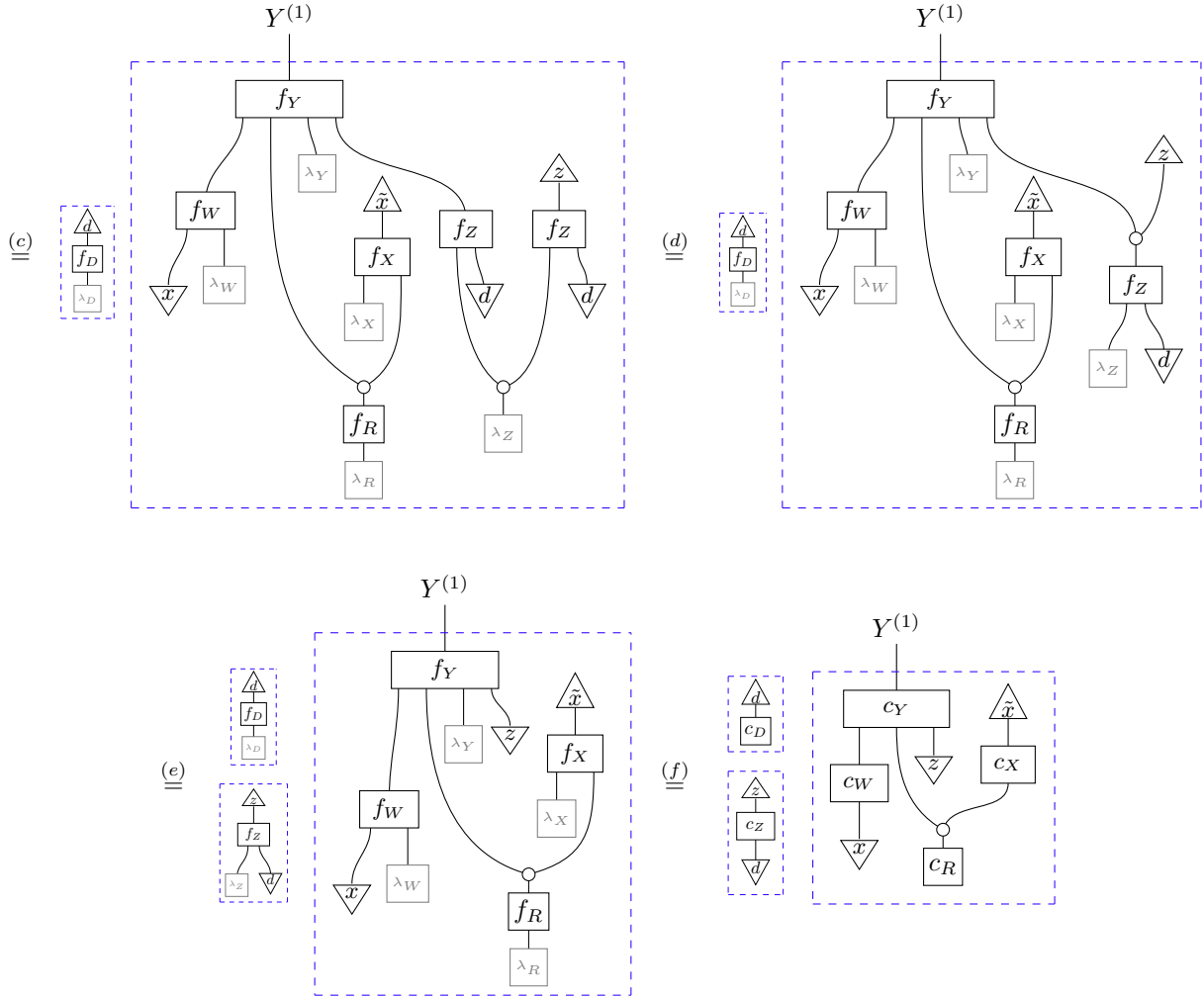
\begin{figure}[H]
	\centering
	\input{Fig_CFid_Example_step_1.tikz}
\end{figure}

\begin{figure}[H]
	\centering
	\hspace*{-1.0cm} $\stackrel{(a)}{=}$ 
	\begin{minipage}{8.3cm}
		\centering
		\input{Fig_CFid_Example_step_2.tikz} 
	\end{minipage}
	$\stackrel{(b)}{=}$ 
	\begin{minipage}{7.9cm}
		\centering
		\input{Fig_CFid_Example_step_3.tikz} 
	\end{minipage}
\end{figure}
\begin{figure}[H]
	\centering
	$\stackrel{(c)}{=}$ 
	\begin{minipage}{8.2cm}
		\centering
		\resizebox{0.1\textwidth}{!}{\input{Fig_CFid_Example_step_4_scalar.tikz}} 
		\input{Fig_CFid_Example_step_4.tikz} 
	\end{minipage}
	$\stackrel{(d)}{=}$ 
	\begin{minipage}{7.2cm}
		\centering
		\resizebox{0.1\textwidth}{!}{\input{Fig_CFid_Example_step_4_scalar.tikz}}
		\input{Fig_CFid_Example_step_5.tikz} 
	\end{minipage}
\end{figure}
\begin{figure}[H]
	\centering
	$\stackrel{(e)}{=}$ 
	\hspace*{0.2cm}
	\begin{minipage}{1.0cm}
		\centering
			\resizebox{0.7\textwidth}{!}{\input{Fig_CFid_Example_step_4_scalar.tikz}} \\[0.2cm]
			\resizebox{1.0\textwidth}{!}{\input{Fig_CFid_Example_step_6_scalar.tikz}}
	\end{minipage}	
	\begin{minipage}{4.7cm}
		\centering
		\input{Fig_CFid_Example_step_6.tikz} 
	\end{minipage}
	$\stackrel{(f)}{=}$ 
	\hspace*{0.2cm}
	\begin{minipage}{1.0cm}
		\centering
			\resizebox{0.85\textwidth}{!}{\input{Fig_CFid_Example_step_4_scalar_merged.tikz}} \\[0.2cm]
			\resizebox{0.85\textwidth}{!}{\input{Fig_CFid_Example_step_6_scalar_merged.tikz}}
	\end{minipage}	
	\begin{minipage}{3.7cm}
		\centering
		\input{Fig_CFid_Example_step_7.tikz} 
		\vspace*{0.7cm}
	\end{minipage}
\end{figure}

Step $(a)$ simply uses that discards `fall through' the mechanisms and remove wires when connected to copy maps, as used throughout Sec.~\ref{Sec_CEI_ExamplesGeneralCase}. 
Step $(b)$ uses that $f_D$ is deterministic, and so two copies of $f_D$ can be moved through a copy map leaving a  single $f_D$ whose output is copied; analogously for the function $f_R$. 
Step $(c)$ uses that the deterministic effect $d$ connects to a copy map and hence separates off $d \circ f_D \circ \lambda_D$ as a scalar (see Sec.~\ref{sec:normalisation}). 
Steps $(d)$ and $(e)$ repeat the same sort of steps -- first $f_Z$ is moved down through the copy and then it is used that the effect $z$ connects to a copy map. 
Finally, step $(f)$ simply replaces the functions and background noises with their respective channels according to the equivalence between Fig.~\ref{Fig_CFid_Example_CBN_ND} and Fig.~\ref{Fig_CFid_Example_SCM_ND}.  

Note that the scalars, floating around in the above diagrams, are important -- the respective normalised scalars are either one or zero and the latter case would have the entire state of the counterfactual evaluate to the zero state (see Rem.~\ref{Rem_CF_Notion}).  
Also note that while wires pertaining to distinct worlds might be labelled distinctly, say $Z^{(1)}$ and $Z^{(3)}$ in the example, they correspond to the same underlying object in the category, which is why `copying through' of functions as done with $f_Z$ is well-defined. 
In order to avoid clutter the label for most wires is suppressed, but wherever it is important to keep track of which world a variable pertains to -- because the counterfactual we wish to calculate makes reference to this particular variable -- that variable is either an open wire and still has its label or else the value it is being conditioned on indicates that world. 
Hence, there is no danger of ambiguity.

Now, on the basis of the network diagram in Fig.~\ref{Fig_CFid_Example_CBN_ND}, observe the following equivalences (straightforward) in Fig.~\ref{Fig_CF_id_main_example_id_components}.
\begin{figure}[H] \label{Fig_CF_id_main_example_id_components}
	\centering
	\begin{subfigure}{6cm}		
		\centering
		\begin{minipage}{3.5cm}
			\centering
			\input{Fig_CFid_Example_CBN_ND_id_2_a.tikz}
		\end{minipage}
		$=$ 
		\begin{minipage}{1.7cm}
			\centering
			\input{Fig_CFid_Example_CBN_ND_id_2_b.tikz}
		\end{minipage}
		\caption{\label{Fig_CF_id_main_example_id_components_a}}
	\end{subfigure}
	\begin{subfigure}{10cm}				
		\centering		
		\begin{minipage}{4.5cm}
			\centering
			\input{Fig_CFid_Example_CBN_ND_id_3_a.tikz}
		\end{minipage}
		$=$ 
		\begin{minipage}{3.7cm}
			\centering
			\input{Fig_CFid_Example_CBN_ND_id_3_b_v2.tikz}
		\end{minipage}
		\caption{\label{Fig_CF_id_main_example_id_components_b}}
	\end{subfigure} 
	\\[0.6cm] %---------------------------------
	\begin{subfigure}{6cm}		
		\centering
		\begin{minipage}{2.0cm}
			\centering
			\input{Fig_CF_id_main_example_id_components_c_LHS.tikz}
		\end{minipage}
		$=$ 
		\begin{minipage}{2.0cm}
			\centering
			\input{Fig_CF_id_main_example_id_components_c_RHS.tikz}
		\end{minipage}
		\caption{\label{Fig_CF_id_main_example_id_components_c}}
	\end{subfigure}
	\begin{subfigure}{10cm}				
		\centering		
		\begin{minipage}{3.0cm}
			\centering
			\input{Fig_CF_id_main_example_id_components_d_LHS.tikz}
		\end{minipage}
		$=$ 
		\begin{minipage}{2.0cm}
			\centering
			\input{Fig_CF_id_main_example_id_components_d_RHS.tikz}
		\end{minipage}
		\caption{\label{Fig_CF_id_main_example_id_components_d}}
	\end{subfigure} 
	\caption{\label{Fig_CF_id_main_example_id_components}}
\end{figure}
In summary we find: 
\begin{equation}
	\begin{minipage}{4.0cm}
		\centering
		\input{Fig_CFid_Example_Result.tikz}
	\end{minipage}
	%--------
	\hspace*{0.1cm} =  \hspace*{0.1cm}
	\begin{minipage}{1.0cm}
		\centering
			\resizebox{0.85\textwidth}{!}{\input{Fig_CFid_Example_step_4_scalar_merged.tikz}} \\[0.2cm]
			\resizebox{0.85\textwidth}{!}{\input{Fig_CFid_Example_step_6_scalar_merged.tikz}}
	\end{minipage}	
	\begin{minipage}{3.4cm}
		\centering
		\input{Fig_CFid_Example_step_7_Y.tikz}
		\vspace*{0.7cm}
	\end{minipage} 
\end{equation}
\begin{equation}
	\hspace*{-0.3cm}
	=  \hspace*{0.1cm} 
	\begin{minipage}{2.0cm}
		\centering
			\resizebox{0.6\textwidth}{!}{\input{Fig_CFid_Example_step_4_scalar_merged_b_explicit.tikz}} \\[0.2cm]
			\resizebox{0.95\textwidth}{!}{\input{Fig_CFid_Example_step_6_scalar_merged_b_explicit.tikz}}
	\end{minipage}	
	\begin{minipage}{4.3cm}
		\centering
		\input{Fig_CFid_Example_step_final_corrected_explicit.tikz}
		\vspace*{0.7cm}
	\end{minipage}
	 \hspace*{-0.1cm}
	= \hspace*{0.2cm}
	%--------
	\begin{minipage}{3.0cm}
		\centering
			\resizebox{0.65\textwidth}{!}{\input{Fig_CFid_Example_step_4_scalar_merged_b.tikz}} \\[0.2cm]
			\resizebox{1.15\textwidth}{!}{\input{Fig_CFid_Example_step_6_scalar_merged_b.tikz}}
	\end{minipage}	
	\begin{minipage}{5.5cm}
		\centering
		\input{Fig_CFid_Example_step_final_corrected.tikz}
		\vspace*{0.7cm}
	\end{minipage}
	\hspace*{-0.2cm}
	\label{Eq_CF_id_main_example_final_solution}
\end{equation}
 
Seeing as the right-hand side of Eq.~\ref{Eq_CF_id_main_example_final_solution} is entirely composed of identifiable expressions, 
we conclude identifiability for this counterfactual.  
Note that this is indeed the same solution as given in Ref.~\cite{ShpitserEtAl_2008_CompleteIdentificationMethodCausalHierarchy}, 
only here with the scalars additionally making explicit conditions under which the counterfactual evaluates to the zero state.
\end{example}

%**********************************************
\subsubsection{The general recipe for simplification -- the \protect\ouralgo\ algorithm \label{Sec_CF_makeCG_algorithm_1}}
%**********************************************

The discussion of Ex.~\ref{Example_CF_identifiability} in the previous section proceeded in two parts. 
First, steps $(a)$ to $(f)$ applied various rewrite rules to transform the counterfactual diagram into an equivalent, but much simpler diagram. 
Second, the simplified diagram then provided the basis for studying the identifiability of the counterfactual given $P_{*}$. 
This two-part approach has a direct analogue in the conventional treatment of counterfactuals in the literature. 
In order to make this correspondence more explicit and, above all, to provide a general recipe in diagrammatic terms, this section presents the following algorithm that abstracts the sort of diagrammatic simplifications as used in Ex.~\ref{Example_CF_identifiability}.   

\vspace*{0.2cm}
\noindent \rule{\textwidth}{0.03cm} \\[0.05cm]
\noindent function \ouralgo\big($D$, $V$, $\pi$\big): \\[-0.15cm]
\noindent \rule{\textwidth}{0.01cm} \\
\noindent INPUT: \textit{Diagram $D$ in $\catC$ given by the network diagram of a parallel worlds model, possibly composed with some effects, 
the set $V$ of variables of the functional model $\modelM$ that underlies the parallel worlds model and $\pi$ a topological ordering\footnote{Given a DAG $G$ with vertices $V$ a \emph{topological order} on $V$ is a total order that is compatible with $G$, i.e. such that $X < Y$ whenever $X$ is an ancestor of $Y$ in $G$.} on $V$ for the DAG $G_{\modelM}$.} \\[0.1cm]
OUTPUT: \textit{Simplified, but equivalent diagram $D$.} \\[-0.5cm]
\begin{enumerate}
	\item \textit{Let all discards `fall through', i.e. make iterative use \rlb{of the defining property of channels (Def.~\ref{Def_Channel})} and drop discarded wires wherever connected to a copy map (Def.~\ref{def:cd_category}) until no discards are left in the diagram.}
	\item \textit{Wherever a sharp effect is connected to a copy map use Eq.~\eqref{eq:sharp-state-eff} to `separate' all the involved wires from each other.} 
	\item \textit{Starting with the lowest root node, iteratively go through the variables $L \in V$ in the order $\pi$ and apply the below steps for the respective variable $L$.} 
	\begin{enumerate} 
		\item[($\alpha$)] \textit{Consider all those $m$ appearances of the functional mechanism $f_L$ from across the different worlds that share their inputs in the sense as on the left-hand side below and rewrite $D$ accordingly:}
		
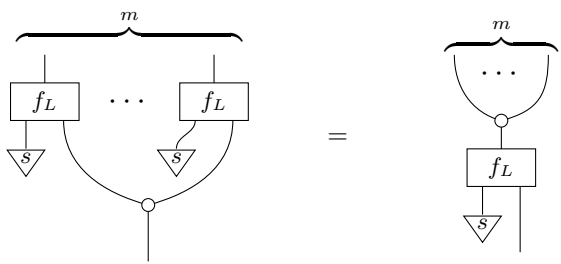
\begin{equation}
			\begin{minipage}{5cm}
				\centering
				\input{Fig_Alg_1_b_LHS.tikz} 
			\end{minipage}			
			= 
			\begin{minipage}{4cm}
				\centering
				\input{Fig_Alg_1_b_RHS.tikz} 
			\end{minipage}	
			\nonumber
		\end{equation}
		\item[($\beta$)] \textit{If sharp effects are then connected to the output of $f_L$ via a copy map, rewrite as: }
		
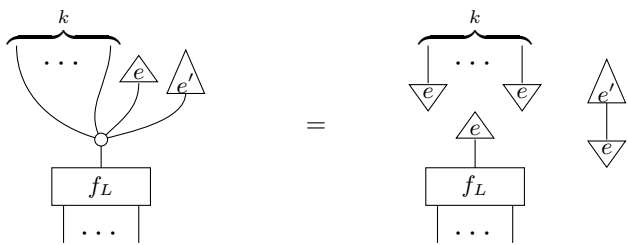
\vspace*{-0.2cm}
		\begin{equation}
			\begin{minipage}{5cm}
				\centering
				\input{Fig_Alg_2_b_LHS.tikz} 
			\end{minipage}			
			= 
			\begin{minipage}{5cm}
				\centering
				\input{Fig_Alg_2_b_RHS.tikz} 
			\end{minipage}	
			\nonumber
		\end{equation}
	\end{enumerate}
	\vspace*{-0.4cm}
	\item \textit{For any $X \in V$ such that $f_X$ appears in $D$ with the state $\lambda_X$ directly fed in, that is, rather than `sharing' it with other occurrences of $f_X$ via a copy map, replace it with $c_X$ as in Eq.~\eqref{Eq_CBN_SCM_Relation}.}
	\item \textit{Output the rewritten diagram $D$.}
\end{enumerate}
\vspace*{-0.4cm}
\rule{\textwidth}{0.03cm} \\

\st{This algorithm takes as input, and then simplifies, the sort of diagram which arises from a counterfactual, though without the normalisation box, since none of these simplifications are to do with normalisation.} So considering a counterfactual diagram $C = \normmor{D}$ for \cfterms\ $W^k = \stdwts$, where $D$ is the corresponding diagram without normalisation, it is $D$ that is rewritten by the algorithm. Note that, in $\MatR$, $D$ represents the state 
$P_{\model{M}^k_{s}} \big(E^{(1)},...,E^{(k)}, \ \allowbreak C^{(1)} = c^{(1)}, \allowbreak ..., \allowbreak C^{(k)} = c^{(k)} \big)$, 
which is not yet a correctly normalised probability distribution over $E^{(1)},...,E^{(k)}$, but a function that returns the corresponding probability for every value assignment of $E$. The counterfactual $C$ is then the corresponding normalised conditional distribution.  

A key idea of the algorithm is to do rewriting following a topological order $\pi$: while to begin with, for some $L$ the different appearances of $f_L$ may not have all their inputs connected to copy maps, this may become the case as one keeps copying through functions according to $\pi$, with copy maps thus `moving up'. 
That the algorithm is sound, meaning that the output is indeed equivalent to the input $D$, is manifest seeing as it only uses equivalences that generally hold in a cd-category $\catC$.  

For illustration of the algorithm, consider again Ex.~\ref{Example_CF_identifiability}. Step (a) of the rewriting in that example is the result of steps 1 and 2 of \ouralgo. 
The result of these steps is reproduced on the left-hand side below, now arranged according to the topological order $R,X,W,D,Z,Y$ indicated by \rlb{the corresponding blue slicing.} 
It is then simplified according to step 3 of \ouralgo, where $(\alpha)$ and $(\beta)$ do nothing to $X,W$ and $Y$ and $(\beta)$ does nothing to $R$. 
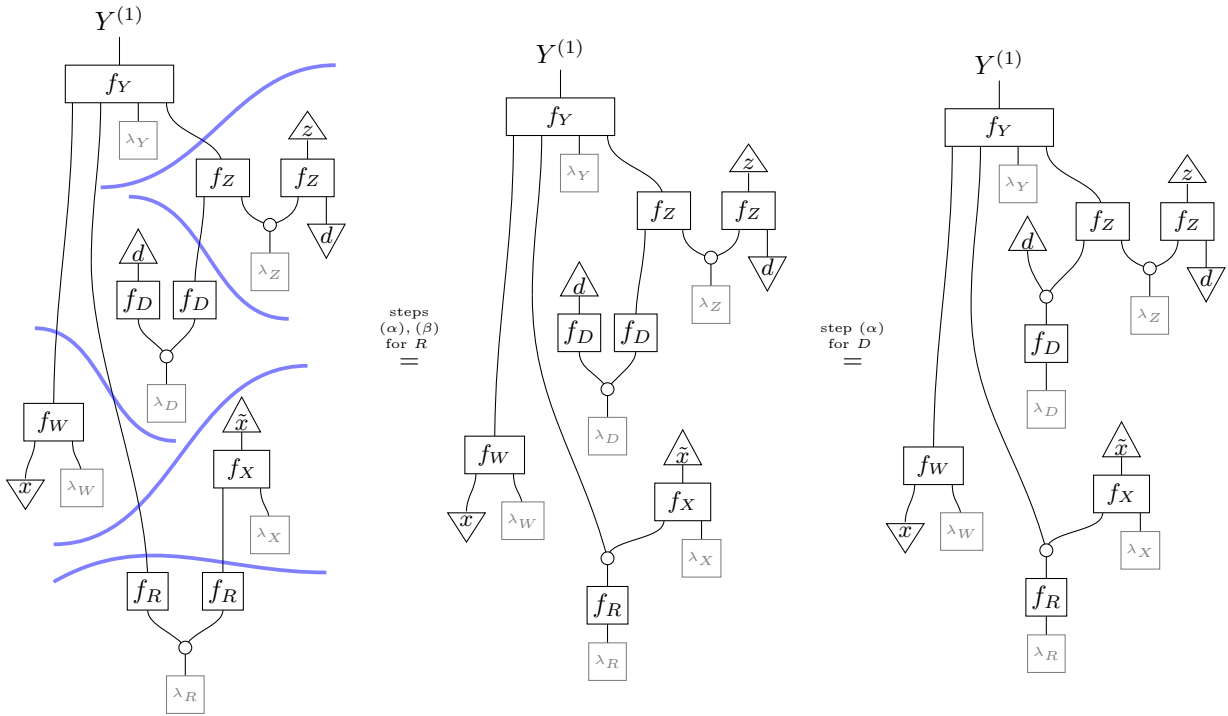
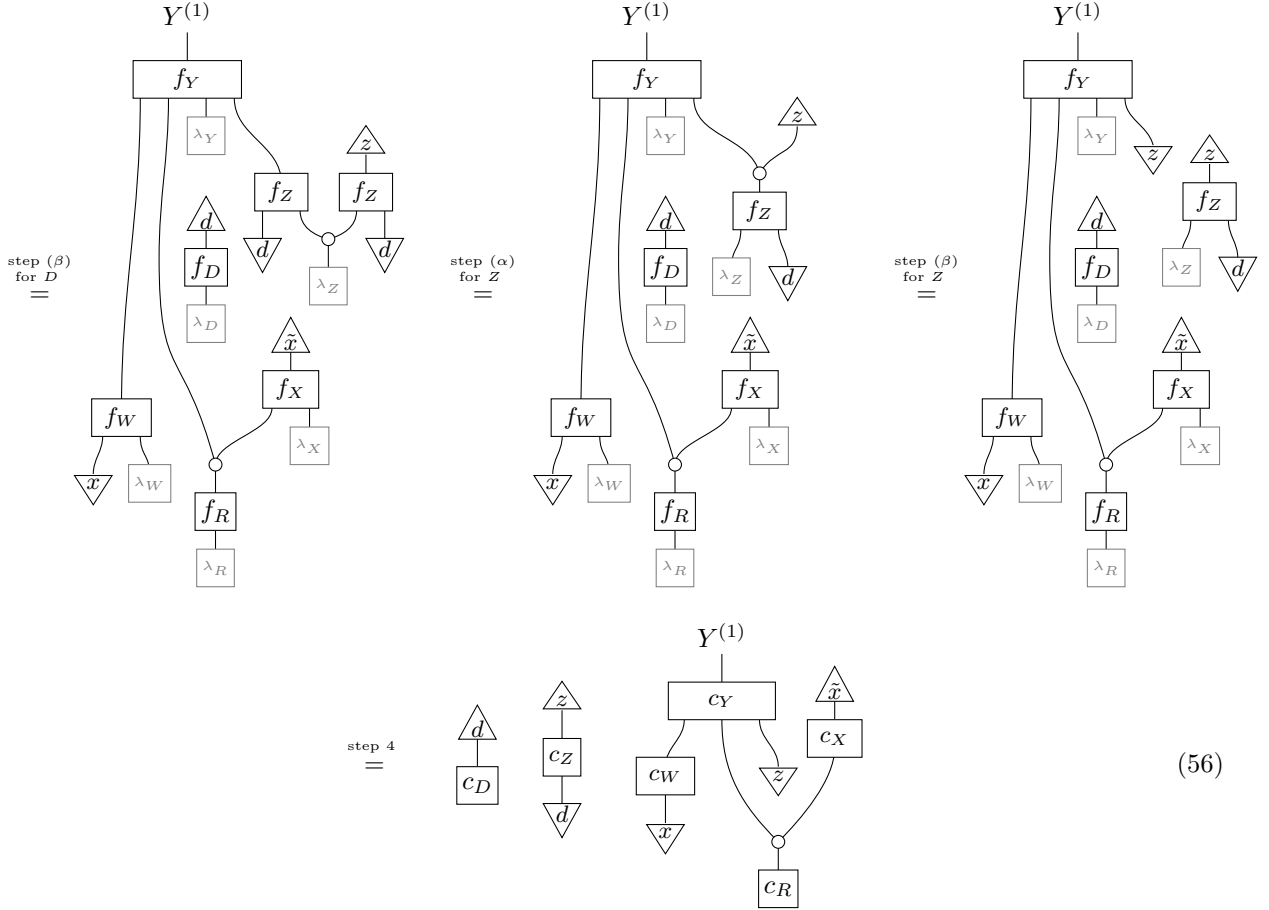
\begin{equation}
	\begin{minipage}{4.0cm} \centering \hspace*{-0.6cm}
	\input{Fig_Alg_Ex_1_sliced_c.tikz} 
	\end{minipage}	
	\stackrel{\begin{minipage}{1.5cm} \centering \tiny steps \\ $(\alpha), (\beta)$ \\ for $R$ \vspace*{0.1cm} \end{minipage}}{=} 
	\begin{minipage}{4.2cm} \centering \hspace*{-0.5cm}
	\input{Fig_Alg_Ex_2.tikz} 
	\end{minipage}
	\stackrel{\begin{minipage}{1.5cm} \centering \tiny step $(\alpha)$ \\ for $D$ \vspace*{0.1cm} \end{minipage}}{=} 
	\begin{minipage}{4.5cm} \centering \hspace*{-0.8cm}
	\input{Fig_Alg_Ex_3.tikz} 
	\end{minipage} \nonumber 
\end{equation}
\begin{equation}
	\stackrel{\begin{minipage}{1.5cm} \centering \tiny step $(\beta)$ \\ for $D$ \vspace*{0.1cm} \end{minipage}}{=} 
	\begin{minipage}{4.2cm} \centering \hspace*{-0.6cm}
	\input{Fig_Alg_Ex_3b.tikz}
	\end{minipage}
	\stackrel{\begin{minipage}{1.5cm} \centering \tiny step $(\alpha)$ \\ for $Z$ \vspace*{0.1cm} \end{minipage}}{=} 
	\begin{minipage}{4.2cm} \centering \hspace*{-0.8cm}
	\input{Fig_Alg_Ex_4.tikz}
	\end{minipage}
	\stackrel{\begin{minipage}{1.5cm} \centering \tiny step $(\beta)$ \\ for $Z$ \vspace*{0.1cm} \end{minipage}}{=} 
	\begin{minipage}{4cm} \centering \hspace*{-0.8cm}
	\input{Fig_Alg_Ex_4_b.tikz}
	\end{minipage} \nonumber
\end{equation}
\vspace*{0.2cm}
\begin{equation} 
	\stackrel{\begin{minipage}{1.5cm} \centering \tiny \rlb{step $4$} \vspace*{0.1cm} \end{minipage}}{=} 
	\begin{minipage}{6cm} \centering 
	\input{Fig_Alg_Ex_5_UsMarginalised.tikz}
	\end{minipage} \label{Eq_AlgoAppliedToExample_finalStep}
\end{equation}
Clearly, adding a normalisation box around the entire output state yields precisely the expression from the right-hand side of $(f)$ in Ex.~\ref{Example_CF_identifiability}. 

Note that step 4 of \ouralgo\ does generally \emph{not} get rid of all the occurrences of the states $\lambda_X$ of the background variables $U_X$, through which the different worlds are connected -- a fact that is closely related to identifiability criteria, and discussed further in the next section. 

\begin{remark} \textnormal{(Comments on \ouralgo):} 
Some further useful observations about \textnormal{\ouralgo}:
	\begin{enumerate}
		\item There is no ambiguity concerning which world a variable pertains to when using the rewrite rules in steps $3(\alpha)$ and $3(\beta)$ -- any variable referenced by the counterfactual of interest is either an open wire, namely $E^{(1)},...,E^{(k)}$, and hence still has its label, or else is being conditioned on and the corresponding value indicates the world it pertains to, namely $c^{(1)}, \allowbreak ..., \allowbreak c^{(k)}$. 
		\item In step 3$(\alpha)$ the wire of the state $s$ and the open input wire may be composite, i.e. stand for a set of variables, in which case it is understood as a product of corresponding sharp states and a product of copy maps, respectively.
		\item The output diagram of \textnormal{\ouralgo} is of the form of a product of scalars and a (product) state and there will be no discards \ $\discard{}$ in the output due to the special structure of network diagrams composed of copy maps and single-output channels. 
	\end{enumerate}
\end{remark}

\begin{remark} \textnormal{(Relation to \makecg\ algorithm):} \label{Rem_RelationToMakeCGAlgorithm}
Recalling Rem.~\ref{Rem_CF_Notion}, given an ADMG $G$ and expressions $\gamma, \delta$, 
the algorithm \texttt{IDC}$^*(G, \gamma, \delta)$ by Shpitser and Pearl \cite{ShpitserEtAl_2008_CompleteIdentificationMethodCausalHierarchy} outputs whether $P(\gamma | \delta)$ is identifiable and if so what it is. 
Suppose $\gamma \wedge \delta$ and $\delta$ are not inconsistent -- if they were, this is accounted for through our normalisation boxes -- 
then the \textnormal{\makecg}$(G, \gamma \wedge \delta)$ subroutine of \texttt{IDC}$^*$ constructs a \emph{parallel worlds graph}, then simplifies it by merging and relabelling nodes based on the assumed underlying functional relationships and outputs a corresponding ADMG $G'$, called a \emph{counterfactual graph}. Identifiabilty is judged on the basis of, \rl{in particular,} $G'$. 
Without spelling out the correspondence in a formal and detailed manner, the relation to \textnormal{\ouralgo} is: 
if $D_{\tilde{\rho}(G')}$ is the network diagram obtained from rootifying the counterfactual graph $G'$ 
then marginalisation over any variable not in $\gamma \wedge \delta$ and composing with sharp effects $c^{(1)}, ..., c^{(k)}$ according to $\delta$ gives a diagram equivalent to the output of \textnormal{\ouralgo} for the corresponding counterfactual diagram.
Intuitively, it is not hard to see that the `copying through' of functions $f_L$ following a topological order implements the same as the `merging of vertices' as part of the \textnormal{\makecg} algorithm. 
\end{remark}

%**********************************************
\subsubsection{The identification algorithm \protect\ouridalgo \label{Sec_CF_id_criteria}}
%**********************************************

The goal of this section is to state an algorithm that generalises the second part of Ex.~\ref{Example_CF_identifiability}, that is, a procedure that takes the diagram simplified by \ouralgo\ and then checks whether it is possible to rewrite it in terms of $P_*$ and, if so, actually does the rewriting. 
One may think of this as the natural translation into diagrammatic terms of the core ideas, especially the necessary and sufficient condition for identifiability, in the algorithmic solution that Shpitser and Pearl present in Ref.~\cite{ShpitserEtAl_2008_CompleteIdentificationMethodCausalHierarchy}. 
To this end, the below first discusses two pedagogical examples from Ref.~\cite{ShpitserEtAl_2008_CompleteIdentificationMethodCausalHierarchy} for non-identifiability, which help build the intuition for the general condition for identifiability that will feature in the algorithm. 
While the algorithm will be stated relative to any suitable cd-category $\catC$ just as our treatment so far, the following two examples from the literature assume $\MatR$ as the ambient category.

\begin{example} 
First, consider again the aspirin and headache example from Sec.~\ref{Sec_CF_IntroductoryExample}. 
Write superscript $(1)$ for the actual world and $x$ and $y$ \rl{instead of 0 and 1} for the conditioning in that world, 
and $(2)$ for the counterfactual world defined by $\Do(X^{(2)} = \tilde{x})$. The counterfactual diagram for $P_{\model{M}^2_{s}}(Y^{(2)} | X^{(1)}=x, Y^{(1)}=y)$ 
(up to normalisation and renaming the same as Fig.~\ref{Fig_CF_Intro_Example_step5}) 
is shown below on the left-hand side together with the output of \textnormal{\ouralgo}.
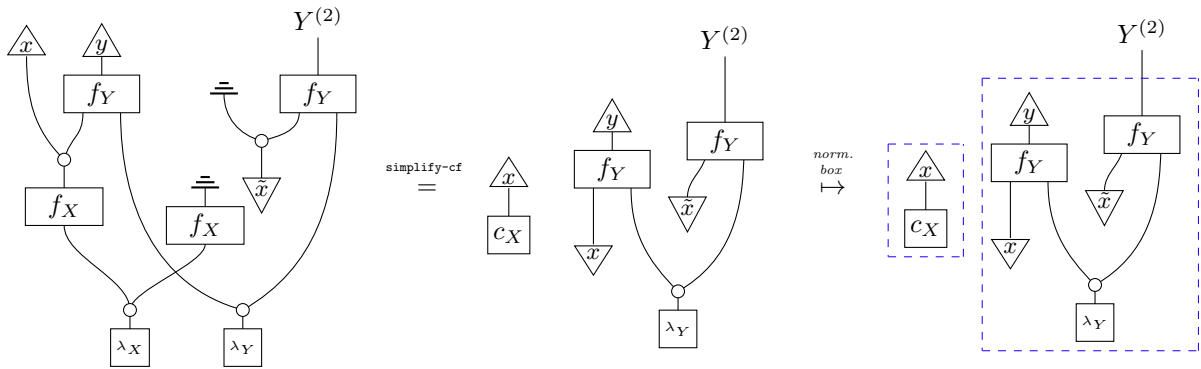
\begin{equation}
	\begin{minipage}{5.0cm} \centering 
	\input{Fig_Alg_Ex_b_1.tikz} 
	\end{minipage}  
	\stackrel{\begin{minipage}{1.2cm} \centering \tiny \textnormal{\ouralgo} \vspace*{0.1cm} \end{minipage}}{=} 
	\begin{minipage}{4.0cm} \centering 
	\input{Fig_Alg_Ex_b_2.tikz} 
	\end{minipage} 
	\stackrel{\begin{minipage}{1.2cm} \centering \tiny norm. \\ box \vspace*{0.1cm} \end{minipage}}{\mapsto} 
	\begin{minipage}{4.2cm} \centering 
	\input{Fig_Alg_Ex_b_3.tikz} 
	\end{minipage}\nonumber
\end{equation}
As argued in Ref.~\cite{ShpitserEtAl_2008_CompleteIdentificationMethodCausalHierarchy} this counterfactual is not identifiable as long as $x \neq \tilde{x}$. 
Looking at the simplified diagram it is obvious why: 
for $x \neq \tilde{x}$ the two boxes $f_Y$ cannot be moved down through the copy map, leading to this outcome of \textnormal{\ouralgo}. 
$P_*$ contains $c_X$ and $c_Y$, but since $U_Y$ is not observed, \rl{$P_*$ does not allow one to infer how $Y$ depends on $U_Y$ through $f_Y$, which would however be needed to identify the sought counterfactual.} 
\end{example}

The intuition in the above example is that whenever a background variable, like $U_Y$, cannot be `simplified away' as was the case in Ex.~\ref{Example_CF_identifiability}, but remains in the diagram output by \ouralgo, this constitutes an `irreducible' correlation across distinct worlds and the counterfactual is not identifiable from $P_*$. 

\begin{example} 
For a second example, consider the ADMG $G$ in Fig.~\ref{Fig_Fig_CF_NonId_Ex_2_1}, the rootification $\tilde{\rho}(G)$ in Fig.~\ref{Fig_Fig_CF_NonId_Ex_2_2} 
and \rl{a corresponding} FCM $\modelM$ in Fig.~\ref{Fig_Fig_CF_NonId_Ex_2_3}. Consider the state $D$ representing $P_{\model{M}^3_s}\big( W_1^{(1)}, W_2^{(1)}, Y^{(2)}, Z^{(3)}\big)$ that arises from an $\modelM$-based parallel worlds model and world terms $W^3$ with world (1) being the original one, i.e. $S_1=\emptyset$, and worlds (2) and (3) defined by $S_2 = \{X\} = S_3$ with $s_2=x$ and $s_3=x'$, respectively.
Note that these world terms do not contain any conditioning -- they are not \cfterms, but would become such by choosing appropriate sharp effects. 
\begin{figure}[H]
	\centering
	\begin{subfigure}{0.22\textwidth}
		\centering
		\input{Fig_CF_NonId_Ex_2_1.tikz} 
		\caption{\label{Fig_Fig_CF_NonId_Ex_2_1}}
	\end{subfigure}
	\begin{subfigure}{0.25\textwidth}
		\centering
		\input{Fig_CF_NonId_Ex_2_2.tikz} 
		\caption{\label{Fig_Fig_CF_NonId_Ex_2_2}}
	\end{subfigure}
	\begin{subfigure}{0.5\textwidth}
		\centering
		\input{Fig_CF_NonId_Ex_2_3.tikz} 
		\caption{\label{Fig_Fig_CF_NonId_Ex_2_3}}
	\end{subfigure}
	\caption{\label{Fig_Fig_CF_NonId_Ex_2}}
\end{figure}

The simplified diagram $D$ output by \textnormal{\ouralgo} is depicted in Fig.~\ref{Fig_Fig_CF_NonId_Ex_2_4}. As argued in Ref.~\cite{ShpitserEtAl_2008_CompleteIdentificationMethodCausalHierarchy} the distribution $P_{\model{M}^3_s}( W_1^{(1)}, W_2^{(1)}, Y^{(2)}, Z^{(3)})$ is not identifiable from $G$ and $P_*$. 
Observe that this time for every (variable associated with a) vertex $A$ of $\tilde{\rho}(G)$, \textnormal{\ouralgo} absorbed $\lambda_A$ into the corresponding channel $c_A$. Nonetheless this is a non-identifiable case. In order to make this intuitive, Fig.~\ref{Fig_Fig_CF_NonId_Ex_2_4} marks in gray all wires corresponding to the `unobserved' variables $R_1,R_2,R_3$ introduced in the rootification $\tilde{\rho}(G)$. 
Note that the set $\{Y, W_1, W_2, Z\}$ forms a c-component in $G$ (see Sec.~\ref{Sec_JKZResult}) with $R_1,R_2,R_3$ being the explicit latent structure. 
While $P(Y, W_1, W_2, Z ; \Do(X))$, which is shown in Fig.~\ref{Fig_CF_NonId_Ex_2_5}, by definition is in $P_*$, the state in Fig.~\ref{Fig_Fig_CF_NonId_Ex_2_4} is not and also is not identifiable from it, because $c_Y$ and $c_Z$ are fed distinct states $x$ and $x'$. 
To identify the sought state one would need $c_Y$ and $c_Z$ independently, which however is in conflict with $R_1$ and $R_3$ being unobserved. 

The same conclusion of non-identifiability applies immediately to any counterfactual arising from $D$ by additionally conditioning on some of the variables $W_1^{(1)}, W_2^{(1)}, Y^{(2)}, Z^{(3)}$.  

\begin{figure}[h]
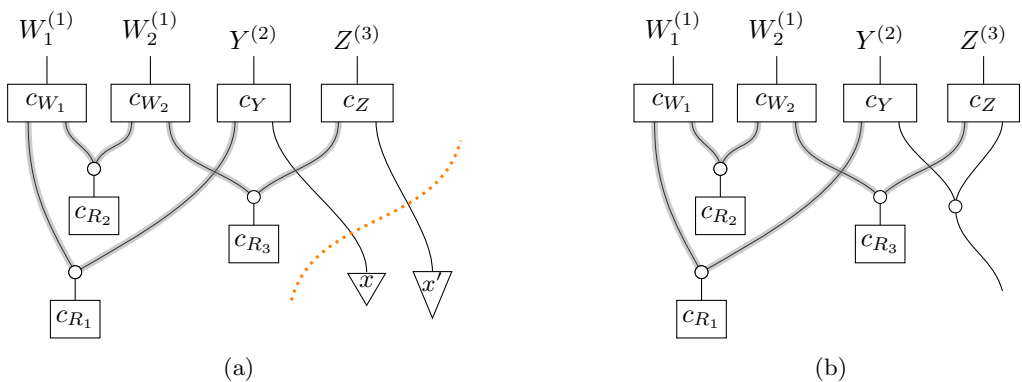

	\centering
	\begin{subfigure}{0.4\textwidth}
		\centering
		\input{Fig_CF_NonId_Ex_2_4.tikz} 
		\caption{\label{Fig_Fig_CF_NonId_Ex_2_4}}
	\end{subfigure}
	\hspace*{1cm}
	\begin{subfigure}{0.4\textwidth}
		\centering
		\input{Fig_CF_NonId_Ex_2_5.tikz} 
		\caption{\label{Fig_CF_NonId_Ex_2_5}}
	\end{subfigure}
	\caption{(a) shows the output $D$ of \textnormal{\ouralgo}\ for $P_{\model{M}^3_s}\big( W_1^{(1)}, W_2^{(1)}, Y^{(2)}, Z^{(3)}\big)$. 
		The dotted orange line delineates an $R$-fragment of $D$ (see Def.~\ref{Def_RFragment} below); 
		(b) shows  $P(Y, W_1, W_2, Z ; \Do(X))$.}
\end{figure}
\end{example}

\st{The previous example suggests it will be helpful to introduce a notion to refer to a part of a network diagram that corresponds to a confounded path in an ADMG.}

\begin{definition} \label{Def_RFragment}
Given an ADMG $G$ with vertices $O$, let $R$ be the set of additional root nodes introduced by the rootification $\tilde{\rho}(G)$ 
and let $D_{\tilde{\rho}(G)}$ be a corresponding network diagram in $\catC$. 
Suppose $D$ is a string diagram composed of only mechanisms that appear in $D_{\tilde{\rho}(G)}$, copy maps, as well as states and effects for objects in $O$.\footnote{Equality of mechanisms is as morphisms in $\catC$, where it does not matter how the wire is labelled so long as they refer to the same objects.}
Then an \emph{$R$-fragment} of $D$ is a maximal fragment of $D$ composed of those and only those mechanisms $\{c_X\}_{X \in S_O \cup S_R}$ 
(and copy maps) for some $S_O \subseteq O$ and $S_R \subseteq R$ such that 
\begin{itemize}
	\item there is a confounded path\footnote{Recall from Sec.~\ref{Sec_ADMGs} that a confounded path in an ADMG is a path consisting of only bi-directed edges.} in $G$ between any pair of vertices of $S_O$ with all vertices on that path inside $S_O$ and the bi-directed edges on that path \rlb{corresponding to root nodes of $\tilde{\rho}(G)$ contained in $S_R$;} 
	\item the wiring \emph{between} those mechanisms in $S_O$ and those in $S_R$ is as in $D_{\tilde{\rho}(G)}$ (though no condition regarding other inputs and outputs of the mechanisms $c_X$ for $X \in S_O$).
\end{itemize} 
\end{definition}

\begin{example} 
The fragment of the diagram in Fig.~\ref{Fig_Fig_CF_NonId_Ex_2_4} obtained from cutting along the dashed orange line, is an $R$-fragment of that diagram  relative to the ADMG $G$ in Fig.~\ref{Fig_Fig_CF_NonId_Ex_2_1} with $R=\{R_1,R_2,R_3\}$ as in Fig.~\ref{Fig_Fig_CF_NonId_Ex_2_2}. 
\end{example} 

\begin{example} 
Consider again Ex.~\ref{Example_CF_identifiability}. For the single additional root node $R = \{ R\}$ in this example, the four $R$-fragments for the output of \textnormal{\ouralgo} from Eq.~\ref{Eq_AlgoAppliedToExample_finalStep}, separated by dashed orange slicing, are as follows:
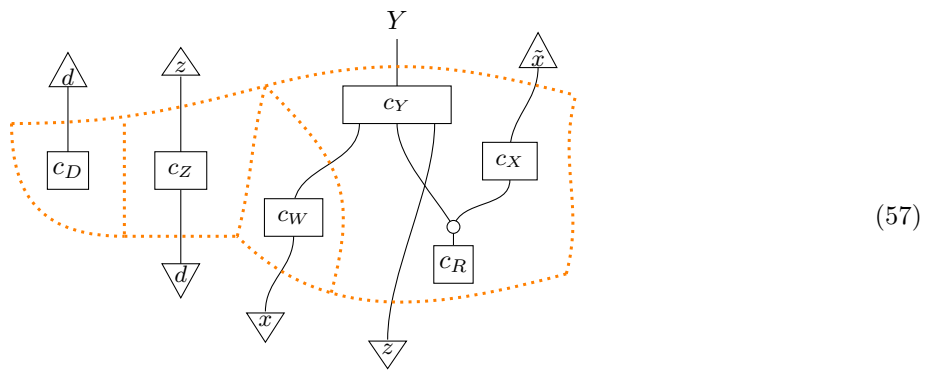
\begin{equation}
%	\begin{minipage}{7cm} \centering \includegraphics[scale=1.5]{figures/Fig_Alg_Ex_5_UsMarginalised_sliced.pdf} \end{minipage}
	\input{Fig_Alg_Ex_5_UsMarginalised_sliced.tikz} 
	\label{Eq_Fig_Alg_Ex_5_UsMarginalised_sliced}
\end{equation}
\end{example}

Given the examples already discussed it now is not difficult to state an algorithm that incorporates the same ideas as in the algorithmic solution in Shpitser \etal\ \cite{ShpitserEtAl_2008_CompleteIdentificationMethodCausalHierarchy} into our diagrammatic treatment. 
We give a proof (sketch) of the soundness of this algorithm below, but it is worth pointing out that this proof is almost trivial -- the algorithm does not require any non-obvious facts or manipulations that would need a separate proof, but relies entirely on the rewrite rules in a cd-category with \diagconditioning. 

\vspace*{0.2cm}
\noindent \rule{\textwidth}{0.03cm} \\[0.05cm]
\noindent function \ouridalgo\big($G$, $W^k$, $P_*(O)$\big): \\[-0.15cm]
\noindent \rule{\textwidth}{0.01cm} \\
\noindent INPUT: \textit{ADMG $G$ with vertices $O$ in $\catC$, \cfterms\ $W^k = \stdwts$ and set $P_*(O)$.}\\[0.1cm]
OUTPUT: \textit{FAIL or the counterfactual $C$ corresponding to $W^k$, assuming $G$, and expressed in terms of $P_*(O)$.} \\[-0.4cm]
\begin{enumerate}
	\item \textit{Let $R$ be the set of additional root nodes introduced by the rootification $\tilde{\rho}(G)$. 
Let $\model{M} = \model{F} \circ \model{L}$ be a corresponding FCM \rl{with endogenous variables $V := O \cup R$ and} background variables $\{U_X\}_{X\in V}$ such that $\pi_O(G_{\model{M}}) = G$ and such that it is compatible with $P_{*}$.   
For any $X \in V$ write $c_X$ for the corresponding probabilistic mechanism obtained from feeding $\lambda_X$ into $f_X$ as in Eq.~\eqref{Eq_CBN_SCM_Relation}. 
Finally, let $C$ be the counterfactual defined by  $W^k$ on the basis of $\model{M}$ and let $D$ be the same diagram up to normalisation, $C=\normmor{D}$.} 
	\item  $D=$\textnormal{\ouralgo}$(D,V,\pi)$ for some topological order $\pi$ for $\tilde{\rho}(G)$. 
	\item \textit{If $\exists X \in V$ s.th. $\lambda_X$ was not absorbed into $c_X$ by \textnormal{\ouralgo}, output \textnormal{FAIL}, otherwise continue.}
	\item \textit{For $F_l \in \{F_1,...,F_m\}$, the set of $R$-fragments of $D$:}
%	\textit{Let $\{F_1,...,F_m\}$ be the set of $R$-fragments of $D$.} \\[0.2cm] 
%	\textit{For $l=1,...,m$:}
		\begin{enumerate}[label=\arabic*.]
			\item \textit{For each $X \in O$ such that $X$ appears as the type of a wire in $F_l$: }
				\begin{enumerate}[label=\arabic*.]
					\item \textit{If mechanism $c_X$ is a component of $F_l$:}
						\begin{enumerate}[label=\alph*.]
							\item \textit{If the output wire of $c_X$ does \emph{not} have some sharp effect $x$ on it, i.e. is fed into some other $R$-fragment, or is an output of $D$, and no other $X$ type wire appears in $F_l$, do nothing.}  
							\item \textit{If the output wire of $c_X$ is composed with some sharp effect $x$ in $D$ and any other wires of type $X$ in $F_l$ all have the state $x$ fed into them, rewrite $D$ according to:} 
								\begin{equation}
									\input{Proof_CF_Id_Thm_Rewrite_1_LHS.tikz} \hspace*{0.5cm} \mapsto \hspace*{0.5cm} \label{Eq_Cf_id_Rewrite_1} \input{Proof_CF_Id_Thm_Rewrite_1_RHS.tikz}
								\end{equation} 
							\item \textit{Else output \textnormal{FAIL}.}   
						\end{enumerate}
					\item \textit{If mechanism $c_X$ is \emph{not} a component of $F_l$: }
						\begin{enumerate}[label=\alph*.]
							\item \textit{If all $X$ type wires, input to $F_l$, are connected via copy maps to the same output of $c_X$ in some other $R$-fragment, do nothing.}  
							\item \textit{If all $X$ type wires, input to $F_l$, are fed the same state $x$, rewrite $D$ according to:}
								
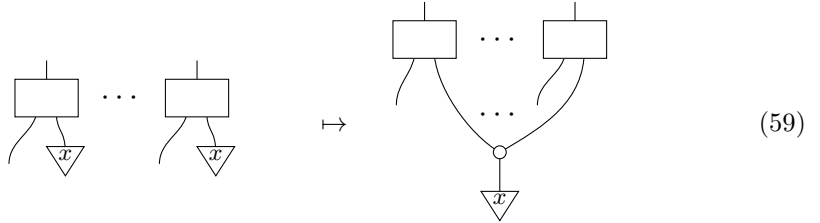
\begin{equation}
									\begin{minipage}{4cm} \centering \input{Proof_CF_Id_Thm_Rewrite_2_LHS.tikz} \end{minipage}
									 \hspace*{0.5cm} \mapsto \hspace*{0.5cm} 
									 \input{Proof_CF_Id_Thm_Rewrite_2_RHS.tikz} \label{Eq_Cf_id_Rewrite_2}
								\end{equation}
							\item \textit{Else output \textnormal{FAIL}. }
						\end{enumerate}
				\end{enumerate}
			\item \textit{Replace the thus rewritten $R$-fragment $\widetilde{F}_l$ -- same as $F_l$ up to more copy maps -- according to}: 
				\begin{equation}
					 \input{CF_id_Algo_channel.tikz} \label{Eq_Cf_id_Rewrite_3} \hspace*{0.3cm} ,
				\end{equation}
				\textit{where $D_l$, $F_l^{\text{in}}$, $C_l$ and $F_l^{\text{out}}$ are the sets of objects of wires going into $\widetilde{F}_l$ with sharp states fed into them, the remaining inputs to $\widetilde{F}_l$, wires coming out of $\widetilde{F}_l$ with sharp effects on them,  
			 and the remaining outputs of $\widetilde{F}_l$, respectively.} 
		\end{enumerate} 
	\item \textit{Output $\normmor{D}$.}
\end{enumerate}
\vspace*{-0.3cm}
\rule{\textwidth}{0.03cm} \\

For illustration consider once more Ex.~\ref{Example_CF_identifiability} and in particular Eq.~\ref{Eq_Fig_Alg_Ex_5_UsMarginalised_sliced}, where the slicing indicates the $R$-fragments of the output of \ouralgo.   
The conditions in steps $(3.)$ and $(4.1.)$ of \ouridalgo\  are all met with no rewrites taking place. 
Observe that the equalities in Figs.~\ref{Fig_CF_id_main_example_id_components_a}-\ref{Fig_CF_id_main_example_id_components_d}, which we used earlier, correspond precisely to the rewrites that \ouridalgo\ would do for this example in step $(4.2.)$.  
The algorithm thus outputs the correct identifying expression from Eq.~\eqref{Eq_CF_id_main_example_final_solution}. 

The key idea in step $(4.)$ of the algorithm, even if spelling it out precisely required some space, is simple: 
check whether within any $R$-fragment $F_l$ any variable $X$ is, intuitively speaking, necessarily set to the same value $x$ everywhere. 
This might be achieved due to being copied out from another $R$-fragment that has the mechanism $c_X$, due to separate, but coinciding states fed into them, or due to being consistently conditioned on this $x$. 
Whatever way it is achieved that this $X$ has agreeing values everywhere, \rlb{if it is the case for all variables $X$ within an $R$-fragment, then that $R$-fragment is identifiable.} 

\begin{theorem} \label{Thm_CF_Algo_Soundness} \textnormal{(Soundness of \ouridalgo):}
	The algorithm \textnormal{\ouridalgo} is sound for the identification of counterfactuals, i.e. whenever it does not output FAIL, the output $C$ in terms of $P_*$ is the correct counterfactual. 
\end{theorem}

\noindent \textit{Proof (sketch):} see App.~\ref{App_proof_Thm_CF_Algo_Soundness}. \hfill $\square$ \\ 

\begin{corollary} \label{Cor_CF_Id_Condition} 
Given ADMG $G$ with vertices $O$ in $\catC$, \cfterms\ $W^k$ and $P_*(O)$, 
then the counterfactual $C$ corresponding to query $W^k$ is identifiable from $G$ and $P_{*}$ if for $D$, the corresponding unnormalised diagram simplified by \textnormal{\ouralgo}, it holds that:
	\begin{enumerate}
		\item[\textnormal{($\alpha$)}] $\forall X\in O \cup R$ any $\lambda_X$ was absorbed into $c_X$.
		\item[\textnormal{($\beta$)}] in any $R$-fragment $F_l$ and for any $X \in O$ the conditions $(4.1.1.a)$ or $(4.1.1.b)$ and $(4.1.2.a)$ or $(4.1.2.b)$ are met, i.e., intuitively speaking, all occurrences of $X$ in $F_l$ have agreeing value.   
	\end{enumerate}
\end{corollary}

The obvious question now of course concerns the \emph{completeness of the algorithm} \ouridalgo. 
Proving completeness requires establishing that whenever the algorithm outputs FAIL, i.e. whenever at least one of the conditions $(\alpha)$ and $(\beta)$ of Cor.~\ref{Cor_CF_Id_Condition} fails, that the counterfactual is not identifiable, that is, constructing an example of two distinct interpretations of the diagram $D$ output by \ouralgo\ such that both are compatible with $P_*$, but imply distinct counterfactuals. 
As such it clearly depends on more structure than $\catC$ being a cd-category with \diagconditioning\ -- one needs `enough maps' to be guaranteed to always be able to construct a counterexample.
However, in the category $\MatR$ it is clear that completeness will hold. 
First, the above algorithm and with it the identifiability criteria in Cor.~\ref{Cor_CF_Id_Condition} only translate into our diagrammatic approach the results from Ref.~\cite{ShpitserEtAl_2008_CompleteIdentificationMethodCausalHierarchy}, where the equivalence between identifiability and the condition in Cor.~\ref{Cor_CF_Id_Condition} -- just in their terms -- is proven. 
Second, also intuitively speaking, it is clear why that should be the case in our formulation for the case of $\catC = \MatR$. 
The above two examples of non-identifiability expose this intuition for both sort of situations, when failing condition $(\alpha)$ or $(\beta)$  of Cor.~\ref{Cor_CF_Id_Condition}. 
The arguments as such that are needed to establish completeness are thus not really diagrammatic in nature. 
Spelling out the precise conditions on the category $\catC$ such that the proof methods for completeness in the literature like those in Ref.~\cite{ShpitserEtAl_2008_CompleteIdentificationMethodCausalHierarchy} go through and can be made obvious in our set-up may not be difficult, but we leave this to future work.   

\begin{remark} \textnormal{(The rationale of distinguishing conditions $(\alpha)$ and $(\beta)$):} \label{Rem_Cor_CF_Id_Condition} 
Note that the distinct treatment of, on the one hand, the latent common cause variables represented as bi-directed edges in the given ADMG and as root nodes $R$ in the corresponding rootification and, on the other hand, the FCM's background noise variables $U_X$ with each only a single child $X$ is our choice and differs from the typical presentation in the literature \cite{ShpitserEtAl_2008_CompleteIdentificationMethodCausalHierarchy}. 
It is a choice for conceptual reasons, not one of any difference in generality. 
It is not difficult to see\footnote{By proof of contradiction and following the topological order $\pi$ in \textnormal{\ouralgo}.} that if condition $(\alpha)$ of Cor.~\ref{Cor_CF_Id_Condition} is not met, i.e. the output of \textnormal{\ouralgo} still explicitly contains states of background noise variables, this must be because for some $X \in O$, there are occurrences of $f_X$ with distinct states fed into some of the parents of $X$ (other than $\lambda_X$). 
Hence, by simply dropping the distinction between the two kinds of unobserved variables, those in $R$ and those in $(U_X)_{X\in R\cup O}$, Cor.~\ref{Cor_CF_Id_Condition} would reduce to condition $(\beta)$ with respect to that entire set of unobserved variables. 
This work prefers to make the distinction, because unobserved common causes `within a world' and those `across worlds' may be regarded as leading to distinct reasons of the failure of identifiability. 
There will always be unobserved noise variables, introducing some stochasticity and making it unrealistic to learn the underlying functions $f_X$. 
Hence, if for a given causal structure some counterfactual is non-identifiable for a failure of condition $(\alpha)$, this may be distinguished from a failure of $(\beta)$ -- the latter could be circumvented if all causally relevant variables (i.e. common causes) within the actual world were observed, whereas the former is more principally unrealistic. 
\end{remark}

\begin{remark} \label{Rem_ADMGsAndNonID} \textnormal{(Role of ADMGs and rootification for non-identifiability):}
Finally, recall the question from Rem.~\ref{Rem_Rootification_issue_CF} whether working with latent projection ADMGs comes with any loss of generality when studying the identifiability of counterfactuals. 
To our knowledge this has not been answered in the literature. 
Provided one assumes the right kind of structure for $\catC$ that allows one to prove sufficiency of the conditions in Cor.~\ref{Cor_CF_Id_Condition}, then in light of footnote~\ref{Footnote_RootificationForCF}, this question may well be easily answerable by inspecting the consequences of working with the rootification method $\rho$ instead of $\tilde{\rho}$ for non-identifiablity. 
\end{remark}

%**********************************************
% \subsection{Natural generalisations of the notion of a counterfactual \label{Sec_CF_Further_Generalisation}}
\subsection{Generalisations of counterfactuals \label{Sec_CF_Further_Generalisation}}
%**********************************************

Let us finish by discussing several natural generalisations of the notion of counterfactual from Def.~\ref{Def_Counterfactual}. These are given by generalising both of the main features of counterfactuals: the interventions which define the parallel worlds, and the potential conditioning \rl{within worlds.} 

\paragraph{General interventions.} 
Firstly, when going from a parallel worlds model to a counterfactual there is, formally speaking, no reason to insist the interventions $\sigma_j$ be do-interventions. As argued above, the do-interventions in Def.~\ref{Def_Counterfactual} are not of \rl{an operational} significance, but only formalise the `small miracle' needed to miraculously have a different value for some variable. As such we can consider counterfactuals defined via an arbitrary parallel worlds model; these allow for any interventions $\sigma_j$ on the deterministic open model $\modelF$ for which $\sigma_j(\modelF)$ remains deterministic, beyond merely do-interventions. 

More broadly still, we could consider a generalisation of parallel worlds model defined just as in \eqref{eq:parallel-worlds}, but now allowing for arbitrary interventions $\sigma_j$. The resulting models $\sigma_j(\modelF)$ may no longer be deterministic so that each world's closed model $\sigma_j(\modelF) \circ \model{L}$ may no longer be an FCM. 

For either kind of generalised counterfactual, each branch can be seen as a hypothetically different world which differs from the actual world in the facts about the mechanisms of some variables, rather than just facts about the values they take. 
\rl{Such generalised notion may also correspond to contemplating, roughly, what would have happened, had one done some intervention $\sigma$, given that one actually did not do that and observed this and that and maybe also given observations from a previously done yet different intervention $\sigma'$.  
Note that the `all else the same' aspect is not lost through such general interventions seeing as that still is reflected in the fact that one starts from an FCM to build the generalised parallel worlds model in the same way with shared $U$ variables.}  

\paragraph{Soft conditioning.}

The second key feature of counterfactuals is the conditioning taking place \rl{within  worlds}. So far we considered only conditioning on event-like `facts' defined via sharp effects $c^{(j)}$ on $C^{(j)}$. However, one can condition a probability distribution $P$ not only on some sharp state $X=x$ to yield $P(Y|X=x)$, but also given `fuzzy facts' as representable by a general state (in $\MatR$ a distribution) or effect (in $\MatR$ a positive function) on $X$. 

A further natural generalisation is thus to consider conditioning via `soft evidence' given by an arbitrary state or effect $c^{(j)}$ over $C^{(j)}$. Such fuzzy conditioning corresponds to updating one's believes in light of some possibly imperfect knowledge of the values obtained in each world. 

However, an underappreciated fact from probability theory, which diagrams help to clarify, is that there are in fact two natural ways to define such `soft conditioning' \cite{jacobs2019mathematics}. Given a morphism $f \colon X \to Y \otimes Z$, we can define the \emph{upper conditional} $f|^e$ of $f$ by an arbitrary effect $e$ on $Z$, and the \emph{lower conditional} $f|_{\rho}$ of $f$ by an arbitrary state $\rho$ of $Z$, \rlb{respectively, as follows}\footnote{In a category without cancellative caps, the dashed box in $f|_\rho$ should be replaced with a more general notion of conditional channel as discussed in Appendix \ref{sec:cond-appendix}.}.
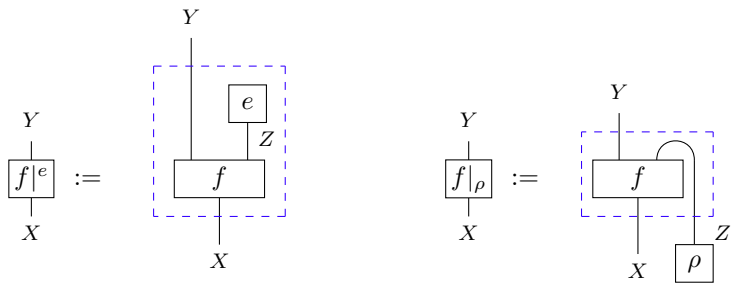
\[
\input{pearl-norm.tikz}
\qquad  \qquad \qquad 
\input{lower-cond.tikz}
\]
In Section \ref{sec:conditioning} we saw that when $z$ is a sharp state and $z^\dagger$ the corresponding sharp effect, both update procedures coincide and are given by the Bayesian conditional $f|_z = f|^{z^\dagger}$. But for a general state $\rho$, letting the effect $e$ be given by composing $\rho$ with a cap, we typically have $f|^e \neq f|_\rho$, owing to the different placement of the normalisation box.
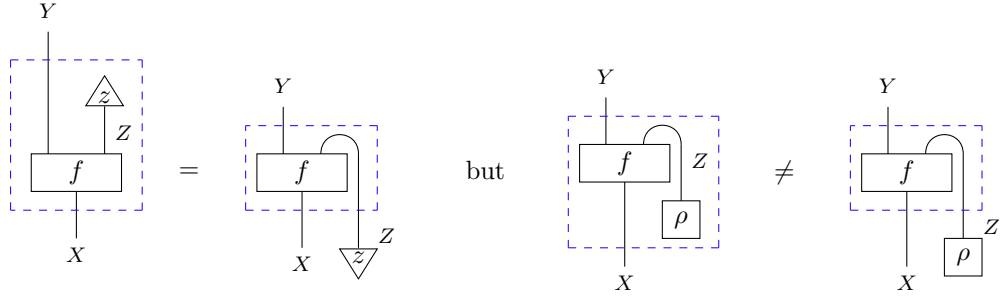
\[
\input{cond-by-z4-again.tikz} 
\qquad \text{ but } \qquad 
\input{pearldiff2.tikz}
\]
Jacobs has distinguished and explored both forms of soft conditioning or \emph{belief updating} in detail (in the case where $f$ is a state, and with $e$ outside the diagrams) \cite{jacobs2019mathematics,jacobs2021learning}, referring to them as `Pearl-style' and `Jeffrey-style' updating, respectively. 

\begin{example}
In $\MatR^+$ the two forms of updating are given by the following formulae. 
\begin{align*}
f|^e(y \mid x) := \frac{ \sum_{z \in Z} e(z) f(y, z \mid x)}{\sum_{y' \in Y, z' \in Z} e(z') f(y', z' \mid x)}
\qquad \qquad 
f|_\rho(y \mid x) := \sum_{z \in Z} \frac{\rho(z) f(y,z \mid x)}{\sum_{y' \in Y} f(y', z \mid x)}
\end{align*}
\end{example}

We arrive now at two corresponding forms of generalised counterfactual, allowing for general interventions and conditioning either via general effects or states.

\begin{definition} 
Let $\modelM = \modelF \circ \model{L}$ be an FCM in $\catC$ with deterministic part $\modelF$ and outputs $O$, and let \rl{$\sigma_1,\dots,\sigma_k$} be arbitrary interventions on $\modelF$, labelling and partitioning the outputs of the resulting models as $O^{(j)} = C^{(j)} \cup D^{(j)} \cup E^{(j)}$ \rl{such that for at least some $j \neq j'$ it holds that $C^{(j)} \neq \emptyset \neq E^{(j')}$.}   A \emph{generalised counterfactual from effects} \rl{$c^{(1)},\dots,c^{(k)}$} is a state of the form:  
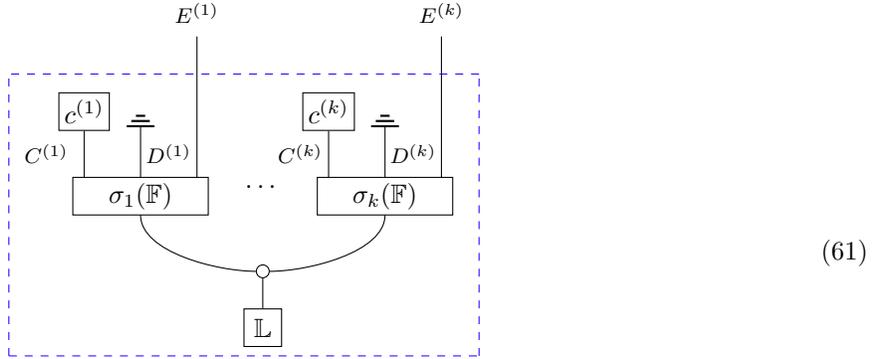
\begin{equation} \label{eq:CF-general-effects}
\input{cf-general-effects.tikz}
\end{equation}
A \emph{generalised counterfactual from states} \rl{$c^{(1)},\dots,c^{(k)}$} is a state of the form: 
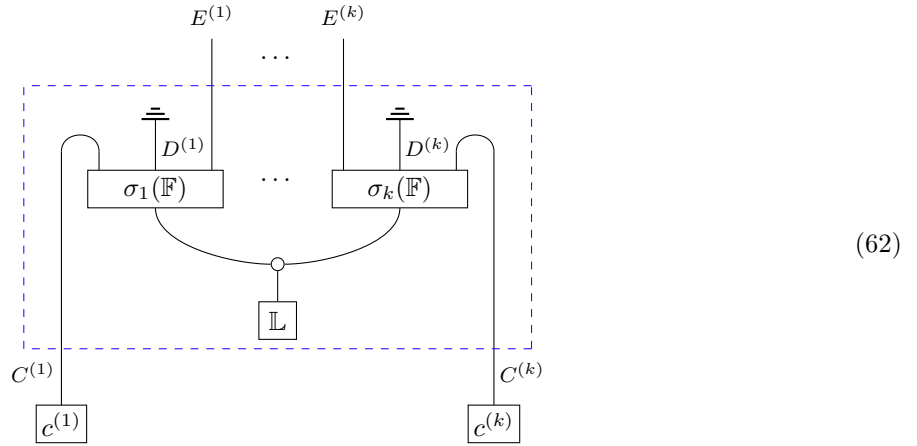
\begin{equation}\label{eq:CF-general-states}
\input{cf-general-states.tikz}
\end{equation}
\end{definition} 

There is much more to be explored in the treatment of such generalised counterfactuals, including the interpretation (and possible exclusion) of general interventions $\sigma_j$ on each world, and choice of which form of conditioning \eqref{eq:CF-general-effects}, \eqref{eq:CF-general-states} has the most natural counterfactual interpretation.  In principle even a mixture of different kinds of conditioning within one counterfactual state is conceivable. We also leave it for future work to explore identifiability criteria for this more general notion of a counterfactual.

% \bibliographystyle{utphys} \bibliography{CauseComp.bib} % Uncomment while working on file standalone, if needed 
%\end{document}
%\documentclass[main.tex]{subfiles} \begin{document}

%*****************************************************************
\section{Outlook} \label{Sec_Conclusions}
%*****************************************************************

We have presented many key notions and tools of the causal model framework in string diagrammatic language, within a general categorical set-up. These include causal models, functional causal models, interventions and counterfactuals, as well as the less standard notion of an open causal model. This extended the works by Fong \cite{Fong_2013_CausalTheories} and Jacobs \etal\ \cite{JacobsEtAl_2019_CausalInferenceByDiagramSurgery, JacobsEtAl_2021_CausalInferencesAsDiagramSurgery_DiagramsToCounterfactuals} and allowed us to treat the problem of the identifiability of causal effects and counterfactuals diagrammatically.  
One of the main take-away messages is that causal reasoning is most naturally and easily done diagrammatically -- it seems causal reasoning wants to be cast in string diagrammatic terms. 

There are a great many directions for future work (despite the length of this manuscript). Here we give a non-exhaustive list. 

\paragraph{Identification of causal effects.} 
A major goal would be to derive a general diagrammatic solution for causal effect identifiability.  
First, this involves finding a diagrammatic condition equivalent to the identifiability of causal effects of the form $P(Y | Z ; \Do(X))$ for $X,Y, Z$ arbitrary disjoint subsets of the variables, as well as deriving the solution for how to identify $P(Y | Z ; \Do(X))$ through diagrammatic reasoning.   
This would generalise the JKZ-result from Ref.~\cite{JacobsEtAl_2021_CausalInferencesAsDiagramSurgery_DiagramsToCounterfactuals}, discussed in Sec.~\ref{Sec_CE_Identifiability}, 
and would be an analogue of the existing algorithms in the conventional literature that are sound and complete for this problem (see, e.g., Ref.~\cite{ShpitserEtAl_2008_CompleteIdentificationMethodCausalHierarchy}). 

As remarked in Sec.~\ref{Sec_CEI_ExamplesGeneralCase} and alluded to in Footnote~\ref{Footnote_Categorical_DSeparation_ForCEid}, one possibility here is to look for a diagrammatic condition following `surgery' on the diagram akin to the categorical generalisation of d-separation presented by Fritz \etal\  
Indeed, one option is to start by using their results, in order to cast and derive the do-calculus rules diagrammatically. 

Secondly, one would want to derive general solutions for the identifiability of $P(Y | Z ; \sigma_X)$ for non-atomic interventions $\sigma_X$, since these  are both natural in the diagrammatic treatment and highly practically relevant. This work only discussed specific examples of such general interventions. One immediate option could be to cast the rules of the $\sigma$-calculus from Ref.~\cite{CorreaEtAl_2020_CalculusForStochasticInterventions} in diagrammatic form. 

\paragraph{\rl{Higher-order map view} of causal models.}

To the end of studying causal effect identifiability, a particularly suitable perspective is likely to be 
\rlb{that of the \emph{split-node representation of causal models}, which is sketched in Appendix ~\ref{Sec:Split_node_models}}. 
Instead of an interpreted network diagram, this uses \emph{combs}, also known as \emph{broken circuits}, or \emph{higher order maps}. 
Combs have already featured in the work of Jacobs \etal\ \cite{JacobsEtAl_2021_CausalInferencesAsDiagramSurgery_DiagramsToCounterfactuals} and also in Ref.~\cite{FriendEtAl_2022_IdentificationCausalInfluenceInQuantumprocesses}, 
as well as in Ref.~\cite{BarrettEtAl_2019_QCMs}, where the term `split-node causal model' is coined. 
Essentially, the idea here is to `break' wires in the network diagram of a causal model, thereby opening a `slot', where one may insert an intervention.  
While mathematically distinct objects, they are closely related -- split-node representations of causal models are in 1-to-1 correspondence with their representation as network diagrams and going between them is easy. 
% In light of the manuscript's length, we refrained from a detailed introduction of the split-node formalism, beyond App.~\ref{Sec:Split_node_models}, simply because this work focused on aspects that \emph{can} easily be treated with network diagrams alone.} 
Here we have refrained from a detailed introduction of the split-node formalism, beyond App.~\ref{Sec:Split_node_models}, since this work focused on aspects that \emph{can} easily be treated with network diagrams alone.  However we do not think of the presentation in this work as fundamentally distinct from a split-node focused one. Rather, the two perspectives go hand in hand and are both useful for distinct purposes. 
It would therefore be valuable to pin down which aspects of the causal model framework, beyond just the definition of a model, may have a natural manifestation in terms of the split-node causal models and where exactly the benefits of each of the two perspectives lie.

\paragraph{Continuous and fuzzy conditioning.} 
As noted in Appendix \ref{sec:cond-appendix}, it would be desirable to extend our approach \rl{of defining conditional channels $P(Y \mid X)$ -- and with it the applications in Sections \ref{Sec_CE_Identifiability} and \ref{Sec_Counterfactuals} --} to a more general categorical form of conditioning appropriate in the continuous and measure-theoretic settings outlined in Appendix \ref{sec:measure-theory}. For this one could hope to specify graphical axioms for the conditional $f|_Z \colon X \otimes Z \to Y$ of a morphism $f \colon X \to Y \otimes Z$ \rl{thereby extending those given here for normalisation and answering an open problem in Ref.~\cite{Fritz_2020_SyntheticApproachToMarkovKernels}.}

A separate manner in which we can generalise these conditionals, mentioned in  Sec.~\ref{Sec_CF_Further_Generalisation}, is to allow for conditioning based on `fuzzy' facts, given by general states and effects rather than merely sharp ones like $Z=z$. Seeing as whether the set $Z$ in $P(Y \mid Z ; \Do(X))$ is empty or not makes a difference to the identifiability of that causal effect, one may naturally wonder what becomes of the known identifiability conditions if one allows for conditioning on such fuzzy facts.

\paragraph{Counterfactuals.} 
An obvious next step, pointed out in Sec.~\ref{Sec_CF_id_criteria}, is to \emph{prove completeness} of the identification algorithm \ouridalgo\ for counterfactuals. This would establish necessity of the diagrammatic identifiability criterion from Cor.~\ref{Cor_CF_Id_Condition} and complete the diagrammatic translation of the results by Shpitser \etal\ from Ref.~\cite{ShpitserEtAl_2008_CompleteIdentificationMethodCausalHierarchy}. 
Once completeness is established, it is natural to check whether the string diagrammatic treatment of counterfactuals actually allows for an easy argument to the effect that there is no loss of generality in studying the identifiability of counterfactuals based on ADMGs as latent projections. This would extend an analogous result for causal effects by Richardson \etal\ from Ref.~\cite{RichardsonEtAl_2022_NestedMarkovForADMGs} (see Rems.~\ref{Rem_Rootification_issue_CF} and \ref{Rem_ADMGsAndNonID}) and fill a gap that seems to currently exist in the conventional literature. 

Beyond these, it would be interesting to study whether our approach might naturally extend to cover so-called \emph{nested counterfactuals} \cite{CorreaEtAl_2021_NestedCounterfactuals, Pearl_2011_AlgorithmizationOfCounterfactuals, Pearl_2001_DirectAndIndirectEffects} by allowing for connections across worlds other than just through the shared background variables $U$. 
Finally, the \emph{generalisation of the notion of a counterfactual} in Sec.~\ref{Sec_CF_Further_Generalisation}, which involves the conditioning with fuzzy facts and parallel worlds defined by interventions more general than do-interventions, deserves further exploration. In particular, one might study how the criteria and methods for the identification of counterfactuals would change in this case.

\paragraph{Notion of causal models.} 
This work restricted itself to acyclic causal structures. 
However, \emph{cyclic causal structures} have been studied in the literature, especially in order to account for feedback loops (see, e.g., Refs.~\cite{Spirtes_1995_DirectedCyclicGraphsForFeedback, Richardson_1997_CharacterizationMCCyclicGraphs, ForreEtAl_2017_MarkovPropertiesCycles, ForreEtAl_2020_CausalCalculusWithCycles}). 
A natural step would be to explore a string diagrammatic representation of such cyclic causal models. 
It would then also be insightful to relate them to (the classical instance of) cyclic causal models as studied in Ref.~\cite{BarrettEtAl_2021_CyclicQCMs}, where the cyclicity is in principle allowed to be of a very different nature -- not one from a feedback loop, that could be unravelled in a time series -- and relevant to the study of `indefinite causal order' in the foundations of quantum theory. 

Another possible direction of exploration concerns the treatment of \emph{latent variable models}. 
This work introduced and worked with the rootification of an ADMG (see Sec.~\ref{Sec_TreatmentLatentvariables}), i.e. working with a DAG such that the corresponding latent projection recovers the given ADMG. In many cases this does not constitute any loss of generality when studying identifiability problems. 
One wonders though whether there is a more direct diagrammatic -- string diagrammatic or otherwise -- representation of models based on an ADMG such that the structural property expressed in the diagram is equivalent to the nested Markov condition with respect to that ADMG, or yet more generally, an mDAG \cite{Evans_2016_GraphsForMarginsOfBNs, Evans_2018_MarginsOfDiscreteBN}. 

\paragraph{Structure of open causal models.} 
In Section \ref{sec:openCMs} we introduced the category $\OCM(\catC)$ of open causal models in a cd-category $\catC$, but noted it had more structure than just a monoidal category, for example including the ability to `reach in' to a morphism to externalise its internal variables. It would be interesting to precisely identify the formal structure of open causal models themselves, as well as `open causal structures' i.e. network diagrams or equivalently open DAGs. The latter are related to the formal structures of open graphs \cite{OpenGraph1,OpenGraph2} and `generalised causal models' in the sense of Fritz and Liang \cite{FritzEtAl_2022_FreeGSMonoidalCategories}. 

\paragraph{Transformations and automated causal reasoning.} 
We introduced a number of interventions and more general transformations of (open) causal models, but there may be many more to explore. More generally, one would hope to develop a formal approach to reasoning about valid transformations of models, noting that each transformation may require certain conditions on an input model (\rl{e.g. to ensure} the result remains acyclic). One approach could be to define a category of transformations of open models, whose objects are the sets of models satisfying certain criteria and morphisms are kinds of transformation. \rlb{The hope would be that this would assist in automatic causal reasoning and construction of open models, by formalising the way in which we can apply the same transformation (e.g. $\Do(X=x)$) to a range of models (e.g. any model including $X$ as a variable), or by formalising notions of coarse-graining and causal abstraction as transformations.}

As mentioned in the introduction, a long term goal of this work is to assist in the development of such methods for automatically learning and discovering causal representations of data, as in so-called `causal representation learning' \cite{SchoelkopfEtAl_2021_TowardCausalRepresentationLearning,Schoelkopf_2019_CausalityForMachineLearning,SchoelkopfEtAl_2022_StatisticalToCausalLearning}. 

\paragraph{Foundational aspects.} 
As mentioned in the introduction, there are various more foundational aspects of causality for which the current work is expected to be useful for. This includes situating causal models within the framework of causal-inferential theories from Ref.~\cite{SchmidEtAl_2020_UnscramblingOmletteOfCausationAndInference}. 
\rl{It also includes exploring the connection to a potential diagrammatic representation of \emph{quantum causal models} \cite{AllenEtAl_2016_QCM, BarrettEtAl_2019_QCMs, LorenzEtAl_2021_CausalAndCompositionalStructure, OrmrodEtAl_2022_CausalStructureWithSectorialConstraints} 
and the exploration of `quantum counterfactuals' \cite{suresh2023semantics} on that basis.}

% \bibliographystyle{utphys} \bibliography{CauseComp.bib} % Uncomment while working on file standalone, if needed 
%\end{document}

\section{Acknowledgements}
\rlb{We would like to thank Bob Coecke, Steve Clark, Matty Hoban, Konstantinos Meichanetzidis, Ilya Shpitser, Robin Evans and Ciarán Gilligan-Lee for helpful discussions and feedback.} 

\addcontentsline{toc}{section}{References}
%\bibliographystyle{utphys}
%\bibliography{CauseComp.bib}

\providecommand{\href}[2]{#2}\begingroup\raggedright\endgroup

\appendix

\section{Measure-theoretic probability} \label{sec:measure-theory}

The framework of cd-categories allows one to consider settings much more general than $\MatR$. In particular they provide a diagrammatic treatment of probability channels between general measurable spaces, such as those defined by continuous probability density functions. Here we briefly illustrate how the measure-theoretic probability can be treated as a cd-category or Markov category, assuming some knowledge of measure theory. For more details see for example \rlb{Refs.~\cite{panangaden1998probabilistic,ChoEtAl_2019_DisintegrationViaStringDiagrams,Fritz_2020_SyntheticApproachToMarkovKernels}.} 

The following example categories are not required for the remainder of the paper where our examples focus on the finite settings $\MatR$ and $\FStoch$, which suffice to illustrate the key notions of causal models. 

\begin{example} \label{ex:KlG} \cite{panangaden1998probabilistic}
In the cd-category $\MeasProb$ the objects are \emph{measurable spaces} $(X, \Sigma_X)$, that is, sets $X$ coming with a $\sigma$-algebra $\Sigma_X$ whose elements are called the \emph{measurable} subsets of $X$. Recall that a sub-probability measure on $X$ is then a function $\mu \colon \Sigma_X \to [0,1]$ such that 
\[
\mu \left(\bigcup _{k=1}^{\infty }E_{k}\right)=\sum _{k=1}^{\infty }\mu (E_{k})
\]
\rl{whenever the $E_k \in \Sigma_X$} are all disjoint. A morphism $f \colon X \to Y$ is given by a \emph{Markov sub-kernel} from $X$ to $Y$. This is a function sending each $x \in X$ to a subprobability measure $f(x)$ on $Y$ such that for each $A \in \Sigma_Y$ the mapping $x \mapsto f(x,A) := f(x)(A)$ is a measurable function $X \to [0,1]$. 

The identity $\id{X}$ sends each $x$ to the point-measure $\delta_x$ with $\delta_x(A) = 1$ iff $x \in A$ and $0$ otherwise. The composition of $f \colon X \to Y$ and $g \colon Y \to Z$ is given by integration:
\[
(g \circ f)(x,A) := \int_{y \in Y} g(y,A) df(x)(y)
\]
The tensor is defined by $X \otimes Y = X \times Y$, equipped with the \emph{product} $\sigma$-algebra $\Sigma_{X \otimes Y}$. For sub-kernels $f \colon X \to Y$ and $g \colon W \to Z$ we define $f \otimes g$ to send each $(x,y) \in X \times Y$ to the \emph{product measure} $f(x) \otimes g(y)$. Here $I = \{\star\}$ is a singleton. 

A state $\omega \colon I \to X$ can be \rl{identified, via $\omega = \omega(\star)$,} with a single sub-probability measure $\omega$ over $X$, while an effect $e \colon X \to I$ similarly amounts to a measurable function $e = e(\star) \colon X \to [0,1]$. In particular, non-zero effects $e$ with discrete values in $\{0,1\}$ can be identified with measurable subsets $M\subseteq X$ via their indicator functions $e=1_M$, with
\[
1_M(x) = 
\begin{cases}
1 & x \in M \\ 0 & \text{otherwise} 
\end{cases}
\]
The discarding effect $\discard{X} = 1_X$ corresponds to the subset $X$, given by $x \mapsto 1$ for all $x \in X$. A scalar is a probability $p \in [0,1]$. 

Here a state $\omega$ is normalised iff it is a probability measure, with $\omega(X) = 1$. More generally, a sub-kernel $f \colon X \to Y$ is a channel iff it is a probability channel, sending each $x \in X$ to a probability measure $f(x)$. The Markov subcategory of channels is often denoted $\MeasStoch$. 

The marginalisation of a state (or process) corresponds to marginalisation in the usual sense of probability theory. That is, for $f \colon X \to Y \times Z$ the marginal $g \colon X \to Y$ is given by $g(x,A) = f(x,A \times Z)$.

Each point $x \in X$ defines a state on $X$ via the probability measure $\delta_x$ with $\delta_x(A) = 1$ iff $x \in A$ \rl{and $\delta_x(A) = 0$ otherwise.} The copy map $\tinycopy$ is given by $x \mapsto \delta_{(x,x)}$ for each $x \in X$. As a result, each state $\delta_x$ is sharp, and any (partial) function $f \colon X \to Y$ may be viewed as a deterministic (partial) channel $f \colon X \to Y$ via $f (x) = \delta_{f(x)}$ whenever $f(x)$ is defined, and $f(x) = 0$ otherwise. 

$\MeasProb$ has normalisation, given for a morphism $f \colon X \to Y$ by 
\[
\normmor{f}(x,A) := \begin{cases} 
\frac{f(x,A)}{f(x,Y)} & f(x,Y) \neq 0 \\ 0 & \text{otherwise} 
\end{cases}
\]
However, $\MeasProb$ lacks caps in general. To obtain caps one may work in the sub cd-category $\SepMet$ whose objects are standard Borel spaces. For any such space $X$, for each $x \in X$ the subset $\{x\}$ is measurable, and so determines a sharp effect on $X$ via the indicator function $x^\dagger = 1_{\{x\}}$. Moreover, the diagonal subset $\Delta = \{(x,x) \mid x \in X\}$ is measurable, giving caps \ $\tinycap = 1_{\Delta}$. Note however that though caps exist here they are not cancellative in general. In particular for a state $\omega$ on $X$ the effect on $X$ given by composing $\omega$ with a cap is $e(x) := \omega(\{x\})$ which can be zero without $\omega = 0$. 
\end{example}

\section{Normalisation} \label{sec:normalisation-appendix}

\subsection{Uniqueness of normalisation}

Normalisation in a cd-category $\catC$ according to Definition \ref{def:norm} is actually unique whenever it exists. For this we introduce the following notions. 

\begin{definition}
For any morphism $f \colon X \to Y$ we say that a morphism $g \colon X \to Y$ \emph{normalises} $f$ when we have 
\begin{equation} \label{eq:norm-on-sup}
\input{normsup.tikz}
\end{equation}
\end{definition}
Thus a partial channel is a morphism which normalises itself. 

\begin{lemma} \label{lem:norm-asymm}
Normalisation is:
\begin{enumerate}
	\item  \label{enum:trans}
	Transitive: if $g$ normalises $f$ and $h$ normalises $g$ then $h$ normalises $f$.
	\item  \label{enum:asym}
	\rl{Antisymmetric:} if $f$ and $g$ normalise each other then $f=g$. 
\end{enumerate}
\end{lemma}
\begin{proof}
Write $e_f$ for the morphism given by copying out an extra wire and composing it with $\discard{} \circ f$. Then $g$ normalises $f$ iff $f = g \circ e_f$. Then applying discarding we see that $e_f = e_g \circ e_f$ also. Then for \eqref{enum:trans} observe that $f = g \circ e_f = h \circ e_g \circ e_f = h \circ e_f$. For \eqref{enum:asym} note that \rl{if $f$ and $g$ normalise each other then} $f$ normalises itself. Now $f =g \circ e_f = f \circ e_g \circ e_f = f \circ e_f \circ e_g = f \circ e_g =g$, using commutativity of copy to swap $e_f$ and $e_g$.
\end{proof}

In general there are many morphisms $g$, which normalise a morphism $f$, since intuitively, $g$ may take any value on inputs $x$ with $f(x) = 0$. The `correct' normalisation of $f$ should send such inputs to $0$ also. Another way to capture this `minimality' is the following.

\begin{definition}
We say that a morphism $f$ has a \emph{minimal normalisation} when there is a partial channel $f'$ which normalises $f$ and is such that whenever $g$ is any other partial channel which normalises $f$ then $g$ normalises $f'$ also. 
\[
\input{normsup1condalt.tikz}
\qquad  \quad \text{ and } \quad \qquad 
\input{normsupcond2alt.tikz}
\]
\end{definition}

It follows from Lemma \ref{lem:norm-asymm} that a minimal normalisation is unique when it exists. By definition if $f$ is already a partial channel then it satisfies the necessary condition and so is equal to its minimal normalisation.

\begin{proposition} \label{prop:norm-is-min-norm}
Let $\catC$ be a cd-category with normalisation. Then $\normmor{f}$ is the minimal normalisation of $f$.
\end{proposition}
\begin{proof}
By \eqref{eq:norm-sup-cond} $\normmor{f}$ indeed normalises $f$, we must verify that it is minimal. But if $g$ is a partial channel which normalises $f$ then $g$ normalises $\normmor{f}$ since:
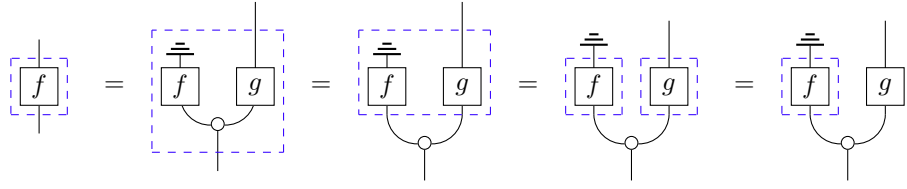
\[
\input{normproof.tikz}
\]
\end{proof}

From the uniqueness of minimal normalisations, we obtain the following.

\begin{corollary}
A normalisation structure on a cd-category $\catC$ is unique whenever it exists.
\end{corollary}

Given Proposition \ref{prop:norm-is-min-norm} it is natural to ask if there are any more elementary conditions which ensure that minimal normalisations satisfy the properties of Definition \ref{def:norm}.

\begin{lemma} \label{lem:disc-cond}
Suppose every morphism in $\catC$ has a minimal normalisation. Then minimal normalisations satisfy \eqref{eq:disc-cond}. 
\end{lemma}

\begin{proof}
For any morphism $f$, it is easy to see that since $\normmor{f}$ is a partial channel then so is $\discard{} \circ \normmor{f}$. Applying discarding to both sides of \eqref{eq:norm-on-sup} shows that $\discard{} \circ \normmor{f}$ normalises $\discard{} \circ f$. Hence it normalises $\normmor{\discard{} \circ f}$ also, that is: 
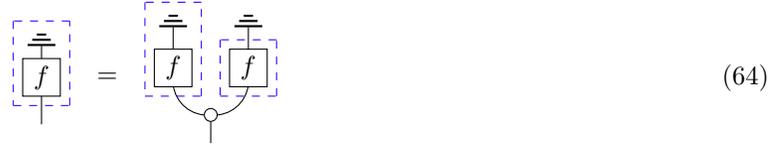
\begin{equation}  \label{eq:normA}
\input{normsupdiscs.tikz}
\end{equation}
Now for all morphisms $f$ we have 
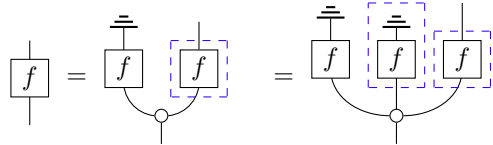
\[
\input{normf2.tikz}
\]
Hence $f$ is normalised by the morphism 
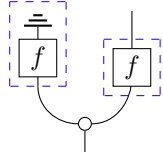
\[
\input{normmor.tikz}
\]
which may be verified to be a partial channel, and more generally that $(g \otimes h) \circ \tinycopy$ will be whenever $g, h $ are. Hence this morphism must also normalise $\normmor{f}$. Writing this out as an equality and applying discarding we obtain:
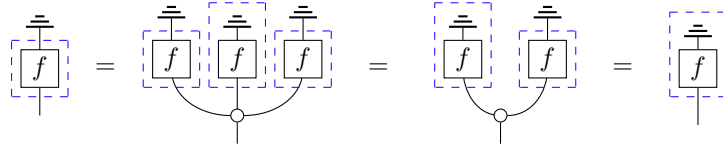
\[
\input{normf3.tikz}
\]
using \eqref{eq:normA} in the last step.
\end{proof}

\begin{corollary} \label{cor:norm-from-min-cond}
A cd-category $\catC$ has normalisation iff every morphism has a minimal normalisation and these together satisfy \eqref{eq:mon-cond} and \eqref{eq:copy-cond}. 
\end{corollary}

Note that the presence of normalisation also ensures the following conditions. 

\begin{lemma}
If $\catC$ has normalisation then $\catC$ satisfies the properties of Lemma \ref{lem:norm-lemma} as well as \emph{discard cancellation}:
\[
\input{disc-cancel.tikz}
\]
for all morphisms $f$ and effects $e$.
\end{lemma}
\begin{proof}
We begin with discard cancellation. Suppose the left-hand equation holds. Then 
\[
\input{disc-cancel2.tikz}
\]
Next we establish conditions \eqref{enum:chancond}, \eqref{enum:Extracopy}, \eqref{enum:detstatecond} of Lemma \ref{lem:norm-lemma}. 

For \eqref{enum:chancond} it is easy to see that when $f$ is a channel $f \circ \normmor{g}$ is a partial channel and normalises $f \circ g$. Hence it must normalise $\normmor{f \circ g}$, \rl{too, since} the latter is the minimal normalisation of $f \circ g$. But then 
\[
\input{normeff.tikz}
\]
 For \eqref{enum:Extracopy} note that since $\discard{}$ is a channel it is a partial channel and \rlb{so $\normmor{\discard{}}= \discard{}$.} Replacing $f$ with $f \otimes \discard{}$ and using \eqref{eq:mon-cond} gives the result. Then \eqref{enum:detstatecond} follows as shown below, using that each effect $x^\dagger$ is deterministic and hence a partial channel. 
\[
\input{norm-det-proof.tikz}
\]
\[
\input{norm-det-proof-2.tikz}
\]
\end{proof}

\subsection{Deriving normalisation}

There are some other natural ways one may derive the presence of normalisation. 

Perhaps the most direct way to understand normalisation is by its action on deterministic states, as in \eqref{eq:norm-on-states}. Recall that we say $\catC$ has zero morphisms when for all objects $X,Y$ there is a morphism $0 \colon X \to Y$, such that $0 \circ f = 0 =f \circ 0$ and $f \otimes 0 = 0$ for all morphisms $f$. 

We say a cd-category has \emph{enough deterministic states} if whenever $f \circ x = g \circ x$ for all deterministic states $x$ we have $f = g$. We say $\catC$ has \emph{scalar cancellativity} if whenever $r \cdot \omega = r \cdot \omega'$ for any states $\omega, \omega'$ and non-zero scalar $r$ then $\omega=\omega'$. Finally, say that $\catC$ has \emph{state normalisation} when every non-zero state $\omega$ there is a normalised state $\normmor{\omega}$, depicted via a dashed box, such that \eqref{eq:state-norm} holds (and which is then unique by scalar cancellativity). 

\begin{proposition} \label{prop:states-to-norm}
Let $\catC$ be a cd-category with  scalar cancellativity, state normalisation and enough deterministic states. Then $\catC$ has normalisation iff for every morphism $f$ there is a morphism $\normmor{f}$ satisfying \eqref{eq:norm-on-states} for all deterministic states $x$. 
\end{proposition}

\begin{proof}
By assumption $\normmor{f}$ is uniquely determined by the fact that $\normmor{f}\circ x = \normmor{f \circ x}$ for all deterministic states $x$, and by Lemma \ref{lem:norm-lemma} \eqref{enum:detstatecond} this condition is necessary. 

Now by the definition of state normalisation, \eqref{eq:norm-sup-cond} holds on all input deterministic states $x$ and hence holds by assumption. By the uniqueness of state normalisation we have $\normmor{\omega} = \omega$ whenever $\omega$ is already normalised, and it follows that $\normmor{f} =\normmor{\normmor{f}}$ for all $f$, making $\normmor{f}$ always a partial channel.

Conversely if $f$ is a partial channel then plugging in a deterministic state $x$ into the definition of a partial channel we see that $f \circ x = (\discard{} \circ f \circ x)(f \circ x)$. By scalar cancellativity either $f \circ x = 0$ or $\discard{} \circ f \circ x = 1$, i.e. $f \circ x$ is normalised. Hence $f \circ x = \normmor{f \circ x} = \normmor{f} \circ x$ for all deterministic states $x$, giving $f = \normmor{f}$. 

We now verify \eqref{eq:mon-cond}, \eqref{eq:disc-cond}, \eqref{eq:copy-cond}. 

\eqref{eq:mon-cond}: One may readily show that in any cd-category every deterministic state of $X \otimes Y$ is of the form $x \otimes y$ for deterministic states $x, y$ of $X, Y$ respectively. Plugging in states of this form to both sides shows that the condition reduces to $\normmor{a \otimes b} = \normmor{a} \otimes \normmor{b}$ for all states $a, b$, which follows from uniqueness of normalisation.

\eqref{eq:disc-cond}: Plugging in deterministic states reduces to the requirement that $\normmor{\discard{} \circ \omega} = \discard{} \circ \normmor{\omega}$ for all states $\omega$, which follows since by state normalisation $\normmor{\discard{} \circ \omega} = 1$ or $0$ in which case $\discard{} \circ \normmor{\omega} =1 $ or $0$ also. 

\eqref{eq:copy-cond}: Plugging in deterministic states $x$ we have $\normmor{f} \circ \tinycopy \circ x = \normmor{f} \circ (x \otimes x) = \normmor{f \circ (x \otimes x)} = \normmor{f \circ \tinycopy \circ x} = \normmor{f \circ \tinycopy} \circ x$.
\end{proof}

Another way to derive normalisation is from the ability to `invert' scalars and effects, as follows. For effects $d, e$ let us write $d \star e$ for the effect
\[
\input{effect-mult.tikz}
\]
Then an effect $e$ is a partial channel iff it is deterministic iff $e = e \star e$, and normalises an effect $d$ iff $ d = d \star e$. 

\begin{definition}
Let $\catC$ be a cd-category. We say that $\catC$ has \emph{effect inverses} when for every effect $e$ there is a unique effect $e^{-1}$ such that 
$e \star e^{-1} = \normmor{e} = \normmor{e^{-1}}$ is the minimal normalisation for $e$ and for $e^{-1}$. 
\end{definition} 

\begin{proposition} \label{prop:eff-norm}
Let $\catC$ be a cd-category with effect inverses and satisfying discard cancellation. Then $\catC$ has normalisation, given for a morphism $f$ by 
\begin{equation} \label{eq:inv-norm}
\input{inv-norm.tikz}
\end{equation}
where $e = \discard{} \circ f$.
\end{proposition}

\begin{proof}
We will first show that \eqref{eq:inv-norm} is the minimal normalisation of $f$. It is easy to check it is a partial channel. Now if $g$ normalises $f \colon X \to Y$ then composing both sides of \eqref{eq:norm-on-sup} with the morphism $X \to X$ given by copying out $X$ and composing with $e^{-1}$, shows that $g$ normalises \eqref{eq:inv-norm} also. Conversely we must show that \eqref{eq:inv-norm} normalises $f$. Applying discard cancellation to the fact that $\normmor{e}$ normalises $e$ tells us that 
\[
\input{fcomp.tikz}
\]
Hence \eqref{eq:inv-norm} indeed forms a minimal normalisation for $f$. By Corollary \ref{cor:norm-from-min-cond} it suffices to check \eqref{eq:mon-cond} and \eqref{eq:copy-cond}. 

We now verify \eqref{eq:mon-cond}. From \eqref{eq:inv-norm} we see that this is equivalent to checking that $(d \otimes e)^{-1} = d^{-1} \otimes e^{-1}$ for all effects $d, e$. Since $(d^{-1} \otimes e^{-1}) \star (d \otimes e) = \normmor{d} \otimes \normmor{e}$, by the uniqueness of inverses it suffices to show that $\normmor{d \otimes e} = \normmor{d} \otimes \normmor{e}$, i.e. establish \eqref{eq:mon-cond} in the case where $f =d $ and $g = e$ are effects. 

Now by definition $\normmor{d} \otimes \normmor{e}$ normalises $d \otimes e$. Conversely if some effect $k$ normalises $d \otimes e$ then we have 
\[
\normmor{d} \otimes \normmor{e}
=
(d^{-1} \otimes e^{-1}) \star (d \otimes e) 
=
(d^{-1} \otimes e^{-1}) \star (d \otimes e) \star k 
=
\normmor{d} \otimes \normmor{e} \star k
\]
and so $\normmor{d} \otimes \normmor{e}$ forms a minimal normalisation, as required. Hence \eqref{eq:mon-cond} holds.

Finally we must verify \eqref{eq:copy-cond}. 
Firstly, observe that $\normmor{f} \circ \tinycopy$ does normalise $f \circ \tinycopy$ since: 
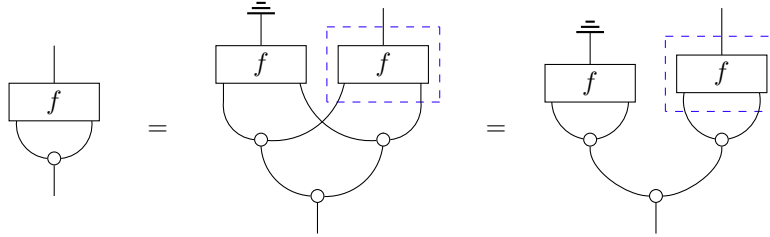
\[
\input{a11.tikz}
\]
Suppose that $g$ is another partial channel which normalises $f \circ \tinycopy$. Then we have 
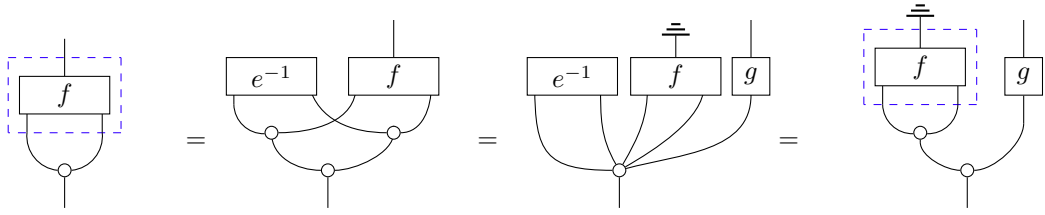
\[
\input{invproof.tikz}
\]
as required. 
\end{proof}

\begin{example}
$\MatR$ satisfies all the requirements of both Proposition \ref{prop:states-to-norm} and Proposition \ref{prop:eff-norm} and so has normalisation as well as inverses for effects. 
\end{example}

\section{\rlb{Conditioning and conditional independence}} \label{sec:cond-appendix}

Let us see more precisely how our treatment of conditioning relates to previous approaches in the literature. The following generalises Lemma \ref{lem:chain-rule}.

\begin{proposition} \label{prop:disint}
Let $\catC$ be a cd-category with \diagconditioning. Then each conditional $f|_Z$ satisfies the following. 
\begin{equation} \label{eq:disint}
\input{disint.tikz}
\end{equation}
\end{proposition}
\begin{proof}
Using the properties of normalisation, we have the following.
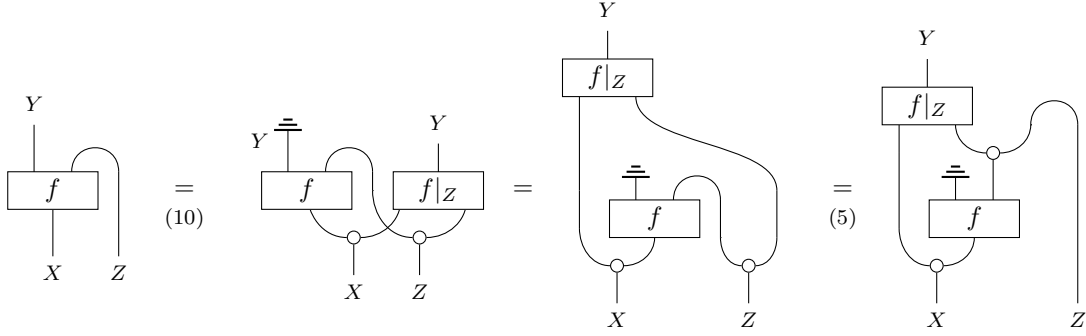
\[
\input{disintproof.tikz}
\]
Applying cancellativity of caps yields the result. 
\end{proof}

Cho and Jacobs defined a normalised joint state $\omega$ has having a \emph{disintegration} when there exists a channel $\omega|_X$ such that \eqref{eq:chainrule} holds \cite{ChoEtAl_2019_DisintegrationViaStringDiagrams}. More generally, Fritz has stated that a Markov category $\catC$ has \emph{conditionals} when for every morphism $f \colon X \to Y \otimes Z$ there is a morphism $f|_Z$ satisfying \eqref{eq:disint} \cite{Fritz_2020_SyntheticApproachToMarkovKernels}. Note that for this $f|_Z$ to exist uniquely as a channel $f$ must have full support over $X, Z$. 

Thus when cancellative caps are present our notion of conditioning generalises conditioning beyond the full support setting, specifying $f|_Z$ uniquely even when it is only a partial channel (and not requiring $f$ to be a channel). However when caps are not cancellative it can differ from the standard notion of probabilistic conditioning, as follows. 

\begin{example}\label{ex:SepMet}
In $\SepMet$ consider a joint probability measure $\omega$ over $X, Y$ with $X = \mathbb{R}^n$, $Y=\mathbb{R}^m$, determined by a density $p(x,y)$ so that 
\[
\omega(A \times B) = \int_{x \in A} \int_{y \in B} p(x,y) dx dy
\]
For $B \in \Sigma_Y$ let us define 
\[
\omega(x,B) := \int_{y \in B} p(x,y) dy
\]
Then $\omega$ does have a conditional $\omega|_X$ in the sense of \eqref{eq:disint} (that is, a disintegration \ref{eq:chainrule} given by 
\[
\omega|_X(x,B) := \frac{\omega(x,B)}{\omega(x,Y)}
\]
whenever the denominator is non-zero, and zero otherwise. However the effect $e$ given by composing $\omega$ with a cap is given by $e(x) := \omega(\{x\} \times B) = 0$ since the singleton $\{x\}$ has Lebesgue measure $0$. Hence the conditional $\omega|_X'$ defined as in \eqref{eq:disintegration} is simply the zero state. 
\end{example}

Thus \eqref{eq:disintegration} is only appropriate as a definition of a conditional channel when caps are cancellative, which fails in the continuous setting. It would be interesting to extend the approach of this work, for example to treat \rl{conditional channels $P(Y | X)$} in more general cd-categories. For this an extension of conditionals from channels to partial channels, and a graphical treatment similar to our treatment of normalisation, would both be desirable.

\color{\rlbcolor}

\subsection{Proof of Prop.~\protect\ref{Prop_Commutativity_of_conditional} \label{App_Proof_Prop_Commutativity_of_conditional}}
\begin{proof} 
	Let $\catC$ be a cd-category with \diagconditioning\ and $\omega$ a state of $X \otimes Y \otimes Z$.
	First note that due to $\omega |_Z$ being a partial channel it follows that 
	\begin{equation}
		\input{Fig_Proof_Commutativity_1.tikz}
	\end{equation}
	\rlb{Then, using Lem.~\ref{lem:chain-rule}, Eq.~\eqref{eq:mult-map} and Def.~\ref{def:norm}}
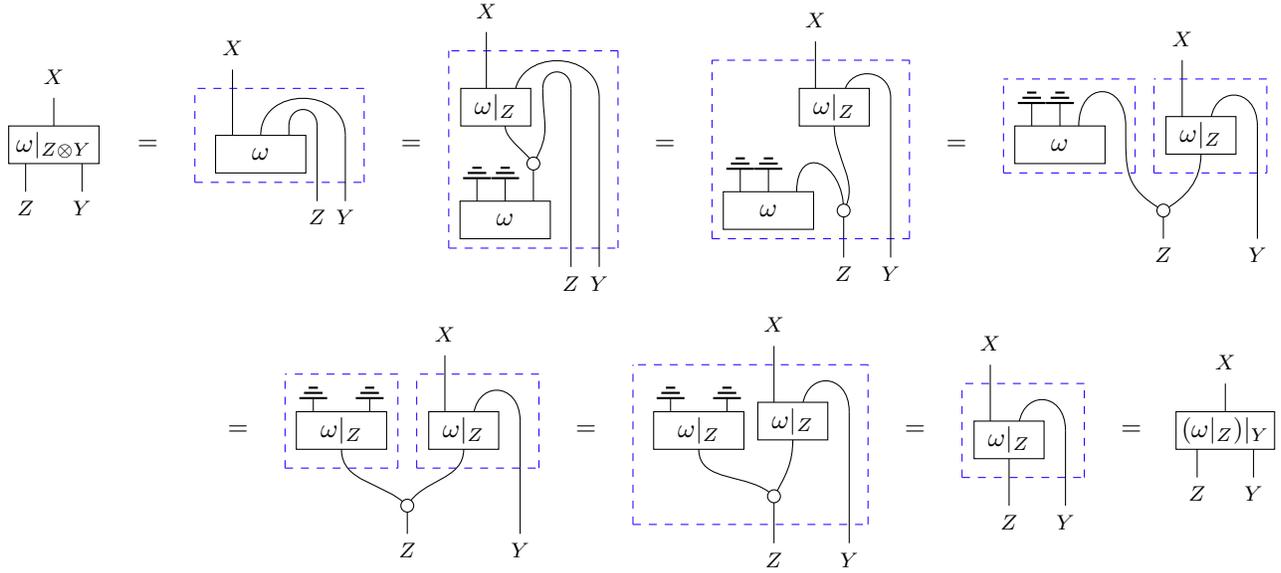
	\begin{eqnarray}
		\input{Fig_Proof_Commutativity_2.tikz} \nonumber \\
		\input{Fig_Proof_Commutativity_3.tikz} \nonumber 
	\end{eqnarray}
\end{proof}

\subsection{Proof of Lem.~\protect\ref{Lem_support_projector} \label{App_Proof_Lem_support_projector}}
\begin{proof} 
        For (1), note that each support effect is a partial channel and hence deterministic. Hence we have 
        \[
    \input{supportidem.tikz}
        \]
	Equation (2) also follows from the fact that $\omega |_X$ is a partial channel. Then 
        (3) follows since: 
	\begin{eqnarray}
		\input{Fig_Proof_support_projector_3.tikz} \nonumber 
	\end{eqnarray}
\end{proof}

\subsection{Proof of Lem.~\protect\ref{Lem_CI_equivalences} \label{App_Proof_Lem_CI_equivalences}}
\begin{proof}
	Let $\catC$ be a cd-category with \diagconditioning\ and $\omega$ a state of $X \otimes Y \otimes Z$.  \\

\noindent ``$(X \bigCI Y | Z)_{\omega} \Rightarrow (1)$": 
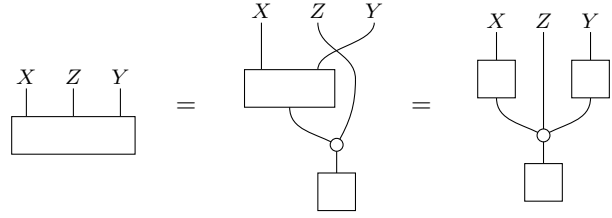
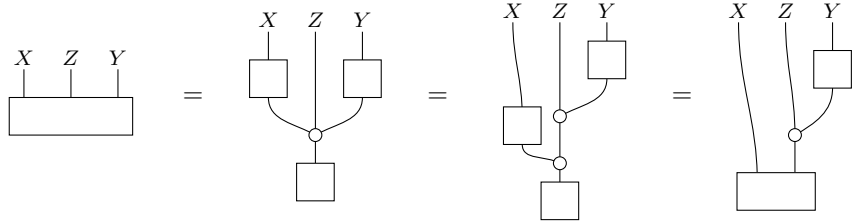
	\begin{eqnarray}
		\input{Fig_Proof_Lem_CI_equivalences_1.tikz} \nonumber 
	\end{eqnarray}

\noindent ``$(1) \Rightarrow (2)$": 
	\begin{eqnarray}
		\input{Fig_Proof_Lem_CI_equivalences_2.tikz} \nonumber 
	\end{eqnarray}

\noindent ``$(2) \Rightarrow (3)$": 
	\rlb{Using $(2)$, Eq.~\eqref{eq:mult-map} and Def.~\ref{def:norm} one finds}
	\begin{eqnarray}
		\input{Fig_Proof_Lem_CI_equivalences_3.tikz} \nonumber 
	\end{eqnarray}

\noindent ``$(3) \Rightarrow (X \bigCI Y | Z)_{\omega}$": 
	\rlb{Using Lem.~\ref{lem:chain-rule}, $(3)$, Eq.~\eqref{eq:mult-map} and Def.~\ref{def:norm} one finds}
	\begin{eqnarray}
		\input{Fig_Proof_Lem_CI_equivalences_4.tikz} \nonumber \\
		\input{Fig_Proof_Lem_CI_equivalences_5.tikz} \nonumber 
	\end{eqnarray}
	
\end{proof}

\subsection{Proof of Thm.~\protect\ref{Thm_SG_axioms} \label{App_Proof_Thm_SG_axioms}}
\begin{proof}
	Let $\catC$ be a cd-category with \diagconditioning\ and  $\omega$ a state of $X \otimes Y \otimes W \otimes Z$. 
	The semi-graphoid axioms read:
	\begin{eqnarray}
		&\text{Symmetry:}& \quad \quad (X \bigCI Y | Z)_{\omega} \quad \Leftrightarrow \quad (Y \bigCI X | Z)_{\omega} \nonumber \\
		&\text{Decomposition:}& \quad \quad (X \bigCI Y\otimes W | Z)_{\omega} \quad \Rightarrow \quad 
							(X \bigCI Y | Z)_{\omega} \ \wedge \ (X \bigCI W | Z)_{\omega} \nonumber \\
		&\text{Weak Union:}& \quad \quad (X \bigCI Y\otimes W | Z)_{\omega} \quad \Rightarrow \quad 
							(X \bigCI Y | Z \otimes W)_{\omega} \nonumber \\
		&\text{Contraction:}& \quad \quad (X \bigCI Y | Z \otimes W)_{\omega} \ \wedge \ (X \bigCI W | Z)_{\omega}
							\quad \Rightarrow \quad  (X \bigCI Y\otimes W | Z)_{\omega} \nonumber					
	\end{eqnarray}
	We now verify each of these axioms through straightforward calculation. \\
	
	\noindent \textit{Symmetry:} Obvious. \\
	
	\noindent \textit{Decomposition:} 
	Suppose $(X \bigCI Y\otimes W | Z)_{\omega}$, then
	\begin{eqnarray}
		\input{Fig_Proof_Thm_SG_axioms_1.tikz} \nonumber 
	\end{eqnarray}
	Hence, $(X \bigCI Y | Z)_{\omega}$. Analogously for $(X \bigCI W | Z)_{\omega}$. \\
	
	\noindent \textit{Weak Union:} \rlb{Suppose $(X \bigCI Y\otimes W | Z)_{\omega}$. Then, using Def.~\ref{def:norm}, Lem.~\ref{Lem_support_projector}, as well as $(X \bigCI W | Z)_{\omega}$, which is implied by the assumption due to the decomposition axiom, one finds}
	\begin{eqnarray}
		\input{Fig_Proof_Thm_SG_axioms_2.tikz} \nonumber  \\
		\input{Fig_Proof_Thm_SG_axioms_3.tikz} \nonumber 
	\end{eqnarray}
	Hence, $(X \bigCI Y | Z \otimes W)_{\omega}$. \\
	
	\noindent \textit{Contraction:} Suppose $(X \bigCI Y | Z \otimes W)_{\omega}$ and $(X \bigCI W | Z)_{\omega}$. 
	Together with the axiom of weak union and similar steps as above in the proof of that axiom, it follows straightforwardly that  
	\begin{eqnarray}
		\input{Fig_Proof_Thm_SG_axioms_4.tikz} \nonumber 
	\end{eqnarray}
	Then compute
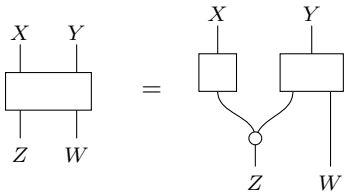
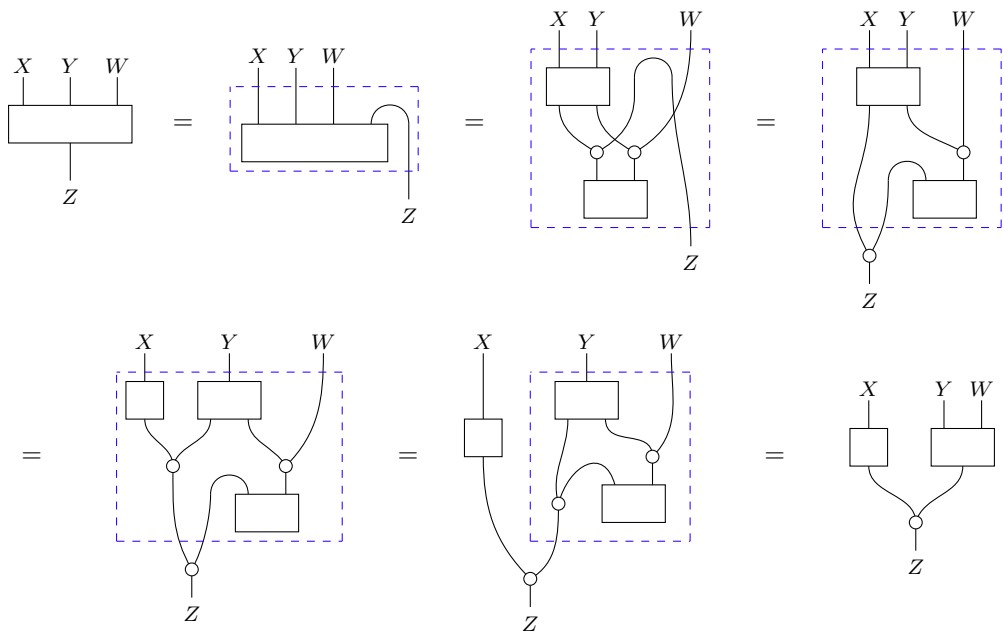
	\begin{eqnarray}
		\input{Fig_Proof_Thm_SG_axioms_5.tikz} \nonumber \\
		\input{Fig_Proof_Thm_SG_axioms_6.tikz} \nonumber
	\end{eqnarray}	
	Hence, $ (X \bigCI Y\otimes W | Z)_{\omega}$. 
\end{proof}

\subsection{Proof of Thm.~\protect\ref{Thm_DSeparation_Theorem} \label{App_Proof_Thm_DSeparation_Theorem}}

For completeness and ease of reference, we first state the definition of d-separation. 

\begin{definition} \textnormal{(Blocked paths and d-separation \cite{Pearl_1988_Probabilistic}):}  \label{Def_DSeparation}
	Given a DAG $G$, a path between nodes $N$ and $N'$ is \emph{blocked} by the set of nodes $W$ if the path contains either 
	\begin{enumerate}[leftmargin=2cm]
		\item a chain $A \rightarrow M \rightarrow  B$ or a fork  $A \leftarrow M \rightarrow  B$ with the middle node $M\in W$ 
		\item a collider $A \rightarrow C \leftarrow  B$ such that neither $C$ nor any descendant of $C$ lies in $W$.
	\end{enumerate}
	For subsets of nodes $Y$, $Z$ and $W$, say that $Y$ and $Z$ are \emph{d-separated} by $W$, and write $(Y \bigCI Z | W)_G$, if for every $N\in Y$ and $N'\in Z$, every path between $N$ and $N'$ is blocked by $W$.
\end{definition}

Note that Thm.~\ref{Thm_SG_axioms} established that the three-place relation $( \_\_ \bigCI \_\_ | \_\_ )_{\omega}$ satisfies the semi-graphoid axioms. 
As is a well-known fact, this implies the equivalence between the local and global Markov condition, where the latter is nothing but the soundness of d-separation for the respective three-place relation \cite{Verma&Pearl_1990_CausalNetworks, Lauritzen_2011_DirectedMarkovProperties}. 
Hence, all that is left to do is to verify that, given a DAG $G$ with vertices $V$, any state $\omega$ on $V$ that factorises according to a network diagram $D_{G,O=V}$, satisfies the local Markov condition, 
i.e. $\forall X \in V$ it holds that $\big(X \bigCI (\Nd(X) \setminus \Pa(X)) \ | \ \Pa(X) \big)_{\omega}$, where $\Nd(X)$ denote the non-descendants of $X$. \\

\noindent \textit{Proof of Thm.~\ref{Thm_DSeparation_Theorem}.} 
Given a DAG $G$ with vertices $V$ suppose $\omega$ is a state on $V$ that factorises according to a network diagram $D_{G,O=V}$.
Let $X \in V$, write $P:=Pa(X)$ and define $D:=V \setminus \{X\} \setminus Nd(X)$ (i.e. the descendants) and partition $Nd(X) \setminus Pa(X) =: A \cup N$, where $A$ are the ancestors of $X$, excluding the parents, and $N$ the remaining non-descendants. 
If `suppressing' the structure within the sets of vertices $D$, $P$, $A$ and $N$, the DAG $G$ can be seen as a subgraph of: 
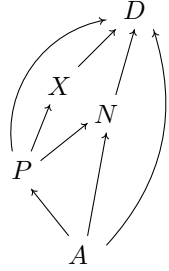
\[
\input{Fig_loc_markov_DAG.tikz} 
\]
Seeing as $\omega$ is equal to an interpretation of $D_{G,O=V}$ with channels, the marginal state of $\omega$ after discarding $D$ can be written as:
\[
\input{Fig_loc_markov_1.tikz}  
\]
Hence, by $(2)$ of Lem.~\ref{Lem_CI_equivalences} indeed $\big(X \bigCI (\Nd(X) \setminus \Pa(X)) \ | \ \Pa(X) \big)_{\omega}$.  
Together with Thm.~\ref{Thm_SG_axioms} this establishes the claim. \hfill $\square$

\color{black}

\subsection{Proof of Lemma~\protect\ref{lem:con-indep-mechs} \label{app:mechcon}}

Thanks to its factorisation over $D_\modelM$ there exist channels $a, b$ such that $\omega$ factorises as below. 
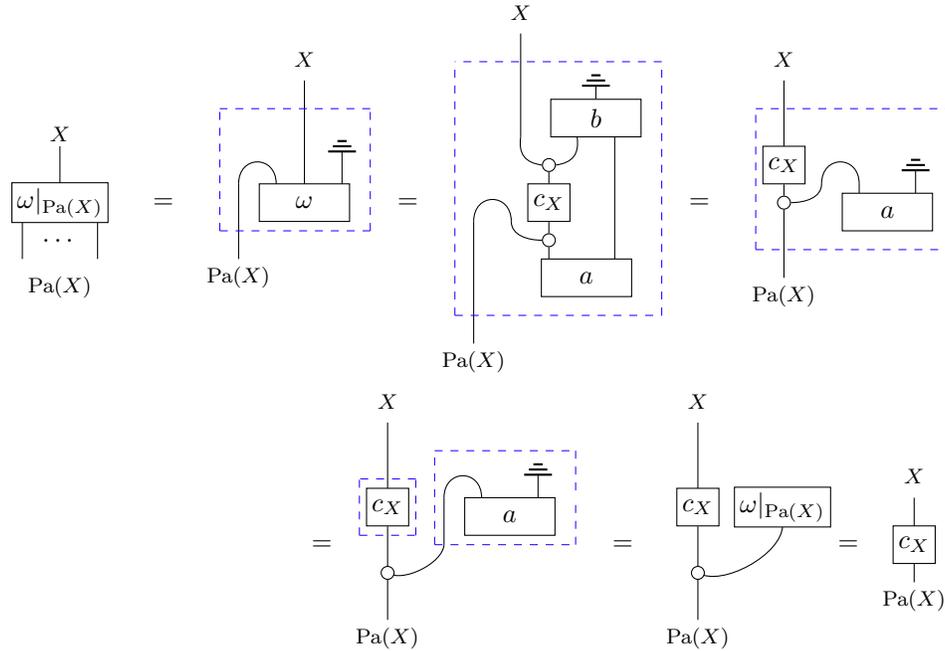
\begin{align*} 
\input{condep-mech.tikz} \\ 
\input{condep-mech2.tikz}
\end{align*}

\section{Trimming interventions} \label{app:trim}

The following result allows us to remove unnecessary inputs from a \rl{channel} to make it faithful, and thus define trimming interventions.

\begin{lemma} \label{lem:trim-lem}
Suppose that $c$ is a \rl{channel} with inputs $S$ and each input object has at least one normalised state. Then there is a unique subset $T \subseteq S$ and \rl{channel} $d=\trim(c)$ such that \eqref{eq:nosignal} holds and $d$ is faithful.
\end{lemma} 
\begin{proof} 
The morphism $d$ in \eqref{eq:nosignal} is automatically unique since it is given by plugging any normalised state into each of the $T$ inputs in $c$. First consider the case where $c$ has three inputs $A, B, C$ and \rl{the output of $c$} is independent from $A, C$ individually. Then we have  
\[
\input{indepc.tikz}
\]
Then for any normalised state $\rho$ of $C$ we obtain
\[
\input{indepc2.tikz}
\]
and so \rl{the output} is independent from $\{A,C\}$, or equivalently $A \otimes C$. Applying this iteratively to a general \rl{channel} $c$ as in the statement we see that \rl{independence} from $T$ holds iff \rl{the output is independent} from $\bigotimes T$, and that if \rl{the output} of $c$ is also independent from $\{A_i\}$, then it is independent from $T \cup \{A_i\}$. Thus \rl{defining $T$ as the union of all inputs $A_i$, for which the output is independent from $\{A_i\}$, concludes the proof.}
\end{proof} 

%*******************************************
\section{Proof of Thm.~\ref{Thm_CF_Algo_Soundness} \label{App_proof_Thm_CF_Algo_Soundness}}
%*******************************************

\begin{proof}
Let $\big(G$, $W^k$, $P_*(O) \big)$ be the input to \ouridalgo\ and $D$ the diagram after step $(2.)$, i.e. the simplified (unnormalised) diagram representing the counterfactual of interest. 

Suppose the condition for continuation in step $(3.)$ is satisfied, i.e. the state $\lambda_X$ of any $U_X$ was absorbed into the corresponding $c_X$. (Note that for any $X \in R$ this can always be done by construction of $D$, seeing as in $\tilde{\rho}(G)$ the $R$ nodes are root nodes.)
In that case $D$ has at most one occurrence of $c_X$ for any $X \in O \cup R$. 
It is not difficult to see that $D$ then has to be a composite of $R$-fragments, copy maps, as well as sharp states and effects concerning variables $X \in O$. 
(See the examples in Fig.~\ref{Fig_Fig_CF_NonId_Ex_2_4} and Eq.~\ref{Eq_Fig_Alg_Ex_5_UsMarginalised_sliced}.) 
Let $\{F_1,...,F_m\}$ be the set of $R$-fragments of $D$, recalling that an $R$-fragment corresponds to a \emph{maximal} fragment such that the conditions in Def.~\protect\ref{Def_RFragment} are satisfied.
Note that the wires ingoing and outgoing of any $R$-fragment of $D$ have type in $O$, i.e. by definition cannot be of the type of $R$ variables. 

Suppose the algorithm does not output FAIL, i.e. the conditions in steps $(4.1.1.a)$ or $(4.1.1.b)$ and $(4.1.2.a)$ or $(4.1.2.b)$ are satisfied for any objects appearing as wire types of a fragment $F_l$.  
In that case the corresponding rewrites in Eqs.~\eqref{Eq_Cf_id_Rewrite_1} and \eqref{Eq_Cf_id_Rewrite_2} clearly lead to an equivalent diagram, since they are just \rlb{basic equivalences in a cd-category.} 

Recall that the definition of an $R$-fragment in Def.~\ref{Def_RFragment} does not specify a condition concerning \rlb{the wiring up of  mechanisms $c_X$ for $X\in O$ beyond the condition that the connectivity between the output of one mechanism and the input of another \emph{within} an $R$-fragment and the wiring up with the states of the variables in $R$ has to be as in $D_{\tilde{\rho}(G)}$.  
Hence, the rewritten fragment $\widetilde{F}_l$, which may now include further copy maps,} indeed still is an $R$-fragment and is such that for any variable $X \in O$ there is at most one wire coming out of, or going into $\widetilde{F}_l$, never both or more than one. 
The sets of objects that are specified in step $(4.2.)$ are thus such that the type of the fragment, suppressing needed swaps, is  
$\widetilde{F}_l: \  D_l  \otimes F_l^{\text{in}}  \ \rightarrow \ C_l \otimes F_l^{\text{out}}$, where we let the symbols denote, both, sets of objects, as well as the corresponding objects for some order of factors. 

In light of the procedure that gave rise to $\widetilde{F}_l$, it is evident that this fragment is now composed of mechanisms that appear in the causal model with network diagram $D_{\tilde{\rho}(G)}$ and, moreover, that it is also wired up precisely in the way as in $D_{\tilde{\rho}(G)}$ -- each  $\widetilde{F}_l$ is also a fragment of $D_{\tilde{\rho}(G)}$. 
It is not difficult to see that $\widetilde{F}_l$ is therefore identical to the channel  
$P\big( C_l, F_l^{\text{out}} \ ; \ \Do(D_l), \Do(F_l^{\text{in}}) \big)$, shown in Eq.~\eqref{Eq_Cf_id_Rewrite_3}. 
The rewriting in step $(4.2.)$ is thus producing an equivalent diagram. 

The diagram $D$ now is a composite of sharp states and effects, copy maps and channels of the form 
$P\big( C_l, F_l^{\text{out}} \ ; \ \Do(D_l), \Do(F_l^{\text{in}}) \big)$ with the latter inside $P_*$ and given by assumption. 
In conclusion, whenever \ouridalgo\ does not output FAIL, the counterfactual $C=norm(D)$ is identifiable \rl{and given 
by the expression that is output by the algorithm.} 
\end{proof}

\section{Split-node representation of causal models \label{Sec:Split_node_models}}

Aside from the representation of causal models in terms of network diagrams, there is a further representation  of causal models that is also of a diagrammatic nature.  Taking inspiration from Ref.~\cite{BarrettEtAl_2019_QCMs}, we refer to this as the \emph{split-node representation of a causal model}. 

Basically, given a causal model $\modelM$ in $\catC$ with ouptut variables $O \subseteq V$, we consider its network diagram but rather than `copying out' each variable $X \in O$, instead break, or \emph{split} the wire coming out of its mechanism $c_X$ to create a `slot' where interventions may be inserted. Here we consider an alternative notion of `intervention' to those given above, related to non-wide local interventions but now allowing for an extra wire which records an \emph{outcome} for the intervention. 

For illustration return to our running example of a CBN with \rlb{the network diagram reproduced on the left-hand side below.} For all variables $X \in O(=\{S,L,A\})$ isolate a copy map corresponding to the output wire of $X$, 
that is, if $\Ch(X)=\emptyset$ introduce a copy map with one wire discarded as for $L$ below, if $|\Ch(X)|=1$, like for $S$ below, then do nothing and otherwise rewrite with two copy maps as for $A$ below. 
Then, remove these copy maps so as to leave two dangling wires in each place, thereby creating the \emph{slots}, indicated by the pink dashed circles on the right-hand side below. 

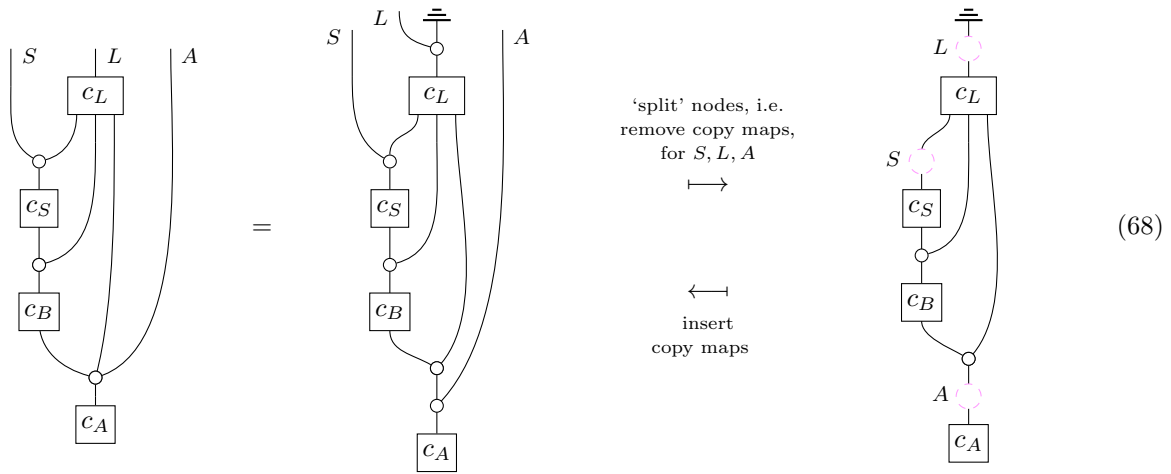
\begin{eqnarray}
	\input{Fig_Example_CBN_DAG_ND.tikz}  
	\hspace*{0.5cm} = \hspace*{0.5cm} 
	\input{Fig_Example_CBN_DAG_ND_CopyMaps_moved.tikz}  
	\hspace*{1cm} 
	\begin{minipage}{2.5cm}	
		\centering
		\rlb{
		{\scriptsize \text{`split' nodes, i.e.} }\\[-0.1cm]
		{\scriptsize  \text{remove copy maps,} }\\[-0.1cm]
		{\scriptsize  \text{for $S,L,A$} }\\
		$\longmapsto$ \\[1cm]
		$\longmapsfrom$ \\
		{\scriptsize \text{insert}}\\[-0.1cm]
		{\scriptsize \text{copy maps} }	
		}	
	\end{minipage}	 
	\hspace*{1cm} 
	\input{Fig_Example_CBN_split_node.tikz}  
	\label{Eq_ND_to_split_node_example}
\end{eqnarray}

One can of course also construct this sort of `string diagram with slots' as on the right-hand side above, in the obvious way, directly from a causal model given in terms of mechanisms $( c_X)_{X\in V}$ and outputs $O\subseteq V$. 

Now, what do the slots mean and what does the overall string diagram with slots mean?  
For a slot labelled $X$, the two wires going into that slot and coming out of it are copies of the corresponding variable $X$, where it is helpful to label them as $X^{\text{in}}$ and $X^{\text{out}}$ for distinction.  
The slot is where an intervention can be inserted and the variables $X^{\text{in}}$ and $X^{\text{out}}$ represent the variable $X$ just before and just after that potential intervention, hence the labels of the variables as coming \emph{into} and going \emph{out} of the slot. 
The slot thus is understood as a locus of potential intervention.  
By an intervention at slot $X$ with outcome $k_X$ we mean a channel $\eta_X$ of this form:
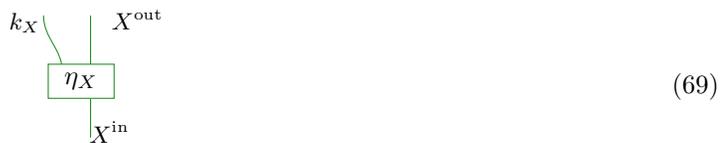
\begin{equation}
	\begin{minipage}{0.5\textwidth}
		\centering
		\input{Fig_split_node_int.tikz}
		\vspace*{-1.0cm}
	\end{minipage}
	\label{Eq_split_node_int}
\end{equation}
This notion of `intervention' (formally outside of Def.~\ref{Def_general_Intervention})  allows one to model situations where one observes $X$, but maybe only with a certain limited precision, or where an agent tosses a coin and depending on the outcome, prepares variable $X$ in a particular state and $k_X$ records the outcome of the coin toss. 

Formally, the overall string diagram with slots defines a \emph{higher-order map}, that is, a function that maps processes in some category to processes in that same category. 
The split-node representation of a causal model maps a choice of interventions at all slots to a state over the outcomes of the respective interventions.
Figure~\ref{Fig_Example_CBN_split_node_gen_int} shows the string diagram obtained from inserting choices of interventions for the slots $A,L$ and $S$ into the diagram on the right-hand side of Eq.~\eqref{Eq_ND_to_split_node_example}. 
Note that the resultant diagram is not a network diagram since it has boxes $\eta_X$ with more than one output, which are not (necessarily) copy maps.
\begin{figure}[H]
	\centering
	\begin{subfigure}{0.4\textwidth}
		\centering
		\input{Fig_Example_CBN_split_node_gen_int.tikz} 
		\vspace*{0.4cm}
		\caption{\label{Fig_Example_CBN_split_node_gen_int}}
	\end{subfigure}
	\begin{subfigure}{0.4\textwidth}
		\centering
		\input{Fig_Example_CBN_split_node_do_int.tikz}  
		\vspace*{0.2cm}
		\caption{\label{Fig_Example_CBN_split_node_do_int}}
	\end{subfigure}
	\caption{ }
\end{figure}

Nonetheless, the channel $\eta_X$ may be a copy map, or contain a copy map as one of its components, as in the special case shown in Fig.~\ref{Fig_Example_CBN_split_node_do_int}, 
which recovers our example of a do-intervention $\Do(S=s)$ from Ex.~\ref{Ex_Do_Intervention}. 
Here $\eta_A$ and $\eta_L$ each are a copy map, reflecting that $A$ and $L$ are simply `observed'. In the split-node model view, `observing' $X$ becomes formally a special case of an intervention with $k_X = X$. 

In terms of the choice of how to represent the data of a causal model, what appears to be just another way of bookkeeping for the subset $O \subseteq V$ -- copy map vs. open slot -- does define a distinct overall object, a higher-order map, and also suggests an attitude whereby one has particular interventions in mind. 
For the formal definition and further technical details on higher-order maps, which in this context of \emph{acyclic} causal models always form so called `combs', we refer the reader to the literature (see, e.g., Refs.~\cite{ChiribellaEtAl_2009_TheoreticalFrameworkCombs, oreshkov2012quantum, uijlen2019categorical, bisio2019theoretical, wilson2021causality, wilson2022mathematical,  WilsonEtAl_2022_QuantumSupermapsCharacterisedbyLocality}). 

\begin{remark} \textnormal{(The needed categorical structure to define higher-order maps):} 
	Given a causal model, the fact that representing it as a string diagram with slots always yields a well-defined higher-order map is guaranteed by just the symmetric monoidal structure of the underlying category -- in particular, there is no need to assume compact closure \cite{WilsonEtAl_2022_QuantumSupermapsCharacterisedbyLocality}.
\end{remark}

This work focuses on aspects of the causal model framework for which it is not particularly instructive or beneficial to employ the higher-order map view. We therefore leave further exploring the benefits of it to future work and use the formalism as laid out in Sec.~\ref{sec:causal-models}.

% \bibliographystyle{utphys} \bibliography{CauseComp.bib} % Uncomment while working on file standalone, if needed 
%\end{document}

\end{document}